\documentclass[11pt,a4paper,oneside,titlepage,onecolumn,openright]{book}
\usepackage[utf8]{inputenc}
\usepackage[top=3cm, bottom=3cm, left=3cm, right=3cm]{geometry}
\usepackage{amsmath}
\usepackage{geometry}
\usepackage[T1]{fontenc}
\usepackage{multirow}
\usepackage{amsfonts}
\usepackage{slashbox}
\usepackage{rotating}
\usepackage[titletoc]{appendix}
\usepackage{amssymb}
\usepackage{graphicx}
\providecommand{\U}[1]{\protect\rule{.1in}{.1in}}
\begin{document}
\newgeometry{tmargin=2.5cm,bmargin=2.5cm,lmargin=2.5cm,rmargin=2.5cm}
\begin{titlepage}
\begin{center}
\begin{table}
\begin{tabular}{ll}
\multirow{2}{*}{\includegraphics[width=0.2 \textwidth]{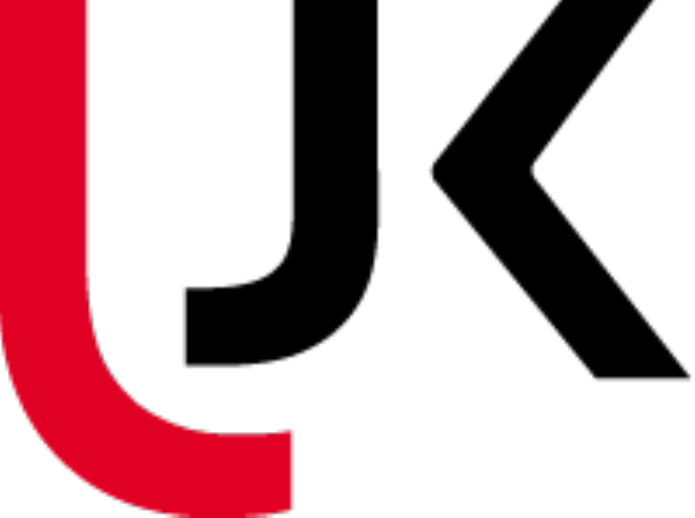}} & \\
& \textsc{\Large Jan Kochanowski University in Kielce}
\\[0.4cm]
&\textsc{\large Faculty of natural sciences}
\end{tabular}
\end{table}

\textsc{\Large}\\[2cm]
\textmd{\Large Doctoral Thesis}\\[1cm]
\textbf{\huge STUDY OF CONVENTIONAL}\\[0.4cm]
\textbf{\huge AND NON-CONVENTIONAL }\\[0.4cm]
\textbf{\huge SCALAR AND VECTOR MESONS}\\[2.0cm]
\textsc{\huge Milena Piotrowska}\\[5cm]
\textmd{\Large Main supervisor: prof. dr hab. Francesco Giacosa}\\[0.4cm]
\textmd{\Large Second supervisor: dr Peter Kovacs}\\[3.5cm]
\textsc{\Large Kielce 2020}

\end{center}
\end{titlepage}

\setlength{\oddsidemargin}{0.46cm}   
\setlength{\evensidemargin}{-0.54cm} 
\newpage
\newpage
\begin{center}
\textbf{\LARGE Abstract}
\end{center}

 Enormous progress in physics, both on experimental and theoretical side, enriched our knowledge about the particles which build matter. Among those particles there are mesons, to which this thesis is entirely devoted. The overwhelming majority of mesons is made of `conventional' quark-antiquark pairs, but nowadays there is mounting evidence for `non-conventional' mesons, such as tetraquarks, glueballs, hybrids, and molecules. The main aim of this thesis is to understand the nature of some scalar and vector mesonic resonances which still remain puzzling and need to be clarified. 
 
After a short introduction concerning mesons in the context of Quantum Chromodynamics (QCD), we present the theoretical formalism which we use throughout this work. Next, we investigate two nonets of conventional (quark-antiquark) excited vector mesons. Within an effective Quantum Field Theoretical (QFT) model we evaluate various decay channels of these states and compare the results with existing experimental data listed in the Particle Data Group (PDG). Moreover, we make predictions for a not-yet observed $s\bar{s}$ state with $1$ $^3D_1$ quantum numbers.

Some non-conventional mesons can be understood by the mechanism of dynamical generation. We show that the inclusion of a single conventional quark-antiquark seed in the relativistic QFT Lagrangian may cause the appearance of an additional associated state as a dynamically generated companion pole. This is a consequence of the strong coupling of the standard $q\bar{q}$ seed to its decay products generating quantum fluctuations which modify the propagator (and thus the spectral function) of the original meson. 

Along this line, we show that light scalar kaon $K^*_0(700)$, which is not yet well understood, emerges as a companion pole of the heavier $q\bar{q}$ meson $K_0^*(1430)$. Moreover, we show that the enigmatic axial-vector resonance $X(3872)$ in the charmonium sector can be interpreted as the (virtual) companion pole of the conventional $c\bar{c}$ state $\chi_{c1}(2P)$.

By applying the same theoretical formalism to the vector charmonium sector, we study the conventional $c\bar{c}$ meson $\psi(4040)$. We find out that an additional companion pole appears on the complex plane. At first sight one may identify it with the puzzling enhancement $Y(4008)$ observed by the Belle Collaboration. However a detailed analysis reveals that an utterly different mechanism, independent of the existence of the companion pole, is responsible for the creation of $Y(4008)$. Hence, the bump associated with $Y(4008)$ should not be interpreted as an independent resonance, but rather as a manifestation of the strong coupling of $\psi(4040)$ to the $D^*D$ channel. 

\newpage
\begin{center}
\textbf{\LARGE Streszczenie} 
\end{center}

Pomimo ogromnego postępu w dziedzinie fizyki cząstek elementarnych wciąż niewyjaśnionych pozostaje wiele kwestii, które mogą przyczynić się do lepszego zrozumienia budowy materii. Spośród ogromnej liczby istniejących cząstek, w niniejszej pracy uwagę skupiamy na konwencjonalnych oraz niekonwencjonalnych mezonach opisanych skalarnymi oraz wektorowymi liczbami kwantowymi.  

Mezony konwencjonalne, które w przyrodzie występują w zdecydowanej większości, składają się z par kwark-antykwark. Znanych jest jednak coraz więcej dowodów potwierdzających istnienie mezonów niekonwencjonalnych takich jak glueballe, hybrydy czy też obiekty czterokwarkowe, których budowa wykracza poza ten prosty schemat. 

Stosując efektywne modele kwantowej teorii pola systematyzujemy wiedzę dotyczącą mezonów konwencjonalnych, wyjaśniamy naturę kilku zagadkowych mezonów niekonwencjonalnych oraz dokonujemy licznych przewidywań teoretycznych. 

Zrozumieliśmy, że dwa nonety mezonów wektorowych, jeden ze wzbudzeniem radialnym zawierającym stany $\{\rho(1450),$ $K^*(1410),$ $\omega(1420),$ $\phi(1680)\}$, drugi ze wzbudzeniem orbitalnym ze stanami $\{\rho(1700),$ $K^*(1680),$ $\omega(1650),$ $\phi (1959)\}$, bardzo dobrze wpisują się w obraz konwencjonalny. Oprócz analizy szeregu kanałów rozpadu dokonujemy przewidywań dla nieodkrytego jeszcze eksperymentalnie stanu $s\bar{s}$ o liczbach kwantowych $1$ $^3D_1$, w niniejszej pracy oznaczonego jako $\phi(1959)$. Wykazaliśmy, że poszukiwany stan rozpada się głównie do kanałów $ \bar{K}K ^ * (892) $ oraz $ \bar{K}K $, ale możliwy jest również rozpad do pary $\gamma \eta$. Taki wynik daje nadzieję na odkrycie mezonu $\phi(1959)$ w trwających eksperymentach  GlueX i CLAS12 w Jefferson Lab. 

Co dotyczy mezonów niekonwencjonalnych, pokazaliśmy, że naturę niektórych z nich można zrozumieć poprzez mechanizm dynamicznej generacji biegunów stowarzyszonych.

Analiza sektora skalarnego potwierdziła istnienie enigmatycznego stanu $K^*_0(700)$, dopełniającego nonet lekkich mezonów poniżej energii $1$ GeV. Nasze wyniki stanowią dodatkowy i niezależny dowód na to, że $K^*_0(700)$ powinien zostać zaakceptowany w spisie cząstek elementarnych w Particle Data Book jako odrębny mezon. Co więcej, wykazaliśmy, że $K_0^*(700)$ można interpretować jako dynamicznie generowany biegun stowarzyszony do cięższego konwencjonalnego mezonu $K^*_0(1430)$. 

Podobna sytuacja ma miejsce w sektorze wektorowym, gdzie uwagę skupiamy na stanie $X(3872)$. Pomimo licznych dowodów eksperymentalnych potwierdzających istnienie tego mezonu, jego natura wciąż pozostaje niewyjaśniona. Nasza analiza pokazuje, że $X(3872)$ jest dynamicznie generowanym, wirtualnym biegunem stowarzyszonym do konwencjonalnego stanu $\chi_{c1}(2P)$, dla którego również poczyniono przewidywania teoretyczne. 

Stosując podobny formalizm, zbadaliśmy zagadkowy stan $Y(4008)$, zaobserwowany w eksperymencie przeprowadzonym przez kolaborację Belle. 
Bardziej szczegółowa analiza pokazuje jednak, że $Y(4008)$ nie jest realnym stanem a jedynie konsekwencją silnego sprzężenia mezonu $\psi(4040)$ z pętlą $DD^*$ poprzez którą ten rozpada się w kanał $\pi^+\pi^- J/ \psi$. Dochodzimy do wniosku, że nie każda wypukłość pojawiająca się w danych eksperymentalnych odpowiada realnej cząstce.

\newpage
\begin{center}
\textbf{\LARGE Acknowledgements} 
\end{center}

I would like to take the opportunity to thank all the people who supported me throughout writing this thesis.

First, I wish to express the deepest gratitude to my supervisor, Francesco Giacosa, for his valuable guidance, constant support and involvement. I am grateful for many helpful and interesting discussions about different aspects of physics. I also thank him for all the positive energy that kept me going in the very intensive PhD time. 

I would also like to thank my co-supervisor, Peter Kovacs, for his useful remarks, fruitful cooperation and careful correction of the thesis. The successful completion of this work would have not been possible without his support. 

I thank Wojciech Broniowski for inspiring me to start the PhD studies and also for all the interesting lectures (physics and beyond) over the last years.

I thank Stanisław Mrówczyński for his kindness, support, and all the physics lectures given with passion.   

I thank Wojciech Florkowski for his help and cordial commitment as the leader of PhD studies at the beginning of my research.

I thank the whole Institute of Physics for the nice and lively working atmosphere. 
 

A special thank goes to my beloved husband, Paweł who was with me at every stage of this thesis. I thank for his patience, and trust in me, as well as for his graphic skills that helped me a lot.

I am grateful to my wonderful parents, my brother Grzegorz, my sister Anna, and all my friends for their huge support.

This thesis was supported by the Polish National Science Centre (NCN) through the OPUS project no. 2015/17/B/ST2/01625.

\newpage
\tableofcontents
\newpage
\chapter{Introduction}
\label{intfr}
\section{Mesons in the framework of the quark model and beyond}
A significant number of states listed in the Particle Data Book (PDG) is identified as strongly interacting and extremely short-lived ($\tau \approx 10^{-22}$ s) resonances. Since many decades their nature and properties are extensively studied by both experimental and theoretical physicists. However, some of these resonances still remain puzzling and need to be clarified. This is what we aim to do in this thesis.

Mesonic resonances emerge as composite objects in the framework of Quantum Chromodynamics (QCD) - the theory describing the quark-gluon strong interactions. Namely, although quarks and gluons are the fundamental degrees of freedom in QCD, they cannot be observed as free objects. Quarks are structureless fermionic (spin-$1/2$) particles that carry an electric and color charges. At present, six types (flavors) of quarks are known to exist in nature: u (up), d (down), s (strange), c (charm), b (bottom), and t (top). Each flavor may appear in one of three possible colors, that are: red, green, and blue. For what concerns the gluons, they are eight massless, bosonic (spin-$1$) color-charged particles. Yet, what the detectors can really observe are the colorless combinations of quarks and gluons. These so-called ``white'' states are denoted as hadrons. 

How are they formed? The quark fields interact strongly with each other via gluon fields and form bound states. Hadrons can be further classified into
\begin{equation}
\textbf{baryons:}\text{ fermionic hadrons with baryon number }  \mathcal{B}=1 \text{ ,} \label{bardef}
\end{equation}
and
\begin{equation}
\textbf{mesons:} \text{ bosonic hadrons with baryon number } \mathcal{B}=0 \text{ .} \label{mesdef}
\end{equation}
The baryon quantum number $\mathcal{B}$, mentioned above, is defined as
\begin{eqnarray}
\mathcal{B}\equiv\frac{1}{3}\left[n(q)-n(\bar{q})\right],
\end{eqnarray} 
where $n(q)$ and $n(\bar{q})$ are the numbers of quarks and antiquarks in a given hadron, respectively. In addition, a conventional meson is a quark-antiquark $(q\bar{q})$ object, while a conventional baryon consists of three quarks $(qqq)$. Conventional mesons and baryons can be described by the quark model \cite{GodfreyIsgur}. Indeed, the vast majority of hadrons listed in the PDG can be understood as conventional objects \cite{pdg}.  

This thesis is entirely devoted to the study of some mesons. As clear from the definitions in Eqs. (\ref{bardef}) and (\ref{mesdef}) we recall that the only conditions that a strongly interacting particle has to fulfill to be a meson is a vanishing baryon number and integer spin. It follows that a meson is not necessarily a bound state of one quark and one antiquark. As various experimental results suggest, there are ,,white'', bosonic particles that do not consist of only one $q\bar{q}$ pair. Still, according to the definition of Eq. (\ref{mesdef}) they are also mesonic states. Such objects are called non-conventional mesons. Among them one can distinguish multiquarks (including the most popular four-quark objects such as tetraquarks and molecules), hybrids, and glueballs. 

\begin{figure}[h!]
\begin{center}
\includegraphics[width=0.4 \textwidth]{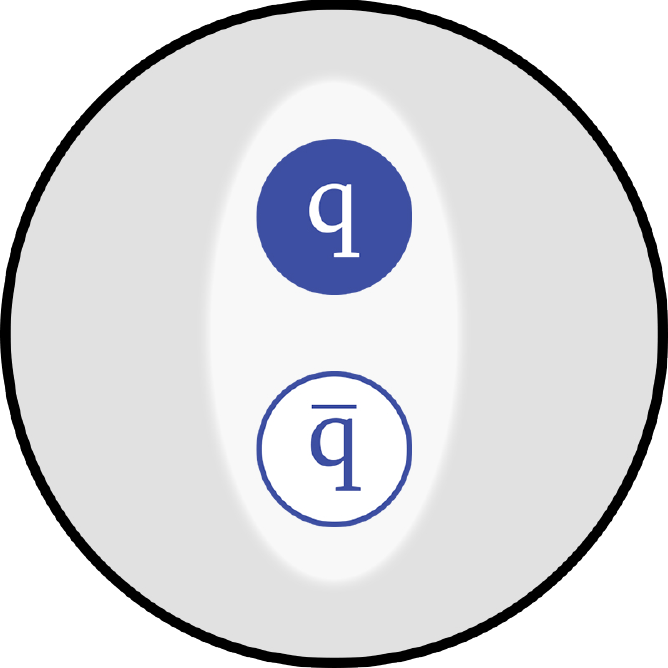}
\caption{\label{convmes} Schematic illustration of a conventional (quark-antiquark) meson.  }
\end{center}
\end{figure}

\begin{figure}[h!]
\begin{center}
\includegraphics[width=1.0 \textwidth]{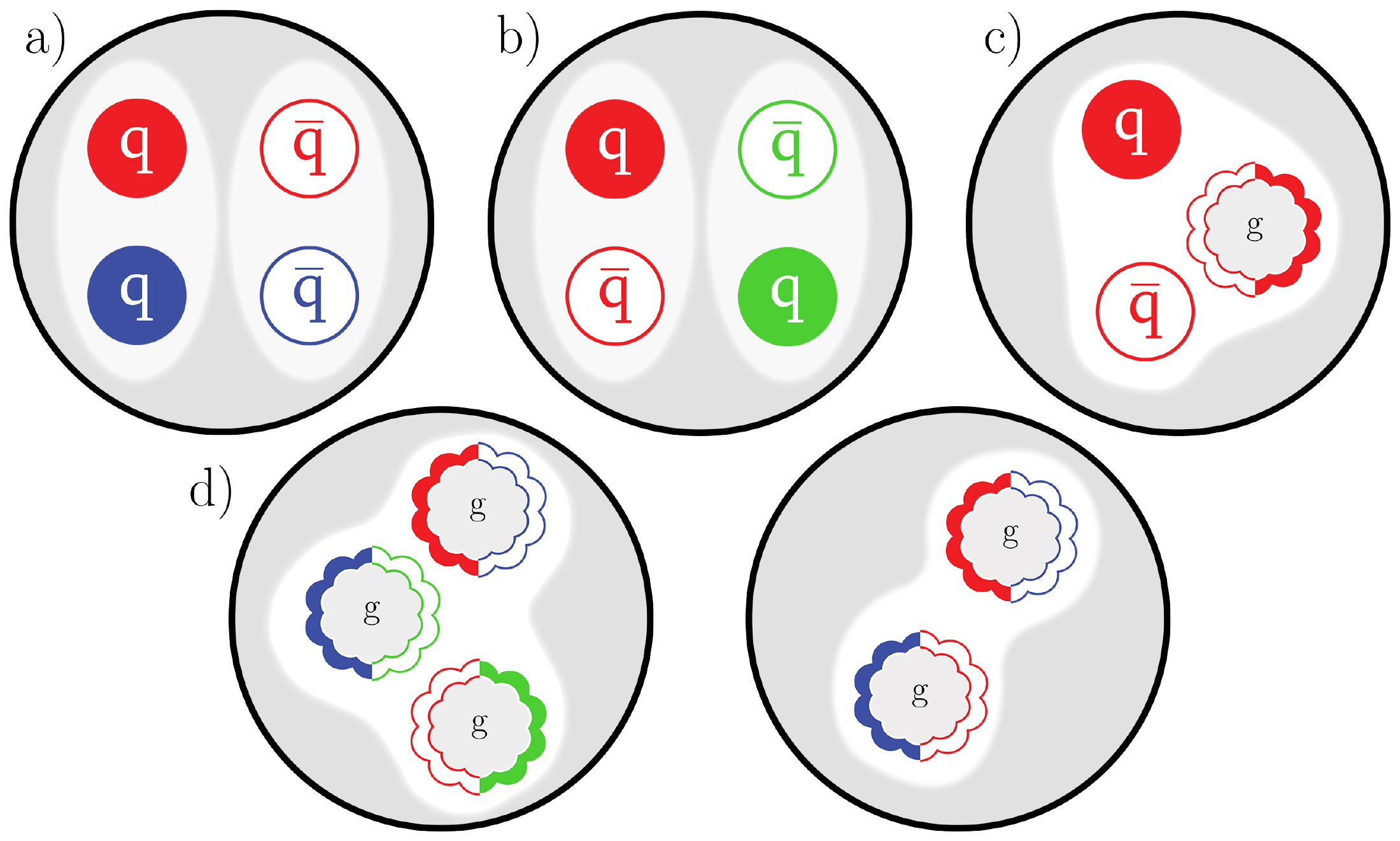}
\caption{\label{nonconvmes} Schematic illustration of non-conventional mesons such as tetraquark (a), molecule (b),  hybrid (c), and glueballs (d).}
\end{center}
\end{figure}

It is worthwhile to make a closer inspection of the internal structure of the non-conventional states. A tetraquark is a complex structure built out of one diquark and one antidiquark ($qq$)-($\bar{q}\bar{q}$). Closely related are the mesonic molecules which consist of two quark-antiquark couples ($q\bar{q}$)-($q\bar{q}$). Moreover, hybrids are objects composed of a $q\bar{q}$ pair with at least one additional gluon ($q\bar{q}$)-($g$). The last objects, the glueballs, are made solely of gluons ($gg$ or $ggg$). The schematic illustration of conventional meson is shown in Figure \ref{convmes}, while the different types of non-conventional mesons are presented in Figure \ref{nonconvmes}. Note, in both figures we depict only the ``constituent'' quarks and gluons. 
Next, we discuss quantum numbers. A conventional meson, where both constituents, the $q$ and the $\bar{q}$ carry spin $\frac{1}{2}$, can have two possible values of the spin S, either $S=0$ for antiparallel orientation of the quark spins, or $S=1$ for the parallel one. Another important quantum number of a meson is its spacial angular momentum $L$. Moreover, $S$ and $L$ combine to give the total angular momentum $J$ with $J=|L-S|, |L-S+1|, \dots, |L+S|$. Having these three numbers and taking into account the principal quantum number $n$, one can characterize conventional mesons by using the non-relativistic spectroscopic notation $n$ $^{2S+1}L_{J}$. In addition, one can describe mesons by parity $P$ ($P(q\bar{q})=(-1)^{L+1}$) and charge conjugation $C$ ($C(q\bar{q})=(-1)^{L+S}$). By using a relativistic notation, one can classify mesons into $J^{PC}$ multiplets \cite{Klempt}. For instance, mesons characterized by $L=S=0$ are pseudoscalars ($J^{PC}=0^{-+}$) but those with $L=0$ and $S=1$ are vectors ($J^{PC}=1^{--}$). The further systematic assignment of various known mesonic states is presented in Table \ref{multiplets}.  
\begin{table}[h]
\renewcommand{\arraystretch}{1.23}
\par
\makebox[\textwidth][c] { 
\par%
\begin{tabular}
[c]{|c|c|c|c|c|c|c|c|cccc|}
\hline
&&&&&&&& \multicolumn{4}{c|}{States} \\\cline{9-12}
Name&S&L&J&P&C&$J^{PC}$&$n$ $^{2S+1}L_{J}$&$I=1$&$I=\frac{1}{2}$&$I=0$&$I=0$\\
\hline
\hline
Pseudoscalar&0&0&0&$-$&+&$0^{-+}$&$1$ $^{1}S_{0}$&$\pi$ &  $K $ & $\eta$ & $\eta'(958)$ 
\\
\hline
Vector&1&0&1&$-$&$-$&$1^{--}$&$1$ $^{3}S_{1}$&$\rho(770)$ & $K^{*}(892)$ & $\phi(1020)$ & $\omega(782)$\\
\hline
Pseudovector&0&1&1&+&$-$&$1^{+-}$&$1$ $^{1}P_{1}$&$b_{1}(1235)$ & $K_{1B} \hspace{0.01cm} ^{\dagger}$ & $h_{1}(1380)$ & $h_{1}(1170)$\\
\hline
Scalar&1&1&0&+&+&$0^{++}$&$1$ $^{3}P_{0}$&$a_{0}(1450)$ & $K_{0}^{*}(1430)$ & $f_{0}(1710)$ & $f_{0}(1370)$\\
\hline
Axial vector&1&1&1&$+$&$+$&$1^{++}$&$1$ $^{3}P_{1}$& $a_{1}(1260)$ & $K_{1A}\hspace{0.01cm}^{\dagger}$ & $f_{1}(1420)$ & $f_{1}(1285)$\\
\hline
Tensor&1&1&2&+&+&$2^{++}$&$1$ $^{3}P_{2}$&$a_{2}(1320)$ & $K_{2}^{*}(1430)$ & $f'_{2}(1525)$ & $f_{2}(1270)$\\
\hline
Pseudotensor&0&2&2&$-$&+&$2^{-+}$&$1$ $^{1}D_{2}$& $\pi_{2}(1670)$ & $K_{2}(1770)^{\dagger}$ & $\eta_{2}(1870)$ & $\eta_{2}(1645)$\\
\hline
Excited vector&1&2&1&$-$&$-$&$1^{--}$&$1$ $^{3}D_{1}$&$\rho(1700)$ & $K^{*}(1680)$ & $*$ & $\omega(1650)$\\
\hline
$2^{--}$ Tensors&1&2&2&$-$&$-$&$2^{--}$&$1$ $^{3}D_{2}$&$*$ & $K_{2}(1820)$ & $*$ & $*$\\
\hline
$3^{--}$ Tensors&1&2&3&$-$&$-$&$3^{--}$&$1$ $^{3}D_{3}$&$\rho_{3}(1690)$ & $K_{3}^{*}(1780)$ & $\phi_{3}(1850)$ & $\omega_{3}(1670)$\\
\hline
Excited vector&0&1&1&$-$&$-$&$1^{--}$&$2$ $^{3}S_{1}$&$\rho(1450)$ & $K^{*}(1410)$ & $\phi(1680)$ & $\omega(1420)$\\
\hline
\end{tabular}
}\caption{\label{multiplets}Systematic assignment of some conventional light mesons according to their quantum numbers: spin (S), angular momentum (L), total spin (J), parity (P), charge conjugation (C), isospin ($I$) and principal number (n). Moreover, (*) stands for states which have not yet been observed in experiment.}%
\end{table}

It should be stressed that the notation $J^{PC}$ is applied also to non-$q\bar{q}$ mesons. Moreover, there are some combinations of quantum numbers, as for example $J^{PC}=0^{+-}$, $J^{PC}=1^{-+}$ and $J^{PC}=2^{+-}$, which cannot be produced by $q\bar{q}$ systems. However, such multiplets can be obtained for non-conventional mesons, such as glueballs and hybrids. 
 \newpage
\section{QCD}
In this section we present some basic information concerning QCD. In particular, we describe the QCD Lagrangian and discuss its most salient consequences: asymptotic freedom and quark confinement. 
\subsection{The QCD Lagrangian}
The strong force binding quarks via gluons into hadrons is described by Quantum Chromodynamics (QCD). Its Lagrangian includes the quark fields $q_f(x)$ and, as a consequence of $SU(3)_c$ gauge invariance, the gluon field $A_{\mu}(x)$ (mediating the strong interaction). It explicitly reads:
\begin{equation}
\mathcal{L}_{QCD}=\bar{q}_f(x)(i\gamma^{\mu}D_{\mu}-m_f)q_f(x)-\frac{1}{4}G^{a}_{\mu \nu}G^{a\text{,} \mu \nu} \text{ , } \label{qcdlag}
\end{equation}
where the sum over $f, \mu, \nu, a$ is understood. The quark fields are represented by four-component Dirac spinors with space-time coordinates $x^{\mu}=(t, x, y, z)$ and each appears in the fundamental representation of the color gauge group $SU(3)_c$. In fact, $q_f(x)$ can be understood as a 12-component spinor.  Moreover, the index $f$ stands for the flavor of the quarks with bare quark mass $m_f$, respectively, and can be: up (u), down (d), strange (s), charm (c), bottom (b) and top (t).  

The coupling between the quark fields $q_f(x)$ and the gluon field $A_{\mu}(x)$ is contained in the gauge covariant derivative, defined as
\begin{equation}
D_{\mu}=\partial_{\mu}-igA_{\mu} \text{ , }
\end{equation}
where $g$ is the gauge coupling constant of QCD. The gluon field which mediates the strong interaction is represented by a $3 \times 3$ matrix $A_{\mu}(x)$, where $\mu=0,1,2,3$ is the Lorentz index. This matrix can be expressed as 
\begin{equation}
A_{\mu}(x)=t^{a}A^{a}_{\mu}(x) \hspace{0,6cm} (a=1,\ldots, 8)
\end{equation}
by using the generators of the SU(3) group $t^{a}=\frac{\lambda^{a}}{2}$ , where $\lambda^{a}$ are the Gell-Mann matrices. 

The last term of the Lagrangian, known also under the name Yang-Mills Lagrangian, stands for the kinetic energy of the gluons as well as their self-interactions. In terms of the gluon field, the strength tensor reads: 
\begin{equation}
G_{\mu \nu}^{a}=\partial_{\mu}A_{\nu}^{a}-\partial_{\nu}A_{\mu}^{a}+gf^{abc}A_{\mu}^{b}A_{\nu}^c \text{ ,}
\end{equation}
where the quantities $f^{abc}$ for $(a, b, c=1,\ldots, 8)$ are the antisymmetric structure constant of SU(3) \cite{Friendly}. 

\begin{figure}[h!]
\begin{center}
\includegraphics[width=0.9 \textwidth]{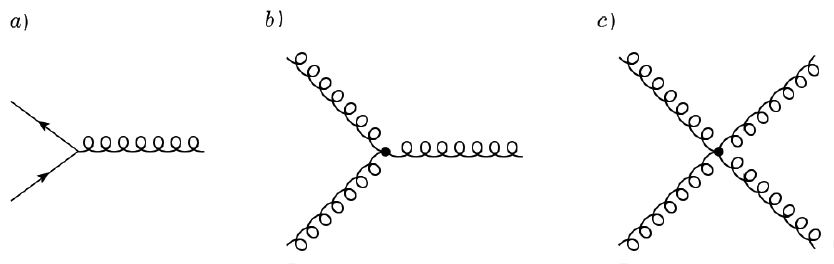}
\caption{\label{vertjk}Feynman diagrams of QCD. The quarks and antiquarks are marked by the straight solid lines, while gluons by the spiral lines.}
\end{center}
\end{figure}

In Figure \ref{vertjk} we present the tree-level Feynman diagrams of the QCD Lagrangian of Eq. (\ref{qcdlag}). 
In panel (a) one can observe the vertex corresponding to the interaction between quarks and gluons. Panels (b) and (c) describe the gluonic self-interactions, with three- and four-gluon vertices, respectively. 

\subsection{The Properties}
The QCD theory has two specific features: confinement and asymptotic freedom. The first property, i.e the confinement, states that quarks and gluons cannot be observed in nature as isolated objects. What really hits the detectors are states which are neutral in color charge. It is not possible to separate the quarks confined in hadrons, since for increasing distance, the strong force between them does not decrease. At most, one may produce additional hadrons when $q\bar{q}$ pairs are created out of the QCD vacuum. Simulations of QCD on the computer (lattice QCD) as well as theoretical approaches and experiments show that confinement is a physical fact, yet a rigorous mathematical proof does not exist currently \cite{nowe1}.

The second property, i.e asymptotic freedom, means that the interaction between the quarks gets weaker as the distance gets shorter. This striking discovery brought the Nobel Prize in 2004 \cite{Politzer}.

\section{Natural units}
In this thesis we use natural units, since it is convenient and rather common for the field of particle physics. According to this convention, two fundamental constants, the velocity of light in vacuum (\textit{c}) and the Planck's constant ($\hslash=\frac{h}{2 \pi}$), are set to unity:
\begin{equation}
c=\hslash=1 \text{ . }
\end{equation}
In this system, $\hslash$ and \textit{c} are the fundamental units of action (and angular momentum at the same time) and velocity, respectively. Moreover, the energy is given in GeV, where $1$ GeV roughly corresponds to the mass of the proton at rest.
As a consequence, the dimensions of all basic quantities (energy, mass, momentum, length, time) are expressed in terms of relevant powers of GeV:
\begin{center}
[energy] = [mass] = [momentum] = [length]$^{-1}$ = [time]$^{-1}$.
\end{center}
\section{Resonances and Breit-Wigner parameterization}
In this section we make some general remarks on resonances. This aspect is important since resonances play a crucial role in our theoretical analysis. The term  ``resonance'' refers to an unstable and short-living state like most mesons are. In fact, such mesons cannot be directly measured in experiments because they live only for an extremely short time ($\tau \approx 10^{-22}$ s). It means that they decay during the propagation from the source to the detector. The existence of resonances is deduced by the study of their decay products (such as pions, photons and leptons) which are much more long-lived/stable.

\begin{figure}[h!]
\begin{center}
\includegraphics[width=0.8 \textwidth]{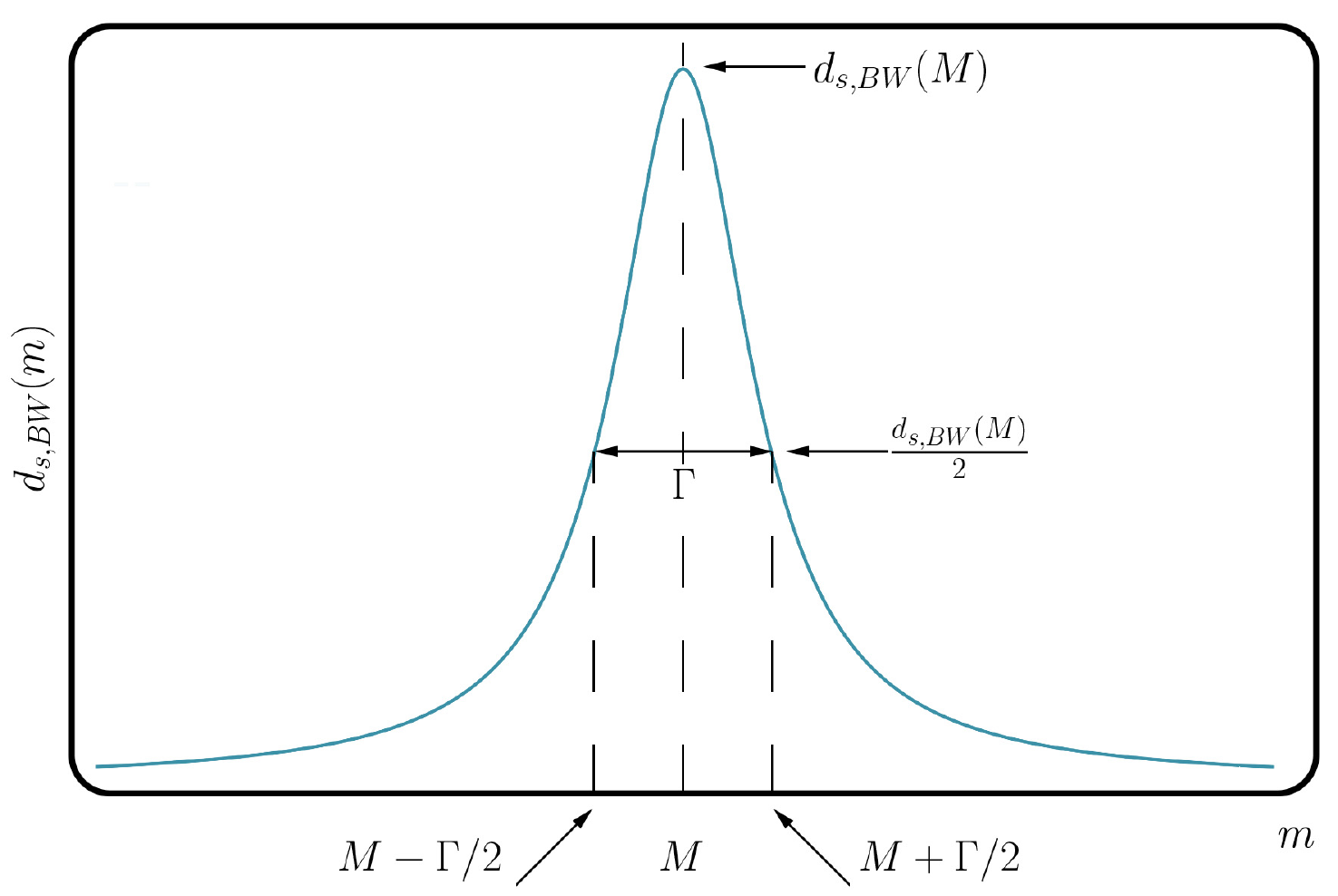}
\caption{\label{bwmp} Shape of the Breit-Wigner distribution.}
\end{center}
\end{figure}

The mean lifetime of a given resonance ($\tau$) is connected with the decay width ($\Gamma$) through the relation:
\begin{equation}
\Gamma= \frac{1}{\tau} \text{ .}
\end{equation} 
One should stress that a resonance does not have a definite value of mass. It rather possesses a mass distribution, which in the case of an isolated resonance can be well approximated by the non-relativistic Breit-Wigner formula (see \textit{e.g.} Ref. \cite{Klempt} and refs. therein):
\begin{equation}
d_{s,BW}(m)=\frac{\Gamma}{2 \pi}\frac{1}{(m-M)^2+\frac{\Gamma^2}{4}}\text{ ,} \label{bwmpmp}
\end{equation}
where $M$ stands for the mass and corresponds to the maximum of the peak. In Figure \ref{bwmp} we present the shape of the Breit-Wigner distribution. 

Interestingly, Eq. (\ref{bwmpmp}) has a pole for the complex value of $m$ given by
\begin{equation}
m_{\text{pole}}=M-\frac{i \Gamma}{2} \text{ . } \label{njk}
\end{equation}
In general, there are many different definitions for the mass of a resonance, which relay on its propagator $\Delta (s)$, where $s$ is the squared four-momentum of the unstable particle (for details see later on). One of them is the so-called Breit-Wigner mass ($m_{BW}$), that is defined as the zero of the real part of the inverse propagator:
\begin{equation}
\text{Re}[\Delta^{-1}(s=m^2_{BW})]=0 \text{ .}
\end{equation}
Another commonly used definition of the mass of the resonance is obtained from the coordinates of the propagator poles (on its second Riemann sheet, see Chapter \ref{chapvec})
\begin{equation}
\Delta^{-1}(s=s_{\text{pole}})=0
\end{equation}
leading to the relation:
\begin{equation}
\sqrt{s_{\text{pole}}}=m_{\text{pole}}- i\frac{\Gamma_{\text{pole}}}{2}, \label{mask}
\end{equation}
see Eq. (\ref{njk}). This means that according to this definition the mass of the resonance corresponds to the real part of the pole while the total decay width can be found by doubling the imaginary part of it. 

One should also remind that the relativistic extension of Breit-Wigner distribution is possible and sometimes useful. It takes the form:
\begin{equation}
d_{s,BW}^{\text{rel}}(m)=\frac{2M}{\pi}\frac{M \Gamma}{(m^2-M^2)^2+M^2 \Gamma^2}\text{ .}
\end{equation}

The latter expression does also have a pole for
\begin{equation}
m_{\text{pole}}=\sqrt{M^2-iM \Gamma} \simeq M-i\frac{\Gamma}{2} \text{ ,}
\end{equation}
in agreement with the previous discussion.

\section{Publications}
This thesis is based on the following: \\
a) regular publications
\begin{enumerate}
\item 
  T.~Wolkanowski, M.~Sołtysiak and F.~Giacosa,
  ``$K_{0}^{\ast}(800)$ as a companion pole of $K_{0}^{\ast}(1430)$,''
  Nucl.\ Phys.\ B {\bf 909} (2016) 418
  [arXiv:1512.01071 [hep-ph]].
 \item  M.~Piotrowska, C.~Reisinger and F.~Giacosa,
  ``Strong and radiative decays of excited vector mesons and predictions for a new $\phi(1930)$ resonance,''
  Phys.\ Rev.\ D {\bf 96} (2017) no.5,  054033
  [arXiv:1708.02593 [hep-ph]].
  \item  M.~Piotrowska, F.~Giacosa and P.~Kovacs,
  ``Can the $\psi(4040)$ explain the peak associated with $Y(4008)$?,''
  Eur.\ Phys.\ J.\ C {\bf 79} (2019) no.2,  98
  [arXiv:1810.03495 [hep-ph]].
 \item  F.~Giacosa, M.~Piotrowska and S.~Coito,
  ``$X(3872)$ as virtual companion pole of the charm-anticharm state $\chi_{c1}(2P)$,''
  Int.\ J.\ Mod.\ Phys.\ A {\bf 34} (2019) no.29,  1950173
  [arXiv:1903.06926 [hep-ph]].
\end{enumerate}
and\\
b) proceedings publications
\begin{enumerate}
\item M.~Soltysiak, T.~Wolkanowski and F.~Giacosa,
  ``Large-$N_{c}$ pole trajectories of the vector kaon $K^{\ast}(892) $ and of the scalar kaons $K_{0}^{\ast}(800)$ and $K_{0}^{\ast}(1430)$,''
  Acta Phys.\ Polon.\ Supp.\  {\bf 9} (2016) 321
  [arXiv:1604.01636 [hep-ph]].
  
Based on the talk given at $11^{\text{th}}$ Workshop on Particle Correlations and Femtoscopy (WPCF 2015) in Warsaw (Poland) (3-7.11.2015)
\item   M.~Soltysiak, T.~Wolkanowski and F.~Giacosa,
  ``A study of the resonances $K_{0}^{*}(800)$ and $K_{0}^{*}(1430)$,''
  J.\ Phys.\ Conf.\ Ser.\  {\bf 742} (2016) no.1,  012014
  [arXiv:1606.02970 [hep-ph]].
  
Based on the talk given at $4^{\text{th}}$ FAIR NExt generation ScientistS (FAIRNESS 2016) in Garmisch-Partenkirchen (Germany) (14-19.02.2016)  
\item   M.~Soltysiak and F.~Giacosa,
  ``A covariant nonlocal Lagrangian for the description of the scalar kaonic sector,''
  Acta Phys.\ Polon.\ Supp.\  {\bf 9} (2016) 467
  [arXiv:1607.01593 [hep-ph]].
  
Based on the talk given at $8^{\text{th}}$ International Winter Workshop ``Excited QCD'' 2016 in Costa da Caparica (Portugal) (6-12.03.2016)
\item   M.~Piotrowska and F.~Giacosa,
 ``Strong decays of excited vector mesons,''
  Acta Phys.\ Polon.\ Supp.\  {\bf 10} (2017) 1015
  [arXiv:1708.03175 [hep-ph]].
  
Based on the talk given at $9^{\text{th}}$ International Winter Workshop ``Excited QCD'' 2017 in Sintra (Portugal) (7-13.05.2017)  
\item  M.~Piotrowska and F.~Giacosa,
 ``A study of the excited radial vector meson $\rho$,''
  PoS Hadron {\bf 2017} (2018) 237
  [arXiv:1712.05617 [hep-ph]].
  
Based on the poster presentation at $17^{\text{th}}$ International Conference on Hadron Spectroscopy and Structure (Hadron 2017) in Salamanca (Spain) (25-29.09.2017)  
  \item  M.~Piotrowska and F.~Giacosa,
  ``Excited vector mesons: phenomenology and predictions for a yet unknown vector $s\bar{s}$ state with a mass of about 1.93 GeV,''
  EPJ Web Conf.\  {\bf 18} (2018) 20209
  [arXiv:1712.01087 [hep-ph]].
  
Based on the talk given at $6^{\text{th}}$ International Conference on New Frontiers in Physics (ICNFP 2017) in Kolymbari, Crete (Greece) (17-26.08.2017)
\item   M.~Piotrowska and F.~Giacosa,
  ``A study of the vector meson $\psi(4040)$,''
  EPJ Web Conf.\  {\bf 199} (2019) 04013
  [arXiv:1810.12702 [hep-ph]].
  
Based on the poster presentation at $15^{\text{th}}$ International Workshop on Meson Physics (MESON 2018) in Cracow (Poland) (7-12.06.2018)  
\item M.~Piotrowska, ``Study of some (non-)conventional mesons in the framework of effective models,'' to be published in Acta Physica Polonica B - Proceedings Supplement
[arXiv:2004.09970 [hep-ph]].

Based on the talk given at $12^{\text{th}}$ Workshop ``Excited QCD'' 2020 in Krynica-Zdrój (Poland) (2-8.02.2020)
  \end{enumerate}

\section{Organization of the thesis}
The structure of the thesis is as follows: In Chapter \ref{intfr} we briefly characterize the mesons, including the classification according to their conventional and non-conventional nature. Moreover, we provide some general information about QCD, the underlying theory of quark-gluon strong interactions. In Chapter \ref{botbds} we discuss in detail the basic aspects concerning the kinematics of the two-body decay. In Chapter \ref{excitki} we perform the phenomenological study of two nonets of excited $q\bar{q}$ vector mesons. In particular, by applying our effective model we make some predictions for the undiscovered $s\bar{s}$ resonance described by $1$ $^3D_1$ quantum numbers. In Chapter \ref{chapvec}, using the example of the conventional vector state $K^*(892)$, we introduce a theoretical approach devoted to the study of the spectral functions and pole positions of resonances. The same model, with small modifications, is also used in Chapter \ref{kappak}, where we investigate the scalar kaonic sector and show that the $K^*_0(700)$ state can be understood as a companion pole of the heavier $q\bar{q}$ resonance $K^*_0(1430)$. Moreover, within the same theoretical setup, in Chapter \ref{chaptpsiy} we consider the conventional vector charmonium state $\psi(4040)$ and describe the puzzling enhancement $Y(4008)$. Later on, in Chapter \ref{Xmain} the idea of dynamical generation is incorporated to explain the nature of $X(3872)$. We summary and present our main conclusions in Chapter \ref{concmp}. 
\section{Presentation of the main results}
\begin{enumerate}
\item By employing an effective QFT model we study two nonets of excited vector mesons, the first one with radial excitations involving the states $\{\rho(1450),$ $K^*(1410),$ $\omega(1420),$ $\phi(1680)\}$  and the second with orbital excitations and the corresponding states $\{\rho(1700),$ $K^*(1680),$ $\omega(1650),$ $\phi (1959)\}$. We explore numerous strong and radiative decay channels and branching ratios related to both nonets and confirm that they accommodate well into the conventional quark-antiquark assignment.  
\item We make predictions for the so far undiscovered state $\phi(1959)$ belonging to the nonet of orbitally excited vector mesons. We obtain that the main decay channels of this resonance are into $\bar{K}K^*(892)+h.c.$ and $\bar{K}K$ meson pairs;  $\phi(1959) \rightarrow \gamma \eta$ is also an interesting channel. It may be possible to confirm the existence of $\phi(1959)$ at the ongoing GlueX and CLAS12 experiments carried at the Jefferson Lab. 
\item We study the non-conventional scalar state $K^*_0(700)$ that very recently has been added to the summary table of PDG, with the annotation that confirmation is needed. We show that this resonance emerges due to the mechanism of dynamical generation. It appears as a companion pole of the heavier conventional $q\bar{q}$ resonance $K_0^*(1430)$ characterized by the same set of quantum numbers. Moreover, we determine the coordinates of the pole of $K^*_0(700)$ on the complex plane. Our results represent an additional and independent proof of the existence of $K^*_0(700)$. We underline that this state should be added to the particle listings as a well-established meson. 
\item We investigate the putative state $Y(4008)$ observed by the Belle Collaboration in order to explain its puzzling nature. Following the idea of dynamical generation we study the system in which a conventional $c\bar{c}$ seed state corresponding to $\psi(4040)$ is included in the Lagrangian. Our analysis reveals that two poles emerge on the complex plane. One of them corresponds to $\psi(4040)$ resonance, but the second one cannot be identified with Y(4008) because of significant discrepancies with the experimental data. Here, a different mechanism takes place. The broad enhancement observed in the spectral function appears in the $J/ \psi \pi^+ \pi^-$ channel when considering the decay of $\psi (4040)$ through an intermediate $DD^*$ loop. We conclude that $Y(4008)$ is not a real state, but rather a manifestation of the strong coupling of the  $DD^*$ to $\psi(4040)$ in a particular channel.  
\item We show that the enigmatic and widely discussed state $X(3872)$ can be understood as a virtual companion pole of the yet undiscovered $\chi_{c1}(2P)$ resonance. It appears as the effect of dressing the $\chi_{c1}(2P)$ seed state  by $DD^*$ loop. The obtained spectral function has a non-trivial shape due to the narrow and sharp peak located at its lowest $D_0D^*_0$ threshold that can be identified with the $X(3872)$ meson. In turn, the $c\bar{c}$ seed state in the spectral function is visible as a broad peak, which for some set of parameters may even disappear. This can possibly explain why $\chi_{c1}(2P)$ could not be found in experiments so far. In the complex plane one always has two well defined poles: one for the $c\bar{c}$ seed state in agreement with quark model predictions and the second (virtual) pole for the $X(3872)$ located just below the lowest $D_0D^*_0$ threshold. 
\end{enumerate}
\chapter{Basics of the two-body decay system}
\label{botbds}
\section{Introduction}
In this chapter we present in detail the kinematic basics of two-body decay systems, which we need for a proper description of the resonances. We start from the general formula of the three-momentum for two-body decays and then we derive the relevant expressions for the decay widths. 
\section{Kinematics of two-body decays} \label{twobody}
\begin{figure}[h!]
\begin{center}
\includegraphics[width=0.4 \textwidth]{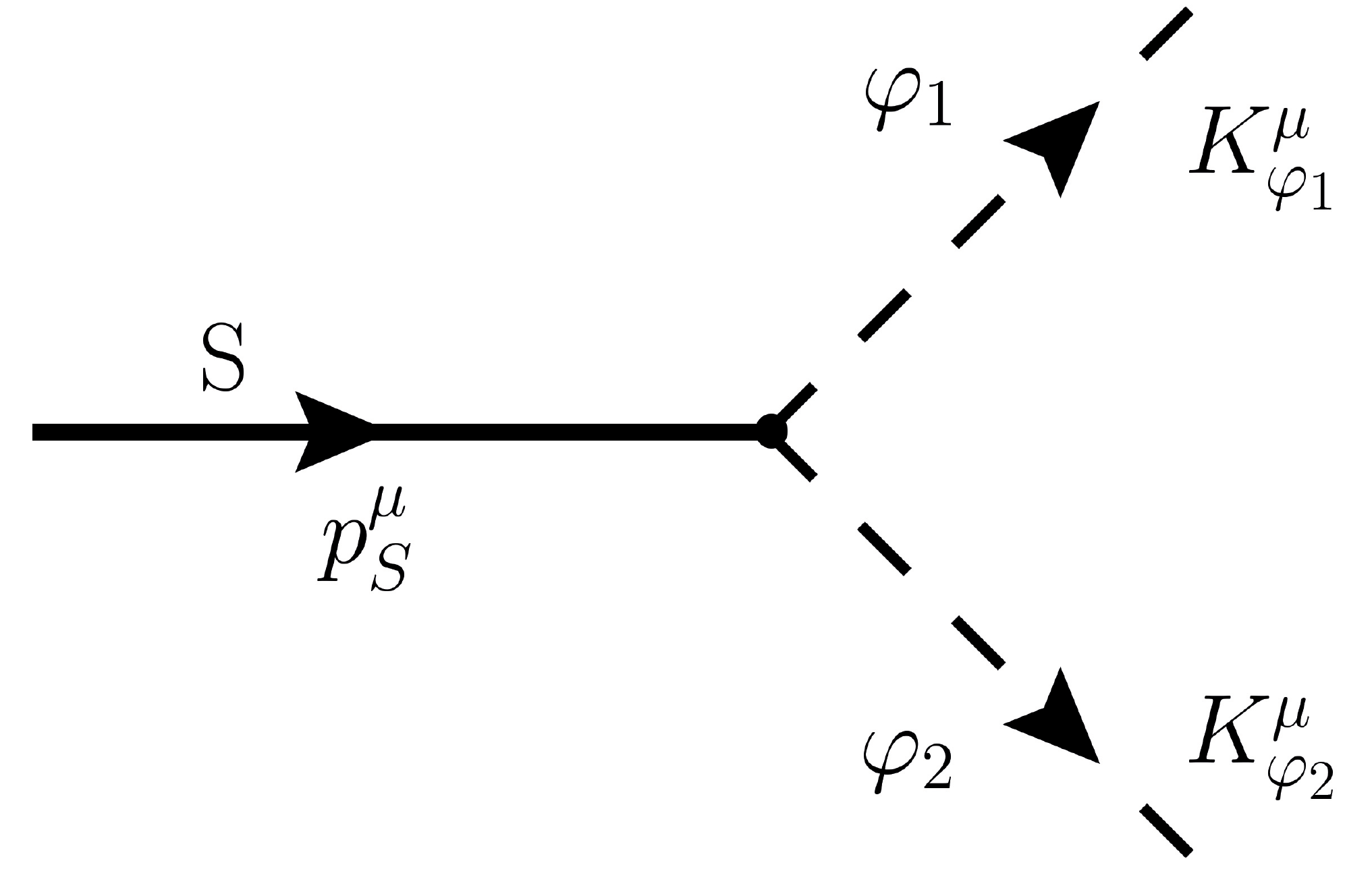}
\caption{\label{twobodyscheme} Schematic illustration of the process $S \rightarrow \varphi_1 \varphi_2$.}
\end{center}
\end{figure}

Let us consider the decay process of the type $S \rightarrow \varphi_1 \varphi_2$ (see Figure \ref{twobodyscheme}) in the rest frame of the decaying particle $S$ with four-momentum $p_S^{\mu}=(M_0, \vec{0})$. The emitted particles $\varphi_1$ and $\varphi_2$ have four-momenta $K_{\varphi_1}^{\mu}\equiv(E_1, \vec{k}_1)=(\sqrt{k_1^2+m_1^2},  \vec{k}_1)$ and $K_{\varphi_2}^{\mu}\equiv(E_2, \vec{k}_2)=(\sqrt{k_2^2+m_2^2},  \vec{k}_2)$, respectively. \\

The four-momentum conservation law 
\begin{equation}
p_S^{\mu}=K_{\varphi_1}^{\mu}+K_{\varphi_2}^{\mu}\text{ , }
\end{equation}
leads to the relations
\begin{equation}
\vec{0}=\vec{k}_1+\vec{k}_2\text{ , }
\end{equation}
and
\begin{equation}
M_0^2=m_1^2+m_2^2+2\vec{k}^2+2 \sqrt{(\vec{k}^2+m_1^2)(\vec{k}^2+m_2^2)}\text{ , }
\label{twoboo}
\end{equation}
where the quantity $\vec{k}$ in Eq. (\ref{twoboo}) is related to $\vec{k}_1$ and $\vec{k}_2$ in the following way:
\begin{equation}
\vec{k}=\vec{k}_1=-\vec{k}_2 \text{ . }
\end{equation}
We then obtain the modulus of the three-momentum of the particle $\varphi_1$ (or $\varphi_2)$ reads as
\begin{equation}
|\vec{k}|=\frac{1}{2M_0}\sqrt{M_0^4+(m_1^2-m_2^2)^2-2M_0^2(m_1^2+m_2^2)}\text{ , }
\end{equation}
valid for $M_0 \geq m_1+m_2$. Namely, for $M_0<m_1+m_2$ the decay cannot take place since it is kinematically forbidden.

For $m_1=m_2=m$, $|\vec{k}|$ simplifies to the formula
\begin{equation}
|\vec{k}|=\sqrt{\frac{M_0^2}{4}-m^2}\text{ . }
\end{equation}

\section{Determination of the decay widths}
\label{decaywidthsall}
\subsection{Vector $\rightarrow$ pseudoscalar + pseudoscalar}
There are numerous examples of this kind of decay:
\begin{equation}
\psi(4040) \rightarrow D^{+}D^{-}, \hspace{0.7cm} \psi(4040) \rightarrow D^{0} \bar{D}^{0}, \nonumber
\end{equation}
\begin{equation}
\rho(1450) \rightarrow \bar{K}K, \hspace{0.7cm} \rho(1450) \rightarrow \pi \pi, \nonumber
\end{equation}
\begin{equation}
K^*(1410) \rightarrow K \pi, \hspace{0.7cm} K^*(1410) \rightarrow K \eta, \nonumber
\end{equation}
\begin{equation}
K^*(1410) \rightarrow K \eta', \hspace{0.7cm} \omega(1420) \rightarrow \bar{K}K, \nonumber
\end{equation}
\begin{equation}
\phi(1680) \rightarrow \bar{K}K, \hspace{0.7cm} \rho(1700) \rightarrow \bar{K}K, \nonumber
\end{equation}
\begin{equation}
\rho(1700) \rightarrow \pi \pi, \hspace{0.7cm} K^*(1680) \rightarrow K \pi, \nonumber
\end{equation}
\begin{equation}
K^*(1680) \rightarrow K \eta, \hspace{0.7cm} K^*(1680) \rightarrow K \eta', \nonumber
\end{equation}
\begin{equation}
\omega(1650) \rightarrow \bar{K}K, \hspace{0.7cm} \phi(1959) \rightarrow \bar{K}K. \nonumber
\end{equation}

\begin{figure}[h!]
\begin{center}
\includegraphics[width=0.38 \textwidth]{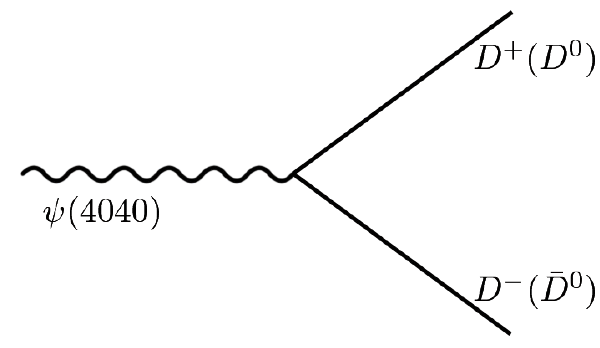}
\caption{\label{jkgmp}Schematic decay of $\psi(4040)$ into $D^+D^-$ $(D^0\bar{D}^0)$.}
\end{center}
\end{figure}

The calculations shown below refer to the decay of the $\psi(4040)$ state into two charged scalar states $(D^{+}, D^{-})$ in the rest frame of the decaying particle (see Figure \ref{jkgmp}). For the neutral case (decay into $D^{0}, \bar{D}^{0}$) the procedure is analogous. \\

The corresponding Lagrangian for the considered process is expressed as\\
\begin{eqnarray}
\mathcal{L}^{charged}_{\psi DD}=ig_{\psi DD} \psi_{\mu}((\partial^{\mu}D^{+})D^{-}-(\partial^{\mu}D^{-})D^{+})\text{ . }
\end{eqnarray}
\newpage
The following notation is used:
\begin{enumerate}
\item Four-momenta of the particles:

\begin{eqnarray}
p^{\mu}_{\psi}=\left(m_{\psi}, \vec{0}\right)\text{ ; }
\end{eqnarray}
\begin{eqnarray}
k^{\mu}_{D^{+}}=\left(E_{D^{+}}, \vec{k}\right)=\left(\sqrt{m^2_{D^{+}}+|\vec{k}|^2}, \vec{k}\right)\text{ ; }
\end{eqnarray}
\begin{eqnarray}
k^{\mu}_{D^{-}}=\left(E_{D^{-}}, -\vec{k}\right)=\left(\sqrt{m^2_{D^{-}}+|\vec{k}|^2},-\vec{k}\right)\text{ . }
\end{eqnarray}
\item Energy:
\begin{eqnarray}
m_{\psi}=E_{D^{+}}+E_{D^{-}}\text{ . }
\end{eqnarray}
For $m_{D^+}=m_{D^-}$ we have
\begin{eqnarray}
E_{D^{+}}=\sqrt{m^2_{D^{+}}+|\vec{k}|^2}=E_{D^{-}}=E=\frac{m_{\psi}}{2}\text{ . }
\end{eqnarray}
\item Momentum:
\begin{eqnarray}
|\vec{k}|=\sqrt{\frac{m^2_{\psi}}{4}-m^2_{D^+}}\text{ . }
\end{eqnarray}
\item Useful relations:
\begin{eqnarray}
k_{\psi} k_{D^+}=m_{\psi}E=k_{\psi} k_{D^-}\text{ ; }
\end{eqnarray}
\begin{eqnarray}
k_{D^+}^2=m^2_{D^+}=k_{D^-}^2\text{ ; }
\end{eqnarray}
\begin{eqnarray}
k_{D^+} k_{D^-}=E^2+|\vec{k}|^2=\frac{m_{\psi}^2}{4}+|\vec{k}|^2\text{ . }
\end{eqnarray}
\end{enumerate}
The total width for the two-body decay reads \cite{peskinQFT}:
\begin{eqnarray}
\Gamma_{\psi \rightarrow D^+D^-}=\frac{|\vec{k}|}{8 \pi m^2_{\psi}}|\mathcal{M}|^2\text{ . }
\end{eqnarray}
In order to obtain the decay  width it is necessary to determine the amplitude of the decay process denoted as $\mathcal{M}$, which can be written as
\begin{eqnarray}
-i\mathcal{M}= ig_{\psi DD} \varepsilon_{\psi,\mu}(i k_{D^+}^{\mu}-i k_{D^-}^{\mu})\text{ , }
\end{eqnarray} 
whose modulus squared reads:
\begin{eqnarray}
|-i\mathcal{M}|^{2}&=&\frac{1}{3} g^{2}_{\psi DD} \sum \limits_{\lambda_1} \varepsilon_{\psi,\mu}(\lambda_{1}) \varepsilon_{\psi, \nu}(\lambda_{1})(k^{\mu}_{D^{+}}-k^{\mu}_{D^{-}})(k^{\nu}_{D^{+}}-k^{\nu}_{D^{-}}) \nonumber\\
&=& \frac{g^2_{\psi DD}}{3}\left(-g_{\mu \nu}+\frac{k_{\psi \mu}k_{\psi \nu}}{m_{\psi}^{2}}\right)\left(k^{\mu}_{D^+}-k^{\mu}_{D^-}\right)\left(k^{\nu}_{D^+}-k^{\nu}_{D^-}\right)\nonumber\\
&=&\frac{g^2_{\psi DD}}{3}\left(\frac{\left(k_{\psi}k_{D^+}-k_{\psi}k_{D^-}\right)^2}{m_{\psi}^2}-\left(k^2_{D^+}+k^2_{D^-}-2k_{D^+}k_{D^-}\right)\right)\nonumber\\
&=&\frac{g^2_{\psi DD}}{3}\left(2k_{D^+}k_{D^-}-2m_{D^+}^2\right)=2\frac{g^2_{\psi DD}}{3}\left(|\vec{k}|^2+\frac{m^2_{\psi}}{4}-m^2_{D^+}\right)\nonumber\\
&=&\frac{4}{3}g^2_{\psi DD}|\vec{k}|^2 \text{ . }
\end{eqnarray}
Finally, the decay width reads
\begin{eqnarray}
\Gamma_{\psi \rightarrow D^+D^-}=\frac{|\vec{k}|}{8 \pi m^2_{\psi}}\frac{4}{3}g^2_{\psi DD}|\vec{k}|^2=\frac{|\vec{k}|^3}{6 \pi m^2_{\psi}}g^2_{\psi DD}\text{ . }
\end{eqnarray}
\subsection{Vector $\rightarrow$ pseudoscalar + vector}
\begin{figure}[h!]
\begin{center}
\includegraphics[width=0.43 \textwidth]{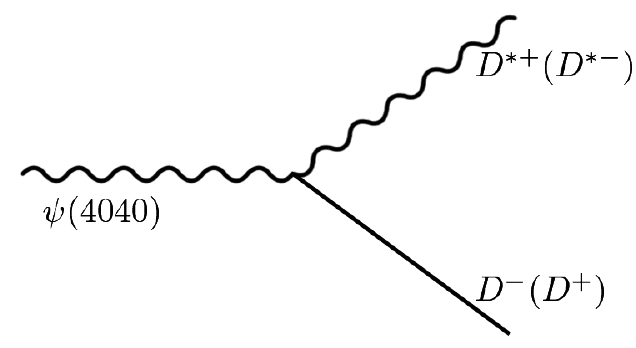}
\caption{\label{decaysv} Schematic decay of $\psi(4040)$ into $D^{*+}D^-$ $(D^{*-}D^+)$.}
\end{center}
\end{figure}

As examples of this kind of decay we mention: 
\begin{equation}
\psi(4040) \rightarrow D^*(2007)^0 \bar{D}^0+ h.c, \hspace{0.7cm} \psi(4040) \rightarrow D^*(2010)^+D^- + h.c. \nonumber
\end{equation}
This subsection contains the calculation of the tree-level decay width into charged particles (see Figure \ref{decaysv}). \\
\newpage
The following relations hold:
\begin{enumerate}
\item Four-momenta:
\begin{eqnarray}
p_{\psi}^{\mu}=\left(E_{\psi}, \vec{0}\right)=\left(m_{\psi},\vec{0}\right)\text{ ; }
\end{eqnarray}
\begin{eqnarray}
k_{D^*}^{\mu}=\left(E_{D^*}, \vec{k}\right)=\left(\sqrt{m^2_{D^*}+|\vec{k}|^2},\vec{k}\right)\text{ ; }
\end{eqnarray}
\begin{eqnarray}
k^{\mu}_{D}=\left(E_{D}, -\vec{k}\right)=\left(\sqrt{m^2_{D}+|\vec{k}|^2}, -\vec{k}\right)\text{ . }
\end{eqnarray}
\item Energy and momentum conservations:
\begin{eqnarray}
m_{\psi}=E_{D^*}+E_{D}\text{ ; }
\end{eqnarray}
\begin{eqnarray}
m_{\psi}=\sqrt{m^2_{D^*}+|\vec{k}|^2}+\sqrt{m^2_{D}+|\vec{k}|^2}\text{ ; }
\end{eqnarray}
\begin{eqnarray}
|\vec{k}|=\frac{\sqrt{ \left( m_{\psi}^2-m^2_{D^*}-m^2_{D}\right)^2-4m^2_{D^*}m^2_{D}}}{2m_{\psi}}\text{ . }
\end{eqnarray}
\item Useful relations:
\begin{eqnarray}
k_{D^*}k_{\psi}=E_{D^*}E_{\psi}-\vec{k}_{D^*}\vec{k}_{\psi}=\sqrt{m^2_{D^*}+|\vec{k}|^2} m_{\psi}\text{ . }
\end{eqnarray}
\end{enumerate}

The Lagrangian for the charged decay mode is:
\begin{eqnarray}
\mathcal{L}_{\psi D^* D}=ig_{\psi D^* D}\tilde{\psi}_{\mu \nu}\left(D^{*\mu \nu +}D^{-}-D^{*\mu \nu -}D^{+}\right)+ig_{\psi D^* D}\tilde{\psi}_{\mu \nu}\left(D^{*\mu \nu 0}\bar{D}^{0}-\bar{D}^{*\mu \nu 0}D^{0}\right)\text{ , }
\end{eqnarray}
where $\tilde{\psi}_{\mu \nu}$ is the dual vector field defined as
\begin{eqnarray}
\tilde{\psi}_{\mu \nu}=\frac{1}{2}\varepsilon_{\mu \nu \rho \sigma}\psi^{\rho \sigma}=\frac{1}{2}\varepsilon_{\mu \nu \rho \sigma}\left(\partial ^{\rho}\psi^{\sigma}-\partial ^{\sigma} \psi^{\rho} \right)=\varepsilon_{\mu \nu \rho \sigma} \partial^{\rho}\psi^{\sigma}\text{ , }
\end{eqnarray}
hence the charged part of the Lagrangian takes the form
\begin{eqnarray}
\mathcal{L}^{charged}_{\psi D^* D}&=&ig_{\psi D^* D}\varepsilon_{\mu \nu \rho \sigma}\partial^{\rho}\psi^{\sigma}\left[\left(\partial^{\mu}D^{*+ \nu}-\partial^{\nu}D^{*+ \mu}\right)D^{-}-\left(\partial^{\mu}D^{*- \nu}-\partial^{\nu}D^{*- \mu}\right)D^{+}\right] \nonumber \\
&=&ig_{\psi D^* D}2\varepsilon_{\mu \nu \rho \sigma}\left[\partial^{\rho}\psi^{\sigma}\left(\partial^{\mu}D^{*+ \nu}D^{-}-\partial^{\mu}D^{*- \nu}D^{+}\right)\right]\text{ . }
\end{eqnarray}
The next step is to determine the amplitude of the decay process into charged particles. The calculation for $\psi(4040)\rightarrow D^{*+}D^{-}$ at tree-level is shown below. According to the Feynman rules the amplitude $-i\mathcal{M}$ takes the form
\begin{eqnarray}
-i\mathcal{M}_{D^{*+}D^-}=i2g_{\psi D^*D}\varepsilon_{\mu \nu \rho \sigma}\varepsilon^{\nu}_{D^{*+}}k^{\mu}_{D^{*+}}\varepsilon_{\psi}^{\sigma}k_{\psi}^{\rho}\text{ , }
\end{eqnarray}
and similarly for $\psi(4040) \rightarrow D^{*-}D^+$
\begin{eqnarray}
-i\mathcal{M}_{D^{*-}D^+}=i2g_{\psi D^*D}\varepsilon_{\mu \nu \rho \sigma}\varepsilon^{\nu}_{D^{*-}}k^{\mu}_{D^{*-}}\varepsilon_{\psi}^{\sigma}k_{\psi}^{\rho}\text{ . }
\end{eqnarray}
Then summarizing:
\begin{eqnarray}
|-i\mathcal{M}_{D^{*+}D^-}|^2&=&4g^2_{\psi D^*D}\frac{1}{3} \sum\limits_{\lambda_1, \lambda_2} \varepsilon_{\mu_1 \mu_2 \mu_3 \mu_4}k^{\mu_1}_{D^{*+}}\varepsilon^{\mu_2}_{D^{*+}}(\lambda_1)k_{\psi}^{\mu_3}\varepsilon^{\mu_4}_{\psi}(\lambda_2) \varepsilon_{\nu_1 \nu_2 \nu_3 \nu_4}k^{\nu_1}_{D^{*+}}\varepsilon^{\nu_2}_{D^{*+}}(\lambda_1)k_{\psi}^{\nu_3}\varepsilon_{\psi}^{\nu_4}(\lambda_2)\nonumber \\
&=& \frac{4}{3}g^2_{\psi D^*D}\varepsilon_{\mu_1 \mu_2 \mu_3 \mu_4} \varepsilon_{\nu_1 \nu_2 \nu_3 \nu_4}k_{D^*}^{\mu_1}k_{D^*}^{\nu_1}k_{\psi}^{\mu_3}k_{\psi}^{\nu_3} \sum\limits_{\lambda_1}\varepsilon^{\mu_2}_{D^*}(\lambda_1)\varepsilon^{\nu_2}_{D^*}(\lambda_1)\sum\limits_{\lambda_2}\varepsilon^{\mu_4}_{\psi}(\lambda_2)\varepsilon^{\nu_4}_{\psi}(\lambda_2) \nonumber \\
&=&\frac{4}{3}g^2_{D^*D}\varepsilon_{\mu_1 \mu_2 \mu_3 \mu_4} \varepsilon_{\nu_1 \nu_2 \nu_3 \nu_4}k^{\mu_1}_{D^*}k^{\nu_1}_{D^*}k^{\mu_3}_{\psi}k^{\nu_3}_{\psi}\left(-g^{\mu_2 \nu_2}+\frac{k_{D^*}^{\mu_2}k_{D^*}^{\nu_2}}{m^2_{D^{*+}}}\right)\left(-g^{\mu_4 \nu_4}+\frac{k_{\psi}^{\mu_4}k_{\psi}^{\nu_4}}{m^2_{\psi}}\right) \nonumber \\
&=&\frac{4}{3}g^2_{\psi D^*D}\varepsilon_{\mu_1 \mu_2 \mu_3 \mu_4}\varepsilon_{\nu_1 \nu_2 \nu_3 \nu_4}g^{\mu_2 \nu_2}g^{\mu_4 \nu_4}k^{\mu_1}_{D^*}k^{\nu_1}_{D^*}k_{\psi}^{\mu_3}k_{\psi}^{\nu_3}+ \nonumber \\
&& -\frac{4}{3}g^2_{\psi D^* D}\varepsilon_{\mu_1 \mu_2 \mu_3 \mu_4}\varepsilon_{\nu_1 \nu_2 \nu_3 \nu_4}k_{D^*}^{\mu_1}k_{D^*}^{\nu_1}g^{\mu_2 \nu_2}\frac{k_{\psi}^{\mu_3}k_{\psi}^{\mu_4}k_{\psi}^{\nu_3}k_{\psi}^{\nu_4}}{m^2_{\psi}}+ \nonumber \\
&&-\frac{4}{3}g^2_{\psi D^* D}\varepsilon_{\mu_1 \mu_2 \mu_3 \mu_4}\varepsilon_{\nu_1 \nu_2 \nu_3 \nu_4}\frac{k_{D^*}^{\mu_1}k_{D^*}^{\mu_2}k_{D^*}^{\nu_1}k_{D^*}^{\nu_2}}{m^2_{D^{*+}}}k_{\psi}^{\mu_3}k_{\psi}^{\nu_3}g^{\mu_4 \nu_4}+ \nonumber \\
&&+\frac{4}{3}g^2_{\psi D^* D}\varepsilon_{\mu_1 \mu_2 \mu_3 \mu_4}\varepsilon_{\nu_1 \nu_2 \nu_3 \nu_4}\frac{k_{D^*}^{\mu_1}k_{D^*}^{\mu_2}k_{D^*}^{\nu_1}k_{D^*}^{\nu_2}}{m^2_{D^{*+}}}\frac{k_{\psi}^{\mu_3}k_{\psi}^{\mu_4}k_{\psi}^{\nu_3}k_{\psi}^{\nu_4}}{m^2_{\psi}} \text{ .}
\end{eqnarray}
The obtained result can be significantly simplified becuse only the first term is nonzero, while the other terms vanish because they involve products of an antisymmetric tensor with a symmetric one. For example, in the second term we have the following quantity: 
\begin{eqnarray}
\varepsilon_{\mu_1 \mu_2 \mu_3 \mu_4}k_{\psi}^{\mu_3}k_{\psi}^{\mu_4}X^{\mu_1 \mu_2}&=&\varepsilon_{0123}k_{\psi}^2k_{\psi}^{3}X^{01}+\varepsilon_{0132}k_{\psi}^3k_{\psi}^{2}X^{01}+ \varepsilon_{1023}k_{\psi}^2k_{\psi}^{3}X^{10}+\nonumber \\
&&+\varepsilon_{1032}k_{\psi}^3k_{\psi}^{2}X^{10}+\ldots = 0\text{ , }
\end{eqnarray}
 where 
 \begin{eqnarray}
 X^{\mu_1 \mu_2}=-\frac{4}{3}g^2_{\psi D^* D}\varepsilon_{\nu_1 \nu_2 \nu_3 \nu_4}k^{\mu_1}_{D^*}k^{\nu_1}_{D^*}g^{\mu_2 \nu_2}\frac{k_{\psi}^{\nu_3}k_{\psi}^{\nu_4}}{m^2_{\psi}}\text{ . }
  \end{eqnarray}
  
Alternatively, one can reach the same conclusion in a general way:
  \begin{eqnarray}
  \varepsilon_{\mu_1 \mu_2 \mu_3 \mu_4}k^{\mu_3}k^{\mu_4}X^{\mu_1 \mu_2}&=& \varepsilon_{\mu_1 \mu_2 \mu_4 \mu_3}k^{\mu_4}k^{\mu_3}X^{\mu_1 \mu_2}=-\varepsilon_{\mu_1 \mu_2 \mu_3 \mu_4}k^{\mu_3}k^{\mu_4}X^{\mu_1 \mu_2} \Rightarrow \nonumber \\
&& \Rightarrow 2 \varepsilon_{\mu_1 \mu_2 \mu_3 \mu_4}k^{\mu_3}k^{\mu_4}X^{\mu_1 \mu_2} = 0\text{ , }
  \end{eqnarray}
  
which vanishes because it contains a product of the antisymmetric Levi-Civita tensor and the symmetric tensor $k^{\mu} k^{\nu}$. Similarly, one can show that the third and the fourth terms vanish.

 After this simplification and renaming the indexes $\mu_2\leftrightarrow \mu_3$ and $\nu_2\leftrightarrow \nu_3$ (this does not change the value of the expression), $|-i\mathcal{M}_{D^{*+}D^-}|^2$ takes the form
 \begin{eqnarray}
 |-i\mathcal{M}_{D^{*+}D^-}|^2&=&\frac{4}{3}g^2_{\psi D^* D} \varepsilon_{\mu_1 \mu_3 \mu_2 \mu_4}\varepsilon_{\nu_1 \nu_3 \nu_2 \nu_4}g^{\mu_2 \nu_2}g^{\mu_4 \nu_4}k_{D^{*+}}^{\mu_1}k_{D^{*+}}^{\nu_1}k_{\psi}^{\mu_3}k_{\psi}^{\nu_3} \nonumber \\
 &=&\frac{4}{3}g^2_{\psi D^* D}\varepsilon_{\mu_1 \mu_3}^{\hspace{0.6cm}\nu_2 \nu_4}\varepsilon_{\nu_1 \nu_3 \nu_2 \nu_4}k_{D^{*+}}^{\mu_1}k_{D^{*+}}^{\nu_1}k_{\psi}^{\mu_3}k_{\psi}^{\nu_3}\text{ . }
 \end{eqnarray}
 In order to make the next steps clear, we use the following general property of $\varepsilon^{\alpha \beta \gamma \delta}$ (four dimensional Levi-Civita in Minkowski space):
 \begin{eqnarray}
 \varepsilon^{\alpha \beta \gamma \delta}\varepsilon_{\rho \sigma \gamma \delta}=-2\delta^{\alpha \beta}_{\rho \sigma}=\left|\begin{array}{cc}
 \delta^{\alpha}_{\rho}& \delta^{\alpha}_{\sigma}\\
 \delta^{\beta}_{\rho}& \delta^{\beta}_{\sigma} \end{array}\right|=-2\left(\delta^{\alpha}_{\rho}\delta^{\beta}_{\sigma}-\delta^{\alpha}_{\sigma}\delta^{\beta}_{\rho}\right)\text{ . }\label{rel}
 \end{eqnarray}
 Consider now that:
 \begin{eqnarray}
 g_{\alpha' \alpha}g_{\beta' \beta}\varepsilon^{\alpha \beta \gamma \delta}\varepsilon_{\rho \sigma \gamma \delta}=\varepsilon_{\alpha' \beta'}^{\hspace{0.5cm}\gamma \delta}\varepsilon_{\rho \sigma \gamma \delta}\text{ . }
 \end{eqnarray}
Then, after using Eq. (\ref{rel}), one has:
 \begin{eqnarray}
 g_{\alpha' \alpha}g_{\beta' \beta}\varepsilon^{\alpha \beta \gamma \delta}\varepsilon_{\rho \sigma \gamma \delta}=-2 g_{\alpha' \alpha}g_{\beta' \beta}\left(\delta^{\alpha}_{\rho}\delta^{\beta}_{\sigma}-\delta^{\alpha}_{\sigma}\delta^{\beta}_{\rho}\right)=-2 \left(g_{\alpha' \rho}g_{\beta' \sigma}-g_{\alpha' \sigma}g_{\beta' \rho}\right)\text{ , }
 \end{eqnarray}
 which leads to the formula
 \begin{eqnarray}
 \varepsilon_{\alpha' \beta'}^{\hspace{0.6cm}\gamma \delta}\varepsilon_{\rho \sigma \gamma \delta}=-2\left(g_{\alpha' \rho}g_{\beta' \sigma}-g_{\alpha' \sigma}g_{\beta'\rho}\right)\text{ . }
 \end{eqnarray}
 Based on the above expressions, the amplitude square $|-i\mathcal{M}_{D^{*+}D^-}|^2$ of the decay $\psi(4040) \rightarrow D^{*+} D^-$ can be expressed as:
 \begin{eqnarray}
 |-i\mathcal{M}_{D^{*+}D^-}|^2&=&-\frac{8}{3}g^{2}_{\psi D^* D}\left(g_{\mu_1 \nu_1}g_{\mu_3 \nu_3}-g_{\mu_1 \nu_3}g_{\mu_3 \nu_1}\right)k_{D^{*+}}^{\mu_1}k_{D^{*+}}^{\nu_1}k_{\psi}^{\mu_3}k_{\psi}^{\nu_3} \nonumber \\
 &=&-\frac{8}{3}g^2_{\psi D^* D}\left(k^2_{D^{*+}}k_{\psi}^2-\left(k_{D^{*+}}k_{\psi}\right)^2\right)
 =\frac{8}{3}g^2_{\psi D^* D}\left(\left(k_{D^*}k_{\psi}\right)^2-m^2_{D^{*+}}m_{\psi}^2\right)\nonumber \\
 &=&\frac{8}{3}g^2_{\psi D^* D}\left(\left(m_{\psi}\sqrt{m^2_{D^{*+}}+|\vec{k}|^2}\right)^2-m^2_{D^{*+}}m_{\psi}^2\right)\nonumber \\
 &=&\frac{8}{3}g^2_{\psi D^* D}\left(m_{\psi}^2\left(m^2_{D^{*+}}+|\vec{k}|^2\right)-m^2_{D^{*+}}m_{\psi}^2\right)=\frac{8}{3}g^2_{\psi D^* D}m_{\psi}^2 |\vec{k}|^2\text{ . }
  \end{eqnarray}
  Note, the expression for the decay channel $\psi(4040) \rightarrow D^{*-}D^+$ is the same. 
  
  Finally, the tree-level decay width for the charged decay channels (two in total) is 
  \begin{eqnarray}
  \Gamma_{D^{*+}D^- + h.c. }=2\frac{|\vec{k}|}{8 \pi m^2_{\psi}}\left[\frac{8}{3}g^2_{\psi D^* D}|\vec{k}|^2 m^2_{\psi}\right]=\frac{2}{3}\frac{|\vec{k}|^3}{\pi}g^2_{\psi D^* D}\text{ . }
  \end{eqnarray}

  \subsection{Vector $\rightarrow$ vector + vector}
  As examples of this kind of decay we mention: 
  
\begin{equation}
\psi(4040) \rightarrow D^*(2007)^0 \bar{D}^*(2007)^0, \hspace{0.7cm} \psi(4040) \rightarrow D^*(2010)^+D^*(2010)^-. \nonumber
\end{equation}

\begin{figure}[h!]
\begin{center}
\includegraphics[width=0.40 \textwidth]{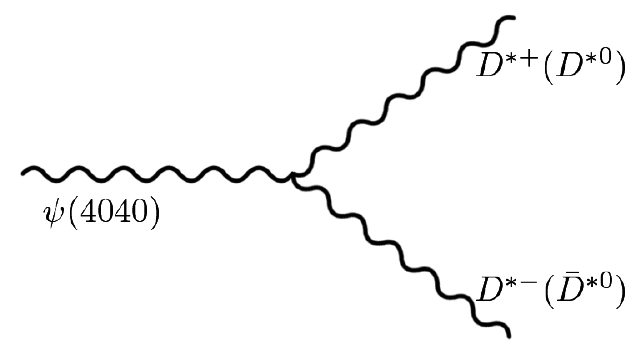}
\caption{\label{nbv}Schematic decay of $\psi(4040)$ into $D^{*+}D^{*-}$ $(D^{*0}\bar{D}^{*0})$.}
\end{center}
\end{figure}

The calculations shown in this subsection refer to the decay of $\psi(4040)$ into $D^{*+}D^{*-}$ particles (see Figure \ref{nbv}).

\newpage
Notation and relations:\\
\begin{enumerate}
\item Four-momenta:
\begin{eqnarray}
p_{\psi}^{\mu}=\left(E_{\psi},\vec{0}\right)=\left(m_{\psi},  \vec{0}\right)\text{ ; }
\end{eqnarray}
\begin{eqnarray}
k_{D^{*+}}^{\mu}=\left(E_{+}, \vec{k}\right)=\left(\sqrt{m^2_{D^{*+}}+|\vec{k}|^2}, \vec{k}\right)\text{ ; }
\end{eqnarray}
\begin{eqnarray}
k_{D^{*-}}^{\mu}=\left(E_{-}, \vec{k}\right)=\left(\sqrt{m^2_{D^{*-}}+|\vec{k}|^2}, -\vec{k}\right)=\left(\sqrt{m^2_{D^{*+}}+|\vec{k}|^2}, -\vec{k}\right)\text{ . }
\end{eqnarray}
because $m_{D^{*+}}=m_{D^{*-}}$.
\item Energy: 
\begin{eqnarray}
m_{\psi}=E_{+}+E_{-}\text{ ; }
\end{eqnarray}
\begin{eqnarray}
E_{+}=\sqrt{m^2_{D^{*+}}+|k|^2}=E_{-}=E=\frac{m_{\psi}}{2}\text{ . }
\end{eqnarray}
\item Momentum:
\begin{eqnarray}
|\vec{k}|=\sqrt{\frac{m_{\psi}^2-4m^2_{D^{*+}}}{2}}\text{ . }
\end{eqnarray}
\item Useful relations:
\begin{eqnarray}
k_{\psi}k_{D^+}=k_{\psi}k_{D^-}=m_{\psi}E=\frac{m_{\psi}^2}{2}\text{ ; }
\end{eqnarray}
\begin{eqnarray}
k_{D^+}k_{D^-}=E^2+|k|^2=\frac{m_{\psi}^2}{4}+|k|^2\text{ ; }
\end{eqnarray}
\begin{eqnarray}
k^2_{D^{*+}}=k^2_{D^{*-}}=m^2_{D^{*+}}\text{ ; }
\end{eqnarray}
\begin{eqnarray}
k^2_{\psi}=m^2_{\psi}\text{ . }
\end{eqnarray}
\end{enumerate}
The Lagrangian
\begin{eqnarray}
\mathcal{L}_{\psi D^* D^*}=ig_{D^* D^*}\left[\psi_{\mu \nu}\left(D^{*+ \mu}D^{*-\nu}-D^{*-\mu}D^{*+\nu}\right)+\psi_{\mu \nu}\left(D^{*0\mu}\bar{D}^{*0\nu}-\bar{D}^{*0 \mu}D^{*0 \nu}\right)\right]\text{ , }
\end{eqnarray}
where $\psi_{\mu \nu}=\partial_{\mu \nu}\psi_{\nu}-\partial_{\nu} \psi_{\mu}$\text{ .} Using Feynman rules:
\begin{eqnarray}
-i\mathcal{M}=2g_{\psi D^*D^*}\left(\varepsilon_{+}^{\mu}\varepsilon_{-}^{\nu}-\varepsilon_{+}^{\nu}\varepsilon_{-}^{\mu}\right)\varepsilon_{\psi \nu}k_{\psi \mu}\text{ , }
\end{eqnarray}
where the following assignements are employed: $\varepsilon_{D^{*+}}^{\mu}\equiv\varepsilon_{+}^{\mu}$ and $\varepsilon_{D^{*-}}^{\mu}\equiv\varepsilon_{-}^{\mu} \text{ . }$ Then:
\begin{eqnarray}
|\mathcal{M}|^2&=&\frac{1}{3} 4g_{\psi D^* D^*}\sum \limits_{\lambda_1, \lambda_2, \lambda_3}[\left(\varepsilon_{+}^{\mu}(\lambda_1)\varepsilon_{-}^{\nu}(\lambda_2)-\varepsilon_{+}^{\nu}(\lambda_1)\varepsilon_{-}^{\mu}(\lambda_2)\right)\varepsilon_{\psi \nu}(\lambda_3)k_{\psi \mu} \cdot \nonumber \\ &\cdot & \left(\varepsilon_{+}^{\alpha}(\lambda_1)\varepsilon_{-}^{\beta}(\lambda_2)-\varepsilon_{+}^{\beta}(\lambda_1)\varepsilon_{-}^{\alpha}(\lambda_2)\right)\varepsilon_{\psi \beta}(\lambda_3)k_{\psi \alpha}] \nonumber \\
&=& \frac{4}{3}g^2_{\psi D^* D^*} \sum \limits_{\lambda_1, \lambda_2, \lambda_3}[\varepsilon^{\mu}_{+}(\lambda_1)\varepsilon^{\alpha}_{+}(\lambda_1)\varepsilon^{\nu}_{-}(\lambda_2)\varepsilon^{\beta}_{-}(\lambda_2)+\nonumber \\
&-& \varepsilon^{\mu}_{+}(\lambda_1)\varepsilon_{+}^{\beta}(\lambda_1)\varepsilon_{-}^{\nu}(\lambda_2)\varepsilon_{-}^{\alpha}(\lambda_2)-
\varepsilon_{+}^{\nu}(\lambda_1)\varepsilon_{+}^{\alpha}(\lambda_1)\varepsilon_{-}^{\mu}(\lambda_2)\varepsilon_{-}^{\beta}(\lambda_2)+ \nonumber \\
&+&\varepsilon_{+}^{\nu}(\lambda_1)\varepsilon_{+}^{\beta}(\lambda_1)\varepsilon_{-}^{\mu}(\lambda_2)\varepsilon_{-}^{\alpha}(\lambda_2)]\varepsilon_{\psi \nu}(\lambda_3)\varepsilon_{\psi \beta}(\lambda_3)k_{\psi \mu}k_{\psi \alpha}\text{ . }
\end{eqnarray}
Going on:
\begin{eqnarray}
|\mathcal{M}|^2&=&\frac{4}{3}g^2_{\psi D^* D^*}\Bigg[\left(-g^{\mu \alpha}+\frac{k_{D^+}^{\mu}k_{D^+}^{\alpha}}{m^2_{{D^{*+}}}}\right)\left(-g^{\nu \beta}+\frac{k_{D^-}^{\nu}k_{D^-}^{\beta}}{m^2_{D^{*+}}}\right)\nonumber \\
&-&\left(-g^{\mu \beta}+\frac{k_{D^+}^{\mu}k_{D^+}^{\beta}}{m^2_{D^{*+}}}\right)\left(-g^{\nu \alpha}+\frac{k_{D^-}^{\nu}k_{D^-}^{\alpha}}{m^2_{D^{*+}}}\right)
-\left(-g^{\nu \alpha}+\frac{k_{D^+}^{\nu}k_{D^+}^{\alpha}}{m^2_{D^{*+}}}\right)\left(-g^{\mu \beta}+\frac{k_{D^-}^{\mu}k_{D^-}^{\beta}}{m^2_{D^{*+}}}\right)\nonumber \\
&+& \left( -g^{\nu \beta} + \frac{k_{D^+}^{\nu}k_{D^+}^{\beta}}{m^2_{D^{*+}}}\right)\left( -g^{\mu \alpha}+\frac{k_{D^-}^{\mu}k_{D^-}^{\alpha}}{m^2_{D^{*+}}}\right)\Bigg]\left(-g_{\nu \beta}+\frac{k_{\psi, \nu}k_{\psi, \beta}}{m^2_{\psi}}\right)k_{\psi \mu} k_{\psi \alpha}\text{ . }
\end{eqnarray}
The previous expression consists of four terms which are considered separately in the following (for simplicity we omit $\cdot$ between Lorentz vectors):
\begin{eqnarray}
|\mathcal{M}|^2_{(1)}&=&\frac{4}{3}g^2_{\psi D^* D^*}k_{\psi \mu}k_{\psi \alpha}\left(-g^{\mu \alpha}+\frac{k^{\mu}_{D^+}k^{\alpha}_{D^+}}{m^2_{D^{*+}}}\right)\left(-g^{\nu \beta}+\frac{k^{\nu}_{D^-}k^{\beta}_{D^-}}{m^2_{D^{*+}}}\right)\left(-g_{\nu \beta}+\frac{k_{\psi,\nu}k_{\psi,\beta}}{m^2_{\psi}}\right)\nonumber \\
&=&\frac{4}{3}g^2_{\psi D^* D^*}\left(-k_{\psi}^2+\frac{\left(k_{\psi} k_{D^+}\right)^2}{m^2_{D^{*+}}}\right)\left(4-\frac{k^2_{\psi}}{m^2_{\psi}}-\frac{k^2_{D^-}}{m^2_{D^{*+}}}+\frac{\left(k_{\psi}k_{D^-}\right)^2}{m^2_{D^{*+}}m^2_{\psi}}\right) \nonumber \\
&=&\frac{4}{3}g^2_{\psi D^* D^*}\left(-k^2_{\psi}+\frac{\left(k_{\psi}k_{D^+}\right)^2}{m^2_{D^{*+}}}\right)\left(2+\frac{\left(k_{\psi}k_{D^-}\right)^2}{m^2_{\psi}m^2_{D^{*+}}}\right)\nonumber \\
&=&\frac{4}{3}g^2_{\psi D^* D^*}\left(-k_{\psi}^2+\frac{\left(k_{\psi}k_{D^+}\right)^2}{m^2_{D^{*+}}}\right)\left(2+\frac{m_{\psi}^2\left(|k|^2+m^2_{D^*_c}\right)}{m^2_{\psi}m^2_{D^{*+}}}\right)\nonumber \\
&=&\frac{4}{3}g^2_{\psi D^* D^*}\left(-k_{\psi}^2+\frac{\left(k_{\psi}k_{D^+}\right)^2}{m^2_{D^{*+}}}\right)\left(3+\frac{|k|^2}{m^2_{D^{*+}}}\right)\text{ ; }
\end{eqnarray}
\begin{eqnarray}
|\mathcal{M}|^2_{(2)}&=&-\frac{4}{3}g^2_{\psi D^* D^*}k_{\psi \mu} k_{\psi \alpha}\left(-g^{\mu \beta}+\frac{k^{\mu}_{D^+}k^{\beta}_{D^+}}{m^2_{D^{*+}}}\right)\left(-g^{\nu \alpha}+\frac{k^{\nu}_{D^-}k^{\alpha}_{D^-}}{m^2_{D^{*+}}}\right)\left(-g_{\nu \beta}+\frac{k_{\psi, \nu}k_{\psi, \beta}}{m^2_{\psi}}\right) \nonumber \\
&=& -\frac{4}{3}g^2_{\psi D^* D^*}\left(-g_{\nu \beta}+\frac{k_{\psi, \nu}k_{\psi, \beta}}{m^2_{\psi}}\right)\left(-k_{\psi}^{\beta}+ \frac{k_{\psi} k_{D^+} k_{D^+}^{\beta}}{m^2_{D^{*+}}}\right)\left(-k_{\psi}^{\nu}+ \frac{k_{\psi} k_{D^-} k_{D^-}^{\nu}}{m^2_{D^{*+}}}\right) \nonumber \\
&=& -\frac{4}{3}g^2_{\psi D^* D^*}\left(k_{\psi, \nu}-\frac{k_{\psi} k_{D^+} k_{D^+, \nu}}{m^2_{D^{*+}}}-\frac{k_{\psi, \nu} k_{\psi}^2}{m^2_{\psi}}+\frac{\left(k_{\psi} k_{D^+}\right)^2 k_{\psi, \nu}}{m^2_{\psi} m^2_{D^{*+}}}\right)\left(-k_{\psi}^{\nu}+ \frac{k_{\psi} k_{D^-} k_{D^-}^{\nu}}{m^2_{D^{*+}}}\right) \nonumber \\
&=&-\frac{4}{3}g^2_{\psi D^* D^*}\Bigg(\frac{\left(k_{\psi} k_{D^+}\right)^2}{m^2_{D^{*+}}}-\frac{\left(k_{\psi} k_{D^+}\right)^2 k^2_{\psi}}{m^2_{\psi} m^2_{D^{*+}}}-\frac{\left(k_{\psi}k_{D^+}\right)\left(k_{\psi} k_{D^-}\right)\left(k_{D^+}k_{D^-}\right)}{m^4_{D^{*+}}}+\nonumber \\
&+& \frac{\left(k_{\psi} k_{D^+}\right)^2 \left(k_{\psi}k_{D^-}\right)^2}{m^2_{\psi} m^4_{D^{*+}}}\Bigg)
= -\frac{4}{3}g^2_{\psi D^* D^*}\Bigg(-\frac{\left(k_{\psi}k_{D^+}\right)\left(k_{\psi} k_{D^-}\right)\left(k_{D^+}k_{D^-}\right)}{m^4_{D^{*+}}}+\nonumber \\
&+&\frac{\left(k_{\psi} k_{D^+}\right)^2 \left(k_{\psi}k_{D^-}\right)^2}{m^2_{\psi} m^4_{D^{*+}}}\Bigg)\text{ ; }
\end{eqnarray}
\begin{eqnarray}
|\mathcal{M}|^2_{(3)}&=& -\frac{4}{3}g^2_{\psi D^* D^*} k_{\psi \mu} k_{\psi \alpha}\left(-g^{\nu \alpha}+\frac{k^{\nu}_{D^+} k^{\alpha}_{D^+}}{m^2_{D^{*+}}}\right)\left(-g^{\mu \beta}+\frac{k^{\mu}_{D^-} k_{D^-}^{\beta}}{m^2_{D^{*+}}}\right)\left(-g_{\nu \beta}+\frac{k_{\psi, \nu} k_{\psi, \beta}}{m^2_{\psi}}\right) \nonumber \\
&=& -\frac{4}{3}g^2_{\psi D^* D^*}\left(-g_{\nu \beta}+\frac{k_{\psi, \nu}k_{\psi, \beta}}{m^2_{\psi}}\right)\left(-k_{\psi}^{\nu}+\frac{k_{\psi} k_{D^+} k_{D^+}^{\nu}}{m^2_{D^{*+}}}\right)\left(-k_{\psi}^{\beta}+\frac{k_{\psi}k_{D^-}k_{D^-}^{\beta}}{m^2_{D^{*+}}}\right) \nonumber \\
&=&-\frac{4}{3}g^2_{\psi D^* D^*}\left(k_{\psi, \beta}-\frac{k_{\psi}k_{D^+}k_{D^+, \beta}}{m^2_{D^{*+}}}-\frac{k_{\psi}^2 k_{\psi, \beta}}{m^2_{\psi}}+\frac{\left(k_{\psi}k_{D^+}\right)^2k_{\psi, \beta}}{m^2_{\psi}m^2_{D^{*+}}}\right)\left(-k_{\psi}^{\beta}+\frac{k_{\psi}k_{D^-}k_{D^-}^{\beta}}{m^2_{D^{*+}}}\right)\nonumber \\
&=& -\frac{4}{3}g^2_{\psi D^* D^*}\Bigg(\frac{\left(k_{\psi}k_{D^+}\right)^2}{m^2_{D^{*+}}}- \frac{k_{\psi}^2\left(k_{\psi}k_{D^+}\right)^2}{m^2_{\psi}m^2_{D^{*+}}}-\frac{\left(k_{\psi}k_{D^+}\right)\left(k_{\psi}k_{D^-}\right)\left(k_{D^+}k_{D^-}\right)}{m^4_{D^{*+}}}+\nonumber \\
&+&\frac{\left(k_{\psi}k_{D^+}\right)^2 \left(k_{\psi}k_{D^-}\right)^2}{m^2_{\psi}m^4_{D^{*+}}}\Bigg)
=-\frac{4}{3}g^2_{\psi D^* D^*}\Bigg(-\frac{\left(k_{\psi}k_{D^+}\right)\left(k_{\psi}k_{D^-}\right)\left(k_{D^+}k_{D^-}\right)}{m^4_{D^{*+}}}+ \nonumber \\
&+&\frac{\left(k_{\psi}k_{D^+}\right)^2 \left(k_{\psi}k_{D^-}\right)^2}{m^2_{\psi}m^4_{D^{*+}}}\Bigg)\text{ ; }
\end{eqnarray}
\begin{eqnarray}
|\mathcal{M}|^2_{(4)}&=&\frac{4}{3}g^2_{\psi D^* D^*}k_{\psi \mu} k_{\psi \alpha}\left(-g^{\mu \alpha}+ \frac{k^{\mu}_{D^-}k^{\alpha}_{D^-}}{m^2_{D^{*+}}}\right)\left(-g^{\nu \beta}+ \frac{k_{D^+}^{\nu}k^{\beta}_{D^+}}{m^2_{D^{*+}}}\right)\left(-g_{\nu \beta}+ \frac{k_{\psi, \nu}k_{\psi, \beta}}{m^2_{\psi}}\right)\nonumber \\
&=&\frac{4}{3}g^2_{\psi D^* D^*}\left(-k^2_{\psi}+\frac{\left(k_{\psi}k_{D^-}\right)^2}{m^2_{D^{*+}}}\right)\left(4-\frac{k^2_{\psi}}{m^2_{\psi}}-\frac{k^2_{D^+}}{m^2_{D^{*+}}}+\frac{\left(k_{\psi}k_{D^+}\right)^2}{m^2_{\psi}m^2_{D^{*+}}}\right)\nonumber \\
&=&\frac{4}{3}g^2_{\psi D^* D^*}\left(-k^2_{\psi}+\frac{\left( k_{\psi}k_{D^-}\right)^2}{m^2_{D^{*+}}}\right)\left(2+\frac{\left(k_{\psi}k_{D^+}\right)^2}{m^2_{\psi}m^2_{D^{*+}}}\right)\nonumber \\
&=&\frac{4}{3}g^2_{\psi D^* D^*}\left(-k_{\psi}^2+\frac{\left(k_{\psi}k_{D^-}\right)^2}{m^2_{D^{*+}}}\right)\left(2+\frac{m_{\psi}^2\left(|k|^2+m^2_{D^{*+}}\right)}{m^2_{\psi}m^2_{D^{*+}}}\right)\nonumber \\
&=&\frac{4}{3}g^2_{\psi D^* D^*}\left(-k_{\psi}^2+\frac{\left(k_{\psi}k_{D^-}\right)^2}{m^2_{D^{*+}}}\right)\left(3+\frac{|k|^2}{m^2_{D^{*+}}}\right)\text{ . }
\end{eqnarray}
Finally:
\begin{eqnarray}
|\mathcal{M}|^2&=&2\frac{4}{3}g^2_{\psi D^* D^*}\left[\frac{\left(k_{\psi}k_{D^-}\right)^2\left(k_{D^+}k_{D^-}\right)}{m^4_{D^{*+}}}-\frac{\left(k_{\psi}k_{D^+}\right)^4}{m_{\psi}^2m^4_{D^{*+}}}-\left(k_{\psi}^2-\frac{\left(k_{\psi}k_{D^+}\right)^2}{m^2_{D^{*+}}}\right)\left(3+\frac{|k|^2}{m^2_{D^*_c}}\right)\right]\nonumber \\
&=&2\frac{4}{3}g^2_{\psi D^* D^*}\left[\frac{m_{\psi}^4}{4m^4_{D^{*+}}}\left(\frac{m^2_{\psi}}{4}+|k|^2\right)-\frac{m^8_{\psi}}{16m^4_{D^{*+}}m^2_{\psi}}-\left(m_{\psi}^2-\frac{m_{\psi}^4}{4m^2_{D^{*+}}}\right)\left(3+\frac{|k|^2}{m^2_{D^{*+}}}\right)\right] \nonumber \\
&=&2\frac{4}{3}g^2_{\psi D^* D^*}\left[\frac{m_{\psi}^4 |k|^2}{4m^4_{D^{*+}}}+\frac{m_{\psi}^2}{m^2_{D^{*+}}}\left(\frac{m_{\psi}^2}{4}-m^2_{D^{*+}}\right)\left(3+\frac{|k|^2}{m^2_{D^{*+}}}\right)\right] \nonumber \\
&=&2 \frac{4}{3} g^2_{\psi D^* D^*}\left[\frac{m_{\psi}^4|k|^2}{4m^4_{D^{*+}}}+\frac{m^2_{\psi}|k|^2}{m^2_{D^{*+}}}\left(3+\frac{|k|^2}{m^2_{D^*_c}}\right)\right]\nonumber \\ &=&2\frac{4}{3}g^2_{\psi D^* D^*}\left[\frac{m_{\psi}^2\left(|k|^2+m^2_{D^{*+}}\right)|k|^2}{m^4_{D^{*+}}}+\frac{3m_{\psi}^2|k|^2}{m^2_{D^{*+}}}+\frac{m^2_{\psi}|k|^4}{m^4_{D^{*+}}}\right]\nonumber \\
&=&\frac{4}{3}g^2_{\psi D^* D^*}\left[4\frac{m_{\psi}^2|k|^4}{m^4_{D^{*+}}}+8\frac{m_{\psi}^2|k|^2}{m^2_{D^{*+}}}\right]\text{ . }
\end{eqnarray}
Then, the decay width reads:
\begin{eqnarray}
\Gamma_{D^{*+} D^{*-}}=\frac{|k|}{8 \pi m^2_{\psi}}\frac{4}{3}g^2_{\psi D^* D^*}\left[4\frac{m_{\psi}^2|k|^4}{m^4_{D^{*+}}}+8\frac{m_{\psi}^2|k|^2}{m^2_{D^{*+}}}\right]=\frac{2|k|^3}{3 \pi m^2_{D^{*+}}}g^2_{\psi D^* D^*}\left[2+\frac{|k|^2}{m^2_{D^{*+}}}\right]\text{ . }
\end{eqnarray}
\subsection{Scalar $\rightarrow$ pseudoscalar + pseudoscalar}
\label{threemomentum1430}
Examples of this kind of decay are: 
\begin{equation}
K_0^*(1430)^- \rightarrow \pi^- \bar{K}^0, \hspace{0.7cm} K^*_0(1430)^- \rightarrow \pi^0 K^-. \nonumber
\end{equation}
\begin{figure}[h!]
\begin{center}
\includegraphics[width=0.40 \textwidth]{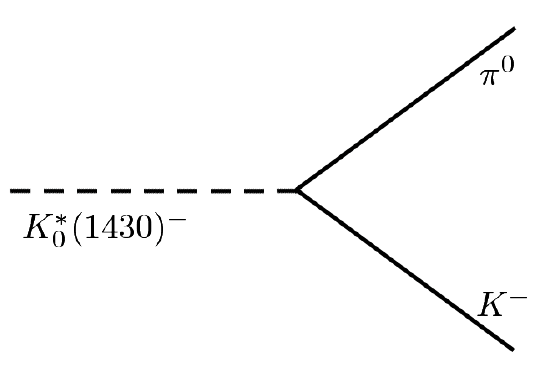}
\caption{Schematic decay of $K_0^*(1430)^-$ into $K^- \pi^0$.}
\end{center}
\end{figure}

The interaction Lagrangian for the resonance $K^*_0(1430)$ decaying into one $\pi$ and one $K$ contains two terms, one involving derivative and one does not: 
\begin{eqnarray}
\mathcal{L}_{int}=aK^{*+}_0K^-\pi ^0+bK_{0}^{*+}\partial_{\mu}K^-\partial^{\mu} \pi ^0+ h.c + \ldots \text{ , }
\end{eqnarray}
where the dots refer to analogous terms involving the other members of the isospin multiplet.

We turn to the notation:
\begin{enumerate}
\item Four-momenta:
\begin{eqnarray}
p^{\mu}_{K_0^*}=\left(m_{K^*_0}, 0\right)\text{ ; }
\end{eqnarray}
\begin{eqnarray}
p_{\pi}^{\mu}=\left(\sqrt{m^2_{\pi}+|\vec{k}|^2},  \vec{k}\right)\text{ ; }
\end{eqnarray}
\begin{eqnarray}
p_k^{\mu}=\left(\sqrt{m^2_{K}+|\vec{k}|^2},  -\vec{k}\right)\text{ . }
\end{eqnarray}
\item Momentum:
\begin{eqnarray}
|\vec{k}|=\frac{\sqrt{m^4_{K_0^*}+\left(m^2_{\pi}-m_K^2\right)^2-2\left(m_{\pi}^2+m_K^2\right)m_{K^*_0}^2}}{2m_{K^*_0}}\text{ . }
\end{eqnarray}
\item Useful relation:
\begin{eqnarray}
k_K k_{\pi}=\frac{m^2_{K^*_0}-m^2_{\pi}-m^2_K}{2}\text{ . }
\end{eqnarray}
\end{enumerate}

\begin{figure}[h!]
\begin{center}
\includegraphics[width=0.99 \textwidth]{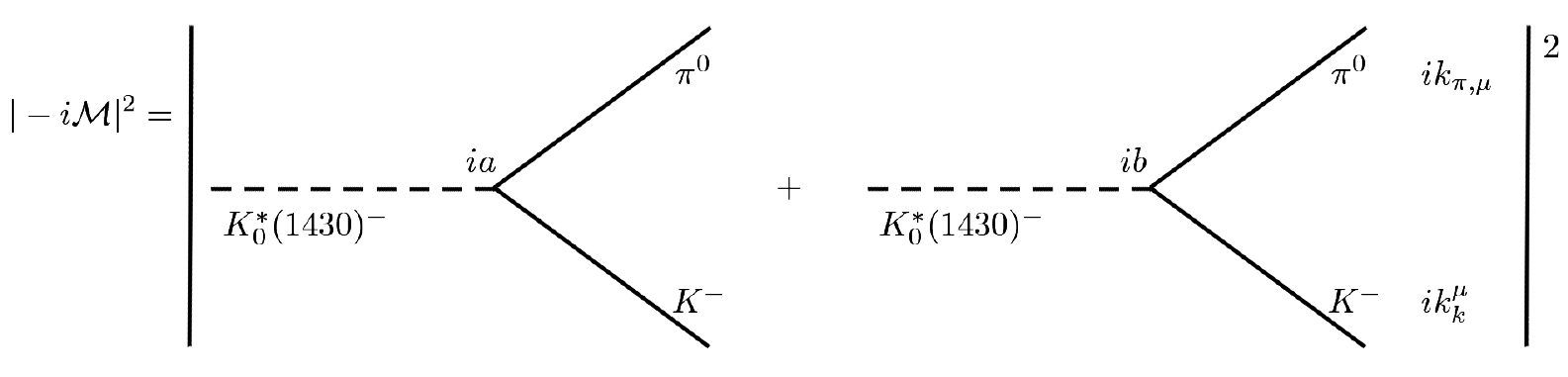}
\caption{Schematic representation of the squared amplitude of the decay of $K_0^*(1430)^-$ into $K^- \pi^0$. Interference of the terms proportional to ``$a$'' and ``$b$'' is made clear.}
\end{center}
\end{figure}
Amplitude calculation:

The amplitude element is given by:
 \begin{eqnarray}
 -i\mathcal{M}=ia+ib\left(ik_{\pi,\mu}\right)\left(ik_{K}^{\mu}\right)=ia-ib\left(k_{\pi}k_{K}\right)=i\left(a-b\frac{m^2_{K^*_0}-m^2_{\pi}-m_K^2}{2}\right)\text{ . }
 \end{eqnarray}
 Finally:
 \begin{eqnarray}
 \Gamma_{K^*_0 \rightarrow K \pi}(m)=3\frac{|\vec{k}|(m)}{8 \pi m^2}\left(a-b\frac{m^2-m_{\pi}^2-m_{K}^2}{2}\right)^2 \text{ . }
 \end{eqnarray}
 
 Moreover, as we shall explain later on in detail (see Chapter \ref{chapvec}), a form factor $F_{\Lambda}(m)$ will be needed. This form factor implies the modification of the previous expression as:
 \begin{eqnarray}
 \Gamma_{K^*_0 \rightarrow K \pi}(m)=3\frac{|\vec{k}|(m)}{8 \pi m^2}\left(a-b\frac{m^2-m_{\pi}^2-m_{K}^2}{2}\right)^2F_{\Lambda}(m) \text{ . }
 \end{eqnarray}
 This is indeed applicable to all previously discussed decays.
 \chapter{Excited vector mesons in the $q\bar{q}$ picture}
 \label{excitki}
 
In this chapter we perform a phenomenological study  of two nonets of excited $q\bar{q}$ vector mesons. The first nonet with the set of states $\{\rho(1450),$ $K^*(1410),$ $\omega(1420),$ $\phi(1680)\}$ is roughly identified with radially excited vector mesons with quantum numbers $n$ $^{2S+1}L_{J}$=$2$ $^{3}S_{1}$. The second nonet with the resonances $\{\rho(1700),$ $K^*(1680),$ $\omega(1650),$ $\phi (???)\}$ corresponds predominantly to orbitally excited vector mesons described by quantum numbers $n$ $^{2S+1}L_{J}$=$1$ $^{3}D_{1}$. After introducing an effective flavor-invariant QFT model, we calculate various decays of the states belonging to these nonets. In particular, we study the strong and radiative  decay channels. The theoretical results for the decay widths as well as many branching ratios are compared with the available experimental data of PDG. A good agreement of theory with data is obtained. In addition, some predictions have been given for the not yet experimentally observed $s\bar{s}$ resonance belonging to $1$ $^{3}D_{1}$ nonet. Since, within our model its mass is estimated to be $\sim 1959$ MeV, throughout this thesis we identify this state with the name $\phi(1959)$. At this point one should point out that in the original paper \cite{PiotrcoitoX} we use the name $\phi(1930)$ when refering to this particular state. This updated value is a consequence of a more detailed evaluation of the mass. The state $\phi(1959)$ decays predominantly into $\bar{K}K^*(892)+h.c.$ and $\bar{K}K$ channels, but it also decays into $\gamma \eta$. This means that it can be found in the ongoing GlueX and CLAS12 experiments at Jefferson Lab which are based on photoproduction.
 \section{Introduction to the excited vector mesons}
Before we proceed with the excited states, some words on the ground-state vector mesons have to be said. This nonet consists of the well-know conventional $q\bar{q}$ resonances $\{\rho(770),$ $K^*(892),$ $\phi(1020),$ $\omega(782)\}$ and is decribed by the following specific quantum numbers: $n$ $^{2S+1}L_{J}$=$1$ $^3S_1$ when using the non-relativistic notation, and $J^{PC}=1^{--}$ for the relativistic one. Taking into account that the pseudoscalar mesons are peculiar due to the phenomenon of spontaneous symmetry breaking and the axial anomaly, the ground-state vector mesons are considered as the lightest ,,almost ideal'' bound states of a constituent quark and a constituent antiquark. 

In contrast to the above arguments, the current knowledge about the excited vector mesons is rather poor and needs to be improved. So far, two groups of excited vector mesons have been seen in experiments: these are the nonets with radial and orbital excitations, respectively. Even if the data are not yet complete, they are good enough to perform a systematic phenomenological analysis. We recall that the nonet of the vector mesons with (predominantly) radial excitation has the quantum numbers $2$ $^3S_1$ (non-relativistic notation) and $1^{--}$ (relativistis notation). The set of states belonging to this nonet is $\{\rho(1450),$ $K^*(1410),$ $\omega(1420),$ $\phi(1680)\}$. For what concerns the nonet of vector mesons with (predominantly) orbital excitation, it has the quantum numbers $1$ $^3D_1$ and $1^{--}$ in non-relativistic and relativistic notation, respectively. The resonances corresponding to this nonet are $\{\rho(1700),$ $K^*(1680),$ $\omega(1650),$ $\phi(???)\}$. Notice that the last state belonging to this nonet is still undiscoved. Moreover, it is important to stress that the physical states which are mentioned above cannot be directly understood as the ideal $q\bar{q}$ configurations described by the non-relativistic notation. In fact, mixing between different configurations is possible. In this respect, the assignment given by the non-relativistic quantum numbers should be regarded as the dominant -but not as the solely- contribution. 
 \section{Theoretical framework}
 
 In this section we introduce our relativistic QFT model described by an effective Lagrangian whose degrees of freedom are mesonic fields corresponding to $q\bar{q}$ states. Our model is constructed in agreement with flavor symmetry. In the following, we first focus on the description of the mesonic fields, then we briefly describe the Lagrangian and the theoretical formulas of the tree-level decay widths (for both the strong and radiative channels). Finally, we show in detail the determination of the model parameters.  
 \subsection{The fields of the theoretical model}
 We start with the introduction of four nonets of mesons, which are presented in forms of matrices. All field components, which correspond to the elements of a matrix, are identified with quark-antiquark states according to the following assignment:
 
 \begin{equation}
\frac{1}{\sqrt{2}}\left(
\begin{array}
[c]{lcr}%
u\bar{u} & u\bar{d} & u\bar{s}\\
d\bar{u} & d\bar{d} & d\bar{s}\\
s\bar{u} & s\bar{d} & s\bar{s}
\end{array}
\right)  \text{ .}%
\end{equation}
To be more precise, the nonet of pseudoscalar mesons $(P)$, the nonet of ground-state vector mesons $(V^{\mu})$, and two nonets of excited vector mesons ($V^{\mu}_E$ and $V^{\mu}_D$) take the following explicit forms:
\begin{equation}
P=\frac{1}{\sqrt{2}}\left(
\begin{array}
[c]{lcr}%
\frac{\eta_{N}+\pi^{0}}{\sqrt{2}} & \pi^{+} & K^{+}\\
\pi^{-} & \frac{\eta_{N}-\pi^{0}}{\sqrt{2}} & K^{0}\\
K^{-} & \bar{K}^{0} & \eta_{S}%
\end{array}
\right)  \text{ ; }V^{\mu}=\frac{1}{\sqrt{2}}\left(
\begin{array}
[c]{lcr}%
\frac{\omega^{\mu}+\rho^{\mu0}}{\sqrt{2}} & \rho^{\mu+} & K_{i}^{\mu\star+}\\
\rho^{\mu-} & \frac{\omega^{\mu}-\rho^{\mu0}}{\sqrt{2}} & K^{\mu\star0}\\
K^{\mu\star-} & \bar{K}^{\mu\star0} & \phi^{\mu}%
\end{array}
\right)  \text{ ; } \label{pv}%
\end{equation}

\begin{equation}
V_{E}^{\mu}=\frac{1}{\sqrt{2}}\left(
\begin{array}
[c]{lcr}%
\frac{\omega_{E}^{\mu}+\rho_{E}^{\mu0}}{\sqrt{2}} & \rho_{E}^{\mu+} &
K_{E}^{\mu\star+}\\
\rho_{E}^{\mu-} & \frac{\omega_{E}^{\mu}-\rho_{E}^{\mu0}}{\sqrt{2}} &
K_{E}^{\mu\star0}\\
K_{E}^{\mu\star-} & \bar{K}_{E}^{\mu\star0} & \phi_{E}^{\mu}%
\end{array}
\right)  \text{ ; }V_{D}^{\mu}=\frac{1}{\sqrt{2}}\left(
\begin{array}
[c]{lcr}%
\frac{\omega_{D}^{\mu}+\rho_{D}^{\mu0}}{\sqrt{2}} & \rho_{D}^{\mu+} &
K_{D}^{\mu\star+}\\
\rho_{D}^{\mu-} & \frac{\omega_{D}^{\mu}-\rho_{D}^{\mu0}}{\sqrt{2}} &
K_{D}^{\mu\star0}\\
K_{D}^{\mu\star-} & \bar{K}_{D}^{\mu\star0} & \phi_{D}^{\mu}%
\end{array}
\right)  \text{ .} \label{ve}%
\end{equation}
For a detailed description of these fields via $q\bar{q}$ microscopic currents, see Ref. \cite{adfran}.
Note, in our relativistic QFT approach it is natural to keep track of the relativistic notation $(J^{PC})$ to describe these multiplets. Nevertheless, we shall also follow the non-relativictic notation which describes the dominant contribution in the (not visible in our approach) mesonic microscopic wave functions. Such treatment (even if only approximate) allows to find the connection with the results of the quark model, thus allowing for a more intuitive understanding of the nature of these states. 

Let us now describe more precisely all these conventional quark-antiquark nonets mentioned above.

(i) The matrix $P$ represents the pseudoscalar meson nonet identified with the particles $\{ \pi, K, \eta \equiv \eta(547), \eta' \equiv \eta'(958)\}$. One should remark that in Eq. (\ref{pv}) two fields, $\eta_N$ regarded as purly nonstrange $\left(\eta_N\equiv\sqrt{\frac{1}{2}}(u\bar{u}+d\bar{d})\right)$ and $\eta_S$ as purly strange $\left(\eta_S\equiv s\bar{s}\right)$, are unphysical. In fact, they mix and generate the physical fields $\eta\equiv\eta(547)$ and $\eta'\equiv\eta(958)$: 
\begin{equation}
\eta_{N}=\eta\cos\theta_{P}-\eta^{\prime}\sin\theta_{P}\text{ and }\eta
_{S}=\eta\sin\theta_{P}-\eta^{\prime}\cos\theta_{P}\text{ .}%
\end{equation}
The value of mixing angle $\theta_{P}$ is estimated by the KLOE Collaboration as $-40.5^{\circ}$ \cite{kloemixang}. Changing its value between the range ($-40^{\circ},$ $-45^{\circ}$), suggested e.g. in Refs. \cite{dick, bass, feldmann}, does not have a significant influence on our results.
The non-relativistic quantum numbers, predominantly corresponding to this nonet, are $(n, L, S)=(1, 0, 0)$, hence  $n$ $^{2S+1}L_{J}=1$ $^{1}S_{0}$. For what concern the relativistic notation, we have $J^{PC}=0^{-+}$.

(ii) The ground-state vector mesons are assembled in the matrix $V^{\mu}$, the associated states to this nonet are $\{ \rho(770), K^{*}(892), \phi(1020), \omega(782) \}$. One should note that also in the vector sector the mixing between two fields $\omega=\sqrt{\frac{1}{2}}(u\bar{u}+d\bar{d})$ and $\phi=s\bar{s}$ takes place. However, the mixing angle is small enough (about $-3^{\circ}$, see Ref. \cite{qqpdg}) to be neglected. Thus, the states $\phi(1020)$ and $\omega(782)$ are interpreted as purely strange and purely non-strange, respectively. 

Notice that the mesons in the ground-state vector nonet are regarded as the lightest ideal quark-antiquark objects. This is due to the fact that efffects of axial anomaly and spontaneous symmetry breaking are not as significant as in the pseudoscalar mesonic sector. The non-relativistic quantum numbers describing this nonet are $(n, L, S)=(1, 0, 1)$, and thus $n$ $^{2S+1}L_{J}=1$ $^{3}S_{1}$. In the relativistic notation we have $J^{PC}=1^{--}$.

(iii) The first group of excited vector mesons is described by matrix $V_E^{\mu}$ and is associated to the states  $\{ \rho(1450), K^{*}(1410), \omega(1420), \phi(1680)\}$. This nonet corresponds  (roughly) to the radial excitation characterized by quantum numbers $(n, L, S)=(2, 0, 1)$, hence $n$ $^{2S+1}L_{J}=2$ $^{3}S_{1}$ when using the non-relativistic notation and $J^{PC}=1^{--}$ in the relativistic notation.  Analogously to the ground-state vector mesons, the effect of isoscalar mixing is negligible for this nonet, hence $\omega(1420)$ is identified with $\sqrt{\frac{1}{2}}(u\bar{u}+d\bar{d})$ state while $\phi(1680)$ with $s\bar{s}$. We also recall that the resonance $K^*(1410)$ was confirmed by lattice studies in Ref. \cite{Prelovsekex}, while the resonances $\omega(1420)$ and $\phi(1680)$ in Refs. \cite{Dudekex, Dudekex2}.

(iv) The second group of excited vector mesons is described by the matrix $V^{\mu}_D$ and contains the resonances  $\{ \rho(1700), K^{*}(1680), \omega(1650), \phi(???) \}$. This nonet corresponds (predominantly) to orbital excitations with $(n, L, S)=(1, 2, 1)$, thus $n$ $^{2S+1}L_{J}=1$ $^{3}D_{1}$ (not relativistically) and $J^{PC}=1^{--}$ (relativistically). In this case the isoscalar mixing can also be omitted. 

It is worth to note that each field described above is invariant under the transformations of flavor symmetry $U(3)_{V}$, parity (P), and charge conjugation (C), see Table \ref{transcpu} in Appendix \ref{ApA} for details. 
\subsection{Estimation of the mass of $\mathbf{\phi(1959)}$}

Let us consider the $s\bar{s}$ state $\phi(???)$ belonging to the $1$ $^{3}D_{1}$ nonet. This state is especially interesting since its existence is up to now not experimentally confirmed. We aim to make some predictions for this state by employing our model. However, in order to do this, one needs to estimate its mass. We observe that the difference of masses of the relevant states from radially and orbitally excited nonets is almost the same in each case, see Table \ref{tablepar} for details. This is due to the same kind of strong dynamics describing these resonances. 

\begin{table}[h!] 
\centering

\renewcommand{\arraystretch}{1.25}
\begin{tabular}[c]{c|c|c|c|c|c}
\hline
\hline
\multicolumn{2}{c|}{\textbf{Radially excited}}&\multicolumn{2}{c|}{\textbf{Orbitally excited}}&\multicolumn{1}{c|}{\textbf{Mass difference}}&\multicolumn{1}{c}{\textbf{Predicted mass}}\\
\multicolumn{2}{c|}{\textbf{vector mesons}}&\multicolumn{2}{c|}{\textbf{vector mesons}}&\multicolumn{1}{c|}{\textbf{$(m_D-m_E)$}}&\multicolumn{1}{c}{of $\phi(???)$}\\
\cline{1-4}
State & Mass $(m_E)$& State & Mass $(m_D)$& [MeV]& $m_{\phi(1680)}+(m_D-m_E)$\\
 & [MeV] &  & [MeV]& & [MeV] \\
\hline
\hline
$\rho(1450)$&$1465 \pm 25$&$\rho(1700)$&$1720 \pm 20$&$255 \pm 32$&$1935 \pm 38$\\
$K^*(1410)$&$1421 \pm 9$&$K^*(1680)$&$1718 \pm 18$&$297 \pm 20$&$1977 \pm 28$\\
$\omega(1420)$&$1425 \pm 25$&$\omega(1650)$&$1670 \pm30$&$245 \pm 39$&$1925 \pm 44$ \\
$\phi(1680)$&$1680 \pm 20$&$\phi(???)$&$?$&-&-\\
\hline
\multicolumn{4}{c|}{\textbf{weighted average}}&$279 \pm 16$&$\mathbf{1959 \pm 20}$\\
\hline
\hline
\end{tabular}
\caption{Estimation of the mass of the putative state $\phi(???)$.} \label{tablepar}
\end{table} 
We have obtained that the average value of the mass difference is $279 \pm 16$ MeV. In our simple estimation one needs to add this value to the mass of $\phi(1680)$. As a consequence, one has:
\begin{equation}
\phi(???)\simeq (m_{\phi(1680)}+279 \pm 16) \text{ MeV}= 1959 \pm 20 \text{ MeV} \text{ ,}
\end{equation}
where the last error of $20$ MeV is obtained by the usual error propagation formula.
From now on, whenever we refer to this putative state, we use the identification:
\begin{equation}
\phi(???)=\phi(1959) \text{ .}
\end{equation}

\subsection{The Lagrangian}

We present the interaction Lagrangian of our model. It is constructed in agreement with $P$, $C$ and $U(3)_{V}$ symmetries, and by coupling the four matrices given by Eqs. (\ref{pv}) and (\ref{ve}). One can write it as
\begin{equation}
\mathcal{L}_{int}=\mathcal{L}_{EPP}+\mathcal{L}_{DPP}+\mathcal{L}_{EVP}%
+\mathcal{L}_{DVP} \text{ ,} \label{lagfull}%
\end{equation}
with:
\begin{equation}
\mathcal{L}_{EPP}=ig_{EPP}Tr\left(\left[\partial^{\mu}P, V_{E,\mu}\right]P\right) \text{ ,} \label{term1a}
\end{equation} 

\begin{equation}
\mathcal{L}_{DPP}=ig_{DPP}Tr\left(\left[\partial^{\mu}P, V_{D,\mu}\right]P\right)\text{ ,} \label{term2b}
\end{equation}

\begin{equation}
\mathcal{L}_{EVP}=ig_{EVP}Tr\left(\tilde{V}_{E}^{\mu \nu}\left\{V_{\mu \nu}, P\right\}\right) \text{ ,} \label{term3c}
\end{equation}

\begin{equation}
\mathcal{L}_{DVP}=ig_{DVP}Tr\left(\tilde{V}_{D}^{\mu \nu}\left\{V_{\mu \nu}, P\right\}\right) \text{ .} \label{term4d}
\end{equation}
Each term of the above Lagrangian describes a different type of decay: $\mathcal{L}_{EPP}$ refers to the process $V_E \rightarrow PP$, $\mathcal{L}_{DPP}$ to $V_D \rightarrow PP$, $\mathcal{L}_{EVP}$ to $ V_E \rightarrow VP$, and finally $\mathcal{L}_{DVP}$ to $V_D \rightarrow VP$. Moreover, the notations $[A,B]=AB-BA$ (in Eqs. (\ref{term1a}) and (\ref{term2b})) and $\{A,B\}=AB+BA$ (in Eqs. (\ref{term3c}) and (\ref{term4d})) stand for the standard commutator and anticommutator relations, respectively. The dual fields are defined by the following terms:
\begin{equation}
\tilde{V}^{\mu \nu}_{E}=\frac{1}{2}\varepsilon^{\mu \nu \alpha \beta}(\partial_{\alpha}V_{E,\beta}-\partial_{\beta}V_{E, \alpha}) \text{ ,}
\end{equation}
\begin{equation}
\tilde{V}^{\mu \nu}_{D}=\frac{1}{2}\varepsilon^{\mu \nu \alpha \beta}(\partial_{\alpha}V_{D,\beta}-\partial_{\beta}V_{D, \alpha}) \text{ .}
\end{equation}
The constants $g_{EPP}$, $g_{DPP}$, $g_{EVP}$ and $g_{DVP}$ are the four free parameters of the model. Two of them ($g_{EPP}$ and $g_{DPP}$) are dimensionless and the other two ($g_{EVP}$ and $g_{DVP}$) have dimension Energy$^{-1}$. In order to determine them we use experimental data listed in the PDG, see details in Sec. \ref{detpar}. The explicit form of the Lagrangian of Eq. (\ref{lagfull}) is shown in Appendix \ref{ApA}.

For completeness, we also explore the radiative decays of both types of excited vector mesons into the $\gamma P$ channel. For this purpose, we shall apply to the vector field strength tensor $V_{\mu \nu}$ the following shift \cite{OConnellpearce}:
\begin{equation} \label{shiftmpp}
V_{\mu \nu} \rightarrow V_{\mu \nu}+\frac{e_0}{g_{\rho}}QF_{\mu \nu} \text{ .}
\end{equation}
In the equation above, the quantity $F_{\mu \nu}$ stands for the field strength tensor of the photons, $g_{\rho}=5.5 \pm 0.5$ is the coupling constant in the $\rho \pi \pi$ channel, $e_{0}=\sqrt{4 \pi \alpha}$ is the proton electric charge and finally, $Q$ is a $3\times3$ diagonal matrix with the quark charges: $Q=diag\{2/3, -1/3, -1/3\}$. We stress that in order to study the radiative decays there is no need to involve any new parameter.
\subsection{Strong and radiative tree-level decay widths}
The usual theoretical expressions for the decay widths at tree-level can be derived from the Feynman rules. In particular, a resonance $R=V_E$ or $R=V_D$ decays into three channels: pseudoscalar-pseudoscalar ($R \rightarrow PP$), ground-state vector-pseudoscalar ($R \rightarrow VP$), and finally photon-pseudoscalar ($R \rightarrow \gamma$P). The corresponding tree-level decay widths take the explicit form:
\begin{equation}
\Gamma_{R\rightarrow PP}=s_{RPP}\frac{|\vec{k}|^{3}}{6\pi m_{R}^{2}}\left(
\frac{g_{RPP}}{2}\lambda_{RPP}\right)  ^{2}\text{ ,} \label{rpp}%
\end{equation}%
\begin{equation}
\Gamma_{R\rightarrow VP}=s_{RVP}\frac{|\vec{k}|^{3}}{12\pi}\left(
\frac{g_{RVP}}{2}\lambda_{RVP}\right)  ^{2}\text{ ,} \label{rvp}%
\end{equation}%
\begin{equation}
\Gamma_{R\rightarrow\gamma P}=\frac{|\vec{k}|^{3}}{12\pi}\left(  \frac
{g_{RVP}}{2}\frac{e_{0}}{g_{\rho}}\lambda_{R\gamma P}\right)  ^{2}\text{ ,}
\label{rgammap}%
\end{equation}
where the quantity
\begin{equation}
|\vec{k}|=\frac{\sqrt{m_{R}^{4}+(m_{A}^{2}-m_{B}^{2})^{2}-2(m_{A}^{2}%
+m_{B}^{2})m_{R}^{2}}}{2m_{R}}%
\end{equation}
stands for the modulus of the three-momentum of one (of the two) outgoing particles (A or B) in the rest frame of decaying resonance with mass $m_R$. Moreover $m_A$ and $m_B$ are the masses of emitted particles A and B, respectively (see Sec. \ref{twobody} for details). In Tables \ref{coef1}, \ref{coef2} and \ref{coef3} we report the isospin/symmetry factors $(s_{RPP}, s_{RVP})$ and Clebsch-Gordan coefficients $(\lambda_{RPP}, \lambda_{RVP}, \lambda_{R \gamma P})$, which are extracted from the explicit expression of the Lagrangian (see Appendix \ref{ApA}).

\begin{table}[h]
\renewcommand{\arraystretch}{1.23}
\par
\makebox[\textwidth][c] { 
\par%
\begin{tabular}
[c]{c|c|c|c}\hline \hline
\multicolumn{2}{c|}{Decay channel} & Symmetry factor &
Clebsch-Gordan coefficient\\
$V_{E}\rightarrow PP$ & $V_{D}\rightarrow PP$ & $s_{EPP}=s_{DPP}$ &
$\lambda_{EPP}=\lambda_{DPP}$\\\hline\hline
$\rho(1450)\rightarrow\bar{K}K$ & $\rho(1700)\rightarrow\bar{K}K$ & $2$ &
$\frac{1}{2}$\\
$\rho(1450)\rightarrow\pi\pi$ & $\rho(1700)\rightarrow\pi\pi$ & $1$ &
$1$\\
\hline
$K^{\ast}(1410)\rightarrow K\pi$ & $K^{\ast}(1680)\rightarrow K\pi$ & $3$ &
$\frac{1}{2}$\\
$K^{\ast}(1410)\rightarrow K\eta$ & $K^{\ast}(1680)\rightarrow K\eta$ & $1$ &
$\frac{1}{2}(\cos\theta_{p}-\sqrt{2}\sin\theta_{p})$\\
$K^{\ast}(1410)\rightarrow K\eta^{\prime}$ & $K^{\ast}(1680)\rightarrow
K\eta^{\prime}$ & $1$ & $\frac{1}{2}(\sqrt{2}\cos\theta_{p}+\sin\theta_{p}%
)$\\\hline
$\omega(1420)\rightarrow\bar{K}K$ & $\omega(1650)\rightarrow\bar{K}K$ & $2$ &
$\frac{1}{2}$\\\hline
$\phi(1680)\rightarrow\bar{K}K$ & $\phi(1959)\rightarrow\bar{K}K$ & $2$ &
$\frac{1}{\sqrt{2}}$\\\hline \hline
\end{tabular}
}\caption{Isospin/symmetry factors and the Clebsch-Gordan coefficients of Eq.
(\ref{rpp}) related to the Lagrangian of Eqs. (\ref{term1a}%
) and (\ref{term2b}). For the explicit expressions see Appendix \ref{ApA}.} \label{coef1}
\end{table}

\begin{table}[h]
\renewcommand{\arraystretch}{1.23}
\par
\makebox[\textwidth][c] {
\begin{tabular}
[c]{c|c|c|c}\hline \hline
\multicolumn{2}{c|}{Decay channel} & Symmetry factor &
Clebsch-Gordan coefficient\\
$V_{E}\rightarrow VP$ & $V_{D}\rightarrow VP$ & $s_{EVP}=s_{DVP}$ &
$\lambda_{EVP}=\lambda s_{DVP}$\\\hline\hline
$\rho(1450)\rightarrow\omega\pi$ & $\rho(1700)\rightarrow\omega\pi$ & $1$ &
$\frac{1}{2}$\\
$\rho(1450)\rightarrow K^{\ast}(892)K$ & $\rho(1700)\rightarrow K^{\ast
}(892)K$ & $4$ & $\frac{1}{4}$\\
$\rho(1450)\rightarrow\rho(770)\eta$ & $\rho(1700)\rightarrow\rho(770)\eta$ &
$1$ & $\frac{1}{2}\cos\theta_{p}$\\
$\rho(1450)\rightarrow\rho(770)\eta^{\prime}$ & $\rho(1700)\rightarrow
\rho(770)\eta^{\prime}$ & $1$ & $\frac{1}{2}\sin\theta_{p}$\\
\hline
$K^{\ast}(1410)\rightarrow K\rho$ & $K^{\ast}(1680)\rightarrow K\rho$ & $3$ &
$\frac{1}{4}$\\
$K^{\ast}(1410)\rightarrow K\phi$ & $K^{\ast}(1680)\rightarrow K\phi$ & $1$ &
$\frac{1}{2\sqrt{2}}$\\
$K^{\ast}(1410)\rightarrow K\omega$ & $K^{\ast}(1680)\rightarrow K\omega$ &
$1$ & $\frac{1}{4}$\\
$K^{\ast}(1410)\rightarrow K^{\ast}(892)\pi$ & $K^{\ast}(1680)\rightarrow
K^{\ast}(892)\pi$ & $3$ & $\frac{1}{4}$\\
$K^{\ast}(1410)\rightarrow K^{\ast}(892)\eta$ & $K^{\ast}(1680)\rightarrow
K^{\ast}(892)\eta$ & $1$ & $\frac{1}{4}(\cos\theta_{p}+\sqrt{2}\sin\theta
_{p})$\\
$K^{\ast}(1410)\rightarrow K^{\ast}(892)\eta^{\prime}$ & $K^{\ast
}(1680)\rightarrow K^{\ast}(892)\eta^{\prime}$ & $2$ & $\frac{1}{4}(\sqrt
{2}\cos\theta_{p}-\sin\theta_{p})$\\\hline
$\omega(1420)\rightarrow\rho\pi$ & $\omega(1650)\rightarrow\rho\pi$ & $3$ &
$\frac{1}{2}$\\
$\omega(1420)\rightarrow K^{\ast}(892)K$ & $\omega(1650)\rightarrow K^{\ast
}(892)K$ & $4$ & $\frac{1}{4}$\\
$\omega(1420)\rightarrow\omega(782)\eta$ & $\omega(1650)\rightarrow
\omega(782)\eta$ & $1$ & $\frac{1}{2}\cos\theta_{p}$\\
$\omega(1420)\rightarrow\omega(782)\eta^{\prime}$ & $\omega(1650)\rightarrow
\omega(782)\eta^{\prime}$ & $1$ & $\frac{1}{2}\sin\theta_{p}$\\\hline
$\phi(1680)\rightarrow K\bar{K}^{\ast}$ & $\phi(1959)\rightarrow K\bar
{K}^{\ast}$ & $4$ & $\frac{1}{2\sqrt{2}}$\\
$\phi(1680)\rightarrow\phi(1020)\eta$ & $\phi(1959)\rightarrow\phi(1020)\eta$
& $1$ & $\frac{1}{\sqrt{2}}\sin\theta_{p}$\\
$\phi(1680)\rightarrow\phi(1020)\eta^{\prime}$ & $\phi(1959)\rightarrow
\phi(1020)\eta^{\prime}$ & $1$ & $\frac{1}{\sqrt{2}}\cos\theta_{p}$\\\hline \hline
\end{tabular}
}\caption{Isospin/symmetry factors and the Clebsch-Gordan coefficients of Eq.
(\ref{rvp}) related to the Lagrangian of Eqs. (\ref{term3c}%
) and (\ref{term4d}). For the explicit expressions see  Appendix \ref{ApA}.}
\label{coef2}
\end{table}

\begin{table}[h]
\renewcommand{\arraystretch}{1.23}
\par
\makebox[\textwidth][c] {%
\begin{tabular}
[c]{c|c|c}\hline \hline
\multicolumn{2}{c|}{Decay channel} & Clebsch-Gordan coefficient\\
$V_{E}\rightarrow\gamma P$ & $V_{D}\rightarrow\gamma P$ & $\lambda_{E\gamma
P}=\lambda_{D\gamma P}$\\\hline\hline
$\rho(1450)\rightarrow\gamma\pi$ & $\rho(1700)\rightarrow\gamma\pi$ &
$\frac{1}{6}$\\
$\rho(1450)\rightarrow\gamma\eta$ & $\rho(1700)\rightarrow\gamma\eta$ &
$\frac{1}{2}\cos\theta_{p}$\\
$\rho(1450)\rightarrow\gamma\eta^{\prime}$ & $\rho(1700)\rightarrow\gamma
\eta^{\prime}$ & $\frac{1}{2}\sin\theta_{p}$\\
\hline
$K^{\ast}(1410)\rightarrow\gamma K$ & $K^{\ast}(1680)\rightarrow\gamma K$ &
$\frac{1}{3}$\\\hline
$\omega(1420)\rightarrow\gamma\pi$ & $\omega(1650)\rightarrow\gamma\pi$ &
$\frac{1}{2}$\\
$\omega(1420)\rightarrow\gamma\eta$ & $\omega(1650)\rightarrow\gamma\eta$ &
$\frac{1}{6}\cos\theta_{p}$\\
$\omega(1420)\rightarrow\gamma\eta^{\prime}$ & $\omega(1650)\rightarrow
\gamma\eta^{\prime}$ & $\frac{1}{6}\cos\theta_{p}$\\\hline
$\phi(1680)\rightarrow\gamma\eta$ & $\phi(1959)\rightarrow\gamma\eta$ &
$\frac{1}{3}\sin\theta_{p}$\\
$\phi(1680)\rightarrow\gamma\eta^{\prime}$ & $\phi(1959)\rightarrow\gamma
\eta^{\prime}$ & $\frac{1}{3}\cos\theta_{p}$\\\hline \hline
\end{tabular}
}\caption{Clebsch-Gordan coefficients of Eq. (\ref{rgammap}) related to Eqs.
(\ref{term3c}) and (\ref{term4d}) combined with the shift given by Eq. (\ref{shiftmpp}).}
\label{coef3}
\end{table}

In general, the simplified (but still not trivial) treatment of the decay width at tree-level can be improved by the inclusion of mesonic loops and studying the pole positions. The loop corrections are important due to the influence of quantum fluctuations on the width ($\Gamma$) and mass (M) of the decaying resonance. For what concerns the decays examined in our model, the ratio $\Gamma/$M is small enough (safely below 1), which means that loop corrections affect only slightly the tree-level results and can be neglected in first approximation, but should be included in future studies \cite{fggp}. \\

\subsection{Determination of the parameters}\label{detpar}
Our model contains four free parameters (two for each nonet): for radial excitations we have $g_{EPP}$ and $g_{EVP}$, while for orbital excitations we have $g_{DPP}$ and $g_{DVP}$. At first sight, it may seem that the best strategy to determine them is to make a full fit to all available experimental data related to both nonets. However, following this procedure turns out not to be the optimal one because some states were observed only in a single experiment and further confirmations would be needed. Moreover, in some cases, there is a large discrepancy between different experimental results describing the same observable. Therefore, instead of doing a fit, in order to determine the four parameters of our model we chose four well known experimental values. 

In order to fix the coupling constants $g_{EPP}$ and $g_{EVP}$, we employ the following two values quoted in the PDG \cite{pdg}:
\begin{align}
\Gamma_{K^{\ast}(1410)\rightarrow K\pi}^{exp}  &  =15.3\pm3.3\text{ MeV} \text{ ,}\\
\Gamma_{\phi(1680)}^{tot,exp}  &  =150\pm50\text{ MeV} \text{ .}%
\end{align}
Such a choice is justified by the good quality of these results: the decay channel $K^*(1410) \rightarrow K \pi$ is precisely known and the resonance $\phi(1680)$ is rather narrow. The decay channels (also included in our model) contributing to the total decay width of $\phi(1680)$ state are: $\phi(1680) \rightarrow K\bar{K}$, $\phi(1680) \rightarrow \phi(1020) \eta$ and $\phi(1680) \rightarrow K^*(892) K$. Moreover, the latter is listed in  the PDG as the dominant one (in agreement with our theoretical results). By minimizing the $\chi^2$ function:
\begin{align}
F_{E}(g_{EPP},g_{EVP})  &  =\left(  \frac{\Gamma_{K^{\ast}(1410\rightarrow
K\pi)}-\Gamma_{K^{\ast}(1410)\rightarrow K\pi}^{\exp}}{\delta\Gamma_{K^{\ast
}(1410)\rightarrow K\pi}^{exp}}\right)  ^{2}\nonumber\\
&  +\left(  \frac{\Gamma_{\phi(1680)\rightarrow K^{\ast}(892)K}+\Gamma
_{\phi(1680)\rightarrow\phi(1020)\eta}+\Gamma_{\phi(1680)\rightarrow\bar{K}%
K}-\Gamma_{\phi(1680)}^{tot,\exp}}{\delta\Gamma_{\phi(1680)}^{tot,exp}%
}\right)  ^{2} \label{fe}%
\end{align}
we obtain:
\begin{equation}
g_{EPP}=3.66\pm0.4 \hspace{0.5cm} \text{ and } \hspace{0.5cm} g_{EVP}=18.4\pm3.8 \text{  [GeV}^{-1}] \text{ .} \label{ge}%
\end{equation}

For consistency, concerning $g_{DPP}$ and $g_{DVP}$, we use two values reported by the experiments ASTON 84 \cite{Aston84} and ASTON 88 \cite{Aston88} about the well known resonance $K^*(1680)$. We stress that both experimental results which we use are compatible with the fit provided by the PDG. 

The first quantity used to fix these parameters is the ratio $K \rho$/$K \pi$
\begin{equation}
\left.  \frac{\Gamma_{K^{\ast}(1680)\rightarrow K\rho}}{\Gamma_{K^{\ast
}(1680)\rightarrow K\pi}}\right\vert _{\exp}=1.2\pm0.4\text{ by ASTON 84
\cite{Aston84}, } \label{ratiogdvp/gdpp}%
\end{equation}
which depends on the ratio $g_{DVP}$/$g_{DPP}$. Notice that the corresponding value reported by the PDG \cite{pdg} reads $0.81^{+0.14}_{-0.09}$ and is comparible with ASTON 84. Additionally, we employ the decay width for the channel $K^*(1680) \rightarrow K \pi$. To this end, we obtain its value from the following two quantities:
\begin{equation}
\left.  \frac{\Gamma_{K^{\ast}(1680)\rightarrow K\pi}}{\Gamma_{K^{\ast}%
(1680)}^{tot}}\right\vert _{\exp}=0.388\pm0.036\text{ and}\left.  \text{
}\Gamma_{K^{\ast}(1680)}^{tot}\right\vert _{\exp}=205\pm50\text{ MeV by
ASTON\ 88 \cite{Aston88}, } \label{aston88}%
\end{equation}
which gives:
\begin{equation} 
\left.  \Gamma_{K^{\ast}(1680)\rightarrow K\pi}\right\vert _{\exp}%
=79\pm21\text{ MeV from ASTON 88 \cite{Aston88}.} \label{k(1680)inkpion}%
\end{equation}
This value is also in good agreement with the results of the PDG, that quotes the ratio\\ $\Gamma_{K^*(1680)\rightarrow K \pi}$/$\Gamma^{tot}_{K^*(1680)}=0.387 \pm 0.026$ and the total decay width $\Gamma^{tot}_{K^*(1680)}=322 \pm 110$ MeV. Next, using Eqs. (\ref{ratiogdvp/gdpp}) and (\ref{k(1680)inkpion}) and minimizing the $\chi^2$ function
\begin{align}
F_{D}(g_{DPP},g_{DVP})  &  =\left(  \frac{\frac{\Gamma_{K^{\ast}%
(1680)\rightarrow K\rho}}{\Gamma_{K^{\ast}(1680)\rightarrow K\pi}}-\left(
\frac{\Gamma_{K^{\ast}(1680)\rightarrow K\rho}}{\Gamma_{K^{\ast}%
(1680)\rightarrow K\pi}}\right)  ^{\exp}}{\delta\left(  \frac{\Gamma_{K^{\ast
}(1680)\rightarrow K\rho}}{\Gamma_{K^{\ast}(1680)\rightarrow K\pi}}\right)
}\right)  ^{2}\nonumber\\
&  +\left(  \frac{\Gamma_{K^{\ast}(1680)\rightarrow K\pi}-\Gamma_{K^{\ast
}(1680)\rightarrow K\pi}^{\exp}}{\delta\Gamma_{K^{\ast}(1680)\rightarrow K\pi
}}\right)  ^{2} \text{ ,} \label{fd}%
\end{align}
we obtain the values for the coupling constants $g_{DPP}$ and $g_{DVP}$ as:
\begin{equation}
g_{DPP}=7.15\pm0.94 \hspace{0.5cm} \text{ and } \hspace{0.5cm} g_{DVP}=16.5\pm3.5 \text{ [GeV}^{-1}] \text{ .} \label{gd}%
\end{equation}

\section{Results of the model}
In this section we present the results for both groups of excited vector mesons. In particular, we focus on strong and radiative decay channels of these nonets. In Sec. \ref{radex} we discuss the vector mesons with radial excitations while in Sec. \ref{orbex} we investigate the mesons with orbital excitations. 
 \subsection{Radially excited vector mesons}
 \label{radex}
 In the following we discuss the strong decays of radially excited vector mesons into two pseudoscalar mesons, and also into one pseudoscalar and one vector meson. Next, the radiative decays of this nonet are described.  
 
 \begin{center}
 \textbf{Strong decays}
 \end{center}
 We first calculate the decay widths of the resonances $\rho(1450)$, $K^*(1410)$, $\omega(1420)$ and $\phi(1680)$ decaying into mesons via strong interactions. The results for the $V_E \rightarrow PP$ channels are shown in Table \ref{Vepp} and for $V_E \rightarrow VP$ channels in Table \ref{Vevp}. Our theoretical results are compared with the experimental data, whenever existent. 
\begin{table}[h]
\renewcommand{\arraystretch}{1.53}
\par
\makebox[\textwidth][c] { 
\par%
\begin{tabular}
[c]{ccc}\hline \hline
Decay process $V_{E}\rightarrow PP$ & Theory [MeV] & Experiment
[MeV]\\\hline\hline
$\rho(1450)\rightarrow\bar{K}K$ & $6.6\pm1.4$ & $<$ $6.7\pm1.0$ by\ DONANCHIE
91 \cite{Donnachie91}\\
$\rho(1450)\rightarrow\pi\pi$ & $30.8\pm6.7$ & $\sim$ $27\pm4,$ seen by CLEGG
94 \cite{Clegg94}\\\hline
$K^{\ast}(1410)\rightarrow K\pi$ & $15.3\pm3.3$ & $15.3\pm3.3$ by PDG\\
$K^{\ast}(1410)\rightarrow K\eta$ & $6.9\pm1.5$ & not listed in PDG\\
$K^{\ast}(1410)\rightarrow K\eta^{\prime}$ & $\approx0$ & not listed in
PDG\\\hline
$\omega(1420)\rightarrow\bar{K}K$ & $5.9\pm1.3$ & not listed in PDG\\\hline
$\phi(1680)\rightarrow\bar{K}K$ & $19.8\pm4.3$ & seen by BUON 82
\cite{Buon82}\\\hline\hline
\end{tabular}
}\caption{\label{Vepp}Results for the partial decay widths of the resonances belonging to the nonet of predominantly radially excited vector mesons decaying into two pseudoscalar mesons.}%
\end{table}
\begin{table}[h!]
\renewcommand{\arraystretch}{1.53}
\par
\makebox[\textwidth][c] { 
\par%
\begin{tabular}
[c]{ccc}\hline\hline
Decay process $V_{E}\rightarrow VP$ & Theory [MeV] & Experiment
[MeV]\\\hline\hline
$\rho(1450)\rightarrow\omega\pi$ & $74.7\pm31.0$ & $\sim84\pm13$ seen by CLEGG
94 \cite{Clegg94}\\
$\rho(1450)\rightarrow K^{\ast}(892)K$ & $6.7\pm2.8$ & possibly seen by COAN
04 \cite{Coan04}\\
$\rho(1450)\rightarrow\rho(770)\eta$ & $9.3\pm3.9$ & $<16.0\pm2.4$ by
Donnachie 91 \cite{Donnachie91}\\
$\rho(1450)\rightarrow\rho(770)\eta^{\prime}$ & $\approx0$ & not listed in
PDG\\\hline
$K^{\ast}(1410)\rightarrow K\rho$ & $12.0\pm5.0$ & $<16.2\pm1.5$ by
PDG\\
$K^{\ast}(1410)\rightarrow K\phi$ & $\approx0$ & not listed in PDG\\
$K^{\ast}(1410)\rightarrow K\omega$ & $3.7\pm1.5$ & not listed in PDG\\
$K^{\ast}(1410)\rightarrow K^{\ast}(892)\pi$ & $28.8\pm12.0$ & $>93\pm8$ by
PDG\\
$K^{\ast}(1410)\rightarrow K^{\ast}(892)\eta$ & $\approx0$ & not listed in
PDG\\
$K^{\ast}(1410)\rightarrow K^{\ast}(892)\eta^{\prime}$ & $\approx0$ & not
listed in PDG\\\hline
$\omega(1420)\rightarrow\rho\pi$ & $196\pm81$ & dominant, $\Gamma
_{tot}=(180-250)$ by PDG\\
$\omega(1420)\rightarrow K^{\ast}(892)K$ & $2.3\pm1.0$ & not listed in
PDG\\
$\omega(1420)\rightarrow\omega(782)\eta$ & $4.9\pm2.0$ & not listed in
PDG\\
$\omega(1420)\rightarrow\omega(782)\eta^{\prime}$ & $\approx0$ & not listed in
PDG\\\hline
$\phi(1680)\rightarrow K\bar{K}^{\ast}$ & $110\pm46$ & dominant, $\Gamma
_{tot}=150\pm50$ by PDG\\
$\phi(1680)\rightarrow\phi(1020)\eta$ & $12.2\pm5.1$ & seen by ACHASOV 14
\cite{Achasov14}\\
$\phi(1680)\rightarrow\phi(1020)\eta^{\prime}$ & $\approx0$ & not listed in
PDG\\\hline\hline
\end{tabular}
}\caption{\label{Vevp}Results for the partial decay widths of the resonances belonging to the nonet of predominantly radially excited vector mesons decaying into one pseudoscalar and one vector meson.}%
\end{table}

 In general, one observes a very good agreement between the theory and experiment. The decay channels for which we obtained large values for the decay widths are observed in experiments, while those with theoretically small decay widths have not been observed. We then conclude that the conventional quark-antiquark assignment of this nonet is well upheld.
 
In PDG one may find other experimental data such as various ratios which can be also studied within our theoretical model. We proceed to calculate some of them. 

Let us first consider the resonance $\rho(1450)$ for which four different branching ratios are experimentally known, see Table \ref{brarho1450} for their values. 
\begin{table}[h!] 
\centering
\renewcommand{\arraystretch}{1.9}
\begin{tabular}[c]{c|c|c|c}
\hline
\hline
\multicolumn{4}{c}{$\mathbf{\rho(1450)}$}\\
\hline
\hline
\multirow{2}{*}{\textbf{Branching ratio}}& \multirow{2}{*}{\textbf{Our model}} & \multicolumn{2}{|c}{\textbf{Experimental results}}\\
\cline{3-4}
&&\textbf{Value}& \textbf{Reference}\\
\hline
\hline
$\frac{\Gamma_{\rho(1450)\rightarrow\pi\pi}}{\Gamma_{\rho
(1450)\rightarrow\omega\pi}}$&$0.41 \pm 0.20$&$\sim0.32$&CLEGG 94 \cite{Clegg94}\\
\hline
$\frac{\Gamma_{\rho(1450)\rightarrow\pi\pi}}{\Gamma_{\rho
(1450)\rightarrow\eta\rho}}$& $3.3 \pm 1.6$&$1.3 \pm 0.4$& AULCHENKO 15 \cite{Aulchenko15}\\
\hline
$\frac{\Gamma_{\rho(1450)\rightarrow KK}}{\Gamma_{\rho(1450)\rightarrow
\omega\pi}}$&$0.088 \pm 0.043$&$<0.08$& DONNACHIE 91 \cite{Donnachie91}\\
\hline
\multirow{3}{*}{$\frac{\Gamma_{\rho(1450)\rightarrow\eta\rho}}{\Gamma_{\rho
(1450)\rightarrow\omega\pi}}$}&\multirow{3}{*}{$\approx 0.12$}&$0.081 \pm 0.020$& AULCHENKO 15 \cite{Aulchenko15}\\
&&$\sim 0.21$&DONNACHIE 91 \cite{Donnachie91}\\
&&$>2$&FUKUI 91 \cite{Fukui91}\\
\hline
\hline
\end{tabular}
\caption{\label{brarho1450}Branching ratios ralated to the $\rho(1450)$ resonance. }
\end{table} 

In the first entry of that table we report the $\pi \pi/ \omega \pi$ ratio. Its value obtained in Ref. CLEGG 94 \cite{Clegg94} is very well explained by our theoretical result.  In the second row, the ratio $\pi \pi/ \eta \rho$ is shown. In this case, a satisfactory agreement is also visible. The next ratio $KK/\omega \pi$, for which only the upper limit is given, seems to be consistent with our theoretical result. Note, the three quantities mentioned above depend on the ratio between two coupling constants $g_{EPP}$ and $g_{EVP}$. The last quantity is the $\eta \rho/ \omega \pi $ rate. The PDG gives for this ratio three different values. Our theoretical value is in very good agreement with the value determined in Ref. AULCHENKO 15 \cite{Aulchenko15} and is also qualitatively compatible with the result obtained in Ref. DONNACHIE 91 \cite{Donnachie91}. For what concerns the value given by Ref. FUKUI 91 \cite{Fukui91} , where only the lower limit is determined, the disagreement with our theory is visible. However, this value is also not compatible with the other experimental results quoted by the PDG. Notice that our theoretical result for the ratio  $\eta \rho/ \omega \pi $ does not depend on any parameter of the model. Hence, we are not able to calculate the theoretical error of this value since within our approach the errors follow solely from the uncertainties related to the decay widths. However, taking into account the errors of the masses of the resonances one may estimate the error of this ratio to be around (10-20)$\%$. The same remark applies to other quantities which do not depend on the coupling constants. 

The next resonance belonging to the nonet of radially excited vector mesons is $K^*(1410)$. This state is very well known from experimental observations as well as from lattice studies \cite{Prelovsekex}. The results are shown in Table \ref{Vevp}, where we remind that the $K \pi$ channel was used to determine the model parameters. A significant discrepancy between theory and experiment is observed for the decay into the $K^*(892) \pi$ channel. Our theoretical value is approximately three times smaller than the PDG quote. 

Let us now look at two ratios related to $K^*(1410)$ resonance which are listed in the PDG, see details in Table \ref{brak1410}. 
\begin{table}[h!] 
\centering
\renewcommand{\arraystretch}{1.9}
\begin{tabular}[c]{c|c|c|c}
\hline
\hline
\multicolumn{4}{c}{$\mathbf{K^*(1410)}$}\\
\hline
\hline
\multirow{2}{*}{\textbf{Branching ratio}}& \multirow{2}{*}{\textbf{Our model}} & \multicolumn{2}{|c}{\textbf{Experimental results}}\\
\cline{3-4}
&&\textbf{Value}& \textbf{Reference}\\
\hline
\hline
 $\frac{\Gamma_{K^{\ast}(1410)\rightarrow\rho K}}{\Gamma_{K^{\ast
}(1410)\rightarrow K^{\ast}(892)\pi}}$&$ \approx 0.42$&$<0.17$ &\multirow{2}{*}{ ASTON 84 \cite{Aston84}} \\
\cline{1-3}
$\frac{\Gamma_{K^{\ast}(1410)\rightarrow\pi K}}{\Gamma_{K^{\ast
}(1410)\rightarrow K^{\ast}(892)\pi}}$&$0.53 \pm 0.26$&$<0.16$&\\
\hline
\hline
\end{tabular}
\caption{\label{brak1410}Branching ratios ralated to the $K^*(1410)$ resonance.}
\end{table} 
In both cases our theoretical results are too large when compared to the experimental data. This disagreement is a natural conseqence of the presence of $K^*(1410) \rightarrow K^*(892) \pi$ channel in both ratios. This particular channel should be verified in the future.

Next, we consider the resonance $\omega(1420)$. The dominant decay, $\omega(1420) \rightarrow K^*(892) K$ is listed in the PDG as the only experimentally measured decay channel: Our theoretical value is well compatible with the experimental result. All the remaining channels of $\omega(1420)$ have not yet been seen in experiments. Hence, our theoretical results for them are predictions. One should take into account that these decay rates are quite small since the predicted decay widths for them are in the order of few MeV. 
 Interestingly, in PDG one can find information about the following quantities:
 \begin{align}
\left.  \frac{\Gamma_{\omega(1420)\rightarrow\omega\eta}}{\Gamma
_{\omega(1420)}^{tot}}\frac{\Gamma_{\omega(1420)\rightarrow e^{+}e^{-}}%
}{\Gamma_{\omega(1420)}^{total}}\right\vert _{\exp}  &  =(1.6_{-0.07}%
^{+0.09})\cdot10^{-8}\text{ by ACHASOV 16B \cite{Achasov16B}}%
\label{omegarhopion}\\
\left.  \frac{\Gamma_{\omega(1420)\rightarrow\rho\pi}}{\Gamma_{\omega
(1420)}^{tot}}\frac{\Gamma_{\omega(1420)\rightarrow e^{+}e^{-}}}%
{\Gamma_{\omega(1420)}^{total}}\right\vert _{\exp}  &  =\left(  0.73\pm
0.08\right)  \cdot10^{-6}\text{ by AULCHENKO 15A \cite{Aulchenko15A}}\text{ ,}
\label{omegarhopionepem}%
\end{align}
 which allow us to determine the ratio
 \begin{equation}
\left.  \frac{\Gamma_{\omega(1420)\rightarrow\omega\eta}}{\Gamma
_{\omega(1420)\rightarrow\rho\pi}}\right\vert _{\exp}=\frac{(1.6\pm
0.08)\cdot10^{-8}}{\left(  0.73\pm0.08\right)  \cdot10^{-6}}=0.021\pm0.001 \text{ ,}
\end{equation}
in good agreement with the corresponding theoretical value of about $0.025$. 

The last state of this nonet is $\phi(1680)$. Also for this resonance only one channel can be found in the PDG, for which a very good agreement with the theoretical result is obtained. According to the PDG, two ratios related to $\phi(1680)$ are experimentally known and can be compared with the results of our model (see Table \ref{phi1680bra}). In both cases a satisfactory agreement is visible. 
\begin{table}[h!] 
\centering
\renewcommand{\arraystretch}{1.9}
\begin{tabular}[c]{c|c|c|c}
\hline
\hline
\multicolumn{4}{c}{$\mathbf{\phi(1680)}$}\\
\hline
\hline
\multirow{2}{*}{\textbf{Branching ratio}}& \multirow{2}{*}{\textbf{Our model}} & \multicolumn{2}{|c}{\textbf{Experimental results}}\\
\cline{3-4}
&&\textbf{Value}& \textbf{Reference}\\
\hline
\hline
$\frac{\Gamma_{\phi(1680)\rightarrow K\bar{K}}}{\Gamma_{\phi
(1680)\rightarrow K^{\ast}(892)K}}$&$0.18 \pm 0.09$&$0.07 \pm 0.01$&BUON 82 \cite{Buon82}\\
\hline
$\frac{\Gamma_{\phi(1680)\rightarrow\eta\phi}}{\Gamma_{\phi
(1680)\rightarrow K^{\ast}(892)K}}$& $\approx0.11$&$0.07 \pm 0.01$& AUBERT 08S \cite{Aubert08S}\\
\hline
\hline
\end{tabular}
\caption{\label{phi1680bra}Branching ratios for the $\phi(1680)$ resonance.}
\end{table}

\begin{center}
\textbf{Radiative decays}
\end{center}
We now turn to the decays of the resonances $\rho(1450)$, $K^*(1410)$, $\omega(1420)$ and $\phi(1680)$ into  one photon and one pseudoscalar meson. In order to evaluate these radiative decays `Vector Meson Dominance' is applied \cite{OConnellpearce}. In this approach one does not need to introduce any new parameter. The results for the partial decay widths are shown in Table \ref{radiaradia}.
\begin{table}[h]
\renewcommand{\arraystretch}{1.53}
\par
\makebox[\textwidth][c] {%
\begin{tabular}
[c]{ccc}\hline\hline
Decay process $V_{E}\rightarrow\gamma P$ & Theory [MeV] & Experiment
[MeV]\\\hline\hline
$\rho(1450)\rightarrow\gamma\pi$ & $0.072\pm0.042$ & not listed\\
$\rho(1450)\rightarrow\gamma\eta$ & $0.23\pm0.14$ & $\sim0.2-1.5,$ see
text\\
$\rho(1450)\rightarrow\gamma\eta^{\prime}$ & $0.056\pm0.033$ & not
listed\\\hline
$K^{\ast}(1410)\rightarrow\gamma K$ & $0.18\pm0.11$ & seen, $<0.0529$ MeV PDG+
Alavi-Harati 02B \cite{Alavi-harati 02B}\\\hline
$\omega(1420)\rightarrow\gamma\pi^0$ & $0.60\pm0.36$ & $1.90\pm0.75,$ see
text\\
$\omega(1420)\rightarrow\gamma\eta$ & $0.023\pm0.014$ & not listed\\
$\omega(1420)\rightarrow\gamma\eta^{\prime}$ & $0.0050\pm0.0030$ & not
listed\\\hline
$\phi(1680)\rightarrow\gamma\eta$ & $0.14\pm0.09$ & seen\\
$\phi(1680)\rightarrow\gamma\eta^{\prime}$ & $0.076\pm0.045$ & not
listed\\\hline\hline
\end{tabular}
}\caption{\label{radiaradia} Results for the partial decay widths of the resonances belonging to the nonet of predominantly radially excited vector mesons decaying into one photon and one pseudoscalar meson.}%
\end{table}
Because of the poor experimental knowledge about the radiative decays of this nonet, most of the results are predictions. However, a general remark can be done: those radiative decays with a (relatively) large theoretical width were actually seen in experiments. 

Even if no direct experimental data on the widths exist in PDG,  for two radiative channels some estimate can be done. First, let us consider the transition $\rho(1450) \rightarrow \gamma \eta$. In PDG one can find that

\begin{equation}
\left.  \Gamma_{\rho(1450)\rightarrow\gamma\eta}\frac{\Gamma_{\rho
(1450)\rightarrow e^{+}e^{-}}}{\Gamma_{\rho(1450)}^{total}}\right\vert _{\exp
}=2.2\pm0.5\pm0.3\text{ eV by AKHMETSHIN 01B \cite{Akhmetshin01B},}%
\end{equation}
and
\begin{equation}
\left.  \Gamma_{\rho(1450)\rightarrow\pi\pi}\frac{\Gamma_{\rho
(1450)\rightarrow e^{+}e^{-}}}{\Gamma_{\rho(1450)}^{total}}\right\vert _{\exp
}=\left\{
\begin{tabular}
[c]{l}%
$\ 0.12$ keV by DIEKMANN 88 \cite{Diekmann88}\\
$0.027_{-0.010}^{+0.015}$ keV by KURDADZE 83 \cite{Kurdadze83}%
\end{tabular}
\ \ \ \ \ \ \right.  .
\end{equation}
Moreover, taking into account that $\Gamma_{\rho(1450) \rightarrow \pi \pi} \approx 84$ MeV (CLEGG 94 \cite{Clegg94}), one gets:
\begin{equation}
\left.  \Gamma_{\rho(1450)\rightarrow\gamma\eta}\right\vert _{\exp}%
\approx\left\{
\begin{tabular}
[c]{l}%
$1.5$ MeV\\
$0.2$ MeV
\end{tabular}
\ \ \ \ \ \ \right.  \text{ .}%
\end{equation}
The second of the estimated values shows a good agreement with our theoretical prediction, see Table \ref{radiaradia}. Note, at present it is not possible to determine the error of the experimental value since in \cite{Clegg94} no information is reported. 

The second transition for which a similar study can be done is $\omega(1420) \rightarrow \gamma \pi^0$. In this case we have
\begin{equation}
\left.  \frac{\Gamma_{\omega(1420)\rightarrow\gamma\pi^{0}}}{\Gamma
_{\omega(1420)}^{total}}\frac{\Gamma_{\omega(1420)\rightarrow e^{+}e^{-}}%
}{\Gamma_{\omega(1420)}^{total}}\right\vert _{\exp}=2.03_{-0.75}^{+0.70}%
\cdot10^{-8}\text{ by AKHMETSHIN 05 \cite{Akhmetshin05}}%
\end{equation}
and
\begin{equation}
\left.  \frac{\Gamma_{\omega(1420)\rightarrow e^{+}e^{-}}}{\Gamma
_{\omega(1420)}^{total}}\right\vert _{\exp}=\left(  23\pm1\right)
\cdot10^{-7}\text{ by HENNER 02 \cite{Henner02}} \label{omegaepem}%
\end{equation}
which upon using $\Gamma_{\omega(1420)}^{total}=215$ MeV \cite{pdg} deliver the width
\begin{equation}
\left.  \Gamma_{\omega(1420)\rightarrow\gamma\pi^{0}}\right\vert _{\exp
}=\left(  1.9\pm0.75\right)  \text{ MeV.}%
\end{equation}
Again, a satisfactory agreement with our theoretical value is obtained, see Table \ref{radiaradia} for comparison. According to our model, this particular channel seems to be the largest radiative decay of the nonet of radially excited vector mesons. 

Furthermore, in PDG one can find information about the upper limit of the width for the transition $\phi(1680) \rightarrow \gamma K$. This value, originally determined in Ref \cite{Alavi-harati 02B}, is  comparable with our predictions, see again Table \ref{radiaradia} for details. 

For what concerns the remaining radiative decays predicted by our model, there is so far no experimental data to which one can compare. Hopefully, ongoing experiments such as GlueX and CLAS12 could deliver new results in the near future.  

\subsection{Orbitally excited vector mesons}
\label{orbex}
Next, we move to the nonet of orbitally excited vector mesons decaying into two pseudoscalar mesons as well as into one pseudoscalar and one vector meson via the strong interactions. Simillarly to the previous section, here radiative decays are also described.
\begin{center}
\textbf{Strong decays}
\end{center}

Let us now discuss the results for the orbitally excited vector mesons. In the first step we calculate the strong decays of the resonances $\rho(1700)$, $K^*(1680)$, $\omega(1650)$ and $\phi(???)=\phi(1959)$. The results for the $V_D \rightarrow PP$ channels are reported in Table \ref{orbOPP} and for $V_D \rightarrow VP$ channels in Table \ref{orbOVP}.
\begin{table}[h]
\renewcommand{\arraystretch}{1.53}
\par
\makebox[\textwidth][c] {%
\begin{tabular}
[c]{ccc}\hline\hline
Decay process $V_{D}\rightarrow PP$ & Theory [MeV] & Experiment
[MeV]\\\hline\hline
$\rho(1700)\rightarrow\bar{K}K$ & $40\pm11$ & $8.3_{-8.3}^{+10}$ MeV, see
text.\\
$\rho(1700)\rightarrow\pi\pi$ & $140\pm37$ & $75\pm30$ by BECKER 79
\cite{Becker79}\\\hline
$K^{\ast}(1680)\rightarrow K\pi$ & $82\pm22$ & $125\pm43$ by PDG\\
$K^{\ast}(1680)\rightarrow K\eta$ & $52\pm14$ & not listed in PDG\\
$K^{\ast}(1680)\rightarrow K\eta^{\prime}$ & $0.72\pm0.02$ & not listed in
PDG\\\hline
$\omega(1650)\rightarrow\bar{K}K$ & $37\pm10$ & not listed in PDG\\\hline
$\phi(1959)\rightarrow\bar{K}K$ & $104\pm28$ & resonance not yet known\\\hline\hline
\end{tabular}
}\caption{\label{orbOPP}Results for the partial decay widths of the resonances belonging to the nonet of predominantly orbitally excited vector mesons decaying into two pseudoscalar mesons.}%
\end{table}
\begin{table}[h!]
\renewcommand{\arraystretch}{1.53}
\par
\makebox[\textwidth][c] {
\begin{tabular}
[c]{ccc}\hline\hline
Decay process $V_{D}\rightarrow VP$ & Theory [MeV] & Experiment
[MeV]\\\hline\hline
$\rho(1700)\rightarrow\omega\pi$ & $140\pm59$ & seen, see text\\
$\rho(1700)\rightarrow KK^{\ast}(892)$ & $56\pm23$ & $83\pm66$ MeV, see
text\\
$\rho(1700)\rightarrow\rho\eta$ & $41\pm17$ & $68\pm42$ MeV, see text\\
$\rho(1700)\rightarrow\rho\eta^{\prime}$ & $\approx0$ & not listed in
PDG\\\hline
$K^{\ast}(1680)\rightarrow K\rho$ & $64\pm27$ & $101\pm35$ by PDG\\
$K^{\ast}(1680)\rightarrow K\phi$ & $13\pm6$ & not listed in PDG\\
$K^{\ast}(1680)\rightarrow K\omega$ & $21\pm9$ & not listed in PDG\\
$K^{\ast}(1680)\rightarrow K^{\ast}(892)\pi$ & $81\pm34$ & $96\pm33$ by
PDG\\
$K^{\ast}(1680)\rightarrow K^{\ast}(892)\eta$ & $0.5\pm0.2$ & not listed in
PDG\\
$K^{\ast}(1680)\rightarrow K^{\ast}(892)\eta^{\prime}$ & $\approx0$ & not
listed in PDG\\\hline
$\omega(1650)\rightarrow\rho\pi$ & $370\pm160$ & $\sim205,$ $154\pm44,$
$\sim273,$ $120\pm18$, see text\\
$\omega(1650)\rightarrow K^{\ast}(892)K$ & $42\pm18$ & not listed in
PDG\\
$\omega(1650)\rightarrow\omega(782)\eta$ & $32\pm13$ & $\sim100,$ $56\pm30,$
see text\\
$\omega(1650)\rightarrow\omega(782)\eta^{\prime}$ & $\approx0$ & not listed in
PDG\\\hline
$\phi(1959)\rightarrow K\bar{K}^{\ast}$ & $260\pm109$ & resonance not yet
known\\
$\phi(1959)\rightarrow\phi(1020)\eta$ & $67\pm28$ & resonance not yet
known\\
$\phi(1959)\rightarrow\phi(1020)\eta^{\prime}$ & $\approx0$ & resonance not
yet known\\\hline\hline
\end{tabular}
}\caption{\label{orbOVP}Results for the partial decay widths of the resonances belonging to the nonet of predominantly orbitally excited vector mesons decaying into one pseudoscalar and one vector meson.}%
\end{table}
When comparing the predictions of our model with the available experimental data a satisfactory agreement is obtained. It should be noted that in some cases slight differences appear. However, besides these small discrepancies, our identification of $\rho(1700)$, $K^*(1680)$, $\omega(1650)$ and $\phi(???)=\phi(1959)$ resonances as the nonet of orbitally excited vector mesons is satisfactory. 

Let us now have a closer look at the results of our model. We start with the $\rho(1700)$ resonance. According to the PDG average, its total decay width reads $0.25 \pm 0.10$ GeV, that fits to our result of $0.42 \pm 0.08$ GeV. From Table \ref{orbOPP} one can see that our theoretical predictions for PP channels are somewhat overestimated  with respect to the experimental data (except $K^*(1680) \rightarrow K \pi$ which is underestimated). For what concerns the decay into $\pi \pi$ (for which we compare with the data reported by BECKER 79 \cite{Becker79}) some older experiments deliver similar results:  

\begin{equation}
\left.  \Gamma_{\rho(1700)\rightarrow\pi\pi}\right\vert _{\exp}=\left\{
\begin{tabular}
[c]{l}%
$56\pm29$ MARTIN 78C \cite{Martin78C}\\
$75\pm32$ FROGGATT 77 \cite{Froggatt77}\\
$63\pm30$ HYAMS\ 73 \cite{Hyams73}%
\end{tabular}
\ \ \ \ \ \ \ \right.  \text{ .}%
\end{equation}
Even if the data are consistent, new measurements and further verification would be advisable. 

Next, we move to the decay into the $K\bar{K}$ channel. The branching ratio
\begin{equation}
\left.  \frac{\Gamma_{\rho(1700)\rightarrow K\bar{K}}}{\Gamma_{\rho(1700)\rightarrow
2(\pi^{+}\pi^{-})}}\right\vert _{\exp}=0.015\pm0.010\text{ DELCOURT\ 81B
\cite{Delcourt81B},}%
\end{equation}
together with
\begin{equation}
\left.  \frac{\Gamma_{\rho(1700)\rightarrow\pi\pi}}{\Gamma_{\rho
(1700)\rightarrow2(\pi^{+}\pi^{-})}}\right\vert _{\exp}=0.13\pm0.05\text{
ASTON 80 \cite{Aston80},}%
\end{equation}
and
\begin{equation}
\left.  \frac{\Gamma_{\rho(1700)\rightarrow\pi\pi}}{\Gamma_{\rho(1700)}^{tot}%
}\right\vert _{\exp}=0.287_{-0.042}^{+0.043}\pm0.05\text{ BECKER 79
\cite{Becker79},}%
\end{equation}
upon using $\Gamma_{\rho(1700)}^{tot}=250$ MeV lead us to 
\begin{equation}
\left.  \Gamma_{\rho(1700)\rightarrow K\bar{K}}\right\vert _{\exp}=8.3_{-8.3}%
^{+10.4}\text{ MeV. }%
\end{equation}
This value, to which we reffer in Table \ref{orbOPP}, is about 5 times smaller than our theoretical prediction. Nevertheless, one should take into account that in both cases large errors occur. 

Similarly, by using the ratio
\begin{equation}
\left.  \frac{\Gamma_{\rho(1700)\rightarrow KK^{\ast}(892)}}{\Gamma
_{\rho(1700)\rightarrow2(\pi^{+}\pi^{-})}}\right\vert _{\exp}=0.15\pm
0.03\text{ DELCOURT\ 81B \cite{Delcourt81B},}%
\end{equation}
one gets:
\begin{equation}
\left.  \Gamma_{\rho(1700)\rightarrow KK^{\ast}(892)}\right\vert _{\exp}%
=83\pm66\text{ MeV }%
\end{equation}
to which we compare in Table \ref{orbOVP}. The obtained value, even if it has a quite large error, is consistent with our calculations. This particular decay channel is also reported by COAN 04 \cite{Coan04} as possibly seen and by DELCOURT 81B \cite{Delcourt81B} and BIZOT 80 \cite{Bizot80} as clearly seen.  

Furthermore, by using the same procedure, we are able to determine the decay width of the mode $\rho(1700) \rightarrow \rho \eta$. To this end, we use the ratio
\begin{equation}
\left.  \frac{\Gamma_{\rho(1700)\rightarrow\rho\eta}}{\Gamma_{\rho
(1700)\rightarrow2(\pi^{+}\pi^{-})}}\right\vert _{\exp}=0.123\pm0.027\text{
DELCOURT\ 82 \cite{Delcourt82},}%
\end{equation}
which delivers the following value
\begin{equation}
\left.  \Gamma_{\rho(1700)\rightarrow\rho\eta}\right\vert _{\exp}%
=68\pm42\text{ MeV.} \label{rhoeta}%
\end{equation}
As one can see from Table \ref{orbOVP} a good agreement with our theoretical value is obtained. 

As a next step, we discuss some ratios related to the $\rho(1700)$ resonance. To do this, we incorporate some available data of decay widths containing dilepton pair $e^+e^-$: 
\begin{equation}
\Gamma_{\rho(1700)\rightarrow MM}  \cdot \frac{\Gamma_{\rho(1700)} \rightarrow e^+e^-}{\Gamma^{\text{tot}}_{\rho(1700)}} \text{ ,} \nonumber
\end{equation}
as well as some ratios containing $2(\pi^+ \pi^-)$: 
\begin{equation}
\frac{\Gamma_{\rho(1700)\rightarrow MM}}{\Gamma_{\rho(1700) \rightarrow 2(\pi^+ \pi^-)}} \text{ .} \nonumber
\end{equation}
In both mentioned quantities we use the notation $MM$ for the decay into meson-meson pairs (which can be $PP$ or $VP$). In Table \ref{brarhoorb} we report the results for the ratios: $PP/PP$ (in the first row), $PP/VP$ (in the second, third and fourth row), and $VP/VP$ (in the last row). 
\begin{table}[h!] 
\centering
 \small{
\renewcommand{\arraystretch}{1.8}
\begin{tabular}[c]{c|c|c|c}
\hline
\hline
\multicolumn{4}{c}{$\mathbf{\rho(1700)}$}\\
\hline
\hline
\multirow{2}{*}{\textbf{Branching ratio}}& \multirow{2}{*}{\textbf{Our model}} & \multicolumn{2}{|c}{\textbf{Experimental results}}\\
\cline{3-4}
&&\textbf{Value}& \textbf{Reference}\\
\hline
\hline
\multirow{3}{*}{$\frac{\Gamma_{\rho(1700)\rightarrow\pi\pi}}{\Gamma_{\rho(1700)\rightarrow KK}%
}$}&\multirow{3}{*}{$\approx3.5$}&$\sim 3.7 $&DIEKMANN 88 \cite{Diekmann88} + BIZOT 80 \cite{Bizot80}\\
&&$0.83 \pm 0.82$& KURDADZE 83 \cite{Kurdadze83} + BIZOT 80 \cite{Bizot80}\\
&&$8.7 \pm 6.7$&ASTON 80 \cite{Aston80} + DELCOURT 81B \cite{Delcourt81B}\\
\hline
\multirow{3}{*}{$\frac{\Gamma_{\rho(1700)\rightarrow\pi\pi}}{\Gamma_{\rho
(1700)\rightarrow\eta\rho}}$}&\multirow{3}{*}{$3.4 \pm 1.1$}&$\sim 18 $&DIEKMANN 88 \cite{Diekmann88} + ANTONELLI 88 \cite{Antonelli88}\\
&&$4.1 \pm 2.7$& KURDADZE 83 \cite{Kurdadze83} + ANTONELLI 88 \cite{Antonelli88}\\
&&$1.1 \pm 0.47$&ASTON 80 \cite{Aston80} + DELCOURT 82 \cite{Delcourt82}\\
\hline
\multirow{2}{*}{$\frac{\Gamma_{\rho(1700)\rightarrow KK}}{\Gamma_{\rho(1700)\rightarrow
\eta\rho}}$}&\multirow{2}{*}{$0.98 \pm 0.33$}&$5.0 \pm 4.7$& BIZOT 80 \cite{Bizot80} + ANTONELLI 88 \cite{Antonelli88}\\
&&$0.12 \pm 0.09$& DELCOURT 81B \cite{Delcourt81B} + ANTONELLI 88 \cite{Antonelli88}\\
\hline
\multirow{3}{*}{$\frac{\Gamma_{\rho(1700)\rightarrow KK}}{\Gamma_{\rho(1700)\rightarrow
K^{\ast}(892)K}}$}&\multirow{3}{*}{$0.71 \pm 0.24$}&$0.11 \pm 0.10 $& BIZOT 80 \cite{Bizot80}\\
&&$0.10 \pm 0.07$& DELCOURT 81B \cite{Delcourt81B}\\
&&$0.052 \pm 0.026$& BUON 82 \cite{Buon82}\\
\hline
\multirow{2}{*}{$\frac{\Gamma_{\rho(1700)\rightarrow K^{\ast}(892)K}}{\Gamma
_{\rho(1700)\rightarrow\eta\rho}}$}&\multirow{2}{*}{$\approx 1.37$}&$43 \pm 21$& BIZOT 80 \cite{Bizot80} + ANTONELLI 88 \cite{Antonelli88}\\
&&$1.22 \pm 0.27$& DELCOURT 81B \cite{Delcourt81B} + DELCOURT 82 \cite{Delcourt82}\\
\hline
\hline
\end{tabular}
\caption{\label{brarhoorb} Branching ratios involving the $\rho(1700)$ resonance.}
}
\end{table}

Let us first consider the $\pi\pi/ KK$ ratio, which is of the $PP/PP$ type. This means that it does not depend on any parameter of our model. The comparison of the theoretical value with three independent experimental determinations is shown in the first entry of Table \ref{brarhoorb}. In all cases a satisfactory agreement is obtained, especially when compared with the value reported by DIEKMANN 88 \cite{Diekmann88} and BIZOT 80 \cite{Bizot80}. 

We move to the $PP/VP$ ratios. In the second entry of Table \ref{brarhoorb} the $\pi \pi/ \eta \rho$ rate is considered. Again, we have three experimental results to which our theory can be compared. Definitely, the determination obtained by KURDADZE 83 \cite{Kurdadze83} and ANTONELLI 88 \cite{Antonelli88} is the closest one to our theoretical value. Also the result obtained by ASTON 80 \cite{Aston80} and DELCOURT 82 \cite{Delcourt82} is not so far from our prediction. On the contrary, the value evaluated by DIEKMANN 88 \cite{Diekmann88} and ANTONELLI 88 \cite{Antonelli88} is much larger than our theory prediction as well as the other experimental determinations.

The second ratio of the type $PP/VP$ is $KK/\eta \rho$. The main information about this decay rate is reported in the third entry of Table \ref{brarhoorb}. Among two available experimental determinations only the first one, determined by  BIZOT 80 \cite{Bizot80} and ANTONELLI 88  \cite{Antonelli88}, is consistent with our theory. This is because large experimental errors appear. The second value, calculated by combining DELCOURT 81B \cite{Delcourt81B} and ANTONELLI 88 \cite{Antonelli88}, is definitely in disagreement with our result. Notice that in this case the experimental results are also not consistent with each other. 

The next ratio of the $PP/VP$ type, listed in the fourth entry of the discussed  table, is $KK/KK^*(892)$. Our theoretical result is compared with three independent experimental values and in all cases a disagreement is visible. This inconsistency can be easily explained by the  fact that this particular ratio depends solely on the ratio of the coupling constants $g_{DPP}/g_{DVP}$ reported in Eq. (\ref{ratiogdvp/gdpp}), which in turn relies on the data of ASTON 84 \cite{Aston84} and ASTON 88 \cite{Aston88}. In this respect, it would be very useful to have new measurements concerning the $KK/KK^*(892)$ ratio. 

The last ratio involving the $\rho(1700)$ resonance and reported in the last entry of Table \ref{brarhoorb} is $K^*(892)K/\eta \rho$. There is a significant discrepancy between the two experimental values. Yet, our theoretical prediction agrees very well with the value determined in DELCOURT 81B \cite{Delcourt81B} and DELCOURT 82 \cite{Delcourt82}. 

Let us now discuss the next state belonging to orbitally excited vector mesons, the $K^*(1680)$ resonance. As we mentioned above, the experimental status of $K^*(1680)$ state is rather well established. Its partial decay widths calculated from our model are in agreement with the experimental ones, see Table \ref{orbOPP} and Table \ref{orbOVP} for details. For this case, two ratios can be studied. Both are reported in Table \ref{bakaorbr}. 
\begin{table}[h!] 
\centering
\renewcommand{\arraystretch}{1.9}
\begin{tabular}[c]{c|c|c|c}
\hline
\hline
\multicolumn{4}{c}{$\mathbf{K^*(1680)}$}\\
\hline
\hline
\multirow{2}{*}{\textbf{Branching ratio}}& \multirow{2}{*}{\textbf{Our model}} & \multicolumn{2}{|c}{\textbf{Experimental results}}\\
\cline{3-4}
&&\textbf{Value}& \textbf{Reference}\\
\hline
\hline
\multirow{2}{*}{$\frac{\Gamma_{K^{\ast}(1680)\rightarrow K\pi}}{\Gamma_{K^{\ast
}(1680)\rightarrow K^{\ast}(892)\pi}}$}&\multirow{2}{*}{$1.01 \pm 0.34$}&$1.30_{-0.14}^{+0.23}$& PDG\\
&&$2.8 \pm 1.1$& ASTON 84 \cite{Aston84}\\
\hline
\multirow{2}{*}{$\frac{\Gamma_{K^{\ast}(1680)\rightarrow K\rho}}{\Gamma_{K^{\ast
}(1680)\rightarrow K^{\ast}(892)\pi}}$}&\multirow{2}{*}{$\approx 0.79$}&$1.05_{-0.11}^{+0.27}$& PDG\\
&&$0.97\pm0.09_{-0.10}^{+0.30}$& ASTON 87 \cite{Aston87}\\
\hline
\hline
\end{tabular}
\caption{\label{bakaorbr}Branching ratios involving the $K^*(1680)$ resonance.}
\end{table} 

We start from the $K \pi/ K^*(892) \pi$ rate. As one can see in the first row of Table \ref{bakaorbr}, our theoretical result is compared with two experimental data, one from PDG average and one determined in ASTON 84 \cite{Aston84}. Definitely, a better agreement is obtained with the former. 

The next ratio, reported in the second entry of the same table, is $K \rho/ K^*(892)\pi$. Again, we have two experimental determinations to compare with. Our theoretical result is consistent with both of them. 

Going further, we study the resonance $\omega(1650)$. According to the PDG its decay into $\rho \pi$ is claimed as the dominant channel, just as our model shows. In PDG one cannot find the numerical value of the width of this particular mode, however it is possible to determine it from available experimental data. To this end, we combine the quantity

\begin{equation}
\left.  \frac{\Gamma_{\omega(1650)\rightarrow\rho\pi}}{\Gamma_{\omega
(1650)}^{tot}}\right\vert _{\exp}=\left\{
\begin{array}
[c]{c}%
\sim0.65\text{ ACHASOV 03D \cite{Achasov03D}}\\
0.380\pm0.014\text{ HENNER 02 \cite{Henner02}}%
\end{array}
\right.
\end{equation}
with the total decay width of $\omega(1650)$ cited by PDG \cite{pdg}
\begin{equation}
\Gamma_{\omega(1650)}^{\text{tot}}=315 \pm 35 \text{ MeV .}
\end{equation}
One gets:
\begin{equation}
\left.  \Gamma_{\omega(1650)\rightarrow\rho\pi}\right\vert _{\exp}=\left\{
\begin{array}
[c]{c}%
\sim205\text{ MeV ACHASOV 03D \cite{Achasov03D}}\\
120\pm18\text{ MeV HENNER 02 \cite{Henner02}.}%
\end{array}
\right.  \text{ .} \label{rhopion1}%
\end{equation}
Our theoretical value of $370 \pm 160$ MeV is consistent with the value based on ACHASOV 03D \cite{Achasov03D}. This is due to the large errors that emerge in our prediction. There is no agreement with the second experimental data, based on HENNER 02.  \cite{Henner02}

Interestingly, there is also another way to extract $\Gamma_{\omega(1650) \rightarrow \rho \pi}$ by using completely different experimental outcomes.  In this case we use the quantities
\begin{equation}
\frac{\Gamma_{\omega(1650)\rightarrow\rho\pi}}{\Gamma_{\omega(1650)}^{tot}%
}\frac{\Gamma_{\omega(1650)\rightarrow e^{+}e^{-}}}{\Gamma_{\omega
(1650)}^{tot}}=1.56\pm0.23\text{ AULCHENKO 15A \cite{Aulchenko15A}}
\label{rad16501}%
\end{equation}
and
\begin{equation}
\left.  \frac{\Gamma_{\omega(1650)\rightarrow e^{+}e^{-}}}{\Gamma
_{\omega(1650)}^{tot}}\right\vert _{\exp}=\left\{
\begin{array}
[c]{c}%
\sim18\text{ ACHASOV 03D \cite{Achasov03D}}\\
32\pm1\text{ HENNER 02 \cite{Henner02}}%
\end{array}
\right. \text{ ,}
\end{equation}
which together deliver the values
\begin{equation}
\Gamma_{\omega(1650)\rightarrow\rho\pi}=\left\{
\begin{array}
[c]{c}%
\sim273\text{ MeV ACHASOV 03D \cite{Achasov03D} + AULCHENKO 15A
\cite{Aulchenko15A}}\\
154\pm44\text{ HENNER 02 \cite{Henner02} + AULCHENKO 15A \cite{Aulchenko15A}}%
\end{array}
\right.  \text{ .}%
\end{equation}
As one can see, the desired value could not be clearly determined. It is however clear that in all cases large numerical values of the width are obtained. This signals that the $\rho \pi$ channel is dominant.

In a similar way we study the decay into the $\omega \eta$ channel. Out of 
\begin{equation}
\left.  \frac{\Gamma_{\omega(1650)\rightarrow\omega\eta}}{\Gamma
_{\omega(1650)}^{tot}}\frac{\Gamma_{\omega(1650)\rightarrow e^{+}e^{-}}%
}{\Gamma_{\omega(1650)}^{tot}}\right\vert _{\exp}=0.57\pm0.06\text{ AUBERT 06D
\cite{Aubert06D}, } \label{rad16502}%
\end{equation}
we have
\begin{equation}
\left.  \Gamma_{\omega(1650)\rightarrow\omega\eta}\right\vert _{\exp}=\left\{
\begin{array}
[c]{c}%
\sim100\text{ MeV, ACHASOV 03D \cite{Achasov03D} +AUBERT 06D \cite{Aubert06D}
}\\
56\pm30\text{ MeV, HENNER 02 \cite{Henner02} + AUBERT 06D \cite{Aubert06D}}%
\end{array}
\right.  \text{ .}%
\end{equation}
Here, only the second determination matches with our value of $32 \pm 13$ MeV. 

Moreover, we are able to establish one additional ratio related to $\omega(1650)$: 
\begin{equation}
\left.  \frac{\Gamma_{\omega(1650)\rightarrow\omega\eta}}{\Gamma
_{\omega(1650)\rightarrow\rho\pi}}\right\vert _{\exp}=0.365\pm0.054\text{ } \text{ .}%
\end{equation}
While, in our model we obtained a much smaller value of about 0.086. 

The last state belonging to the group of orbitally excited vector mesons is $\phi(1959)$. We remind that this resonance is up to now not experimentally confirmed. Therefore, all the results related to $\phi(1959)$ are predictions. For completeness, all of them are reported in Table \ref{predictionsphi}. 
\begin{table}[h!] 
\centering
\renewcommand{\arraystretch}{1.3}
\begin{tabular}[c]{c|c}
\multicolumn{2}{c}{$\mathbf{\phi(1959)}$}\\
\hline
\hline
\multicolumn{2}{c}{BASIC INFORMATIONS}\\
\hline
\hline
Quark content & $\approx s\bar{s}$\\
Mass & $1959 \pm 20$ MeV\\
\hline
\hline
\multicolumn{2}{c}{QUANTUM NUMBERS}\\
\hline
\hline
n & 1 (predom.)\\
L & 2 (predom.)\\
S & 1 $\uparrow \uparrow$ (predom.)\\
Old spectroscopy notation & $n$ $^{2S+1} L_J=1$ $^3D_1$\\
$J^{PC}$ notation & $1^{--}$\\
\hline
\hline
\multicolumn{2}{c}{DECAYS}\\
\hline
\hline
Decay mode & Decay width\\
\hline
$\phi(1959) \rightarrow K \bar{K}$& $104 \pm 28$ MeV\\
$\phi(1959) \rightarrow KK^*$ & $260 \pm 109$ MeV\\
$\phi(1959) \rightarrow \phi(1020) \eta$& $67 \pm 28$ MeV\\
$\phi(1959) \rightarrow \phi(1020) \eta'$& $\approx 0$ MeV\\
$\phi(1959) \rightarrow \gamma \eta$ & $0.19 \pm 0.12$ MeV\\
$\phi(1959) \rightarrow \gamma \eta'$ & $0.13 \pm 0.08$ MeV\\
\hline
\hline
\end{tabular}
\caption{\label{predictionsphi}Summary of basic informations concerning the not yet discovered meson $\phi(1959)$.}
\end{table} 

Even if no experimental data exist, we can compare some of our theoretical results with the predictions of the quark model. In the case of the decay into $KK$ channel the quark model predicts the value of about $100$ MeV, just as in our model where we have $104$ MeV. However, a large discrepancy appears when we consider the decay into $K^*K$.  The prediction of the quark model gives $\sim 50$ MeV \cite{GodfreyIsgur}, much smaller than our result of $260 \pm 109$ MeV. 

\begin{figure}[h!]
\begin{center}
\includegraphics[width=0.4 \textwidth]{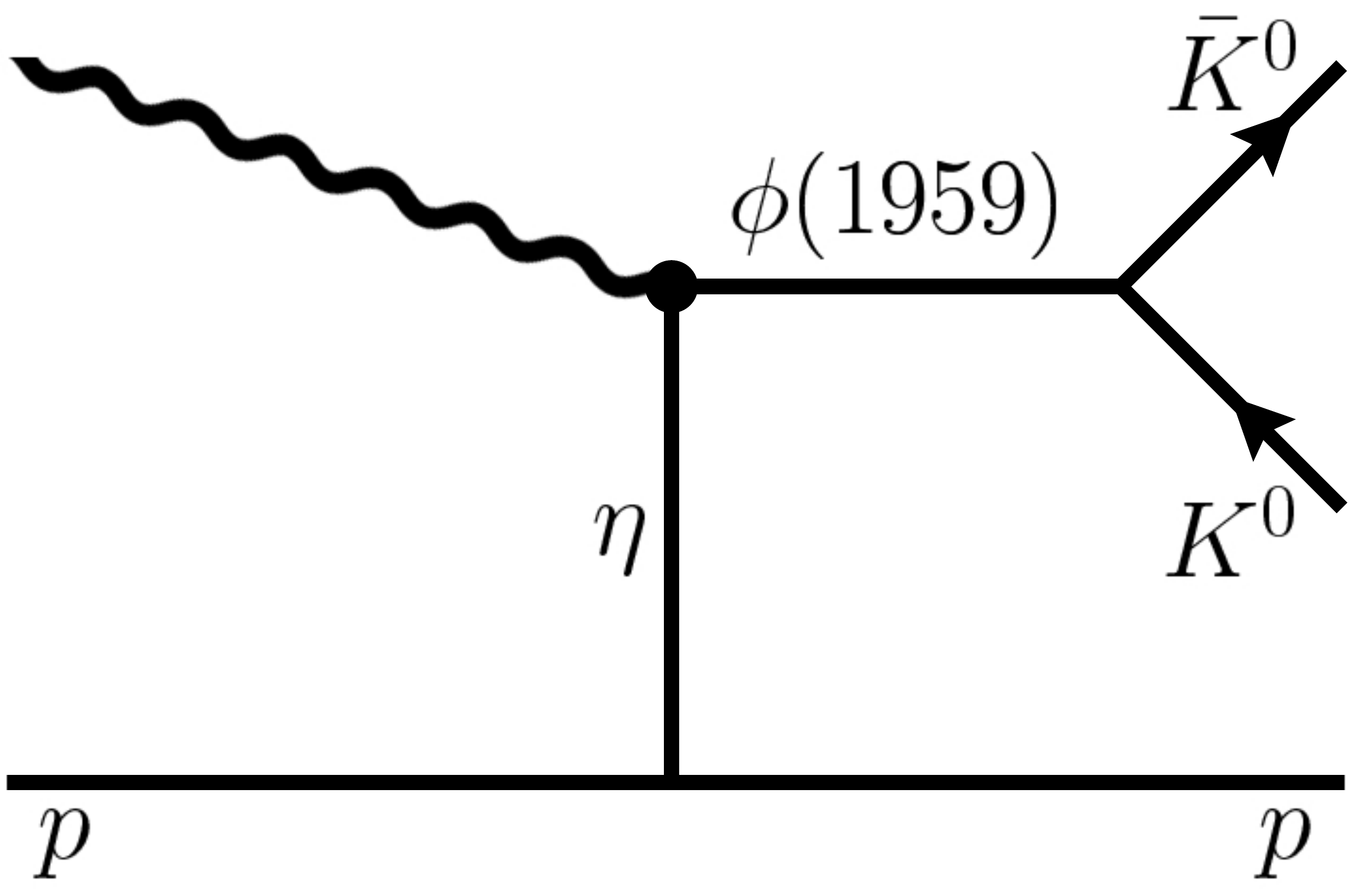}
\caption{\label{expprop} Feynman diagram representing the process of Eq. (\ref{gluexclas}).}
\end{center}
\end{figure}

The lack of the experimental discovery of $\phi(1959)$ could be explained by the very broad total width of this state. According to our model it is about $430$ MeV. Hopefully, $\phi(1959)$ may be nevertheless measured at the ongoing GlueX and CLAS12 experiments at Jefferson Lab. In particular, in Figure \ref{expprop} we present the diagram of the processes 
\begin{equation} \label{gluexclas}
\gamma+ p \rightarrow K^0+\bar{K}^0+p, \hspace{0.5cm} \gamma+ p\rightarrow K^++K^-+p
\end{equation}
which can be examined at GlueX and CLAS12. In our approach all the mesonic vertices are included ($\phi(1959)\gamma \eta$ and $\phi(1959)KK$). The study of the barionic part is doable by using hadronic models that contain the baryons as well as their interactions with mesons. In this context, the extended Linear Sigma Model can be employed \cite{GallasGiacosalinear, sigmaolgiam}. Similarly, the process
\begin{equation}
\gamma+p \rightarrow K^-+K^{*+}(892)+p \rightarrow K^++K^-+\pi^0+p
\end{equation}
can also be studied. 
Moreover, the state $\phi(1959)$ should also appear in the data provided by BaBar, where the process $e^+e^- \rightarrow K^+K^-$ was measured \cite{Leesbabar}. In the study of this particular reaction all vector mesons $\rho(1450)$, $\omega(1420)$, $\phi(1680)$, $\rho(1700)$, $\omega(1650)$ and $\phi(1959)$ should be incorporated since interference effects may occur, see details in Ref. \cite{Sauli}.

\begin{center}
\textbf{Radiative decays}
\end{center}

The results of our approach for the radiative decays of orbitally excited vector mesons are reported in Table \ref{radorbs}. 
\begin{table}[h]
\renewcommand{\arraystretch}{1.53}
\par
\makebox[\textwidth][c] {
\begin{tabular}
[c]{ccc}\hline\hline
Decay process $V_{D}\rightarrow\gamma P$ & Theory [MeV] & Experiment
[MeV]\\\hline\hline
$\rho(1700)\rightarrow\gamma\pi$ & $0.095\pm0.058$ & not listed\\
$\rho(1700)\rightarrow\gamma\eta$ & $0.35\pm0.21$ & not listed\\
$\rho(1700)\rightarrow\gamma\eta^{\prime}$ & $0.13\pm0.08$ & not
listed\\\hline
$K^{\ast}(1680)\rightarrow\gamma K$ & $0.30\pm0.18$ & not listed\\\hline
$\omega(1650)\rightarrow\gamma\pi$ & $0.78\pm0.47$ & not listed\\
$\omega(1650)\rightarrow\gamma\eta$ & $0.035\pm0.021$ & not listed\\
$\omega(1650)\rightarrow\gamma\eta^{\prime}$ & $0.012\pm0.007$ & not
listed\\\hline
$\phi(1959)\rightarrow\gamma\eta$ & $0.19\pm0.12$ & resonance not yet
known\\
$\phi(1959)\rightarrow\gamma\eta^{\prime}$ & $0.13\pm0.08$ & resonance not yet
known\\\hline\hline
\end{tabular}
}\caption{\label{radorbs} Results for the partial decay widths of the resonances belonging to the nonet of predominantly orbitally excited vector mesons decaying into one photon and one pseudoscalar meson.}%
\end{table}
All these transitions have not yet been even seen in experiments, hence there is no data to compare to. In general, the order of magnitude for the partial decay widths of radiative decays of orbitally excited vector mesons is similar to that for the nonet of radially excited vector mesons. The largest width is $\omega(1650) \rightarrow \gamma \pi$, in accordance with the dominant character of the $\omega(1650) \rightarrow \rho \pi$ decay. It would be very useful if experimental data on these decays could be obtained in the future. 

\section{Conclusions}

In this chapter we considered two vector meson nonets, one with (predominantly) radial excitations including the states $\{\rho(1450)$, $K^*(1410),$ $\omega(1420),$ $\phi(1680)\}$ and one with (predominantly) orbital excitations involving the states $\{\rho(1700)$, $K^*(1680),$ $\omega(1650),$ $\phi(???)\equiv\phi(1959)\}$. In our flavor-invariant approach an effective QFT Lagrangian was employed. The four free parameters of the model were determined by using experimental data reported in the PDG. 

For both nonets we evaluated strong and radiative decays. In total, we reported 48 partial decay widths. Moreover, numerous ratios related to both nonets have been studied. Theoretical results were compared with the experimental data. Since not all considered channels are experimentally known, some theoretical outcomes are predictions.

In general, a qualitatively good agreement between theory and experiment is observed. Theoretically dominant channels are detected in experiments, while those with small decay widths are usually not observed. In some cases there is some disagreement between our results and the data. One should, however, note that numerous discrepancies occur between the data reported by different experiments measuring the same quantity. 

A key point of our study is the investigation of the putative state $\phi(???)$ in the $1$ $^3D_1$ nonet. Within our framework we estimated its mass to be $1959 \pm 20$ MeV and named it as the $\phi(1959)$. We were able to make predictions for the strong and radiative decay modes of this state. We conclude that the dominant channels are $\phi(1959) \rightarrow K^*(892)K$ and $\phi(1959) \rightarrow KK$. Moreover, according to our theory, the radiative decay into $\gamma \eta$ is also possible. This means that $\phi(1959)$ can be searched in photoproduction based studies, as for instance at the ongoing GlueX and CLAS12 experiments at the Jefferson Lab.

In summary, both theoretical and experimental studies of the excited vector mesons are important for a better understanding of spectroscopy in the low energy regime. Although the general picture is consistent, new experimental data, with focus on radiative decays, would be advisable. Finally, the experimental confirmation of the existence of $\phi(1959)$ would be a neat proof of the (predominantly) $q\bar{q}$ assignment for the discussed excited vector states.

\chapter{Vector meson $\mathbf{K^*(892)}$}
\label{chapvec}
In this chapter we concentrate on the vector state $K^*(892)$, which we study by using an effective QFT approach at the resummed one-loop level. The validity of the one-loop resummation is in general satisfactory in hadronic models, as it was shown in Ref. \cite{oneloopapprox}. In the following we investigate the spectral function of $K^*(892)$, whose shape turns out to be well approximated by a standard relativistic Breit-Wigner function, and we show that the propagator contains only a single (relevant) pole on the complex plane. Moreover, in order to understand the nature of this resonance, we employ a large-$N_c$ study which confirms that $K^*(892)$ is a conventional quark-antiquark meson. 
\section{Introduction}
 
\begin{figure}[h!]
\begin{center}
\includegraphics[width=0.75 \textwidth]{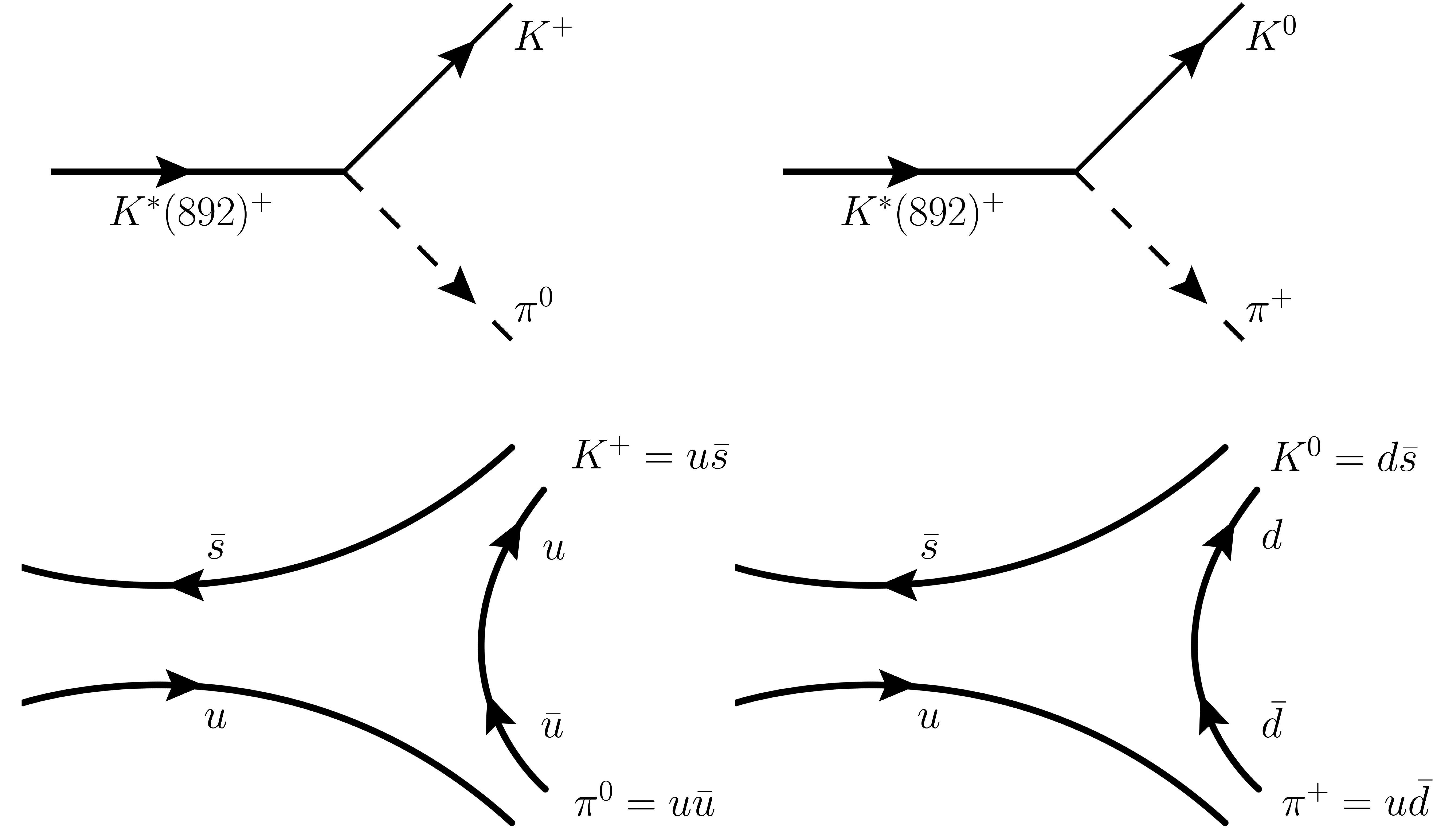} \label{diagrams892b}
\caption{The Feynman diagrams and the corresponding quark-line diagrams of the resonance $K^*(892)$ decaying into one pion and one kaon. (Note, the quark-line graphs are solely illustrative, since we use mesonic d.o.f. in our calculations).}
\end{center}
\end{figure}

 According to established knowledge from experimental and theoretical studies, the resonance  $K^*(892)$ is a ground-state vector meson with quantum numbers $I(J^P)=\frac{1}{2}(1^-)$. The Particle Data Group quotes its mass as $891.66 \pm 0.26$ MeV and the total decay width as $50.8 \pm0.9$ MeV \cite{pdg}. As we shall show, here we confirm that the $K^*(892)$ is a standard quark-antiquark object corresponding well to the four expected charge combinations $K^{*+}\equiv u\bar{s}$, $K^{*-} \equiv s\bar{u}$, $K^{*0} \equiv d\bar{s}$, $\bar{K}^{*0} \equiv s\bar{d}$. This relatively short living state decays predominantly into one pion and one kaon, see Figure \ref{diagrams892b} for an illustrative presentation of Feynman diagrams and corresponding quark-line diagrams.
 The main idea is to apply an effective QFT Lagrangian in which a seed state corresponding to the $K^*(892)$ resonance is coupled to its decay products (one kaon and one pion). The theory is then studied at the one-loop resummed level. 
 
 In this way, we present the main features of our formalism in the case of a well understood $q\bar{q}$ state. Later on, we shall apply it to other mesonic resonances which still need a better understanding. 

\section{Theoretical framework}

We present our model starting from the Lagrangian constructed as
\begin{eqnarray}
\mathcal{L}_v=cK^*(892)^+_{\mu} \partial^{\mu}K^-\pi^0+ \sqrt{2}cK^*(892)^+_{\mu} \partial^{\mu}K^0 \pi^-+\ldots \text{ , } \label{LagK892}
\end{eqnarray}
where $K^*(892)$ is the vector kaonic field, c is the dimensionless coupling constant, and dots stand for the sum over analogous interaction terms for the other members of the isospin multiplet ($K^{*-}, K^{*0}, \bar{K}^{*0}$). As it was shown in Chapter \ref{botbds}, by using the Feynman rules one can obtain the theoretical expression for the total decay width of mesons. Hence, for the unstable $K^*(892)$ resonance with mass m (kept as `running', see below) decaying into one pion with mass $M_{\pi}$ and one kaon with mass $M_{K}$ we have:
\begin{eqnarray}
\Gamma_{K^*(892) \rightarrow K \pi}(m)=3\frac{\left|\vec{k}\right|}{8 \pi m^2}\frac{c^2}{3}\left[-M_{\pi}^2+\frac{\left(m^2+M^2_{\pi}-M_K^2\right)^2}{4m^2}\right]F_{\Lambda}(m) \text{ , }\label{GammaK892}
\end{eqnarray}
where the quantity $|\vec{k}|$ 

\begin{eqnarray}
k(m)=\left|\vec{k}\right|=\frac{\sqrt{m^4+\left(M^2_K-M^2_{\pi}\right)^2-2\left(M_K^2+M^2_{\pi}\right)m^2}}{2m} \label{threemomentum892}
\end{eqnarray}
is the  absolute value of three-momentum of one of the decay products in the reference frame in which $K^*(892)$ is at rest. We refer to Sec. \ref{twobody} and to  Refs. \cite{Friendly, peskinQFT, Mrowczynski} for an explicit derivation of Eq. (\ref{threemomentum892}). The on-shell decay width can be obtained by fixing the mass m as the nominal PDG mass of $K^*(892)$: 
\begin{equation}
\Gamma^{\text{on}\hspace{0.1cm}\text{shell}}=\Gamma_{K^*}(m \simeq 0.892 \hspace{0.1cm} \text{GeV})=50.8 \text{ MeV} \text{ .}
\end{equation}
 Another important quantity, included in Eq. (\ref{GammaK892}), is the form factor (or vertex function) which is choosen as a Gaussian function given by
\begin{eqnarray}
F_{\Lambda}(m)=e^{-2\left|\vec{k}\right|^2/ \Lambda^2}\text{ . } \label{formfactorK892}
\end{eqnarray}
This type of form factor is rather conventional in hadron physics and emerges in various studies on the subject, including microscopic models, such as the $^3P_{0}$ mechanism used in quark models \cite{3p0old, 3p0new}. This mechanism describes the production of mesons from a $q\bar{q}$ pair emerging from the QCD vacuum, as we show in the bottom part of Figure \ref{diagrams892b}. Obviously, the Gaussian function is an important part of our effective model and such a choice is model-dependent and definitely not unique. One should note that the form of the vertex function does not change the qualitative picture of the results as long as it is smooth and goes to zero sufficiently fast. However, the results depend strongly on the numerical value of cutoff $\Lambda$. This parameter acts as an energy scale that takes into account the extended structure of mesons.

Notice that we employed here an effective QFT model, hence one should not interpret the $\Lambda$ parameter as the maximal value of the momentum $k$. For $k$ larger than $\Lambda$ the decay into that particular channel is naturally suppressed. However there are no restrictions for the momentum $k$, which can have values even larger than $\Lambda$ (mathematically, $k$ may range from 0 to $\infty$, yet physically the model is limited by the investigated energy range). Technically, one could include the form factor directly in the Lagrangian of Eq. (\ref{LagK892}) by making it nonlocal \cite{nonlocalmilena}.

\begin{figure}[h!]
\begin{center}
\includegraphics[width=0.8 \textwidth]{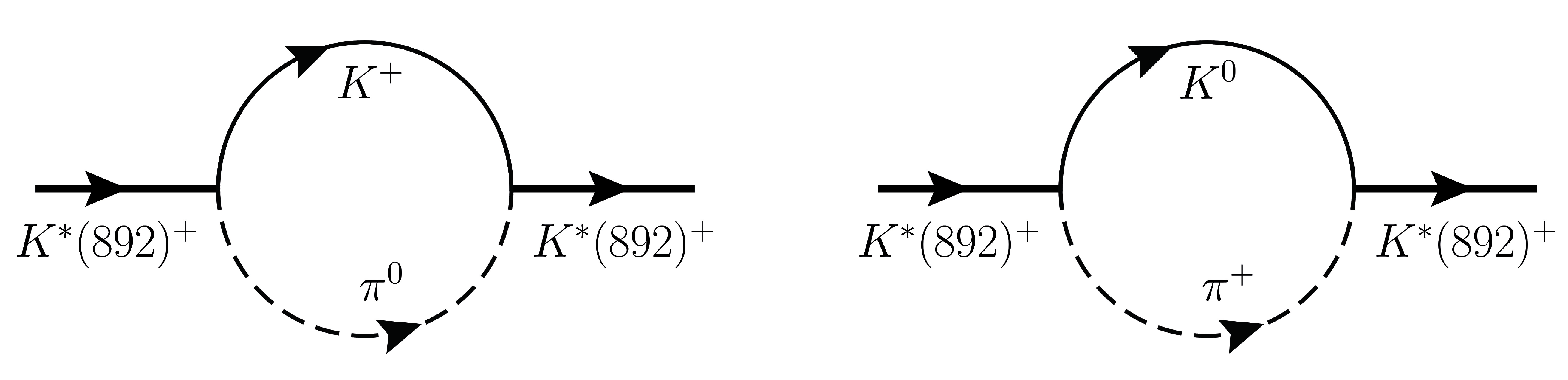} \label{loopk892}
\caption{Schematic illustration of $\pi K$ loops.}
\end{center}
\end{figure}

We further introduce the scalar part of the propagator of the field $K^*$ through the equation
\begin{eqnarray}
\Delta_{K^*}(p^2=m^2)=\frac{1}{m^2-M_0^2+\Pi(m^2)+i\varepsilon}\text{ , }  \label{propK892}
\end{eqnarray}
where $M_0$ refers to the bare mass of the $K^*(892)$ state. Without interactions the function $\Pi(m^2)$ vanishes. Yet, the propagator changes upon turning on the interactions, when intermediate  mesonic loops dress the original state. This mechanism is commonly named as `dressing'. In general, $\Pi(m^2)$ is the contribution to the self-energy of the vector kaonic field $K^*$ and $\Pi(m^2)$ contains all the one-particle irreducible diagrams. Here, we consider as an approximation only one-loop diagrams, see Figure \ref{loopk892} for an illustrative example of $\pi K$ loops. 

Moreover, in terms of the real part $\operatorname{Re}(\Pi(m^2))$ and imaginary part $\operatorname{Im}(\Pi(m^2))$, the one-loop contribution can be expressed as 
\begin{equation}
\Pi(m^2)=\operatorname{Re}(\Pi(m^2))+i\operatorname{Im}(\Pi(m^2)) \text{ .}
\end{equation}

Based on the optical theorem \cite{peskinQFT}, the imaginary part reads 
\begin{equation}
\operatorname{Im}(\Pi(m^2))=m\Gamma_{K^*(892) \rightarrow K \pi}(m) \text{ ,}
\end{equation}
where:
\begin{equation}
\Gamma_{K^*(892) \rightarrow K \pi}= \Gamma_{K^{*+}(892) \rightarrow K^+ \pi^0}+\Gamma_{K^{*+}(892) \rightarrow K^0 \pi^+} \stackrel{\text{isospin symmetry}}{=} 3 \Gamma_{K^*(892) \rightarrow K^+ \pi^0}
\end{equation}

Note, we work here in the isospin-symmetric limit, hence we consider only one channel denoted as:
\begin{equation}
K^*(892) \rightarrow K \pi \text{ .} \nonumber
\end{equation}

Furthermore, in order to calculate the real part we use the corresponding dispersion relation (valid under the conditions that $s=m^2$ is real and exceeds the value $M_{\pi}+ M_{K}$). For the decay channel $K^*(892) \rightarrow K \pi$ we have \cite{complexmmpp1, complexmmpp2}
\begin{eqnarray}
\operatorname{Re}(\Pi(m^2))=-\frac{1}{\pi} \mathcal{P} \int \limits_{M_{K}+M_{\pi}}^{\infty}2m'\frac{m' \Gamma_{K^*(892)\rightarrow K \pi}(m')}{m^2-m'^2+ i \varepsilon}dm' \text{ . }
\end{eqnarray}

Once, the propagator is defined, one can introduce the spectral function, given by
\begin{eqnarray}
d_{K^*}(m)=\frac{2m}{\pi}\left|\operatorname{Im} \Delta_{K^*}(p^2=m^2)\right| \text{ ,} \label{sfK892}
\end{eqnarray}
which must be correctly normalized:
\begin{eqnarray}
\int\limits_0^{\infty}d_{K^*}(m)dm=1\text{ . } \label{normalK892}
\end{eqnarray}
The quantity $d_{K^*}(m)dm$ can be understood as the probability that the mass of the state $K^*(892)$ is in the range between $m$ and $m+dm$. Note, Eq. (\ref{normalK892}) naturally follows from our formalism and is not imposed as an `ad hoc' constraint. The numerical verification of Eq. (\ref{normalK892}) represents an important unitarity test of our approach. 

When $s$ is not real (or real but smaller than $(M_{\pi}+ M_{K})^2$) the loop $\Pi(s)$ can be expressed as:
\begin{equation}
\Pi(s=z^2)=-\frac{1}{\pi} \int \limits_{M_{K}+M_{\pi}}^{\infty}2m'\frac{m' \Gamma_{K^*(892) \rightarrow K \pi}(m')}{s-m'^2+ i \varepsilon}dm'=-\frac{1}{\pi} \int \limits_{(M_{K}+M_{\pi})^2}^{\infty}\sqrt{s'}\frac{\Gamma_{K^*(892) \rightarrow K \pi}(s')}{s-s'+i \varepsilon}ds' \text{ ,}
\end{equation}
where $s=z^2$ is a complex number.

When $\Pi(s)$ is continued to its second Riemann sheet ( II RS), it reads:
\begin{equation}
\Pi_{II}=\Pi(s)+2iI(s) \text{ ,}
\end{equation} 
where
\begin{equation}
I(s)=\sqrt{s} \Gamma(s) \text{,}
\end{equation}
and $\Gamma (s)$ is given in Eq. (\ref{GammaK892}) upon setting $s=z^2$.

We emphasize that the relation $\Pi(m^2 \rightarrow \infty) \rightarrow 0$ holds in all directions of the whole complex plane on the I RS. Namely, the functions $\Pi(m^2 \rightarrow \infty)$ and $e^{-2k(s=z^2)/\Lambda^2}$ are completely different from each other in the I RS. The former does not have any singularity (besides a cut), while the latter has no cut and an essential singularity for $z=\infty$. For what concerns other Riemann sheets, the $\Pi_{II}(z^2)$ develops singular points.

The propagator in the I RS is given by Eq. (\ref{propK892}) upon consdering $m^2 \rightarrow z^2=s$. It can be also continued to the II RS as
\begin{equation}
\Delta_{K^*,II}(s=z^2)=\frac{1}{s-M_s+\Pi_{II}(s=z^2)+i \varepsilon} \text{ .}
\end{equation}
One searches poles of $\Delta_{K^*,II}(s)$ on the II RS as
\begin{equation}
\Delta_{K^*,II}^{-1}(s=z^2)=0 \text{ ,}
\end{equation}
where the solution(s) $s=s_{\text{pole}}$ is (are) such that (see Chapter \ref{intfr}):
\begin{equation}
\sqrt{s_{\text{pole}}}=m_{\text{pole}}-\frac{i \Gamma_{\text{pole}}}{2} \text{ .}
\end{equation}

\section{Results}
Our model contains three free parameters: the bare mass $M_0$ of $K^*(892)$ included in Eq. (\ref{propK892}), the cutoff $\Lambda$ employed in Eq. (\ref{formfactorK892}), and the coupling constant $c$ entering the Lagrangian of Eq. (\ref{LagK892}). For illustrative purposes, we fix $\Lambda=0.5$ GeV in agreement with other approaches \cite{a0980revisited}. Then, $M_0$ is determined under the requirement that the maximum of the spectral function corresponds to the PDG mass value of $K^*(892)$, obtaining $M_0=0.89166$ GeV. Finally, the coupling constant $c$ is evaluated under the requirement to obtain the PDG value of the decay width of $K^*(892)$ state; the value is $c=15.73$. (Note, we omit the anyhow small experimental errors in this illustrative study of $K^*(892)$).

\begin{figure}[h!]
\begin{center}
\includegraphics[width=0.65 \textwidth]{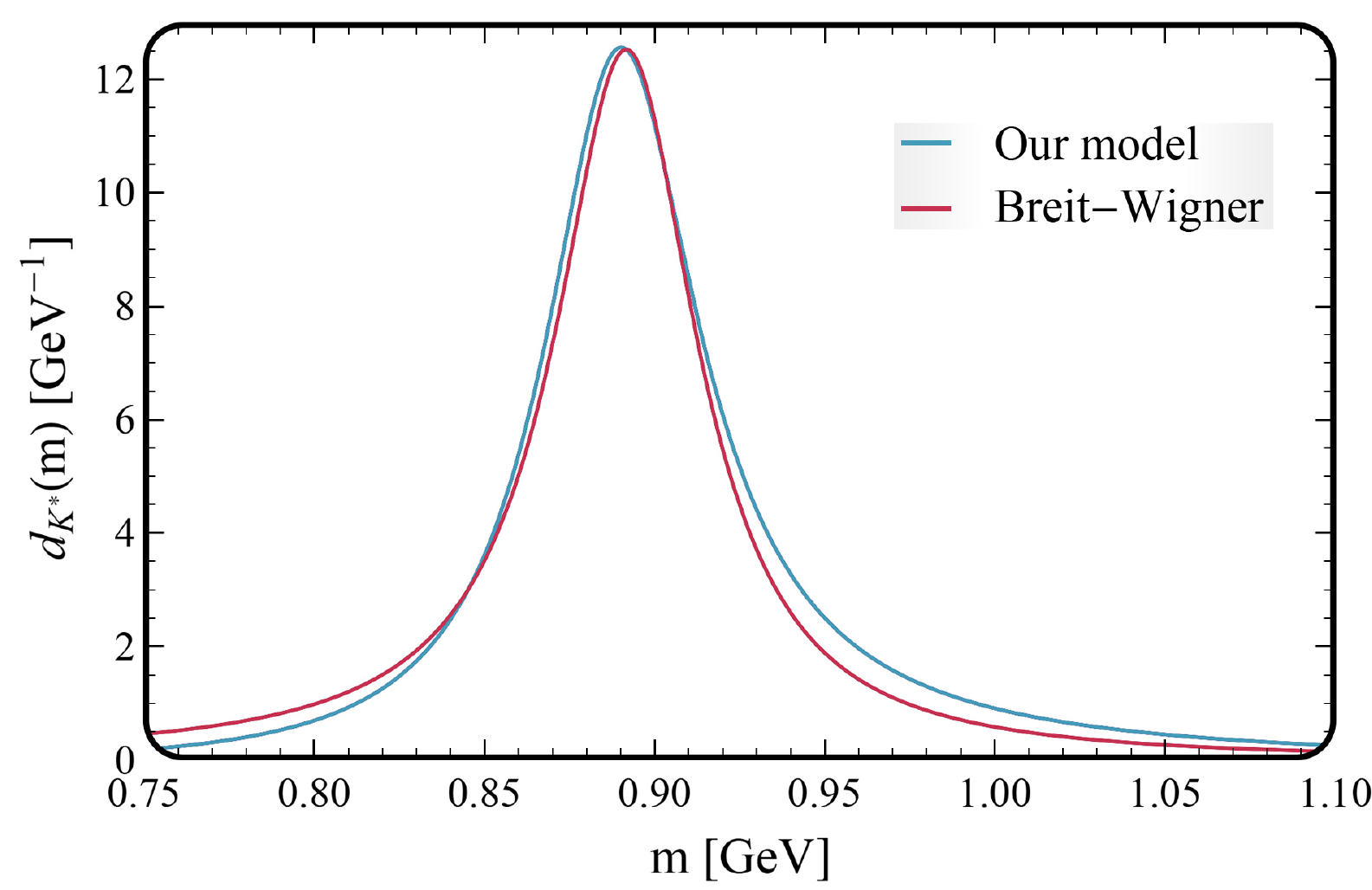} 
\caption{\label{sfblue}The shape of the spectral function of $K^*(892)$ defined in Eq. (\ref{sfK892}) compared with the Breit-Wigner function, plotted for the parameters listed in the PDG \cite{pdg}. The used parameters are: $\Lambda=0.5$ GeV, $M_0=0.89166$ GeV and $c=15.73$.}
\end{center}
\end{figure}

By using these parameters, we calculate the spectral function of Eq. (\ref{sfK892}) for the vector kaon $K^*(892)$. The plot of the normalized spectral function is presented in Figure \ref{sfblue}. 
As expected, only a unique peak (close to $0.9$ GeV) corresponding to $K^*(892)$ is observed. One can approximate well the shape of the spectral function by using a relativistic Breit-Wigner distribution. Moreover, as expected, we find that the propagator of Eq. (\ref{propK892}) has a unique (relevant) pole 
\begin{equation}
\Delta^{-1}_{K^*, II}(s)=s-M_0^2+\Pi_{II}(s)=0 \text{ ,}
\end{equation}
on the second Riemann sheet. This pole corresponds to the seed state $K^*(892)$ and its coordinates are
\begin{equation}
 \sqrt{s_{\text{pole}}}=(0.89-0.028i) \hspace{0.1cm} \text{GeV} \hspace{0.1cm},
\end{equation}
which means that 
\begin{equation}
m^{\text{pole}}_{K^*(892)}=0.89 \text{ GeV} \hspace{0.6cm} \text{and} \hspace{0.6cm} \Gamma^{\text{pole}}_{K^*(892)}=56 \text{ MeV} \text{ .}
\end{equation}

In order to prove the conventional $q\bar{q}$ nature of $K^*(892)$ we analyze the large-$N_c$ behavior of its spectral function and its pole position. For this purpose we employ a new dimensionless parameter $\lambda$, defined as
\begin{equation} \label{scalmmnc}
\lambda=\frac{3}{N_c}\text{ , }
\end{equation}
with $N_c$ being the number of colors. In nature $N_c=3$ but it is convenient to study QCD for large values of $N_c$, since various simplifications occur in this limit \cite{lebed, largencwitten}. 

\begin{figure}[h] 
\begin{center}
\begin{minipage}[b]{7.8cm}
\centering
\includegraphics[width=7.5cm]{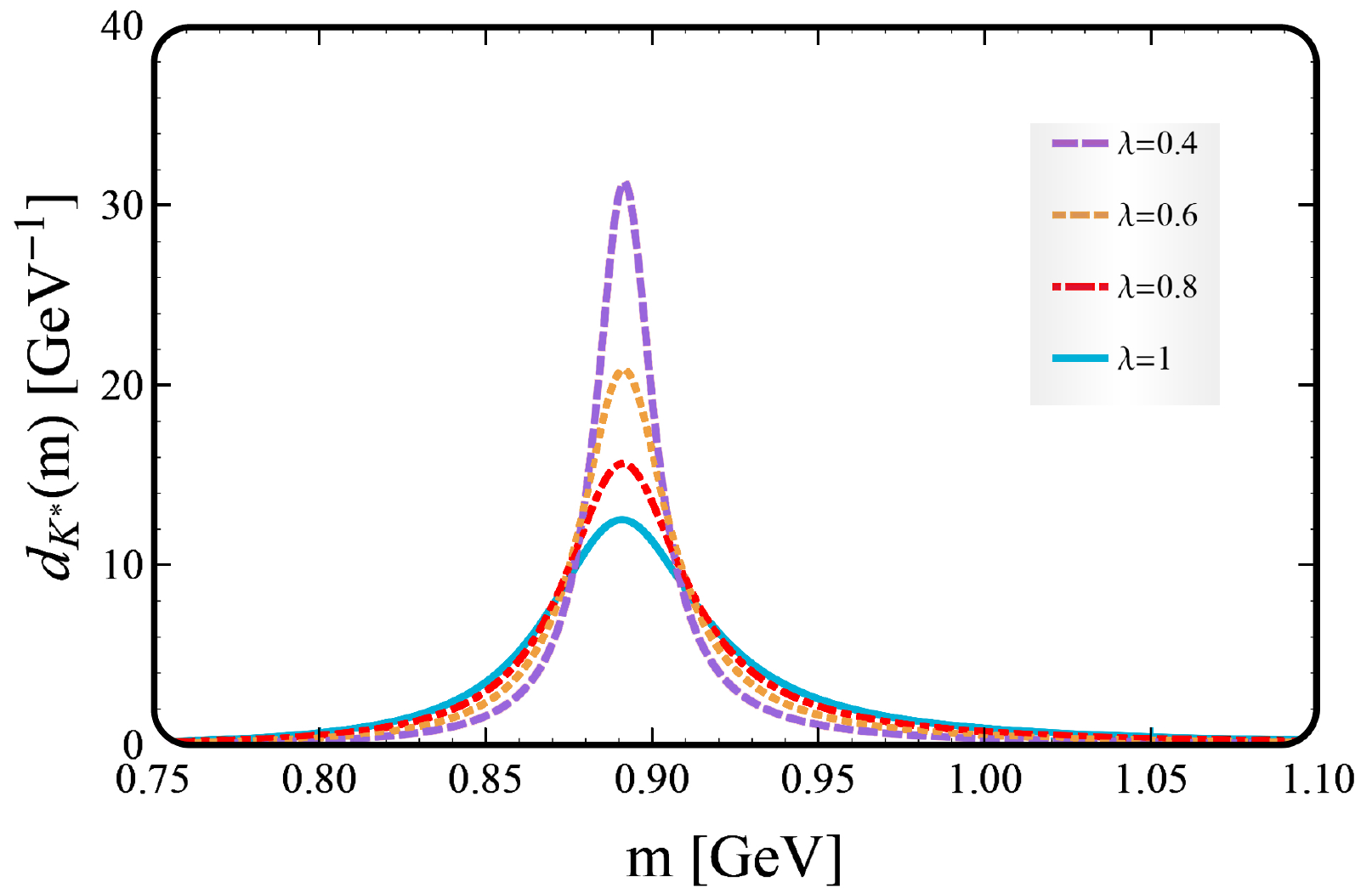}\\\textit{a)} 
\end{minipage}
\begin{minipage}[b]{7.8cm}
\centering
\includegraphics[width=7.75cm]{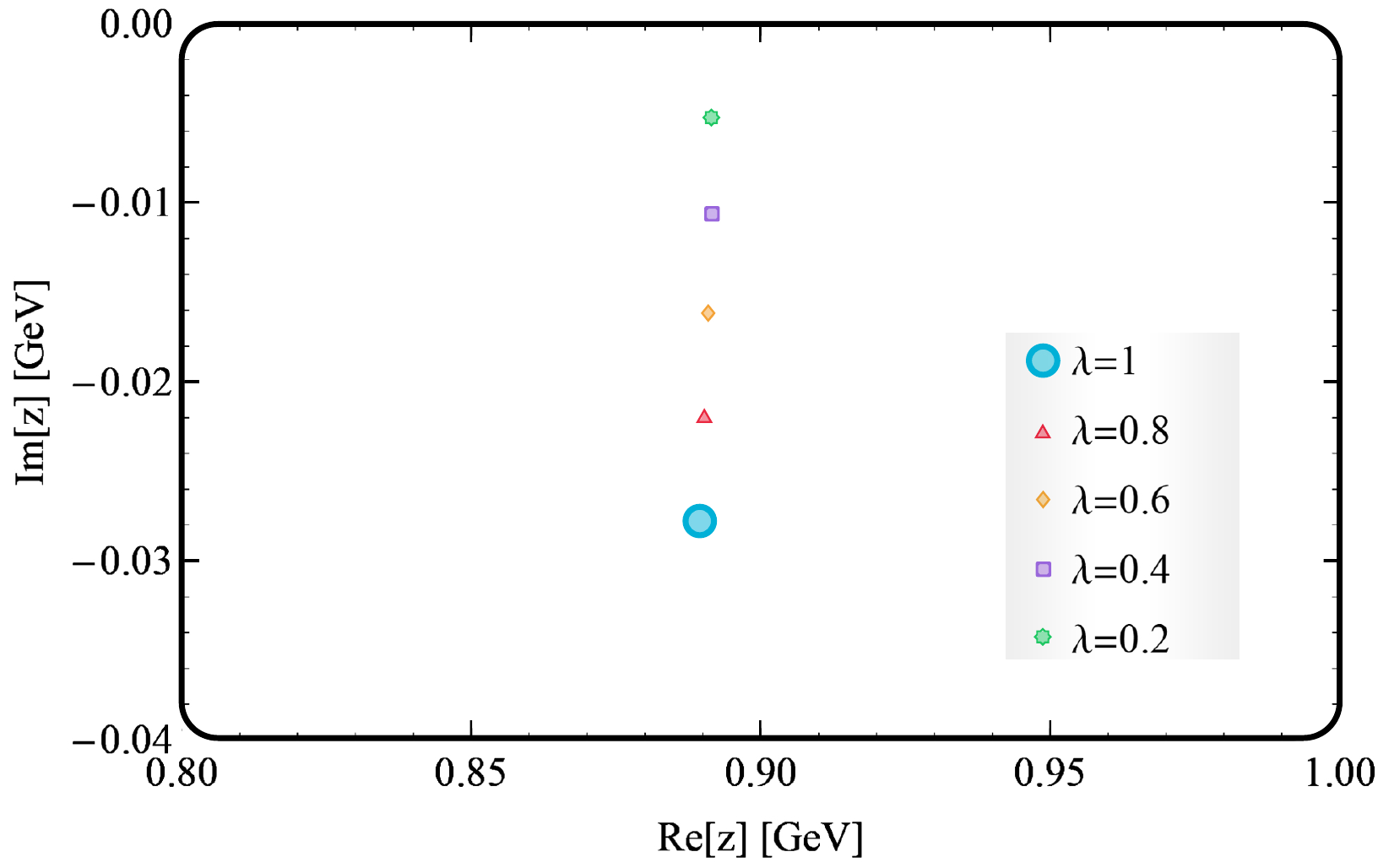}\\\textit{b)}
\end{minipage}
\end{center} 
\caption{\label{nclimitk892}Panel (a) shows the behavior of the spectral function upon variations of $\lambda$ parameter. Panel (b) shows the motion of the pole on the complex plane upon considering the large-$N_c$ limit. The blue point indicates the coordinates of the propagator pole for the physical case, thus $N_c=3$.}
\end{figure}

Then, we impose the scaling of the coupling constant c in the following way
\begin{equation}
c\rightarrow \sqrt{\lambda}c\text{ ,}
\end{equation}
in agreement with large-$N_c$ rules, see Ref. \cite{lebed, largencwitten} and references therein. Thus, for $\lambda=0 \hspace{0.1cm} (N_c \rightarrow \infty)$ the state $K^*_0(892)$ becomes stable. Note, for a large value of $N_c$, $\Gamma$ scales as $1/N_c$. On the other hand, for $\lambda=1$ we obtain the physical results of our model. We chose some intermediate values of the  parameter $\lambda$ and repeat the calculations. The results for the spectral function and the pole trajectories are depicted in Figure \ref{nclimitk892}. 

For what concerns the spectral function, it is visible that for decreasing $\lambda$ the peak related to $K^*(892)$ resonance grows and becomes narrower. Finally, as expected, for $\lambda=0$ the Dirac delta function is obtained: $d_{K^*}=\delta(m-M_0)$. Moreover, the results for the pole in the large-$N_c$ limit as a function of $\lambda$ reveal its movement on the complex plane. The pole assigned to $K^*(892)$ approaches the real energy axis with decreasing $\lambda$ (increasing $N_c$). This behavior is typical for a conventional quark-antiquark object. Moreover, the imaginary part scales as $1/N_c$, as expected for a pole describing a $q\bar{q}$ state.
\section{Concluding remarks}

In this chapter we have presented the formalism of our effective QFT unitarized approach at one loop. In this context we have studied the well-known quark-antiquark $K^*(892)$ resonance. We investigated its spectral function, which can be nicely approximated by Breit-Wigner distribution peaked at around $0.9$ GeV. Moreover, we have found the coordinates of a single pole that appears on the complex plane on the II RS. The large-$N_c$ analysis confirms the quark-antiquark nature of the $K^*(892)$ resonance. 

In the next chapters we will use similar models to describe more controversial and difficult resonances. 

\chapter{Scalar mesons $\mathbf{K^*_0(700)}$ and $\mathbf{K^*_0(1430)}$}
\label{kappak}
This chapter is devoted to the investigation of the existence and the nature of the light resonance $K^*_0(700)$ by employing a QFT approach at the resummed one-loop level. Within our model, a single  (bare) quark-antiquark scalar kaonic seed state, which roughly corresponds to the well-established resonance $K^*_0(1430)$, is included in the relativistic Lagrangian containing both non-derivative and derivative terms. The spectral function in the scalar kaonic sector up to 1.8 GeV cannot be approximated by an ordinary Breit-Wigner shape, due to a significant deformation in the low-energy region. We show that, besides the expected pole of $K^*_0(1430)$, a dynamically generated pole corresponding to the light $K^*_0(700)$ naturally emerges through kaon-pion loops. The performed fit to the experimental $K \pi$ phase-shift data in the $I(J^P)=\frac{1}{2}(0^+)$ channel shows that the scattering data can be correctly described when the poles for both resonances $K^*_0(1430)$ and $K^*_0(700)$ are simultaneously present. 

\section{Introduction to the light scalar mesons}

The resonance $\kappa \equiv K^*_0 (700)$ \cite{pdg} (previously named as $K^*_0 (800)$ \cite{pdgold}) is the lightest scalar state with quantum numbers $I(J^P)=\frac{1}{2}(0^+)$. The $K_0^* (700)$, also denoted as $\kappa$, has been very recently added to the PDG summary table as the last missing state of the nonet of light scalar mesons below 1 GeV (but the remark  ``needs confirmation'' is still present in the detailed entry in the particle listing). Despite numerous theoretical works where the pole position of light $\kappa$ is determined (see Refs. \cite{Ishida97B, Descotes, Zhou06, Pelaez17, Bugg10}), its nature is not yet fully understood. Together with other well-established light scalar mesons of that nonet, such as the broad $f_0 (500)$ and the narrow states $a_0(980)$ and $f_0 (980)$, the $K^*_0 (700)$ is a good candidate to be a non-conventional state. It has been suggested to interpret them as predominantly tetraquark ($qq$-$\bar{q}\bar{q}$) objects \cite{Jaffe, Jaffe2005, Maiani2004, Giacosa2006, GiacosaPagliara, FariborzJora, Fariborz:2003, FariborzAzizi, Napsuciale} or/and as generated dynamically molecular-like states \cite{CloseTornqvist, Pelaez:2004xp, OllerOset, lightestnonet, JaminOller, AlbaladejoOller, PelaezRiosprl, Pelaez:2003dy, MorganPennington, vanBeveren:1986ea, Tornqvistzp, TornqvistRoos, BoglionePenningtonprl, Boglioneprd, OllerOsetnp, OllerOsetPelaezprl,OllOse}. For what concerns the experimental evidence, this resonance has been observed by studying $\pi K$ scattering phase shifts \cite{Aston}. Moreover, quite recently, the BES Collaboration reported the presence of the light $\kappa$ in the decay channel $J/\psi \rightarrow \bar{K}^{*0}(892)K^+ \pi^-$ \cite{Guokappa, Ablikimkappa}. For other experimental signals, see Ref. \cite{evidenceskappa} and references therein. In Ref. \cite{lattice} one can also find the lattice results of the $\pi K$ scattering.\\
The second resonance described by the same set of quantum numbers, thus $I=\frac{1}{2}$ and $J=0$, is the heavier state $K^*_0(1430)$. From both theoretical and experimental side this state is rather well-established. The PDG \cite{pdg} reports for $K^*_0(1430)$ a mass  of $(1425 \pm 50)$ MeV and a total width of $(270 \pm 80)$ MeV. According to the general consensus, this resonance belongs to the nonet of scalar ($J=0$) quark-antiquark states with $L=S=1$. The other members of this nonet are $f_0(1370)$, $f_0(1500)/f_0(1710)$, and $a_0(1450)$. (Note, for $f_0(1500)/f_0(1710)$ mixing with the scalar glueball is possible, see Refs. \cite{gluint1, gluint2, gluint3, gluint4, gluint5, gluint6}).

\section{The model}

For the scalar kaonic sector we use an effective relativistic Lagrangian describing the decay (and interaction) of a single $q\bar{q}$ seed state (denoted as $K^*_0$) into a kaon-pion pair. It is constructed as the sum of two types of interaction terms, one containing a derivative and one without derivative (see Chapter \ref{botbds} for details): 
\begin{eqnarray}
\mathcal{L}_{int}=aK^{*-}_0 \left( K^+ \pi^0 + \sqrt{2} K^0 \pi^+\right)+bK^{*-}_0\left(\partial_{\mu}K^+ \partial^{\mu} \pi^0+\sqrt{2} \partial_{\mu}K^0 \partial^{\mu} \pi^+\right)+ h.c. + \ldots \hspace{0.2cm}. \label{intLag}
\end{eqnarray}
The dots in the above equation represent the sum over the other members of the isospin multiplet. Moreover, the quantities $a$ and $b$ are the coupling constants with dimensions Energy and Energy$^{-1}$, respectively. The simultaneous presence of derivative and non-derivative interactions in the Lagrangian of Eq. (\ref{intLag}) is consistent with other low-energy effective approaches of QCD, such as chiral Perturbation Theory based on the nonlinear realization of chiral symmetry \cite{Gasser, EckerGasser, Scherer} and chiral effective models  based on its linear realization 
 \cite{KoRudaz, urban, ParganlijaKovacs}. As usual, the theoretical formula for the decay width of the unstable $K^*_0$ state with mass $m$ (kept as `running') can be derived from Feynman rules, see Chapter \ref{botbds} for details, and reads as
\begin{eqnarray}
\Gamma_{K^*_0}^{tree}(m)=3\frac{k(m)}{8 \pi m^2}\left(a-b\frac{m^2-M_K^2-M^2_{\pi}}{2} \right)F_{\Lambda}(m)\text{ . } \label{widthpionkaon}
\end{eqnarray}
The quantity $k(m)$ in Eq. (\ref{widthpionkaon}) expressed explicitly as
\begin{eqnarray}
k(m)=\frac{\sqrt{m^4+\left(M^2_{\pi}-M_K^2\right)^2-2\left(M^2_{\pi}+M^2_K\right)m^2}}{2m} \label{threemomentumpionkaon}
\end{eqnarray}
refers to the absolute value of the three-momentum of the kaon (with mass $M_K$) and the pion (with mass $M_{\pi}$) in the reference frame where the decaying $K^*_0$ state is at rest. As explained previously, the quantity $F_{\Lambda}(m)$ is the form factor which, similarly to the case of $K^*(892)$ in the vector kaonic sector, is chosen to be an exponential function of the type
\begin{eqnarray}
F_{\Lambda}(m)=e^{-2k^2(m)/\Lambda^2}\text{ , } \label{vertex}
\end{eqnarray}
where a cutoff parameter $\Lambda$ refers now to the kaonic scalar sector. Upon setting $m \simeq 1.43$ GeV \cite{pdg} in Eq. (\ref{widthpionkaon}) we find the (on-shell) tree-level decay width ($\Gamma_{K^*_0}^{\text{on}\hspace{0.1cm}\text{shell}}=\Gamma_{K^*_0}^{\text{tree}}(m \simeq 1.43 \hspace{0.1cm} \text{GeV})$).

According to some phenomenological models, one can identify it as the physical decay width of the $K^*_0(1430)$ resonance \cite{ParganlijaKovacs}. It shall be here stressed that the bare $q\bar{q}$ seed state $K^*_0$ entering in the Lagrangian of Eq. (\ref{intLag}) roughly corresponds to the well-established $K^*_0(1430)$ state, in accordance with many phenomenological approaches exploring the scalar sector \cite{GodfreyIsgur, FariborzJora, ParganlijaKovacs, AmslerCloseglue, CloseKirk, Giacosagutsche}.

At this point, we turn to the mathematical formalism of our theoretical approach. Similarly to Eq. (\ref{propK892}) we introduce the propagator of the field $K^*_0$ 

\begin{eqnarray}
\Delta_{K^*_0}(p^2=m^2)=\frac{1}{m^2-M_0^2+\Pi(m^2)+i \varepsilon} \text{ . } \label{propagatorscalar}
\end{eqnarray}
The parameter $M_0$ in the above expression stands for the bare mass of the $q\bar{q}$ ($K_0^*(1430): u\bar{s}, \bar{u}s, d\bar{s}, \bar{d}s$) seed state, while the quantity $\Pi (m^2)$ is the self-energy function of the scalar kaonic field $K^*_0$. 

\begin{figure}[h!]
\begin{center}
\includegraphics[width=0.5 \textwidth]{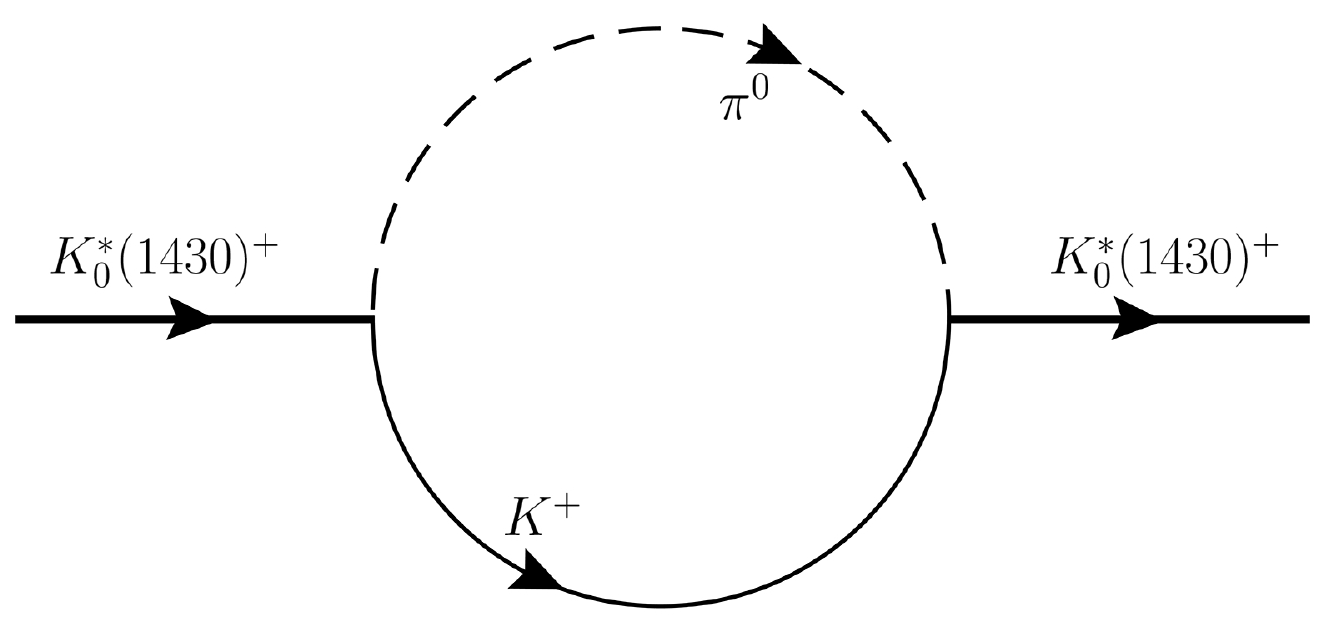}
\caption{ \label{pikaon1430}An illustrative example of one kaon and one pion circulating in the loop, giving the contribution to $\Pi(m)$.}
\end{center}
\end{figure}

Again, the function $\Pi(m^2)$ can be express in terms of real part $\operatorname{Re}\Pi(m^2)$ and imaginary part $\operatorname{Im}\Pi(m^2)$. By applying the optical theorem one gets: 
 
\begin{equation}
\operatorname{Im}\Pi_{K^*_0}(m^2)=m\left(\Gamma_{K_0^{*+}(1430) \rightarrow K^+ \pi^0}+\Gamma_{K_0^{*+}(1430) \rightarrow K^0 \pi^+}\right) \stackrel{\text{isospin symmetry}}{=} m\left(3 \Gamma_{K_0^*(1430) \rightarrow K^+ \pi^0}\right)\text{ .}
\end{equation}
Similarly to $K^*(892)$ also here we work in the isospin limit, therefore we have only one channel (as an example in Figure \ref{pikaon1430} the $K^+ \pi^0$ loop is depicted).
 
In order to calculate the real part we use the corresponding dispersion relation \cite{complexmmpp1, complexmmpp2}:
\begin{eqnarray}
\operatorname{Re}(\Pi_{K^*_0}(m^2))=-\frac{1}{\pi} \mathcal{P} \int \limits_{M_{K}+M_{\pi}}^{\infty}2m'\frac{m' \Gamma_{K^*_0(1430)\rightarrow K \pi}(m')}{m^2-m'^2+ i \varepsilon}dm' \text{ . }
\end{eqnarray}
Furthermore, the propagator is directly connected to spectral function by the following relation
\begin{eqnarray}
d_{K^*_0}(m)=-\frac{2m}{\pi} \text{Im} \Delta_{K^*_0}(p^2=m^2)\text{ . } \label{sftheoretical}
\end{eqnarray}
It is crucial to remind, that the spectral function fulfills again the normalization  condition 
\begin{eqnarray}
\int \limits^{\infty}_{0}d_{K^*_0}(m)dm=1\text{ , } \label{normal1430700}
\end{eqnarray}
see Ref. \cite{Giacosa:2012hd} for a rigorous mathematical proof.
The quantity $d_{K^*_0}(m)dm$ can be identified as the probability that the mass of the resonance lies in the range between $m$ and $m+dm$. Note, the normalization of Eq. (\ref{normal1430700}) demonstrates the conservation of unitarity. 

As it will be discussed in the next subsection, in our approach we use the $K\pi$ phase-shift data for the $I(J^P)=\frac{1}{2}(0^+)$ scattering channel to provide a determination of the model parameters. Within our approach, we considered only the scalar kaonic resonances in the energy region below $1.8$ GeV. Here, the Feynman diagram corresponding to the $\pi K$ s-channel scattering dominates:
\begin{equation}
\mathcal{A}_{\mathcal{F}}=\frac{(ig)^2F^2_{\Lambda}}{s-M^2+ \Pi(s)} \text{ .}
\end{equation}
The phase shift reads \cite{PDG2014kin}
\begin{equation}
\frac{e^{2i\delta_{\pi K}}-1}{2i}=\frac{|k|}{8 \pi \sqrt{s}}\mathcal{A}_{\mathcal{F}} \text{ ,}
\end{equation}
leading to:
\begin{eqnarray}
\delta_{\pi K}(m)=\frac{1}{2} \text{arccos}\left[1-\pi \Gamma^{tree}_{K^*_0}(m)d_{K^*_0}(m)\right]\text{ . } \label{phaseshift}
\end{eqnarray}

Some comments are needed:

1) Eq. (\ref{phaseshift}) is obtained under the assumption that the $K^*_0$ propagation dominates in the s-channel, thus the contributions coming from the u-channel and t-channel meson exchanges are omitted here. For a wider discussion on the validity of such assumption we refer to Refs. \cite{Ruppschannel, Haradaschannel, Isgurschannel}. There is also shown that this approximation does not affect significantly the pole positions of the resonances.  \\

2) Moreover, the approximation of considering only the s-channel and neglecting the u- and t-channels is motivated by the fact that the scaterring data, which we use to perform a fit, starts at a safe distance ($\sim 200$ MeV) above the $K \pi$-threshold. This is especially important, since at threshold all contributions are relevant due to the small overall interaction strength (a consequence of chiral symmetry).\\

3) Within our effective model we investigate the scattering process in the $I=\frac{1}{2}$ channel. Experimentally, the phase shift for $I=\frac{3}{2}$ channel (for which no s-channel is present) is negative (\textit{i.e} repulsion occurs in this case). When comparing the $I=\frac{1}{2}$ and $I=\frac{3}{2} $ channels, the latter is at least four times smaller. This shows that the increased intensity in the former channel is due to the exchange of the scalar kaon in the s-wave.
\section{Results}
\label{resk01430r}
\begin{figure}[h!]
\begin{center}
\includegraphics[width=0.65 \textwidth]{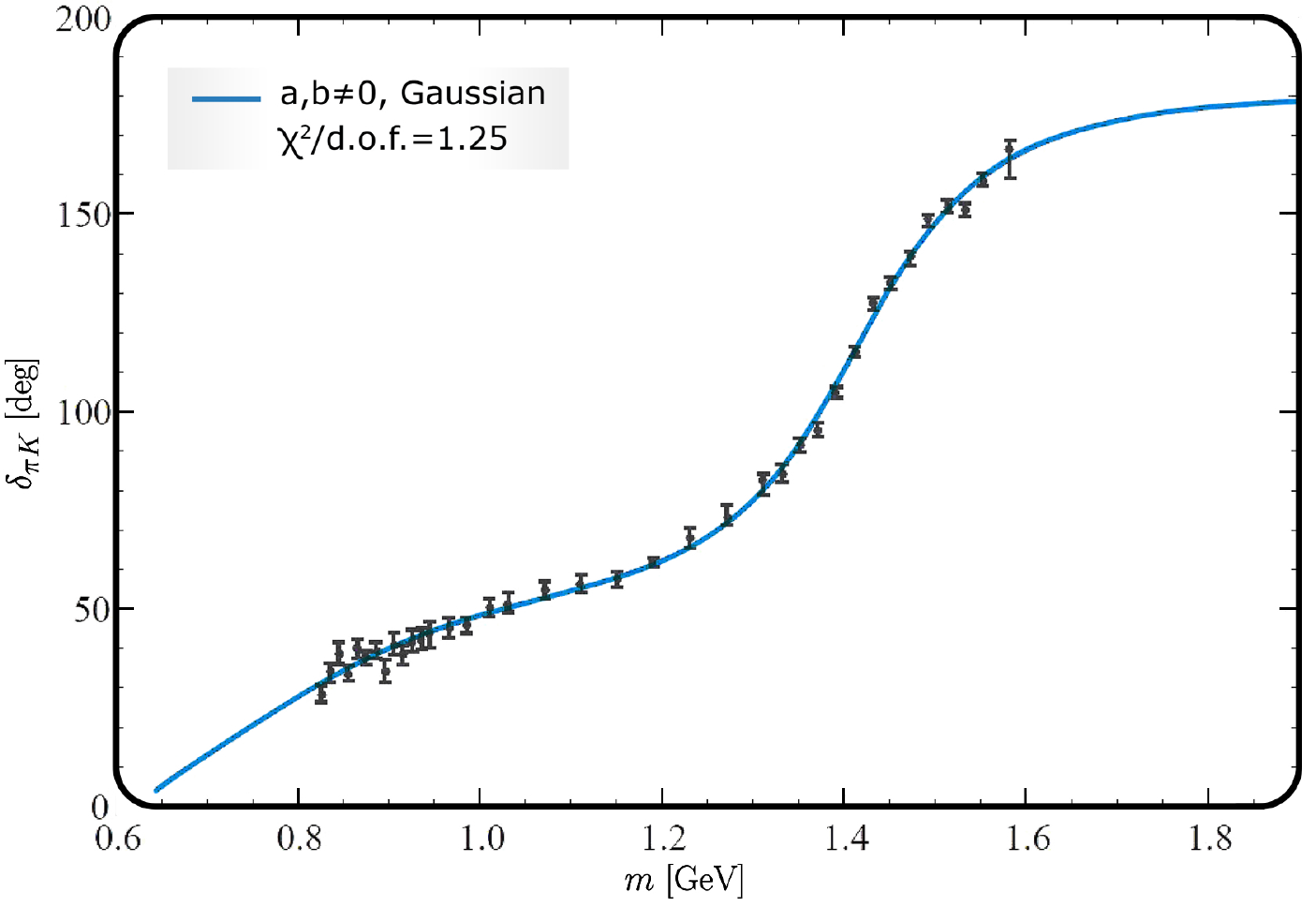}
\caption{\label{resoffit} The solid (blue) curve represents the results of our fit for the kaon-pion phase shift of Eq. (\ref{phaseshift}) performed using the four free parameters of the model $a, b, M_0$ and $\Lambda$ (for their numerical values, see Table \ref{fittingparameters}). The black points stand for the experimental data of Ref \cite{Aston88}. The value of $\chi^2_0/d.o.f.=1.25$ explains the good description of the data.}
\end{center}
\end{figure}

Our model contains four free parameters: the two coupling constants $a$ and $b$ included in Lagrangian of Eq. (\ref{intLag}), the cutoff $\Lambda$ entering Eq. (\ref{vertex}), and finally the bare seed state mass $M_0$ coming from the propagatar of Eq. (\ref{propagatorscalar}). All the model parameters are determined by fitting the expression of Eq. (\ref{phaseshift}) to the phase shift experimental data of Ref. \cite{Aston88}. In order to obtain the parameter's errors, we construct the inverted Hessian matrix out of the $\chi^2$ function and calculate the square roots of its diagonal elements. The results of the fit are illustrated in Figure \ref{resoffit} and the numerical values of all best-fitting parameters as well as their errors are reported in Table \ref{fittingparameters}.

 The statistical analysis of $\chi^2$ gives a satisfactory value $\chi^2_0/d.o.f.=1.25$, thus indicating a good description of the experimental data by the results of our model. 

\begin{table}[h]
\renewcommand{\arraystretch}{1.53}
\par
\makebox[\textwidth][c] {
\begin{tabular}
[c]{c|cccc}\hline \hline
Parameter&a [GeV]& b [ GeV$^{-1}$]& $M_0$ [GeV]& $\Lambda$ [GeV]\\ \hline
Value&$1.6 \pm 0.22$& $-11.16 \pm 0.82$&$1.204 \pm 0.008$& $0.496 \pm 0.008$ \\ \hline \hline
\end{tabular}
}\caption{The numerical values of model parameters and their errors obtained by the fit to the phase-shift data.} \label{fittingparameters}
\end{table}

The poles are searched in the II RS. We therefore refer to Chapter \ref{chapvec} for a treatment of this mathematical aspect. These poles (in unit of GeV) are:

\begin{equation}
K^*_0(1430)\hspace{0.1cm}: \hspace{0.1cm} (1.413 \pm 0.002)-(0.127 \pm 0.003) i \hspace{0.1cm} \text{ [GeV]} \text{ ,}
\end{equation}
\begin{equation}
K^*_0(700)\hspace{0.1cm}: \hspace{0.1cm} (0.746 \pm 0.019)-(0.262 \pm 0.014)i \hspace{0.1cm} \text{ [GeV]} \text{ .}
\end{equation}
We emphasize that a pole related to the resonance $\kappa=K^*_0(700)$ naturally appears in our calculation as a dynamically generated state.

 The obtained coordinates of the pole of the propagator provide us useful information about the resonances. According to the relation given by Eq. (\ref{mask}) one can determine the physical mass of the resonance (to be identified with the real part of the pole) as well as its decay width (to be identified with the doubled imaginary part of the pole). Hence, for both resonances $K^*_0(1430)$ and $K^*_0(700)$ we calculate their masses and decay widths out of the coordinates of the corresponding complex poles. The results of our model in comparison with the PDG values are presented in Table \ref{Tablemasswidth}.

 \begin{table}[h!] 
\centering

\renewcommand{\arraystretch}{1.25}
\begin{tabular}[c]{c|cc|cc}
\hline
\hline
\multicolumn{1}{c|}{}&\multicolumn{2}{c|}{\textbf{Mass [GeV]}}&\multicolumn{2}{c}{\textbf{Decay width [GeV]}}\\
\cline{2-5}
\multicolumn{1}{c|}{STATE}&OUR MODEL&PDG&OUR MODEL&PDG\\
\hline
$K^*_0(1430)$&$(1.413 \pm 0.002)$ &$(1.425 \pm 0.050)$ &$(0.254 \pm 0.006)$ &$(0.270 \pm 0.080)$\\

$K^*_0(700)$&$(0.746 \pm 0.019)$ &$0.630-0.730$ &$(0.524 \pm 0.028)$ &$0.520-0.680$\\
\hline
\hline 
\end{tabular} 
\caption{The numerical values of the masses and decay width for both $K^*_0(700)$ and $K^*_0(1430)$ resonances obtained from our theoretical model compared to the values reported in the PDG. The PDG values for $K^*_0(700)$ correspond to the position of the pole. In PDG one can also find Breit-Wigner mass $m_{\kappa, BW}=824 \pm 30$ MeV and decay width $ \Gamma_{\kappa, BW}= 478 \pm 50$ MeV.}
\label{Tablemasswidth}
\end{table} 

All the theoretical values determined in our model are compatible with those listed in the PDG.  

\begin{figure}[h]
\begin{center}
\begin{minipage}[b]{7.8cm}
\centering
\includegraphics[width=7.7cm]{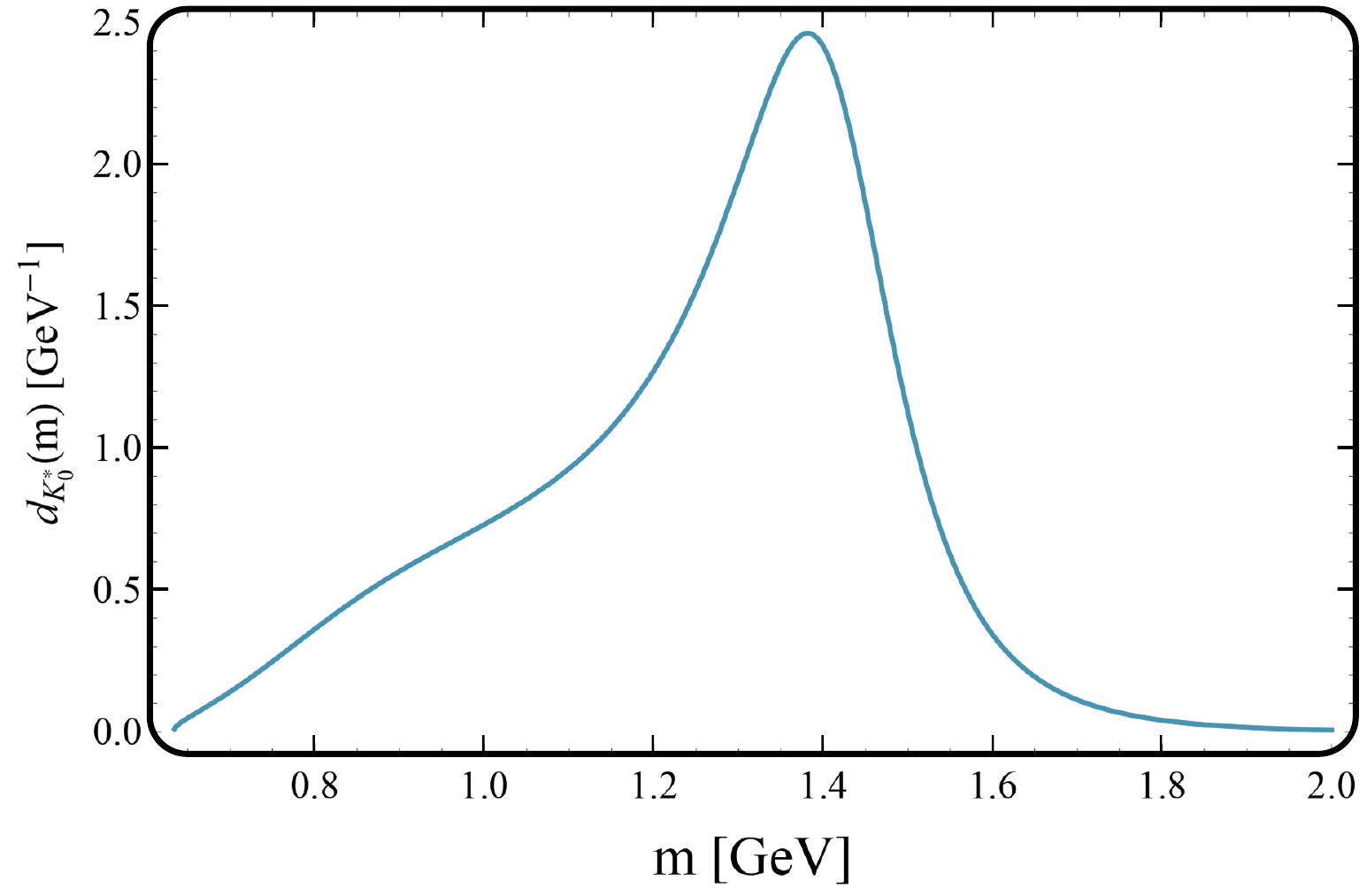}\\\textit{a)} \label{sfscalar}
\end{minipage}
\begin{minipage}[b]{7.8cm}
\centering
\includegraphics[width=7.55cm]{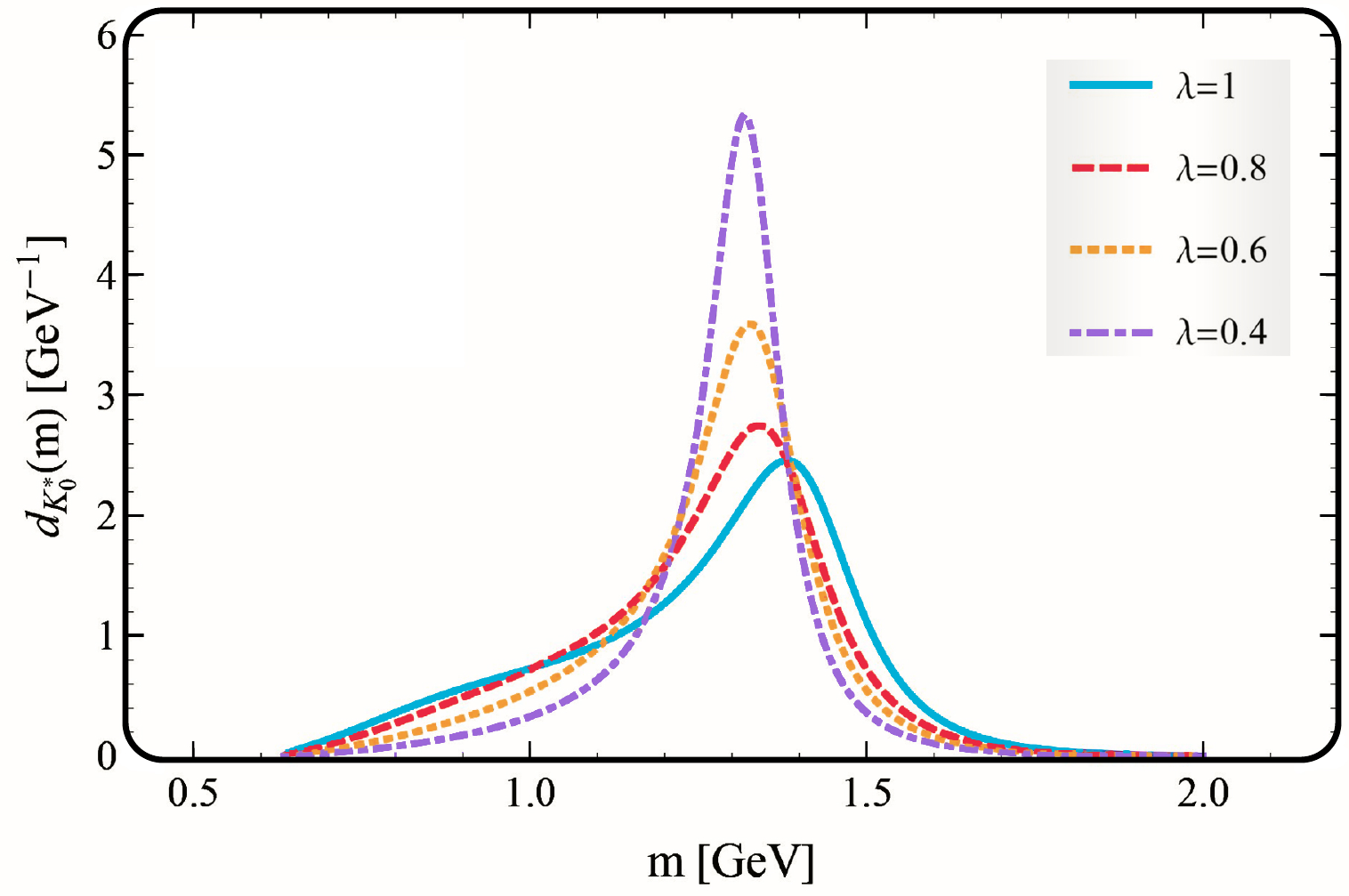}\\\textit{b)}
\end{minipage}
\end{center}
\caption{Panel (a) shows the normalized scalar kaonic spectral function of Eq (\ref{sftheoretical}) plotted for the model parameters reported in Table \ref{fittingparameters}. It is clearly visible that, besides the peak corresponding to $K^*_0(1430)$, an additional broad structure in the low-energy sector is present. Panel (b) shows the shape of the scalar kaonic spectral function in dependence of the scaling parameter $\lambda=3/N_c$. The spectral function becomes more peaked and the enhancement decreases with decreasing $\lambda$.}
\end{figure}

In panel (a) of Figure \ref{sfscalar} we present the normalized scalar kaonic spectral function defined in Eq. (\ref{sftheoretical}) and plotted for the model parameters of Table \ref{fittingparameters}. We stress that the resonance $K^*_0(1430)$ is clearly visible in the scalar spectral function as the unique peak in the energy region close to 1.4 GeV, but there is no peak for light $\kappa$. However, even if only one peak is observed, it turns out that one cannot approximate the shape of the spectral function by the standard Breit-Wigner type, since a broad left hand-side enhancement, identified  with $K^*_0(700)$, is present. The existence of a second, additional companion pole corresponding to the broad distortion in the spectral function is a natural consequence of our theoretical model. 

Interestingly, a similar situation holds in case of broad $f_0(500)$ state in the isoscalar sector: even if its pole is commonly accepted, no corresponding peak is present in the scalar-isoscalar spectral function. In contrast, one can distinguish two scalar resonances $f_0(980)$ and $a_0(980)$ which are quite narrow. Despite the fact that their couplings to kaons are known to be large, these states are situated just at the KK threshold, thus their decays into two kaons are kinematically suppressed. We summarize that the common origin of all those resonances as well as the light $\kappa$ can be explained by mesonic quantum fluctuations. Just as $K^*_0(700)$, the resonances $f_0(500)$, $f_0(980)$ and $a_0(980)$ can be understood as dynamically generated companion poles, see Ref. \cite{pealaezrev4} and refs. therein for details.

In order to understand better the nature of both resonances $K^*_0(1430)$ and $K^*_0(700)$ we study the behavior of the spectral function and position of the poles in the limit in which the number of colors $N_c$ is large. Such theoretical procedure is very useful since for large $N_c$ values conventional mesons become stable. As it was already presented in Eq. (\ref{scalmmnc}) a dimensionless scalling parameter $\lambda$, related to the $N_c$ reads:
\begin{equation}
\lambda=\frac{3}{N_c}\text{ , } \label{nc}
\end{equation}
which enters into our model by replacing the coupling constants in the following way:
\begin{equation}
a \rightarrow \sqrt{\lambda}a, \hspace{0.5cm} b \rightarrow \sqrt{\lambda}b \hspace{0.1cm}.
\end{equation}
As a consequence, by setting $\lambda=0$ $(N_c \rightarrow \infty)$ we obtain the spectral function which corresponds to the non-interacting scalar kaonic seed state, thus a delta function peaked at the seed mass. For the $\lambda=1$ $(N_c=3)$ we reobtain our physical results. By changing the parameter $\lambda$ in the range from 1 to 0, one can successively decrease the interaction and check in a controlled manner the behavior of the spectral function. In panel (b) of Figure \ref{sfscalar} we present the spectral function of $K^*_0(1430)$ plotted for different values of $\lambda$. One can observe that, the smaller $\lambda$ (thus the larger $N_c$), the interaction gets smaller and the peak related to $K^*_0(1430)$ becomes narrower and higher, reducing to a Dirac delta function in the large-$N_c$ limit. Moreover, the broad enhancement identified with $K^*_0(700)$ state becomes less pronounced and completely dissapears for increasing $N_c$ (thus for decreasing interaction). 

\begin{figure}[h!]
\begin{center}
\includegraphics[width=0.65 \textwidth]{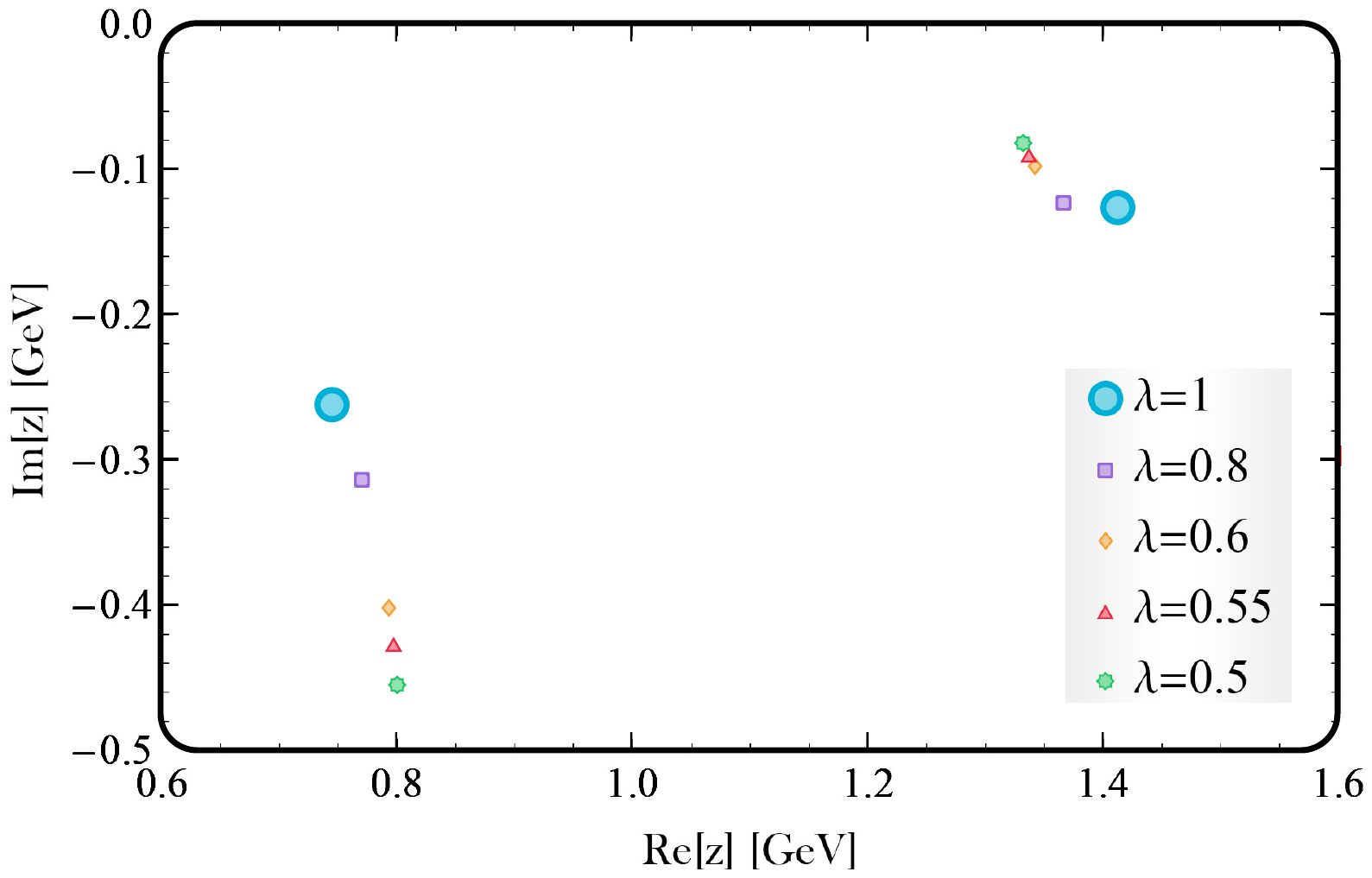} 
\caption{\label{polemovementkappa} The movement of both the seed state and companion pole for five different values of $\lambda$ parameter defined in Eq. (\ref{nc}). The pole positions corresponding to the physical case (thus $\lambda=1$) we indicate by big blue dots.}
\end{center}
\end{figure}

Furthermore, we repeat the same procedure in order to study the behavior of the position of the poles in the complex plane. The results are shown in Figure \ref{polemovementkappa}. 
The large-$N_c$ study reveals that the poles for both $K^*_0(1430)$ and $K^*_0(700)$ move. The former goes towards the real axis, which is the characteristic behavior for a conventional $q\bar{q}$ mesonic state. On the contrary, the latter goes away from it, and dissapears on the II RS for $\lambda\leqslant 0.24$, thus for $N_c\geqslant 13$. We conclude that the pole corresponding to $K^*_0(700)$ is generated dynamically and does not survive in the limit of large-$N_c$. A similar threshold behavior has also been found in Refs. \cite{Pelaez:2004xp, Zhou:2010ra, Guo:2011pat, Guo:2012yt}.

\section{Modification of the model}

\begin{figure}[h]
\begin{center}
\begin{minipage}[b]{7.8cm}
\centering
\includegraphics[width=7.7cm]{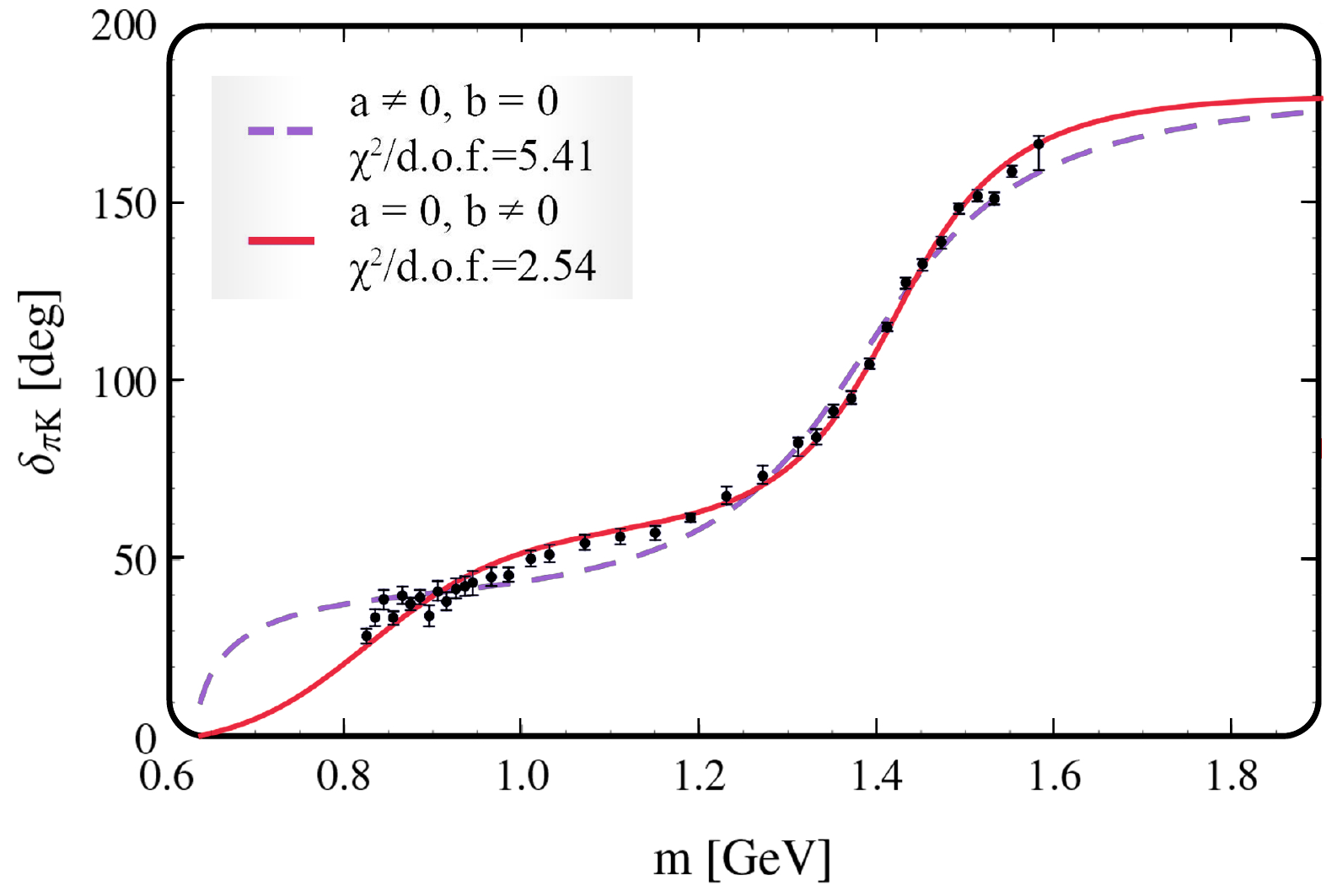}\\\textit{a)}
\end{minipage}
\begin{minipage}[b]{7.8cm}
\centering
\includegraphics[width=7.7cm]{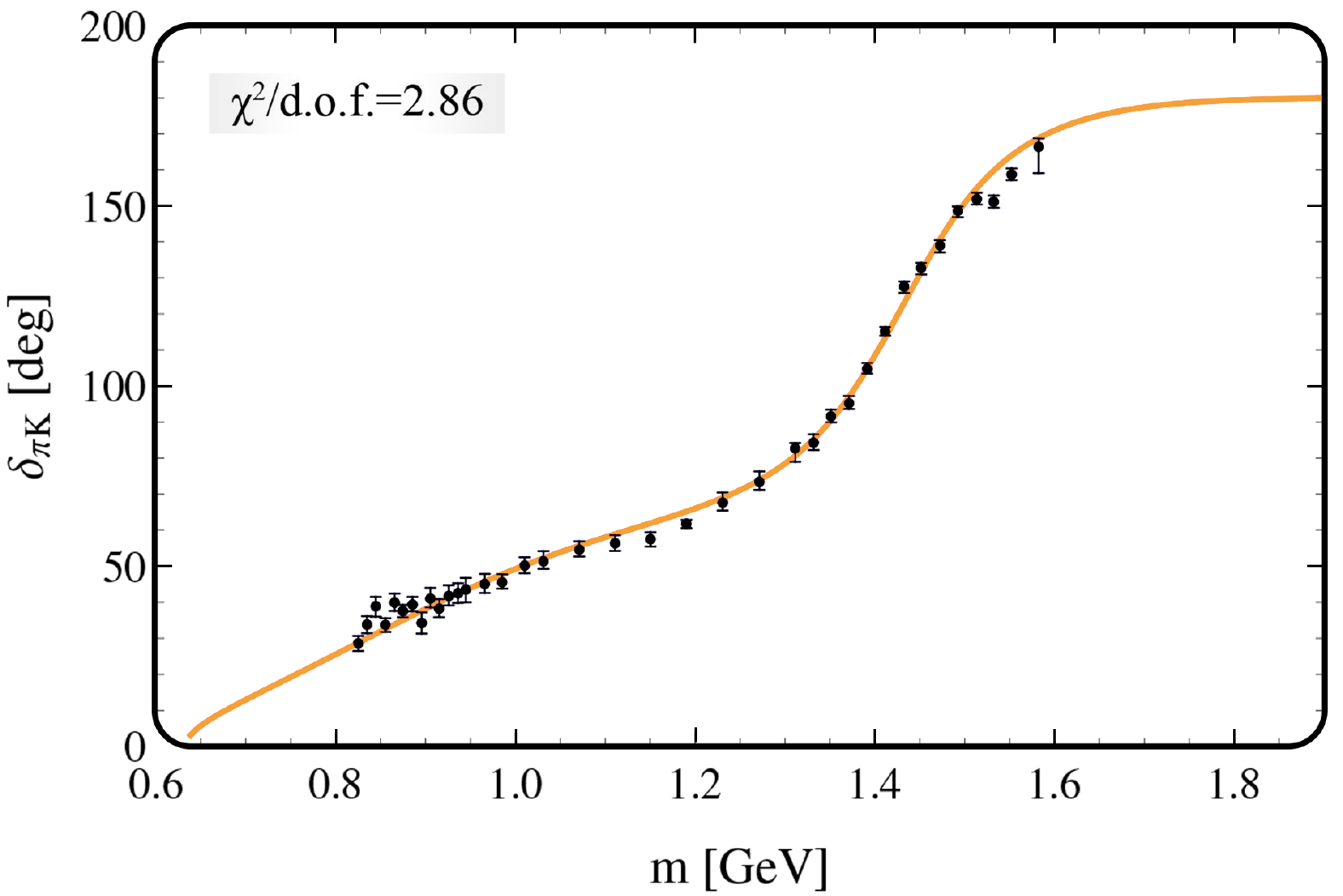}\\\textit{b)}
\end{minipage}
\end{center}
\caption{Panel (a): the solid (red) curve corresponds to the case when only derivative terms are retained in the Lagrangian of Eq. (\ref{intLag}), while the dashed (purple) curve corresponds to the non-derivative case. The black points are the experimental data of Ref. \cite{Aston88}. Panel (b): the solid (orange) curve presents the fitting results for the modified vertex function in Eq. (\ref{vertdipolar}).} \label{3fits}
\end{figure}

We explore various scenarios of our model in order to understand more precisely the results discussed above. As a first step, we consider two simplified forms of the Lagrangian of Eq. (\ref{intLag}), each one containing only one type of interaction term (with derivative or without). In order to obtain the Lagrangian with the non-derivative term only, we set $b=0$ in Eq. (\ref{intLag}). Similarly, in order to obtain the Lagrangian involving only the derivative term we set $a=0$ in the same expression. Then, we perform two additional fits to phase-shift data. The fitting results are illustrated in Figure \ref{3fits}.

In Table \ref{variations} we summarize the outcomes for these alternative scenarios. In the first entry the main results of Sec. \ref{resk01430r} are reported. The results for the non-derivative case ($b=0$) are reported in the second entry. As it is shown in the fifth column, the value $\chi_0^2/d.o.f$ has sizably increased: no satisfactory description can be achieved in this case. In the third entry we present the scenario in which only the derivative term is retained in the Lagrangian $(a=0)$. Even if the value of $\chi_0^2/d.o.f$ is clearly worsened, an qualitative acceptable description of data is still obtained. 

In order to evaluate the validity of performed fits we use the statistical test for the goodness of the fit (see last column of Tab. \ref{variations}). To this end, we introduce the following quantity \cite{taylormmpp3}:
\begin{equation}
p(\chi^2 > \chi^2_0)=\frac{1}{2^{d/2}\Gamma (d/2)} \int \limits^{\infty}_{\chi^2_0}dx \hspace{0.1cm} x^{\frac{d}{2}-1} e^{-x/2} \text{ ,}
\end{equation}
(with $d=d.o.f.$) which stands for the probability to obtain a $\chi^2$ larger than $\chi^2_0$ when performing a new scattering experiment (obviously, without changing the theoretical function for the fit). The very small value of this probability indicates that $(i)$ our theoretical model is not completely correct (the most reasonable conclusion) or $(ii)$ our theoretical model is correct, but quite unluckily, statistical fluctuations have an influence on the experimental results. For instance, when this probability is smaller than $5 \%$, then the thoretical model can be excluded at $95 \%$ confidence level. To be more quantitative, for our original model (with both derivative and non-derivative terms in the Lagrangian and Gaussian form factor) we have $p(\chi^2>\chi^2_0)=15.3 \%$ (with $d.o.f.= 37-4=33$). This result confirms that one cannot reject our theoretical model. The model containing only non-derivative interaction terms gives a very small value of this probability, which is expected because the experimental data are not satisfactorily covered by the theoretical function. For what concern the model with only derivative interaction terms the corresponding value of $p(\chi^2 > \chi^2_0)$ is also very small (here, $d.o.f.=37-3=34$). However, in this case the result may seem surprising at a first sight because the form of the theoretical function match the data. Yet, the results of the statistical analysis implies that both modified models can be excluded with a good level of accuracy.

\begin{sidewaystable}
\renewcommand{\arraystretch}{1.53}
\par
\makebox[\textwidth][c] {
\newcommand{\minitab}[2][l]{\begin{tabular}{#1}#2\end{tabular}}
\begin{tabular}
[c]{llcccc}\hline\hline
$\hspace{0.2cm}$ Scenario & $\hspace{0.4cm}$ Parameters & Poles for $K^*_0(1430)$& Poles for $K^*_0(700)$&$\chi^2_0/d.o.f.$&$p(\chi^2> \chi^2_0)$\\ \hline \hline
\multirow{4}{*}{\minitab[c]{$a, b \neq 0$,\\ Gaussian}}& $a=1.60 \pm 0.22$&\multirow{4}{*}{\minitab[c]{$\hspace{0.45cm}(1.413 \pm 0.002$)\\$-i(0.127 \pm 0.003)$}}&\multirow{4}{*}{\minitab[c]{$\hspace{0.45cm}(0.746 \pm 0.019)$\\$-i(0.262 \pm 0.014)$}}&\multirow{4}{*}{1.25}&\multirow{4}{*}{0.15}\\
&$b=-11.16 \pm 0.082$&&& & \\
&$\Lambda=0.496 \pm 0.008$&&&&\\
&$M_0=1.204 \pm 0.008$&&&&\\ \hline
\multirow{3}{*}{\minitab[c]{$b=0$,\\ Gaussian}}&$a=4.06 \pm 0.04$& \multirow{3}*{\minitab[c]{$\hspace{0.45cm} (1.385 \pm 0.002$) \\$-i(0.146 \pm 0.003)$}}& \multirow{3}{*}{$-$}&\multirow{3}{*}{5.41}&\multirow{3}{*}{$1.72 \cdot 10^{-22}$}\\
&$\Lambda=0.902 \pm 0.015$&&&&\\
&$M_0=1.299 \pm 0.002$&&&\\
\hline
\multirow{3}{*}{\minitab[c]{$a=0$,\\ Gaussian}}&$b=-17.10 \pm 0.17$& \multirow{3}*{\minitab[c]{$\hspace{0.45cm} (1.419 \pm 0.001$) \\$-i(0.112 \pm 0.002)$}}& \multirow{3}{*}{\minitab[c]{$\hspace{0.45cm}(0.820 \pm 0.003)$\\$-i(0.187 \pm 0.002)$}}&\multirow{3}{*}{2.54}&\multirow{3}{*}{$1.92 \cdot 10^{-6}$}\\
&$\Lambda=0.453 \pm 0.002$&&&&\\
&$M_0=1.142 \pm 0.002$&&&\\ \hline
\multirow{4}{*}{\minitab[c]{$a, b \neq 0$,\\ $F(m)=e^{\frac{-k^4(m)}{\Lambda^4}}$}}& $a=2.32 \pm 0.09$&\multirow{4}{*}{\minitab[c]{$\hspace{0.45cm}(1.433 \pm 0.002$)\\$-i(0.112 \pm 0.003)$}}&\multirow{4}{*}{\minitab[c]{$\hspace{0.45cm}(0.863 \pm 0.008)$\\$-i(0.339 \pm 0.017)$}}&\multirow{4}{*}{2.86}&\multirow{4}{*}{$7.98 \cdot 10^{-8}$}\\
&$b=-3.40 \pm 0.026$&&& & \\
&$\Lambda=0.652 \pm 0.006$&&&&\\
&$M_0=1.248 \pm 0.003$&&&&\\
\hline \hline
\end{tabular}
}\caption{Poles for both resonances $K^{*}_0(1430)$ and $K^*_0(700)$, as well as parameters a, $\Lambda$ and $M_0$ are given in units of GeV while b in unit of GeV$^{-1}$.} \label{variations}.
\end{sidewaystable}

For completeness, in the third and fourth columns and for each scenario, we report the position of the poles for both $K^*_0(1430)$ and $K^*_0(700)$ resonances. 

Another important issue requiring a more detailed discussion is the choice of the form factor. As already mentioned in Chapter \ref{chapvec}, the Gaussian form factor originally employed in our model is relatively easy to use and rather standard in numerous works on the subject. However, if any other function (on the positive real axis) goes to zero smoothly but sufficiently fast, can be used as a cutoff function. As a test, we modified our model by using the following form factor:
\begin{equation}
F_{\Lambda}(m)=e^{-2k^4(m)/ \Lambda^4}, \label{vertdipolar}
\end{equation}
where $k(m)$ is given in Eq. (\ref{threemomentumpionkaon}).
The results of the fit are presented in panel (b) of Figure \ref{3fits} and in the last row of Table \ref{variations}. Also for this scenario we observe that the theoretical model is in qualitative agreement with the experimental data. Yet, the statistical test shows that one should reject this scenario. Correspondingly, the value of the pole for $\kappa$ obtained in this model is not compatible with the result of our preffered model (first line of Table \ref{variations}) and with other values quoted by the PDG. We conclude that changing the cutoff function does not provide a satisfactory description of experimental data.

Moreover, another form factor, the so-called Fermi function 
\begin{equation}
F_{\Lambda}(m)=[(1+e^{-\alpha \Lambda^2})/(1+e^{\alpha(k^2(m)-\Lambda^2)})]
\end{equation}

has been tested for different values of the parameter $\alpha$. Also this choice turns out to be not sufficient for the description of the data. The obtained fit does not pass the statistical analysis of the $\chi^2$.

The study of the variations of the model leads us to the conclusion that our original model with the Gaussian form factor and the Lagrangian containing simultaneously both a (dominant) derivative and a (subdominant) non-derivative terms describes accurately the experimental data in contrast to the other scenarios tested here.

\section{Concluding remarks}

In this chapter within our effective model we have studied the $I(J^P)=\frac{1}{2}(0^+)$ channel with two resonances: a conventional $q\bar{q}$ state $K^*_0(1430)$ and a non-conventional state $K^*_0(700)$. In our approach, in agreement with other effective approaches of low-energy QCD, both derivative and non-derivative interactions are taken into account. It was shown that, starting from a single quark-antiquark kaonic field, it is possible to describe both states $K^*_0(1430)$ and $K^*_0(700)$, since the propagator develops two poles, which are needed to reproduce correctly the experimental kaon-pion phase-shift data. We have explored the spectral function, which turns out to be quite different from a Breit-Wigner form, since strong deformations, corresponding to the light $\kappa$, emerge in the low-energy region. 

In the large-$N_c$ limit one observes that the enhancement, corresponding to $K^*_0(700)$ (as well as its dynamically generated pole) disappears when $N_c$ is large enough. At the same time, the pole linked to $K^*_0(1430)$ is present and move toward the real axis. This leads us to conclusion that $K^*_0(1430)$ behaves as a typical $q\bar{q}$ state, while $K^*(700)$ is a non-conventional dynamically generated state (a companion pole within our framework). 

We underline that the inclusion of the derivative interaction term into the Lagrangian is essential for our results since it turns out to be the dominant one for the description of the experimental $K\pi$ phase-shift data. Nevertheless, to achive a satisfactory fit, also the contribution of the non-derivative interaction is needed. A reliable fit cannot be obtained by using only one type of interaction term. Moreover, the Gaussian form factor turns out to be the best choice for the description of data. 

\chapter{Charmonium vector states $\mathbf{\psi(4040)}$ and~$\mathbf{Y(4008)}$ }
\label{chaptpsiy}
In this chapter we investigate the well-established conventional $c\bar{c}$ resonance $\psi(4040)$, which is considered as mainly a $\psi(3S)$ state corresponding to quantum numbers  $3$ $^{3}S_{1}$. To this end we use an effective QFT model in which the decays of $\psi(4040)$ into the $DD$, $DD^*$, $D^*D^*$, $D_sD_s$ and $D_sD_s^*$ channels are included. The evaluated spectral function deviates sizably from the relativistic Breit-Wigner shape. This is due to the dynamically generated enhancement below $4$ GeV, which emerges mostly through $DD^*$ loops. Moreover, two poles are present on the complex plane: one for the standard charm-anticharm seed state, identified with $\psi(4040)$, and the second corresponding to a broad enhancement appearing in the spectral function. At first glance, one can be tempted to identify this additional pole with the puzzling state $Y(4008)$, observed by the Belle experiment in the $\pi^+ \pi^- J/\psi$ channel. Yet, the imaginary part of the dynamically generated pole is too small when compared to $Y(4008)$. Moreover, a deeper analysis reveals that a different mechanism, completely independent of the presence of the dynamically generated pole, is at work: a broad enhancement peaked at about $4$ GeV emerges in the decay chain $e^+e^- \rightarrow \psi(4040) \rightarrow D^*D \rightarrow \pi^+ \pi^- J/\psi$ as a consequence of the $D^*D$ loop. Therefore, the enigmatic state $Y(4008)$ is not a genuine resonance (or pole), but rather an enhancement appearing when the resonance $\psi(4040)$ decays via $DD^*$ loops into the $\pi^+ \pi^- J/\psi$ channel. 

\section{Introduction to the charmonium states}

In the past decades there has been a huge progress in the understanding of the charmonium sector, mainly due to the data collected by $B-$factories experiments. The new observations provided valuable informations about new resonances. Some of them appear as conventional charmonium mesons, (thus, standard c$\bar{c}$ states) whereas others, named as $X, Y$ and $Z$ states, exhibit unusual behavior, suggesting their non-conventional nature, typical for hybrids, molecules, multiquarks or glueballs, see Refs. \cite{rev, rev2016, pillonireview, nielsen}. 

In this chapter, we investigate the vector charmonium sector covering the energy region around $4$ GeV where the well-known resonance $\psi(4040)$ is located. According to the PDG \cite{pdg}, this conventional $c\bar{c}$ state is described by the vector quantum numbers $J^{PC}=1^{--}$.  Moreover, it has spin 1, the principal number 3 and the angular momentum 0, which in the non-relativistic spectroscopic notation corresponds to $n$ $^{2S+1}L_{J} =$  $3$ $^{3}S_{1}$, see Refs. \cite{GodfreyIsgur, eichten4040, Eichten:1979mp, segoviarevmp}. The PDG reports for $\psi(4040)$ a mass of $(4039 \pm 1)$ MeV and a total width of $(80 \pm 10)$ MeV. 

Close to $\psi(4040)$, the puzzling and still not confirmed state $Y(4008)$, with mass $4008 \pm 40^{+114}_{-28}$ MeV and total width $\Gamma=226 \pm 44 \pm 87$ MeV, was observed by the Belle Collaboration. $Y(4008)$ appears as a broad enhancement in the cross-section of the reaction $e^+e^- \rightarrow  J/ \psi \pi^+ \pi^-$, measured by using the initial state radiation (ISR) production process \cite{belle1}. Later on, the same group confirmed this observation in updated measurements \cite{belle2}. Nevertheless, the subsequent experiments at BESIII \cite{besiii} and BaBar \cite{babar4008} could not confirm its existence even if the same ISR-technique was used. Taking into account the above discussion and knowing that there was a limited statistic of the data at Belle, the existence of $Y(4008)$ is rather questionable. Despite all that, this observation initiated an extensive theoretical discussions trying to explain its nature. The most common scenarios concerned the non-conventional character of $Y(4008)$. It was therefore proposed to interpret it as a $\bar{D}^*D^*$ molecule \cite{molecular, molecular2}, as a  tetraquark \cite{tetraquark1, tetraquark2}, and also as an interference effect with the background \cite{interference}. Another possibility, suggesting that $Y(4008)$ is a $\psi(3S)$ charmonium state \cite{charmstate1, charmstate2}, is not preffered, since $\psi(4040)$ seems to be a much better candidate for this assignment. The status of the enigmatic $Y(4008)$ enhancement has not yet been clarified, hence it becomes important to improve our understanding by studying the nearby energy regime.

Following the idea presented in Chapter \ref{kappak}, where we have found that the light $K^*_0(700)$ appears as a companion pole of the heavier $K^*(1430)$ resonance, we investigate if one can describe simultaneously both states $\psi(4040)$ and $Y(4008)$ by using a similar mechanism. It is also interesting that two poles have been observed when studying $\psi(3770)$ charmonium resonance \cite{scoito3770}. In analogy to that, we construct a QFT model in which a unique $c\bar{c}$ seed state corresponding to $\psi(4040)$ is coupled to the $DD$, $D^*D$, $D^*D^*$, $D_s D_s$, $D_s D_s^*$ channels. As we shall demonstrate in this chapter, there are many similarities between this analysis and that one for the scalar sector, but significant differences and new aspects do emerge.

\section{The model Lagrangian}

In our model the two-body decays of a single charm-anticharm seed state identified with $\psi(4040)$ resonance are described by an effective relativistic interaction Lagrangian which contains three types of terms:
\begin{equation}
\mathcal{L}_{\psi(4040)}=\mathcal{L}_{VPP}+\mathcal{L}_{VPV}+\mathcal{L}%
_{VVV} \text{ .} \label{lagpsi4040}%
\end{equation}
The first term refers to the decay channels into two pseudoscalar mesons (with corresponding products such as $DD$ and $D_sD_s$). The second one stays for the decays into one pseudoscalar and one vector meson ($DD^*$ and $D_s^*D_s$ channels). Finally, the last one couples the vector seed state to two vector mesons (decays into $D^*D^*$). Each term of the Lagrangian can be written explicitly as:
\begin{equation}
\mathcal{L}_{VPP}=ig_{\psi DD}\psi_{\mu}\left[  \left(  \partial^{\mu}%
D^{+}\right)  D^{-}+\left(  \partial^{\mu}D^{0}\right)  \bar{D}^{0}+\left(
\partial^{\mu}D_{s}^{+}\right)  D_{s}^{-}\right]  +h.c.\text{ },\label{lag1psi4040}%
\end{equation}%
\begin{equation}
\mathcal{L}_{VPV}=ig_{\psi D^{\ast}D}\tilde{\psi}_{\mu\nu}\left[
\partial^{\mu}D^{\ast+\nu}D^{-}+\partial^{\mu}D^{\ast0\nu}\bar{D}^{0}%
+\partial^{\mu}D_{s}^{\ast+\nu}D_{s}^{-}\right]  +h.c.\text{ },\label{lag2psi4040}%
\end{equation}%
\begin{equation}
\mathcal{L}_{VVV}=ig_{\psi D^{\ast}D^{\ast}}\left[  \psi_{\mu\nu}\left(
D^{\ast+\mu}D^{\ast-\nu}+D^{\ast0\mu}\bar{D}^{\ast0\nu}+D_{s}^{\ast+\mu}%
D_{s}^{\ast-\nu}\right)  \right]  +h.c.\text{ }.\label{lag3psi4040}%
\end{equation}
The three quantities in the above expressions, denoted as $g_{\psi DD}$, $g_{\psi D^*D}$ and $g_{\psi D^*D^*}$ stand for the coupling constants (with dimensions Energy$^0$, Energy$^{-1}$ and Energy$^{0}$, respectively) whose numerical values are determined from the experimental data reported in the PDG \cite{pdg}. Moreover, $\psi_{\mu\nu}$ and $\tilde{\psi}_{\mu\nu}$ are defined in the usual way:
\begin{equation}
\psi_{\mu\nu}= \partial_{\mu}\psi_{\nu}-\partial_{\nu}\psi_{\mu} \text{ ,}
\end{equation}
\begin{equation}
\tilde{\psi}_{\mu\nu}=\frac{1}{2}\varepsilon_{\mu \nu \rho \sigma} \psi^{\rho \sigma} \text{ .}
\end{equation}
For completeness, all the decay processes studied in our model as well as the masses of the decay products are reported in Table \ref{tableprocmass}. One should stress that the decay channel into $D_s^*D^*_s$ is kinematically forbidden and is not considered in our analysis. 
\begin{table}[htbp] 
\centering
\renewcommand{\arraystretch}{1.35}
\begin{tabular}{c|c||c} 
\hline
Term of the Lagrangian& Decay processes & Masses of the decay products [MeV]\\ \hline
\multirow{ 3}{*}{$\mathcal{L}_{VPP}$} & $\psi(4040) \rightarrow D^+ D^-$ &   \\
& $\psi(4040) \rightarrow D^0\bar{D}^0$& $m_{D^+}=m_{D^-}=1869.65 \pm 0.05$ \\ 
&$\psi(4040) \rightarrow D_s^+D_s^-$&$m_{D^0}=m_{\bar{D}^{0}}=1864.83 \pm 0.05$\\
\cline{1-2}
\multirow{ 3}{*}{$\mathcal{L}_{VPV}$} & $\psi(4040) \rightarrow D^{*0} \bar{D}^0+ h.c.$ & $m_{D^{*0}}=m_{\bar{D}^{*0}}=2006.85 \pm 0.05$ \\
& $\psi(4040) \rightarrow D^{*+} D^- + h.c.$& $ m_{D^{*+}}=m_{D^{*-}}=2010.26 \pm 0.05$ \\ 
&$\psi(4040) \rightarrow D_s^{*+}D_s^- +h.c.$&$m_{D_s^+}=m_{D_s^-}=1968.34 \pm 0.07$\\
\cline{1-2}
\multirow{ 2}{*}{$\mathcal{L}_{VVV}$} & $\psi(4040) \rightarrow D^{*+} D^{*-}$ & $m_{D_s^{*+}}=m_{D_s^{*-}}=2112.2 \pm 0.4$ \\
& $\psi(4040) \rightarrow D^{*0} \bar{D}^{*0}$&  \\ 
\hline
\end{tabular}
\caption{Decay processes described by each term of the Lagrangian of Eq. (\ref{lagpsi4040}) with the masses of the products of these decays. The numerical values for the masses are taken from PDG \cite{pdg}.}
\label{tableprocmass}
\end{table}

As it was presented in Sec. \ref{decaywidthsall}, an explicit calculation from the Feynman rules led us to the following theoretical formulas for the decay widths. Depending on the type of decay they are:
\begin{equation}
\Gamma_{\psi\rightarrow D^{+}D^{-}+h.c}(m)=\frac{\left[  k(m,m_{D^{+}%
},m_{D^{-}})\right]  ^{3}}{6\pi m^{2}}g_{\psi DD}^{2}F_{\Lambda}%
(m)\hspace{0.1cm}, \label{deeec1}
\end{equation}%
\begin{equation}
\Gamma_{\psi\rightarrow D^{\ast+}D^{-}+h.c}(m)=\frac{2}{3}\frac{\left[
k(m,m_{D^{\ast+}},m_{D^{-}})\right]  ^{3}}{\pi}g_{\psi D^{\ast}D}%
^{2}F_{\Lambda}(m)\hspace{0.1cm}, \label{deeec2}
\end{equation}%
\begin{equation}
\Gamma_{\psi\rightarrow D^{\ast+}D^{\ast-}}(m)=\frac{2}{3}\frac{\left[
k(m,m_{D^{\ast+}},m_{D^{\ast-}})\right]  ^{3}}{\pi m_{D^{\ast+}}^{2}}g_{\psi
D^{\ast}D^{\ast}}^{2}\left[  2+\frac{\left[  k(m,m_{D^{\ast+}},m_{D^{\ast-}%
})\right]  ^{2}}{m_{D^{\ast+}}^{2}}\right]  F_{\Lambda}(m)\hspace{0.1cm}. \label{deeec3}
\end{equation}
One can notice that each decay width given above is a function of $m$ being the `running' mass of the unstable $\psi$ state. Upon setting it to the nominal mass of $\psi(4040)$ quoted by the PDG \cite{pdg}, thus $m=m_{\psi(4040)}=4.039$ GeV, the on-shell decay widths are obtained. The momentum $k(m)$ was already defined in Eq. (\ref{threemomentum892}).

Next, we turn to the form factor $F_{\Lambda}$ which regularizes our model. Analogously to the previous studies for $K^*(892)$ in the vector sector and light $\kappa$ in the scalar kaonic sector, here we also use the Gaussian function
\begin{equation}
F_{\Lambda}\equiv F_{\Lambda}^{\text{Gauss}}(m)=e^{-2\frac{m^{2}}{\Lambda^{2}%
}}\text{ .}\label{gaussianpsi4040}%
\end{equation}
As it was already mentioned in this thesis, this choice is not unique. For completeness, a dipolar form factor
\begin{equation}
F_{\Lambda}\equiv F_{\Lambda}^{\text{Dipolar}}(m)=\left(  1+\frac{m^{4}%
}{\Lambda^{4}}\right)  ^{-2}\label{dipolarpsi4040}%
\end{equation}
has been also used. We will show later that the results do not depend strongly on the form of the vertex function. Nevertheless, one should be careful with the parameter $\Lambda$ entering Eq. (\ref{gaussianpsi4040}) and Eq. (\ref{dipolarpsi4040}), since its numerical value is important. In our case, $\Lambda$ is between the range from $0.4$ GeV to $0.8$ GeV. This is a typical value for mesonic objects since we have obtained $\Lambda$ to be $\sim 0.5$ GeV for the light $K^*_0(700)$ when performing a fit to experimental data. Moreover, in Ref. \cite{scoito3770} it was found that $\Lambda \approx 0.3$ GeV, but also for a value of around $0.4$ GeV the results are in agreement with data. However, in case of the $^3P_0$ model a value of $0.8$ GeV was found. Although, this microscopic formalism is rather common to study masses and decays, but not necessarily to compute mesonic loops. Since the cutoff $\Lambda$ is not clearly known, we tested different values of this parameter for both types of the form factor. As we shall see later in Appendix \ref{apppsi4040}, only the results for $\Lambda$ up to $0.6$ GeV are physically acceptable. 

On the other hand, mathematical consistency does not imply that our model is physically valid at arbitrarily high momenta, since only one resonance, the $\psi(4040)$ is considered. Our approach can be employed to describe the energy region around $4$ GeV. To go beyond it, one needs to take into account further resonances, such as $\psi(4160)$, and even $\psi(4415)$. For completeness, we performed the calculations for two resonances $\psi(4040)$ and $\psi(4160)$ being simultaneously present. We conclude that there are no significant differences, then we shall omit $\psi(4160)$ in the following. 

\begin{figure}[h!]
\begin{center} 
\includegraphics[width=0.55 \textwidth]{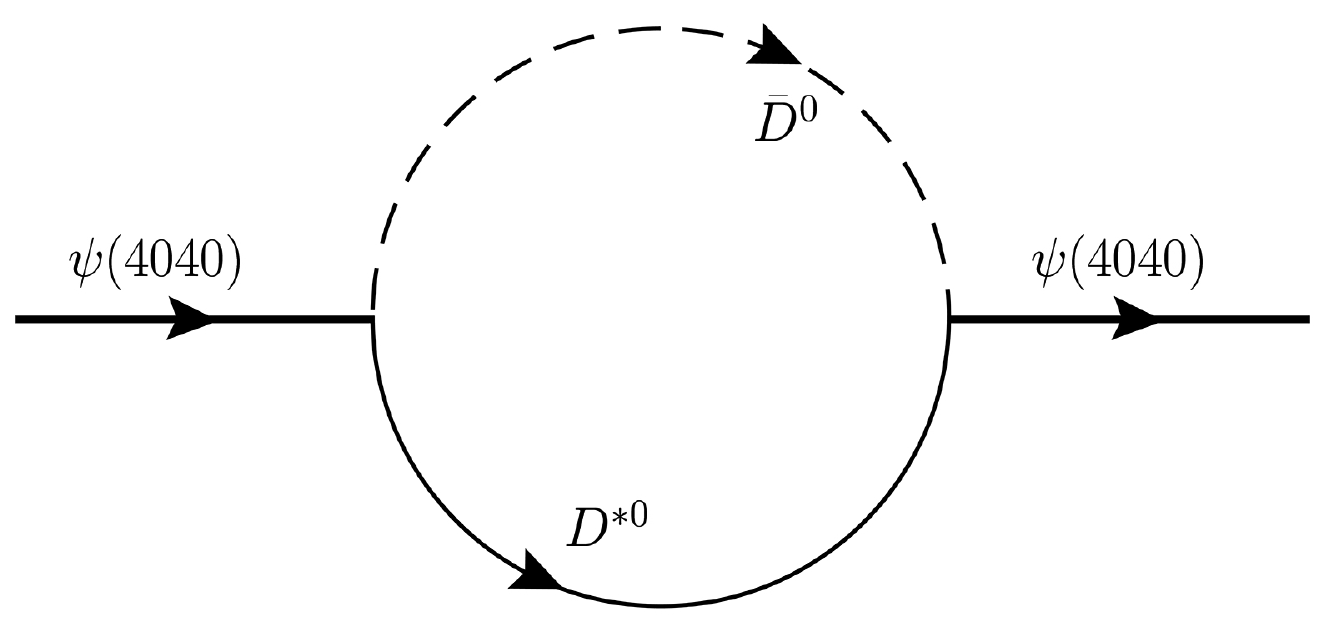} 
\caption{\label{dstardzeroloop} Schematic illustration of one-loop contribution shown for the case of $\bar{D}^0$ and $D^{*0}$.}
\end{center}%
\end{figure}

Following the same formalism used in previous chapter to describe the $K^*_0(700)$, we introduce the propagator for a vector field $\psi_{\mu}$, whose scalar part is given by
\begin{equation}
\Delta_{\psi}(p^{2}=m^{2})=\frac{1}{m^{2}-M_{\psi}^{2}+\Pi(m^{2}%
)+i\varepsilon}\text{ ,}\label{proppsi4040}%
\end{equation}
where $M_{\psi}$ is the bare mass of the $c\bar{c}$ seed state, identified with $\psi(4040)$. The quantity $\Pi(m^2)$ in the denominator is the self-energy, which is a complex function. Explicitly, at the one-loop level, $\Pi(m^2)$ can be written as sum of all possible one-loop contributions:
\begin{align}
\Pi(m^{2}) &  =\Pi_{D^{+}D^{-}}(m^{2})+\Pi_{D^{0}\bar{D}^{0}}(m^{2}%
)+\Pi_{D_{s}^{+}D_{s}^{-}}(m^{2})+\Pi_{D^{\ast0}\bar{D}^{0}+h.c}%
(m^{2})\nonumber\\
&  +\Pi_{D^{\ast+}D^{-}+h.c}(m^{2})+\Pi_{D_{s}^{\ast+}D_{s}^{-}+h.c}%
(m^{2})+\Pi_{D^{\ast0}\bar{D}^{\ast0}}(m^{2})+\Pi_{D^{\ast+}D^{\ast-}}%
(m^{2})+... \text{ .}
\end{align}
The dots in the above equation represent all the other contributions coming from small decay channels. As an illustrative example, in Figure \ref{dstardzeroloop} we present the contribution of the $D^{*0}\bar{D}^0$ loop.

Moreover, one can factorize out the coupling constants. For instance, in the $D^+D^-$ case:
\begin{align}
\Pi_{D^+D^-}(m^2)=g^2_{\psi DD}\Sigma_{D^+D^- +h.c.}(m^2) \text{ .}
\end{align}
Similar definitions can be written for the other channels. At one-loop the quantity $\Sigma_{D^+D^- +h.c.}(m^2)$ is independent of the coupling constant $g_{\psi DD}$. 

In order to find the self-energy $\Pi(m^2)=\operatorname{Re}(\Pi(m^2))+i \operatorname{Im}(\Pi(m^2))$ it is convenient to consider its real and imaginary part separately. The real part $\operatorname{Re}(\Pi(m^2))$ can be computed by using the dispersion relation. For example, for the decay into $D^+D^-$ one has
\begin{equation}
\operatorname{Re}(\Pi_{D^{+}D^{-}}(m^{2}))=-\frac{1}{\pi} \mathcal{P} \int
\limits_{2m_{D^{+}}}^{\infty}2m^{\prime}\frac{m^{\prime}\Gamma_{\psi
(4040)\rightarrow D^{+}D^{-}}(m^{\prime})}{m^{2}-m^{\prime2}+ i \varepsilon}\mathrm{dm}%
^{\prime}\text{ .}%
\end{equation}
Similar expressions hold for all the other decay channels.
For what concerns the imaginary part $\operatorname{Im}(\Pi(m^2))$, the optical theorem \cite{peskinQFT}, implies that
\begin{align}
\operatorname{Im}(\Pi(m^{2})) &  =m\Big(\Gamma_{\psi(4040)\rightarrow
DD}(m)+\Gamma_{\psi(4040)\rightarrow D_{s}D_{s}}+\Gamma_{\psi(4040)\rightarrow
D^{\ast}D}(m)\nonumber\\
&  +\Gamma_{\psi(4040)\rightarrow D_{s}^{\ast}D_{s}}(m)+\Gamma_{\psi
(4040)\rightarrow D^{\ast}D^{\ast}}(m)\Big)\text{ ,}%
\end{align}
where:
\begin{equation}
\Gamma_{\psi(4040)\rightarrow DD}(m)=\Gamma_{\psi(4040)\rightarrow D^{+}D^{-}%
}(m)+\Gamma_{\psi(4040)\rightarrow D^{0}\bar{D}^{0}}(m)\text{ ,}\label{ddpsi4040}%
\end{equation}%
\begin{equation}
\Gamma_{\psi(4040)\rightarrow D^{\ast}D}(m)=\Gamma_{\psi(4040)\rightarrow
D^{\ast+}D^{-}+h.c}(m)+\Gamma_{\psi(4040)\rightarrow D^{\ast0}\bar{D}^{0}%
+h.c}(m)\text{ ,}\label{dsdpsi4040}%
\end{equation}%
\begin{equation}
\Gamma_{\psi(4040)\rightarrow D^{\ast}D^{\ast}}=\Gamma_{\psi(4040)\rightarrow
D^{\ast+}D^{\ast-}}+\Gamma_{\psi(4040)\rightarrow D^{\ast0}\bar{D}^{\ast0}%
}\text{ .}\label{dsdspsi4040}%
\end{equation}

Since the spectral function is at the center of our discussion, it is worth to recall its definition,
\begin{equation}
d_{\psi}(m)=\frac{2m}{\pi}\left\vert \operatorname{Im}\Delta_{\psi}%
(p^{2}=m^{2})\right\vert \text{ .}\label{spectralfunctpsi4040}%
\end{equation}
The quantity $d_{\psi}(m)dm$ stands for the probability that the mass of $\psi(4040)$ is in the range from $m$ to $m+dm$. An important feature of our model is the normalization

\begin{equation}
\int\limits_{0}^{\infty}d_{\psi}(m)\mathrm{dm}=1\text{ .}\label{normpsi4040}%
\end{equation}
Notice that the integration is formally performed up to $m \rightarrow \infty$. However, the numerical calculations show that the condition given by Eq. (\ref{normpsi4040}) is already satisfied for value $m=10$ GeV in the upper limit of the integral, since this is far enough above $4$ GeV.

Note, we work from now on in the isospin symmetric limit, hence $D^0\bar{D}^0\equiv D^+D^-$.

At this point we define the partial spectral functions as:
\begin{align}
d_{\psi\rightarrow DD}(m)  &  =\frac{2m}{\pi}\left\vert \Delta_{\psi}%
(m^{2})\right\vert ^{2}m\Gamma_{\psi(4040)\rightarrow DD}(m)\text{ ,} \label{qqdd}\\
d_{\psi\rightarrow D_{s}D_{s}}(m)  &  =\frac{2m}{\pi}\left\vert
\Delta_{\psi}(m^{2})\right\vert ^{2}m\Gamma_{\psi(4040)\rightarrow D_{s}D_{s}%
}(m)\text{ ,}\\
d_{\psi\rightarrow DD^{\ast}}(m)  &  =\frac{2m}{\pi}\left\vert \Delta_{\psi
}(m^{2})\right\vert ^{2}m\Gamma_{\psi(4040)\rightarrow DD^{\ast}}(m)\text{
,} \label{qqs}\\
d_{\psi\rightarrow D_{s}^{\ast}D_{s}}(m)  &  =\frac{2m}{\pi}\left\vert
\Delta_{\psi}(m^{2})\right\vert ^{2}m\Gamma_{\psi(4040)\rightarrow D_{s}%
^{\ast}D_{s}}(m)\text{ ,}\\
d_{\psi\rightarrow D^{\ast}D^{\ast}}(m)  &  =\frac{2m}{\pi}\left\vert
\Delta_{\psi}(m^{2})\right\vert ^{2}m\Gamma_{\psi(4040)\rightarrow D^{\ast
}D^{\ast}}(m)\text{ ,} \label{qqdsds}%
\end{align}
according to which the total spectral function of Eq. (\ref{spectralfunctpsi4040}) reads:

\begin{equation}
d_{\psi}(m)=d_{\psi\rightarrow DD}(m)+d_{\psi\rightarrow D_{s}D_{s}}(m)+d_{\psi\rightarrow DD^{\ast}}(m)+d_{\psi\rightarrow D_{s}^{\ast}D_{s}}(m)+d_{\psi\rightarrow D^{\ast}D^{\ast}}(m)\text{ . } \label{cytow}
\end{equation}
For example, the quantity $d_{\psi \rightarrow DD^*}(m)dm$ in Eq. (\ref{qqs}) is the probability that $\psi(4040)$, has a mass between $m$ and $m+dm$ and decays into $DD^*$ \cite{duecan}. The partial spectral functions are useful to study the processes corresponding to different decay channels. Moreover, the quantity $d_{\psi \rightarrow DD^*}(m)dm$ is proportional to the cross section of the process $e^+e^- \rightarrow DD^*$.

The extension to the complex plane is done by following the steps of Chapter \ref{chapvec}. Yet, care is needed because a new feature appears since we have more than a single decay channel. 

In general, the loop function of each channel can be continued to the II RS, thus resulting in $2^N$ possible Riemann sheets. For instance, for the $j$-th channel the loop as function of $s=z^2$ (complex number) reads
\begin{equation}
\Pi_j(s=z^2)=-\frac{1}{\pi} \int \limits_{s^2_{th, j}}^{\infty}\sqrt{s'}\frac{\Gamma_{\psi(4040),j}(s')}{s-s'+i \varepsilon}ds' \text{ ,}
\end{equation}
where, $s_{th, j}$ is the threshold of that channel. For instance, for $j=D^0\bar{D}^0$, $s_{th}=(2m_{D^0})^2$, and so on. 

On the II RS of the $j$-th channel $\Pi_{j}(s=z^2)$ reads
\begin{equation}
\Pi_{j, II}(s=z^2)=\Pi_{J}(s=z^2)+2i\sqrt{s}\Gamma_{\psi(4040),j}(s) \text{ .}
\end{equation}

Usually, out of the $2^N$ possible RS, only $N$ of them are taken into consideration. If for instance $\operatorname{Re}(s)$ is larger than $s_{th,1},\ldots, s_{th,n}$ (but smaller than $s_{th,n+1}, \ldots, s_{th,N}$) then
\begin{equation}
\Pi_{(n+1)RS}(s=z^2)=\Pi_{1,II}(s=z^2)+ \ldots +\Pi_{n, II}(s=z^2)+\Pi_{n+1}(s=z^2)+\ldots+\Pi_{N}(s=z^2) \text{ .}
\end{equation}
In our particular case, thus for $\psi(4040)$, we will be interested in $\operatorname{Re}(s)\sim 4.04$ GeV. Then, the thresholds of the channels $DD$, $D^*D$, $D_sD_s$ and $D^*D^*$ are on the left while $D_s^*D_s$ is on the right. We then search for the poles on the $4+1=5$ RS.
\section{Parameters of the model}
There are five free parameters in our model: the couplings $g_{\psi DD}$, $g_{\psi D^*D}$ and $g_{\psi D^*D^*}$ of Eqs. (\ref{lag1psi4040}), (\ref{lag2psi4040}) and (\ref{lag3psi4040}), respectively, the energy scale $\Lambda$ in Eq. (\ref{gaussianpsi4040}), and $M_{\psi}$ the bare mass of $\psi(4040)$ appearing in the propagator of Eq. (\ref{proppsi4040}).
 
For what concerns $\Lambda$, we vary its numerical value within the interval from $0.38$ GeV to $1$ GeV (but as we show in Appendix \ref{apppsi4040} only for $\Lambda$ up to $0.6$ GeV one obtains reasonable results for the pole(s)). Once $\Lambda$ is fixed we can determine the coupling constants. To this end one needs three experimental data. The first one is the total decay width of the $\psi(4040)$, which is given by the PDG \cite{pdg} as
\begin{equation}
\Gamma_{\psi(4040)}^{\text{tot,exp}}=80\pm10\hspace{0.2cm}\text{MeV .}%
\label{totdecwidth4040}
\end{equation}
The corresponding theoretical formula for the total decay width of $\psi(4040)$ reads [for $m=4.039$ GeV (on shell)]:

\begin{equation}
\Gamma_{\psi(4040)}^{\text{tot,theory}}=\Gamma_{\psi(4040)\rightarrow
DD}^{\text{on}\hspace{0.1cm}\text{shell}}+\Gamma_{\psi(4040)\rightarrow
D_{s}D_{s}}^{\text{on}\hspace{0.1cm}\text{shell}}+\Gamma_{\psi
(4040)\rightarrow D^{\ast}D}^{\text{on}\hspace{0.1cm}\text{shell}}%
+\Gamma_{\psi(4040)\rightarrow D_{s}^{\ast}D_{s}}^{\text{on}\hspace
{0.1cm}\text{shell}}+\Gamma_{\psi(4040)\rightarrow D^{\ast}D^{\ast}%
}^{\text{on}\hspace{0.1cm}\text{shell}}\text{ .}%
\end{equation}

Moreover, we choose the following two ratios quoted by the PDG \cite{pdg} (see also Refs. \cite{babarpsi4040, wangpsi4040, belle3psi4040, cleopsi4040}):
\begin{equation}
\left.  \frac{\mathcal{B}(\psi(4040)\rightarrow D\bar{D})}{\mathcal{B}%
(\psi(4040)\rightarrow D^{\ast}\bar{D})}\right\vert _{\exp}=0.24\pm
0.05\pm0.12\text{ ,}%
\end{equation}%
\begin{equation}
\left.  \frac{\mathcal{B}(\psi(4040)\rightarrow D^{\ast}\bar{D}^{\ast}%
)}{\mathcal{B}(\psi(4040)\rightarrow D^{\ast}\bar{D})}\right\vert _{\exp
}=0.18\pm0.14\pm0.03\text{ . }%
\end{equation}
Each ratio above has two errors, the first one is statistical, while the second is systematic. 

Using Eqs. (\ref{deeec1})-(\ref{deeec3}), we are able to  determine the coupling constants and their errors. To this end we minimize the $\chi^2$ function $F_E$ which depends on the parameters $g_{\psi DD}$, $g_{\psi D^*D}$ and $g_{\psi D^*D^*}$:
\begin{align}
F_{E}(g_{\psi DD},g_{\psi D^{\ast}D},g_{\psi D^{\ast}D^{\ast}})  &  =\left(
\frac{\frac{\Gamma_{\psi\rightarrow DD}^{theory}(g_{\psi DD})}{\Gamma
_{\psi\rightarrow D^{\ast}D}^{theory}(g_{\psi D^{\ast}D})}-\left.
\frac{\mathcal{B}(\psi(4040)\rightarrow D\bar{D})}{\mathcal{B}(\psi
(4040)\rightarrow D^{\ast}\bar{D})}\right\vert _{\exp}}{\delta\left.
\frac{\mathcal{B}(\psi(4040)\rightarrow D\bar{D})}{\mathcal{B}(\psi
(4040)\rightarrow D^{\ast}\bar{D})}\right\vert _{\exp}}\right)  ^{2}%
+\nonumber\\
\left(  \frac{\frac{\Gamma_{\psi\rightarrow D^{\ast}D^{\ast}}^{theory}(g_{\psi
D^{\ast}D^{\ast}})}{\Gamma_{\psi\rightarrow D^{\ast}D}^{theory}(g_{\psi
D^{\ast}D})}-\left.  \frac{\mathcal{B}(\psi(4040)\rightarrow D^{\ast}\bar
{D}^{\ast})}{\mathcal{B}(\psi(4040)\rightarrow D^{\ast}\bar{D})}\right\vert
_{\exp}}{\delta\left.  \frac{\mathcal{B}(\psi(4040)\rightarrow D^{\ast}\bar
{D}^{\ast})}{\mathcal{B}(\psi(4040)\rightarrow D^{\ast}\bar{D})}\right\vert
_{\exp}}\right)  ^{2}  &  +\left(  \frac{\Gamma_{\psi(4040)}^{tot,theory}%
(g_{\psi DD},g_{\psi D^{\ast}D},g_{\psi D^{\ast}D^{\ast}})-\Gamma_{\psi
(4040)}^{tot,exp}}{\delta\Gamma_{\psi(4040)}^{tot,exp}}\right)  ^{2} \text{ .}%
\end{align}
Finally, the last parameter $M_{\psi}$ (the bare mass) is determined by imposing the requirement that the spectral function reaches the maximum at $4.039$ GeV (the PDG average).

It should be stress that a change of $\Lambda$ implies the variation of all the other parameters. However, an unambiguous determination of the numerical value of this parameter turns out to be problematic. It is not possible to perform a reliable fit to the available scattering data, at least for two reasons: (i) the experimental errors are still quite large and (ii) an unknown background contribution also needs to be considered. Thus, varying $\Lambda$ in a reasonable range is the best strategy at present. The numerical values of all the model parameters are presented in the Appendix \ref{apppsi4040} where for each value of $\Lambda$ there are different values for coupling constants and for the bare mass. One can see that the dependence on $\Lambda$  is much stronger for $g_{\psi DD}$ than for $g_{\psi D^*D}$ and $g_{\psi D^*D^*}$. This can be explained by the fact that the $DD$ threshold is the most distant one from the on-shell mass value of $\psi(4040)$ resonance and, moreover, the corresponding on-shell momentum $k_{\psi DD}=\sqrt{\frac{m^2_{\psi(4040)}}{4}-m^2_{D^+}}\simeq0.76$ GeV is of the order of $\Lambda$. Note, the contribution from the $DD$ channel to the total width of $\psi(4040)$ is rather small and does not influence much the general picture. The results for the partial decay widths analysis performed for different values of $\Lambda$ parameter and for two form factors can be found in Table \ref{partialdec4040} of Appendix \ref{apppsi4040}. Even if the results are consistent with each other, new experimental data for the channel $\psi(4040) \rightarrow D_s^-D_s^+$ would be needed in order to improve our model.

\section{Results}

Let us discuss now the main outcomes of our study, namely the shape of the total spectral function of the $\psi(4040)$ resonance (with later analysis extended to the partial spectral functions for $DD$, $DD^*$ and $D^*D^*$ channels). The determination of the pole(s) position, and moreover the characteristics of the process $e^+e^- \rightarrow J/ \psi \pi^+\pi^-$ play a crucial role for the generation of a peak corresponding to $Y(4008)$.

\subsection{Spectral function and pole position(s)}
Since we did not determine the precise value for the cutoff, but only limited it to a certain range, we choose $\Lambda=0.42$ for presenting the results and preparing the plots. Such a choice is motivated by the fact that this particular value generates a pole whose imaginary part is $40$ MeV, just the half of the PDG width. Once $\Lambda$ is given, one can determine the other parameters of the model, see Table \ref{tableparameterspsi4040} for their numerical values.
\begin{table}[htbp] 
\centering
\renewcommand{\arraystretch}{1.35}
\begin{tabular}{c|c|c|c|c|c} 
\hline
\hline
Parameters& $\Lambda$ [GeV]&$M_{\psi}$ [GeV]&$g_{\psi_{DD}}$&$g_{\psi D^*D}$ [GeV$^{-1}$] & $g_{\psi D^*D^*}$\\
\hline
Value&$0.42$&$4.01$&$39.6 \pm 5.0$&$3.43 \pm 0.80$&$1.90 \pm 0.95$\\
\hline
\hline
\end{tabular}
\caption{The numerical values of model parameters obtained for $\Lambda=0.42$ and for a Gaussian form factor.}
\label{tableparameterspsi4040}
\end{table}
The partial decay widths are reported in Table \ref{tablepartsmall}.
\begin{table}[htbp] 
\centering
\renewcommand{\arraystretch}{1.35}
\begin{tabular}{c|c|c|c|c|c} 
\hline
\hline
Decay channel& $DD$ &$D_sD_s$&$D^*D$&$D_s^*D_s$ & $D^*D^*$\\
\hline
Partial decay width [MeV]&$5.7 \pm 1.4$&$46 \pm 12$&$24 \pm 11$&$0$&$4.3 \pm 4.3$\\
\hline
\hline
\end{tabular}
\caption{The partial decay widths for different decay channels calculated for the parameters of Table \ref{tableparameterspsi4040}.}
\label{tablepartsmall}
\end{table}
(Note, the value of the partial decay for the decay into $D_s^*D_s$ is zero because it is subtreshold).

\begin{figure}[h] 
\begin{center}
\begin{minipage}[b]{7.8cm}
\centering
\includegraphics[width=7.7cm]{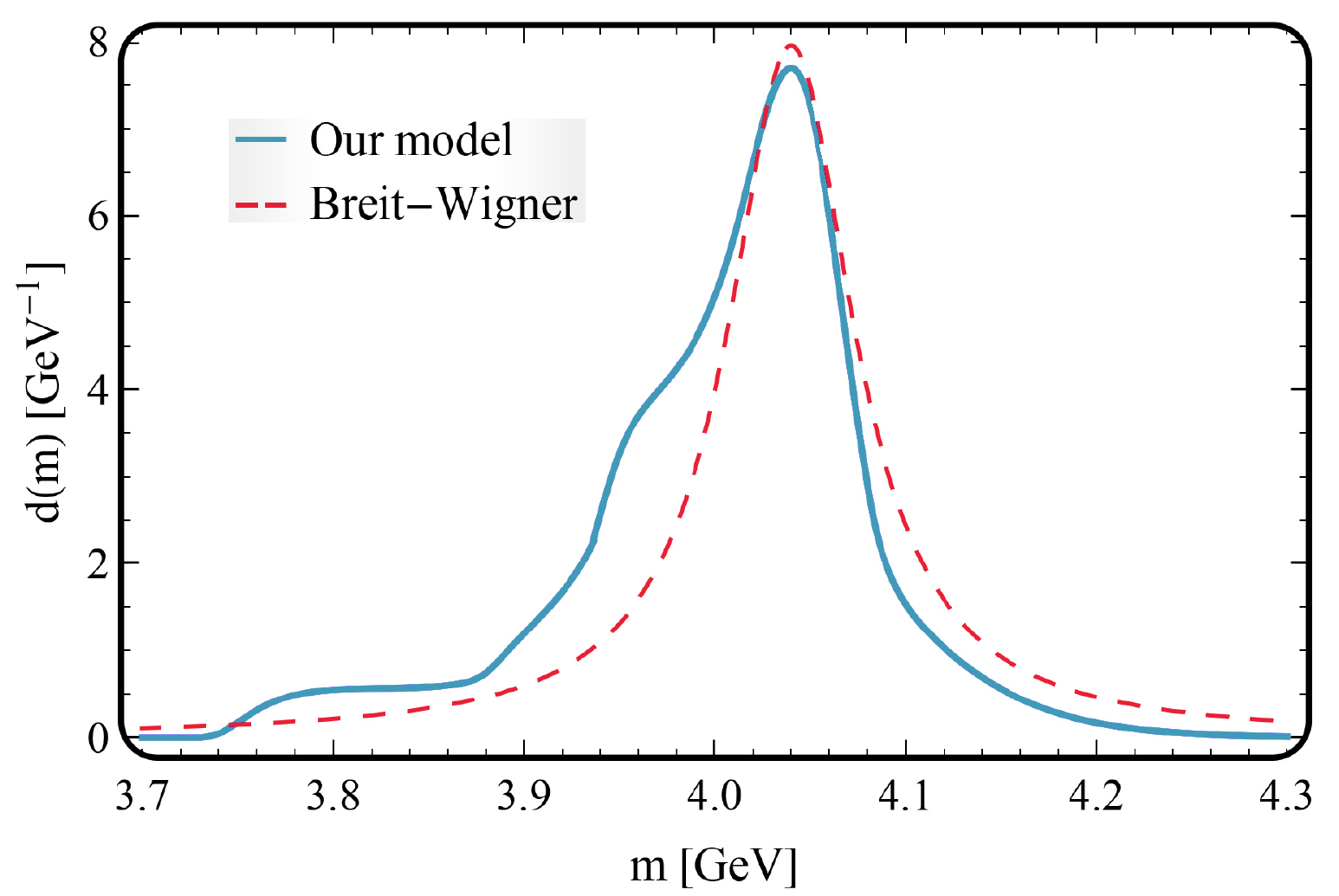}\\\textit{a)} 
\end{minipage}
\begin{minipage}[b]{7.8cm}
\centering
\includegraphics[width=7.75cm]{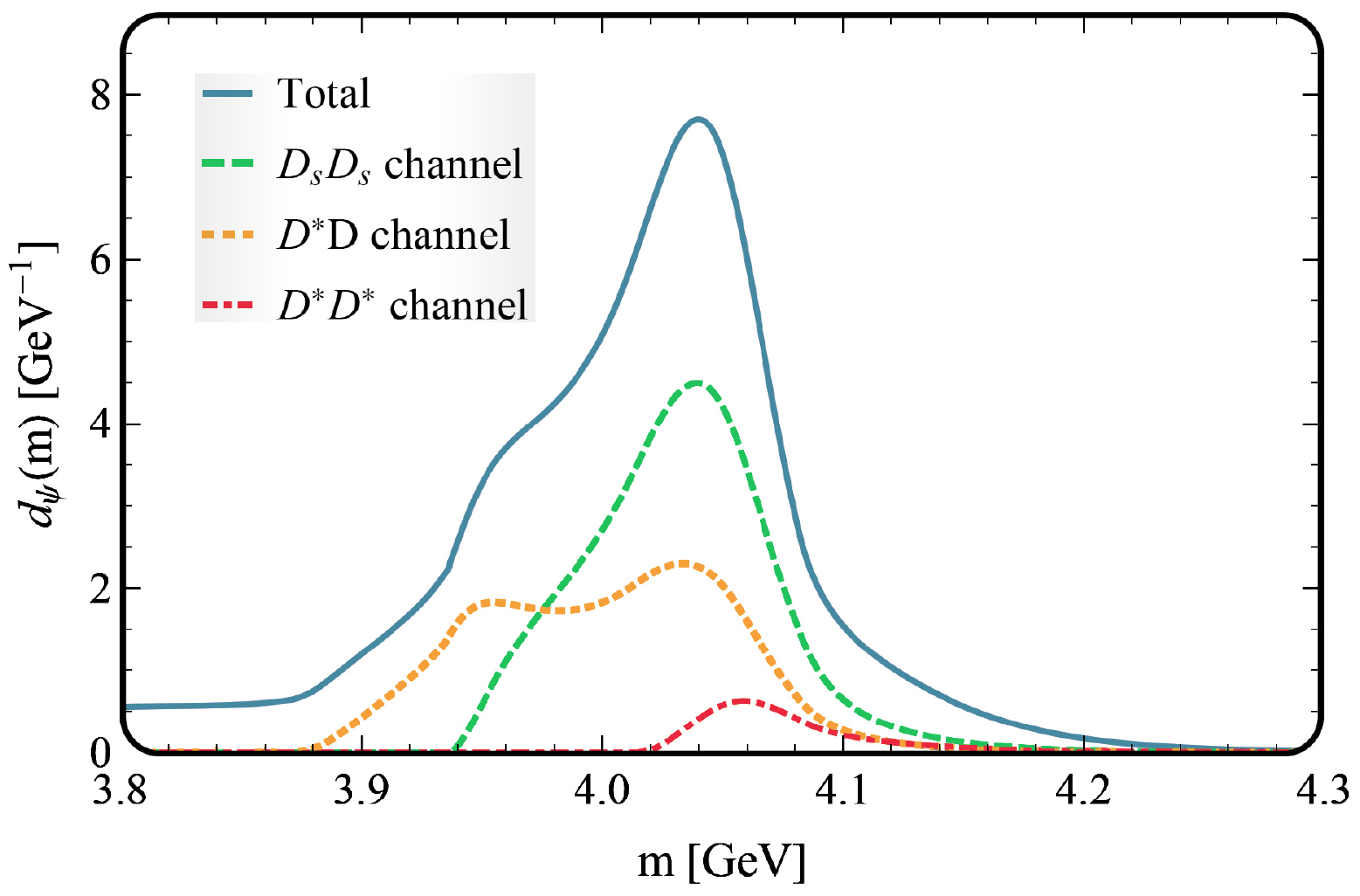}\\\textit{b)}
\end{minipage}
\end{center} 
\caption{Panel (a): Total spectral function of $\psi(4040)$ resonance (blue line), defined in Eq. (\ref{spectralfunctpsi4040}), plotted for the parameters of Table \ref{tableparameterspsi4040} compared with the Breit-Wigner shape (red line) plotted for the parameters quoted by the PDG \cite{pdg}. Panel (b) Contributions to the whole spectral function from $DD$ (green line), $DD^*$ (orange line) and $D^*D^*$ (red line) decay channels, defined in Eq. (\ref{qqdd}), Eq. (\ref{qqs}) and Eq. (\ref{qqdsds}), respectively.}
\label{partial4040mp}
\end{figure}

In the panel (a) of Figure \ref{partial4040mp} we compare the full spectral function defined in Eq. (\ref{spectralfunctpsi4040}) to the ordinary Breit-Wigner shape (plotted for the parameters given by the PDG, thus $M_{\psi}=4.039$ GeV and $\Gamma_{\psi(4040)}=80$ MeV). 
According to the expectation, a $c\bar{c}$ seed state corresponding to the $\psi(4040)$ resonance is visible in the spectral function as a unique peak with the maximum at around $4.04$ GeV. Although the Breit-Wigner distribution describes correctly the region nearby the peak, there are differences close to $3.9$ GeV, where a significant enhancement is present. In order to understand the origin of this puzzling distortion we study the partial spectral functions, thus the contributions of particular channels to the whole spectral function, see Eqs. (\ref{spectralfunctpsi4040}-\ref{cytow}). The results for the most significant channels, $D_sD_s$, $DD^*$ and $D^*D^*$, are presented in panel (b) of Figure \ref{partial4040mp}. One can see that the enhancement is mostly influenced by the $D^*D$ loops.

We have found that there are two complex poles. The first one situated on the V RS with coordinates 
\begin{equation}
(4.053 \pm 0.04)-(0.040 \pm 0.010) i  \text{ GeV}\text{ , }
\end{equation}
is the seed pole related to the peak in the spectral function, while the second located on the IV RS with coordinates 
\begin{equation}
(3.934 \pm 0.006)-(0.030 \pm 0.001)i \text{ GeV} \text{ , }
\end{equation}
is a dynamical ``companion'' pole which corresponds to the broad deformation. Notice that for $\sqrt{s}\simeq 3.93$ GeV, the channel $D^*D^*$ is on the right, hence we consider the IV RS.

We stress that one should not consider these values as fixed results, since the cutoff is not precisely determined. However, we checked that the dynamically generated pole emerges  even if $\Lambda$ is sizably larger, see results in the Appendix \ref{apppsi4040}. Notice that only a single conventional $c\bar{c}$ seed state (identified with $\psi(4040)$) is included in the Lagrangian, therefore the corresponding pole is expected and delivers the results for the mass and decay width of $\psi(4040)$ compatible with the experimental data \cite{pdg}. If the low enhancement in the spectral function of Figure \ref{partial4040mp} is real, one should find it in the cross-section of the reaction $e^+e^- \rightarrow \psi(4040) \rightarrow DD^*$, which in turn is proportional to the partial spectral function $d_{\psi \rightarrow DD^*}(m)$. Nevertheless, because of the large errors of the data, this is not yet visible. Moreover, it is at first tempting to assign it to the questionable state $Y(4008)$. However, this assigment is rather controversial since the corresponding value of the decay width extracted from the pole coordinates, namely $\Gamma \simeq 60$ MeV, is more than three times smaller than the experimental value $(\sim 200$ MeV) measured by Belle \cite{belle1, belle2}. One should also recall that the $Y(4008)$ state has not been seen in the $D^*D$ channel, but solely in the ISR process with the $\pi^+ \pi^- J/ \psi$ final state. This aspect will be elaborated later on in this chapter.

As it was already done for the case of the vector $K^*(892)$ and for the scalars $K^*_0(1430)$ and $K^*_0(700)$, we also perform a large-$N_c$ analysis in the charmonium sector in order to understand the nature of both $\psi(4040)$ and the second pole. To this end, we again introduce a rescaling parameter $\lambda\in [0,1]$, which eneters in our model through the coupling constants:
\begin{equation}
g_{\psi DD}\rightarrow\sqrt{\lambda}g_{\psi DD}\text{ , }g_{\psi D^{\ast}%
D}\rightarrow\sqrt{\lambda}g_{\psi D^{\ast}D}\text{ , }g_{\psi D^{\ast}%
D^{\ast}}\rightarrow\sqrt{\lambda}g_{\psi D^{\ast}D^{\ast}}\text{ }.
\end{equation}
Then, varying  $\lambda$, one can modify the interaction in a controlled manner and then monitor the behavior of the spectral function and positions of the poles. We recall that the connection of $\lambda$ parameter with $N_c$ is given by $\lambda=3/N_c$. Therefore, for $\lambda=1$, the physical results are reobtained.

\begin{figure}[h] 
\begin{center}
\begin{minipage}[b]{7.8cm}
\centering
\includegraphics[width=7.7cm]{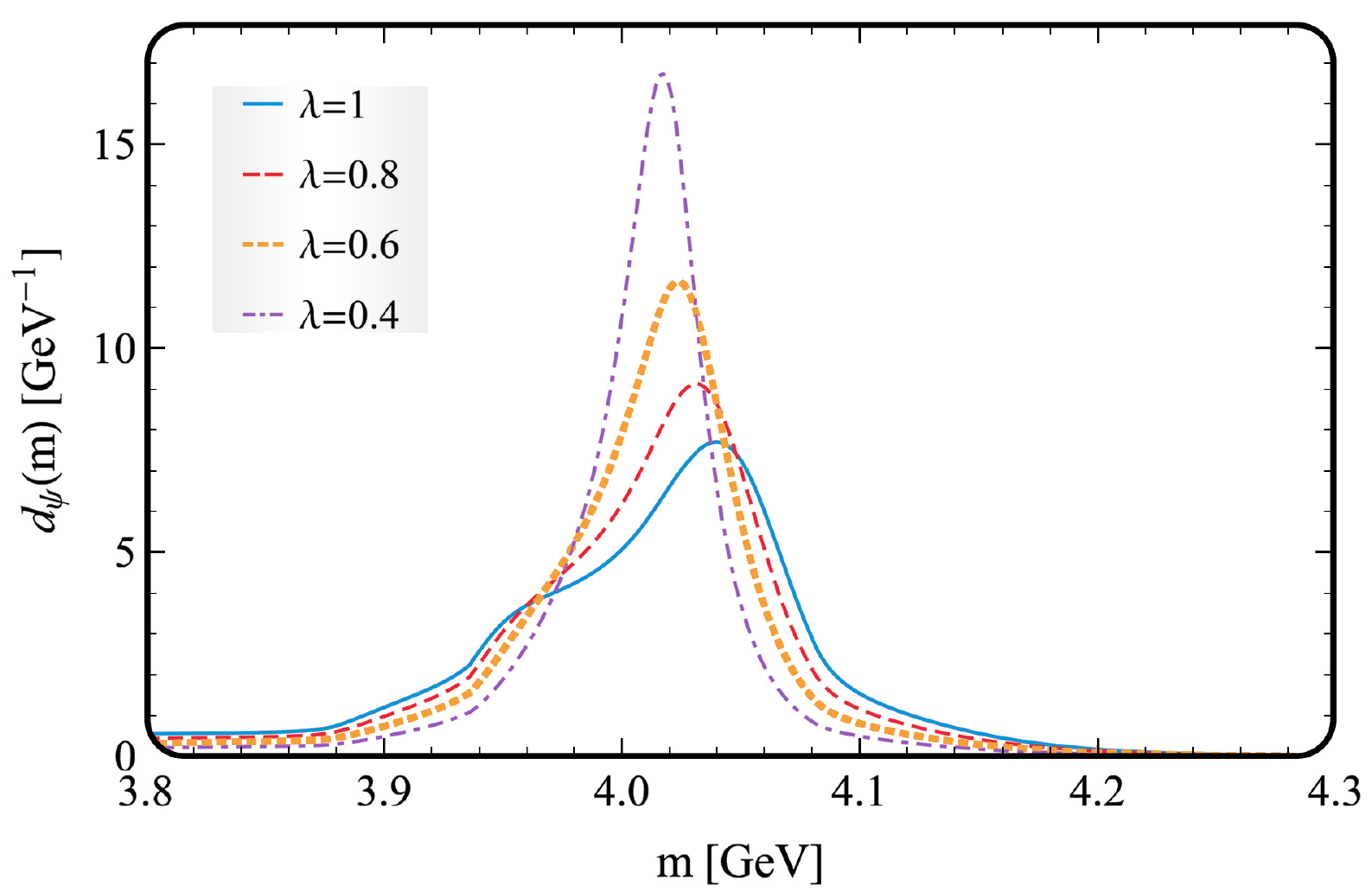}\\\textit{a)} 
\end{minipage}
\begin{minipage}[b]{7.8cm}
\centering
\includegraphics[width=7.75cm]{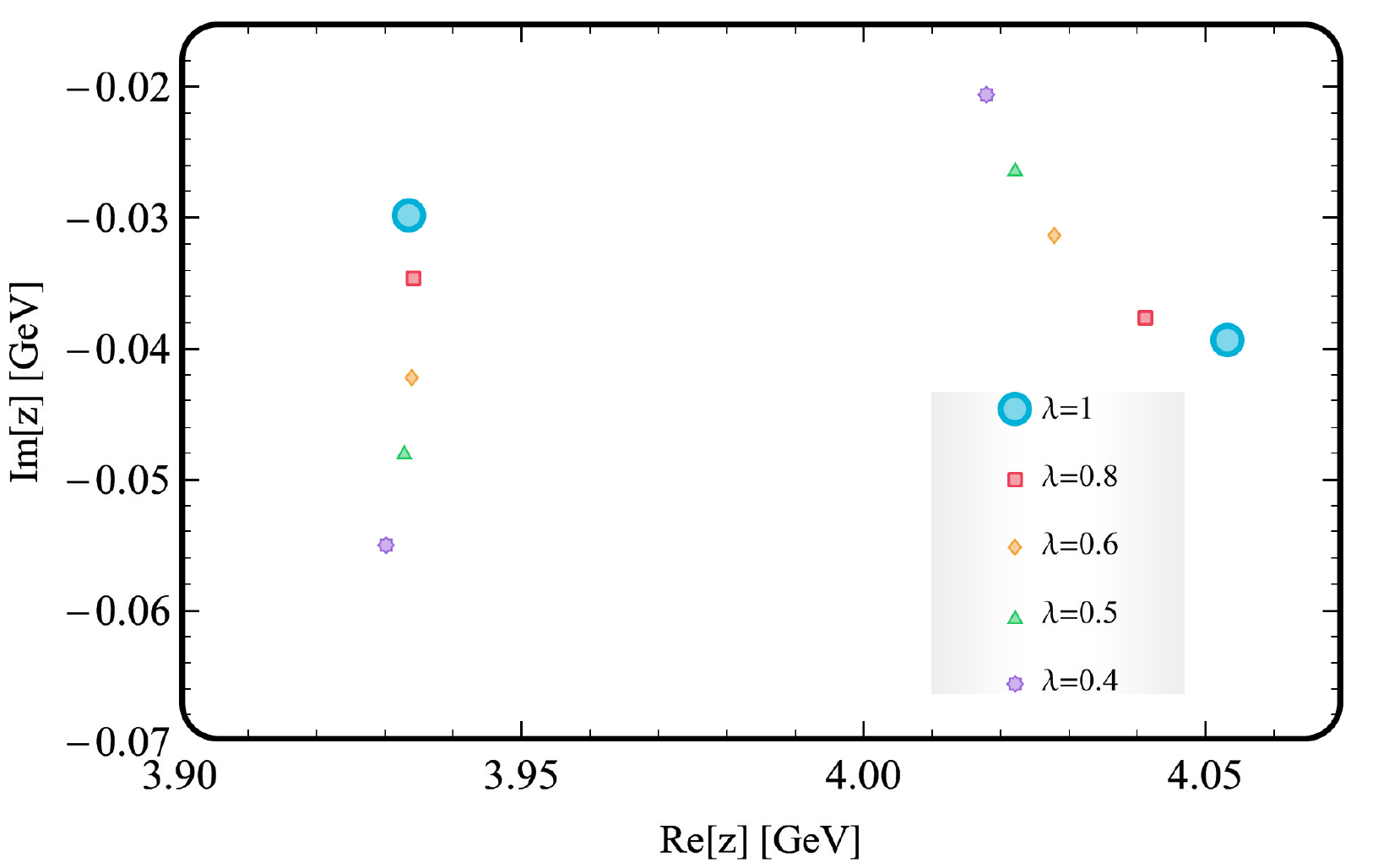}\\\textit{b)}
\end{minipage}
\end{center} 
\caption{Panel (a): Study of the shape of the spectral function of Eq. (\ref{spectralfunctpsi4040}) in large-$N_c$ limit, thus upon changing the value of $\lambda$ parameter. Panel (b): The trajectories of both the seed state and companion poles in the complex plane. The coordinetes of the physical poles, thus for $\lambda=1$ are indicated by the big blue dots. The results of both panels are plotted for the parameters of Table \ref{tableparameterspsi4040}.}
\label{partial24040mp}
\end{figure}

Let us first consider the shape of the spectral function of Eq. (\ref{spectralfunctpsi4040}) plotted for $\lambda=0.4,$ $0.6$ and $0.8$, see panel (a) of Figure \ref{partial24040mp}. For completeness, also the physical result for $\lambda=1$ is shown. 
One observes that the decay width of the $\psi(4040)$ resonance vanishes as $N_c$ becomes infinite, thus for $\lambda \rightarrow 0$, while the enhancement disappears at the same time. Furthermore, we report the coordinates of the poles for $\lambda=0.4,$ $0.5,$ $0.6,$ $0.8$ in panel (b) of Figure \ref{partial24040mp}. The pole of $\psi(4040)$ and the second pole show opposite behavior on the complex plane. For decreasing $\lambda$ the former is moving in the direction of the real energy axis, while the latter is moving away from it and finally fades away.  The behavior of $\psi(4040)$ is typical for a conventional quark-antiquark (here $c\bar{c}$) meson, while the second pole has a non-conventional nature. 

Next, we also check how the mixing of $\psi(4040)$ with $\psi(4160)$ resonance (being the nearest quarkonium state) influences our results. To this end, we constructed the mixed propagator (for the mathematical formalism see Refs. \cite{nonorthm, achasovmmk}), and recalculated the spectral function by considering the chain $\psi(4040) \rightarrow DD^* \rightarrow \psi(4160) \rightarrow DD^* \rightarrow \psi(4040)$. This process is particularly important since $\psi(4160)$ is mainly coupled to $DD^*$ channel. We observed that only small differences of the spectral function appear in the energy region we are intersted in. Such result indicates that the inclusion of even more $c\bar{c}$ states does not significantly affect the overall picture presented in this chapter. 

\subsection{Decay into $\mathbf{J/\psi \pi^+ \pi^-}$}

The $Y(4008)$ state was observed in the reaction $e^+e^- \rightarrow \gamma +J/ \psi+ \pi^+ +\pi^-$, hence a deeper  insight into this particular channel is important. Because the photon $\gamma$ originates from Initial State Radiation (ISR), we can recast the reaction into two parts: 
\begin{equation}
e^+ e^- \rightarrow \gamma(e^+ e^-)_{\text{off-shell}} \text{ and } (e^+ e^-)_{\text{off-shell}} \rightarrow J/ \psi+ \pi^+ +\pi^- \text{ . }
\end{equation}
 
Since, at the fundamental level, both processes are similar to each other, for simplicity we neglect that $e^+ e^-$ pair is off-shell, and focus only on $e^+e^- \rightarrow J/ \psi \pi^+ \pi^-$. It is particularly interesting to study the case in which this process is realized through the intermediate state $\psi(4040)$:
\begin{equation}
e^{+}e^{-}\rightarrow\psi(4040)\rightarrow J/\psi+\pi^{+}+\pi^{-}\text{ .}%
\label{epluseminus}
\end{equation}
Let us now consider two typical decay mechanisms related to this process. In the first scenario, two gluons are emitted:
\begin{equation}
\psi(4040)\equiv c\bar{c}\rightarrow c\bar{c}+gg\rightarrow J/\psi
+f_{0}(500)\rightarrow J/\psi+\pi^{+}+\pi^{-} \text{ . } \label{gluonemiss}
\end{equation}
The resonance $f_0(500)$ has been choosen to mediate this reaction, since it is the lightest one among the states carrying the vacuum quantum numbers and it lies in the allowed kinematic regime. For what concerns the $f_0(980)$ and $f_0(1370)$ states, the former is already at the border while the latter is too heavy. However, there is no problem to consider other $f_0$ resonances, instead of the one proposed here. This decay mode can be formulated by
\begin{equation}
\mathcal{L}_{\psi jf_{0}}^{direct}=g_{\psi jf_{0}}^{direct}\psi_{\mu}j^{\mu
}f_{0}\text{ ,}%
\end{equation}
where the constant $g_{\psi jf_{0}}^{direct}$ couples the $\psi(4040)$ to the $J/ \psi$ and the $f_0(500)$ (denoted by $j^{\mu}$ and $f_0$, respectively). This term would create a peak with a maximum at $4.04$ GeV, which, however, is not visible in the experiment. Indeed, this would correspond to an `ordinary' decay of $\psi(4040)$, which generates the peak located at the Breit-Wigner mass of this resonance. From this, it follows that the coupling $g_{\psi jf_{0}}^{direct}$ should be rather small. Therefore, this particular channel will be omitted in the following.

In the second scenario, a different mechanism takes place, 
\begin{equation}
\psi(4040)\rightarrow DD^{\ast}\rightarrow J/\psi+f_{0}(500)\rightarrow
J/\psi+\pi^{+}+\pi^{-}. \label{secondmechan}
\end{equation}
This process involves an additional vertex, which is described by the following four-body interaction: 

\begin{equation}
\mathcal{L}_{DD^{\ast}jf_{0}}=\lambda_{DD^{\ast}jf_{0}}\left[  \partial^{\mu
}D^{\ast+\nu}D^{-}+\partial^{\mu}D^{\ast0\nu}\bar{D}^{0}\right]  j_{\mu\nu
}f_{0}\text{ ,} \label{lscatf0mp}%
\end{equation}
where $j_{\mu \nu}=\partial_{\mu}j_{\nu}-\partial_{\nu}j_{\mu}$.

Then, for the $J/ \psi \pi^+ \pi^-$ channel one has the corresponding spectral function:
\begin{equation}
d_{\psi(4040)\rightarrow J/\psi\pi^{+}\pi^{-}}(m)=\frac{2m}{\pi}\left\vert
\Delta_{\psi}(m^{2})\right\vert ^{2}m\Gamma_{\psi(4040)\rightarrow J/\psi
\pi^{+}\pi^{-}}(m) \text{ , }\label{dpsijmp}%
\end{equation}
with
\begin{align}
&  \Gamma_{\psi(4040)\rightarrow J/\psi f_{0}(500)}(m)\nonumber\\
&  =\left\vert \lambda_{DD^{\ast}jf_{0}}g_{\Psi D^{\ast}D%
}\left[  \Sigma_{D^{0}D^{\ast0}}(m^{2})+\Sigma_{D^{+}D^{\ast-}}(m^{2})\right]
\right\vert ^{2}\frac{k}{8\pi m^{2}}\left(  3+\frac{k^{2}}{m_{J/\psi}^{2}%
}\right)  e^{-2\frac{k^{2}}{\Lambda^{2}}}%
\end{align}
(the quantity $k=k(m, m_{J/ \psi}, m_{f_0(500)})$ is the momentum of $J/ \psi$). Here, the $D^*D$ loop is responsible for the $J/ \psi \pi^+ \pi^-$ production. 

As the large-$N_c$ limit shows \cite{lebed, largencwitten}, both mechanisms presented above are suppressed. Then, the decay into $J/ \psi f_0(500)$ is rather small. The resonance $f_0(500)$ is (often) interpreted as a combination of non-ordinary states with an admixture of $q\bar{q}$ component, which in turn, may play an important role when $N_c$ becomes large. In the following, we consider the process of Eq. (\ref{secondmechan}) and introduce the $g_{f_0(500)-\bar{q}q}$ to be the coupling of $f_0(500)$ resonance to $q\bar{q}$. Based on the rules presented in Ref. \cite{largencwitten} one can write the scaling of the amplitude as
\begin{equation}
A_{\psi (4040)\rightarrow J/\psi \text{-}gg\rightarrow J/\psi \text{-}%
f_{0}(500)}\propto \frac{1}{\sqrt{N_{c}}}\frac{1}{\sqrt{N_{c}}}\left( \frac{1%
}{\sqrt{N_{c}}}\right) ^{4}N_{c}^{2}g_{f_{0}(500)\text{-}\bar{q}q\text{ }}=%
\frac{g_{f_{0}(500)\text{-}\bar{q}q\text{ }}}{N_{c}}\text{ ,} \label{ncnc}
\end{equation}%
where the factor $\frac{1}{\sqrt{N_c}}$ corresponds to the coupling of a single conventional meson to quarks. Moreover, the two gluons, emitted from $c\bar{c}$ transform into light quarks: this is proportional to $g_{QCD}^4$, where $g_{QCD} \propto \frac{1}{\sqrt{N_c}}$. The factor $N_c^2$, is a consequence of two closed quark lines. For the regular contribution of $q\bar{q}$ to $f_0(500)$, we have $g_{f_{0}(500)\text{-}\bar{q}q} \propto \frac{1}{\sqrt{N_c}}$, which implies the amplitude $1/N_c^{3/2}$. Therefore, for the two pions in the final state one has
\begin{equation}
A_{\psi(4040) \rightarrow J/ \psi-gg \rightarrow J/ \psi \pi^+ \pi^-}\propto N_c^{-2} \text{ .} \nonumber
\end{equation}
For the other non-conventional parts forming the $f_0(500)$, the corresponding amplitude would be even more strongly suppressed. We stress that the main term of the width for the decay into $J/ \psi \pi^+ \pi^-$ goes as $N_c^{-4}$. Such behavior explains the small contribution of this decay. 

\begin{figure}[h!]
\begin{center}
\includegraphics[width=0.55 \textwidth]{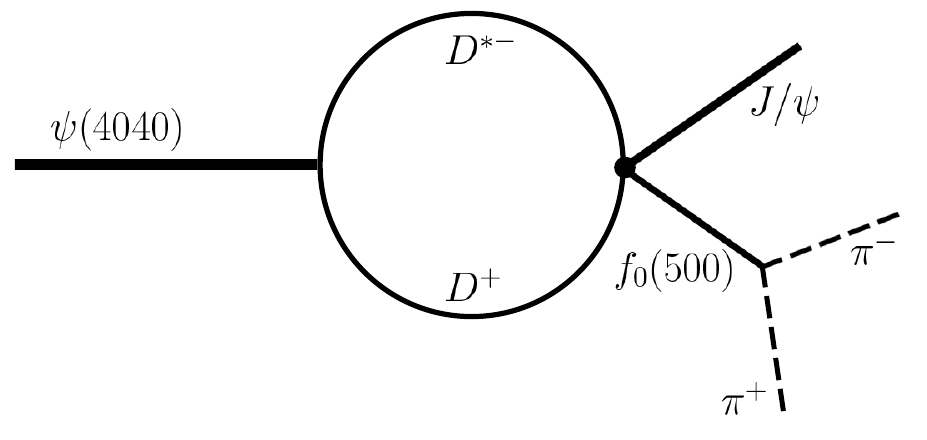}
\caption{\label{abcd} Schematic illustration of the decay of $\psi(4040)$ into $\pi^+ \pi^- J/ \psi$ through the intermediate $D^{*-}D^+$ loop.}
\end{center} 
\end{figure}

\begin{figure}[h] 
\begin{center}
\begin{minipage}[b]{7.8cm}
\centering
\includegraphics[width=8.1cm]{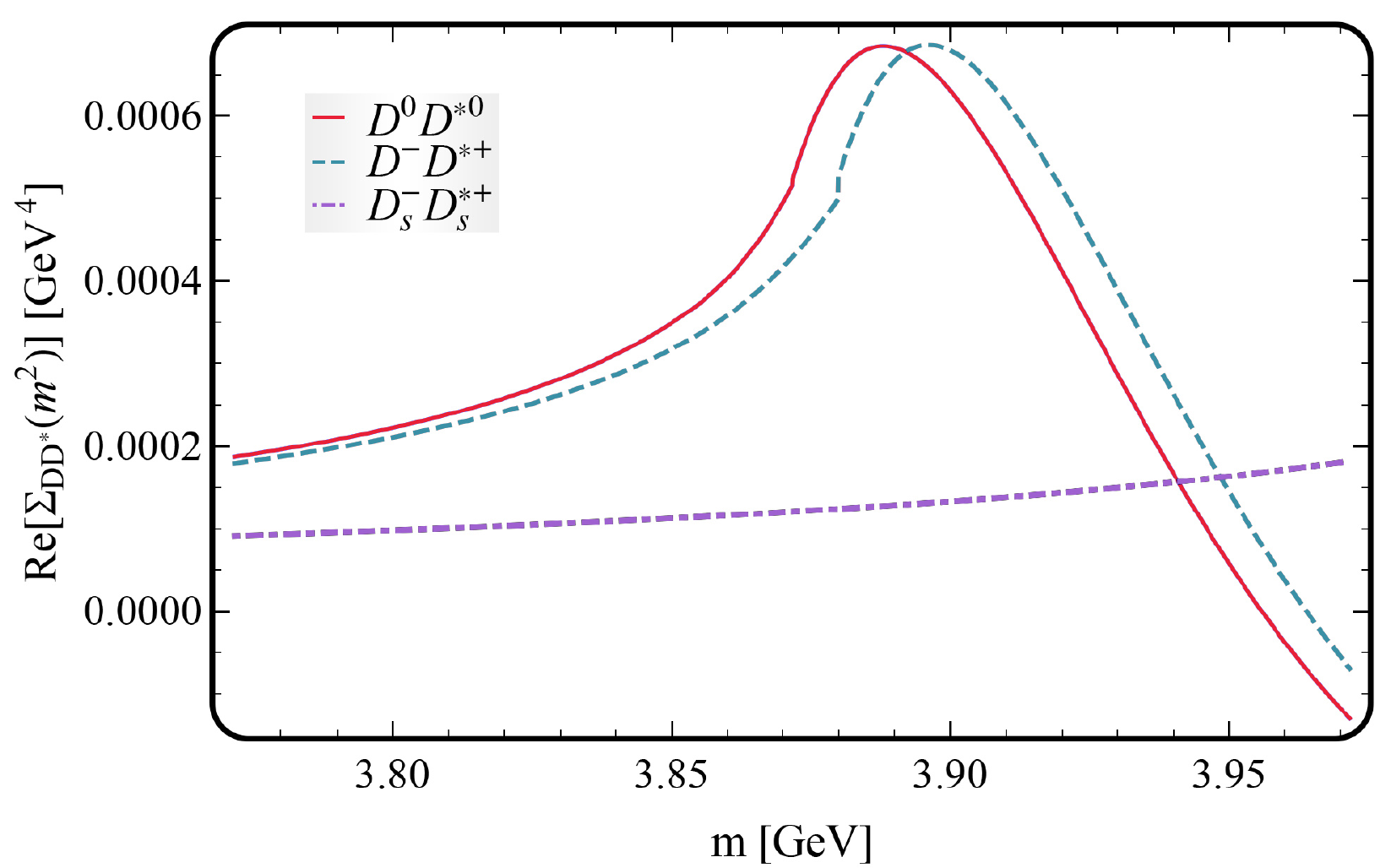}\\\textit{a)} 
\end{minipage}
\begin{minipage}[b]{7.7cm}
\centering
\includegraphics[width=7.5cm]{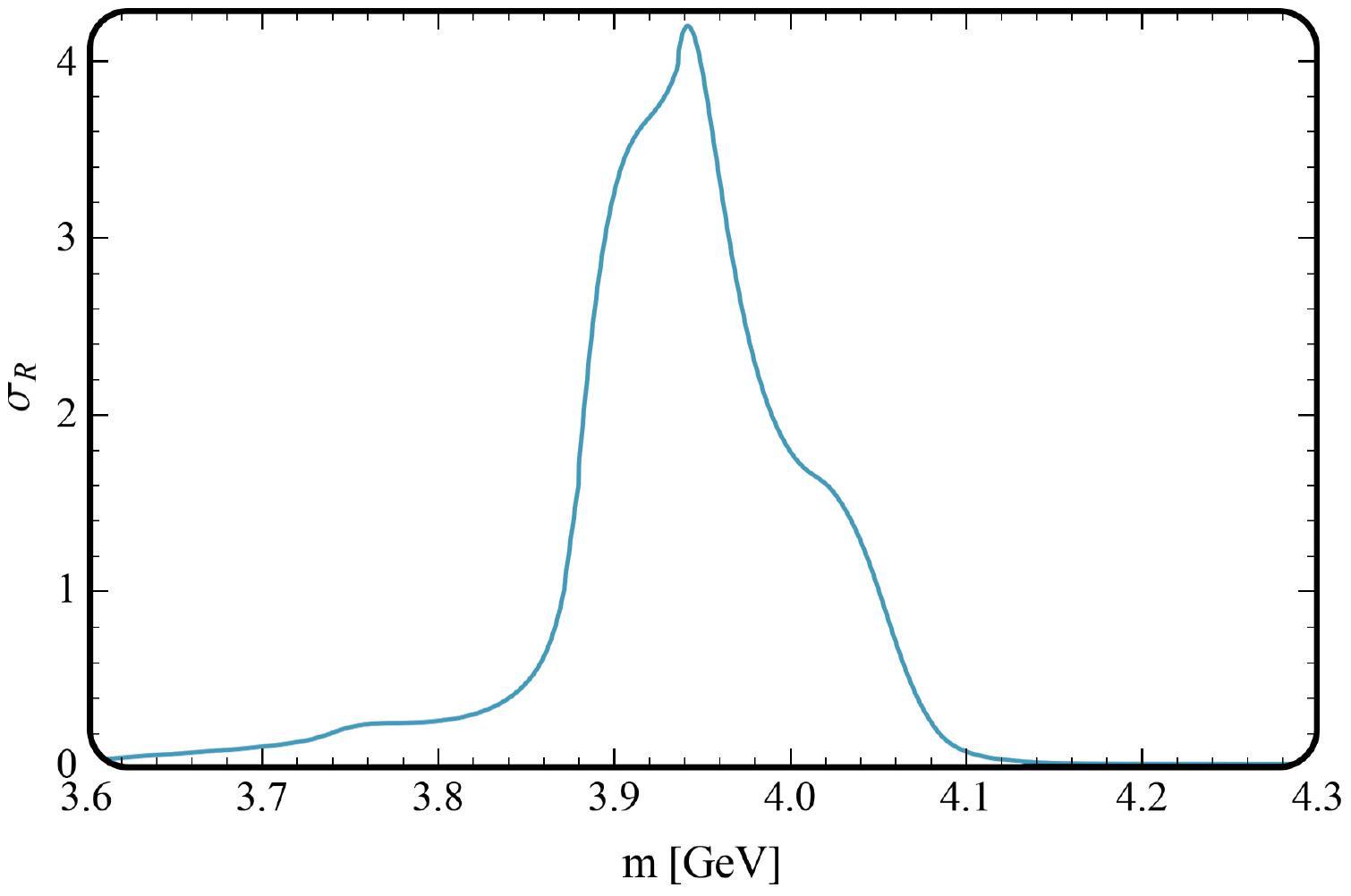}\\\textit{b)}
\end{minipage}
\end{center} 
\caption{Panel (a): Plot of the real part contribution coming from the $D^*D$ loops, as function of $m$. It is visible, that both Re $\Sigma_{D^{*+}D^-}$ and Re $\Sigma_{D^{*0}D^0}$ have their maximum near to $3.9$ GeV. Panel (b): The shape of the ratio $\sigma_R(m)$ of Eq. (\ref{ratiompmp}) representing the normalized cross-section for the process $e^+ e^- \rightarrow \psi(4040) \rightarrow J/ \psi \pi^+ \pi^-$. A distorted and broad structure pronounced close to 3.95 is observed.   }
\label{cross4040mp}
\end{figure}

Let us then consider Figure \ref{abcd}, where the $\psi(4040)$ decays through the $DD^*$ loop. 
One can see that the amplitude referring to this process scales as
\begin{equation}
A_{\psi (4040)\rightarrow j/\psi \text{-}DD^{\ast }\rightarrow j/\psi \text{-%
}f_{0}(500)}\propto \frac{1}{\sqrt{N_{c}}}\frac{1}{\sqrt{N_{c}}}g_{f_{0}(500)%
\text{-}\bar{q}q\text{ }}=\frac{g_{f_{0}(500)\text{-}\bar{q}q\text{ }}}{N_{c}%
}\text{ },
\label{amplittt}
\end{equation}%
just as in Eq. (\ref{ncnc}). Nevertheless, this is expected to be the dominant mechanism because of the following points: first of all there is a strong coupling of the $c\bar{c}$ seed state, thus the $\psi(4040)$ resonance couples to $D^*D$, which turns out to be the main decay channel. Moreover, when looking into panel (a) of Figure \ref{cross4040mp}
where the real part of the $DD^*$ loops has been plotted, one can easily see that the maximum of the peak is located at $m_D+m_{D^*}$, thus around $3.9$ GeV. 

In summary, the contributions from two functions enter into Eq. (\ref{dpsijmp}): $\vert \Delta_{\psi}(m^2)\vert^2$, having the maximum of the peak at $4.04$ GeV and $\Gamma_{\psi(4040)\rightarrow J/ \psi f_0(500)}(m)$, with the peak\footnote{In general, also for other channels, for instance the one into the $DD$, the coupling to $J/ \psi f_0(500)$ would occur. However, the coupling of this particular decay mode to the seed state $\psi(4040)$ is much smaller when compared to the dominant $DD^*$ channel. In fact, $\vert \Sigma_{DD}(m^2)\vert^2$ has its maximum at the value of $m_D+m_D$, thus the overlap with $\vert \Delta_{\psi}(m^2) \vert^2$ is much smaller in this case.} at $3.9$ GeV. 

Using the variable $m=\sqrt{s}$, one has
\begin{equation}
\sigma_{e^{+}e^{-}\rightarrow\psi(4040)\rightarrow J/\psi\pi^{+}\pi^{-}%
}\left(  m\right)  =\frac{2\pi}{m}g_{\psi e^{+}e^{-}}^{2}d_{\psi
(4040)\rightarrow J/\psi\pi^{+}\pi^{-}}(m)\text{ .}%
\end{equation}
In panel (b) of Figure \ref{cross4040mp} we plot the normalized cross-section of the process $e^{+}e^{-}\rightarrow\psi(4040)\rightarrow
J/\psi\pi^{+}\pi^{-}$, defined as
\begin{equation}
\sigma_{R}(m)=\frac{\sigma_{e^{+}e^{-}\rightarrow\psi(4040)\rightarrow
J/\psi\pi^{+}\pi^{-}}\left(  m\right)  }{\sigma_{e^{+}e^{-}\rightarrow
\psi(4040)\rightarrow J/\psi\pi^{+}\pi^{-}}\left(  m=m_{D^{0}}+m_{D^{\ast0}%
}\right)  } \text{ ,} \label{ratiompmp}%
\end{equation}
which is independent on the coupling $\lambda_{DD^*jf_0}$. (When setting $m=m_{D^0}+m_{D^{*0}}$ we have $\sigma_{R}=1$, by construction). The obtained form of the cross-section is rather atypical. It is definitely not a standard Breit-Wigner function. The experimental signal was measured with a quite poor accuracy, thus one could interpret it as a broad resonance peaked at around $4$ GeV. We then suggest that the $Y(4008)$ is not a standard resonance but a manifestation of the conventional state $\psi(4040)$ which strongly couples to $D^*D$ loops. More precisely, the puzzling `state' $Y(4008)$ may appear in the experiment as a consequence of the decay of $\psi(4040)$ via the $DD^*$ loop into the $\pi^+ \pi^- J/\psi$ channel. Note, this result does not exclude the earlier statement concerning the existence of a dynamically generated pole on the complex plane, but that pole, even if existent, should not be assigned to $Y(4008)$. 

\begin{figure}[h!]
\begin{center}
\includegraphics[width=0.63 \textwidth]{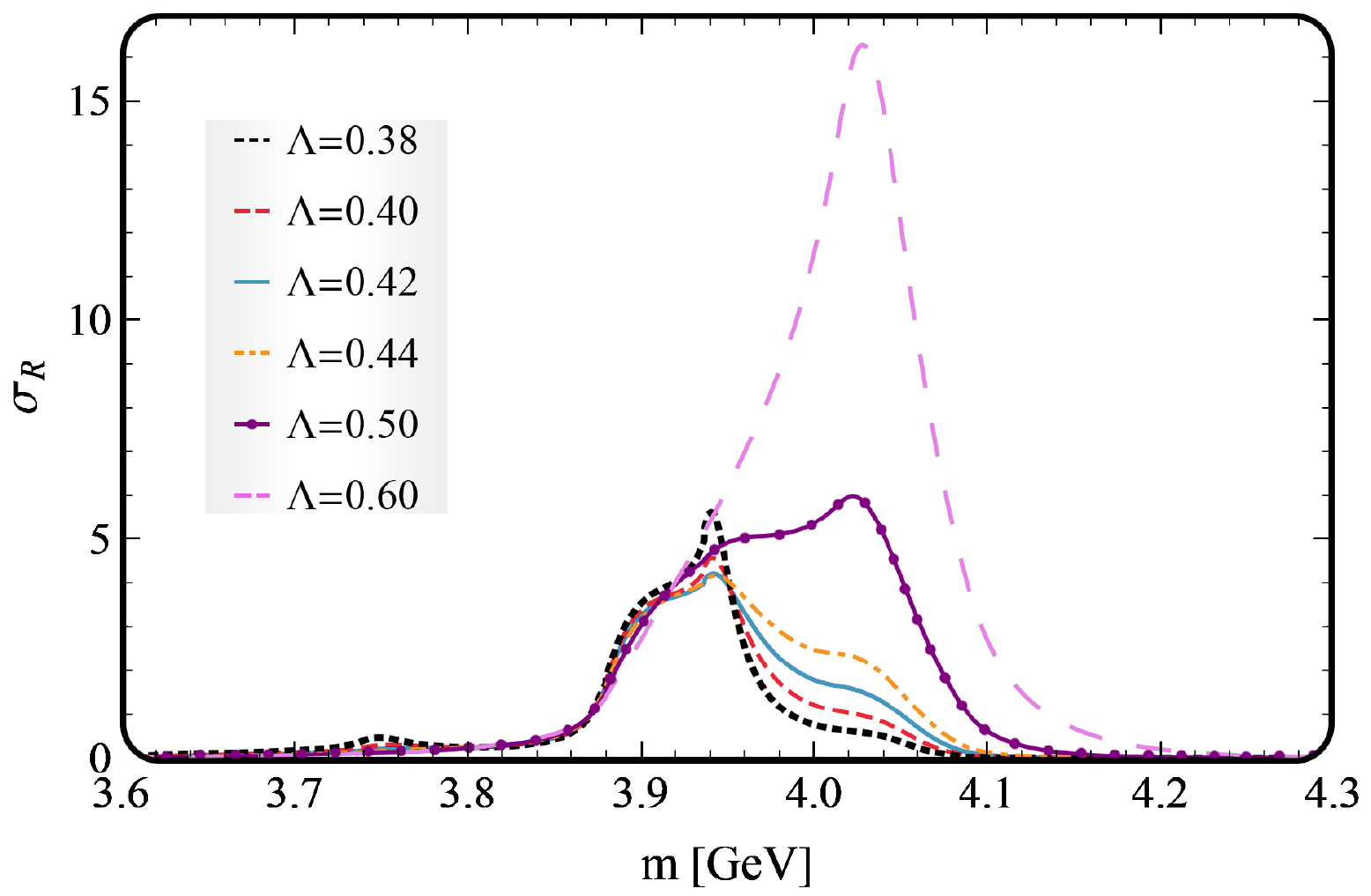} 
\caption{\label{app4040mp} Study of the form of the cross-section for the process $e^+e^- \rightarrow \psi(4040) \rightarrow J/ \psi \pi^+ \pi^-$, given by the Eq. (\ref{ratiompmp}), upon testing different values of $\Lambda$ parameter.}
\end{center} 
\end{figure}

For completeness, in  Figure \ref{app4040mp} we also report the results of the cross-section of Eq. (\ref{ratiompmp}) for different values of the cutoff parameter between $0.38$ GeV to $0.6$ GeV.
A broad structure peaked at around $3.95$ GeV is visible in the cross-section for all values from $0.38$ to $0.5$ GeV. Yet, for $\Lambda=0.5$ one can already find an additional peak near to $4.04$ GeV. Concerning higher values of the cutoff, for instance $\Lambda=0.6$ GeV, only a unique peak at around $4.04$ GeV is observed, which corresponds neatly to $\psi(4040)$. From this it follows that only for $\Lambda \lesssim 0.5$ GeV one can generate a signal consistent with $Y(4008)$.

\section{Concluding remarks}

In this chapter we have explored the energy regime nearby the vector charmonium resonance $\psi(4040)$, thus around $4$ GeV. Within our effective approach only a unique $c\bar{c}$ seed state, assigned to $\psi(4040)$, was included in the original formalism. We have shown that the shape of the spectral function deviates from the Breit-Wigner distribution due to the presence of significant enhancement below $4$ GeV. Moreover, two poles appeared in the complex plane. Their coordinates (being an illustrative results consistent with phenomenology) are: $(4.053 \pm 0.04)-(0.040 \pm 0.010) i $ GeV and $(3.934 \pm 0.006)-(0.030 \pm 0.001)i$ GeV. The former corresponds to the peak in the spectral function, thus to $\psi(4040)$, while the latter corresponds to an additional enhancement on the left side. The performed large-$N_c$ study reveals that $\psi(4040)$ resonance may be regarded as a conventional (predominantly) $c\bar{c}$ meson while the companion pole is generated  dynamically by the mesonic quantum fluctuations. 

Quite importantly, one should not assign this additional pole to $Y(4008)$. Namely, we have considered the reaction:
\begin{equation}
e^+e^- \rightarrow \psi(4040) \rightarrow D^*D \rightarrow J/ \psi + f_0(500) \rightarrow J/ \psi \pi^+ \pi^- \nonumber
\end{equation}
and we have shown that the peak around $3.9$ GeV appears in the cross-section as a consequence of the presence of the $DD^*$ loop. Thus, the strong coupling of the conventional $\psi(4040)$ resonance to $D^*D$ is the reason for the presence of such a broad structure in the spectral function, which can be assigned to $Y(4008)$, even if no real resonance is there. 

\chapter{Axial-vector states $\mathbf{\chi_{c1}(2P)}$ and~X(3872)}
\label{Xmain}
In this chapter we focus on the conventional charmonium state $\chi_{c1}(2P)$, which decays mainly into the $DD^*$ channel. To this end we exploit a QFT model containing originally only a $c\bar{c}$ seed state identified with $\chi_{c1}(2P)$ described by quantum numbers $J^{PC}=1^{++}$ or $n$ $^{2S+1}L_{J}$=$2$ $^{3}P_{1}$. 

We show that two peaks are present in the spectral function: a very broad one related to the original seed state and a very high and narrow one, whose maximum is situated nearby the $D^0D^{*0}$ energy threshold, related to the $X(3872)$ state. Moreover, we search for the propagator poles in the complex plane. Besides the pole related to the seed state $\chi_{c1}(2P)$ in the III RS, there is a second one, a virtual pole on the II RS which we identify with $X(3872)$. Within our study, the $X(3872)$ state appears naturally as a dynamically generated companion pole of the conventional seed state. 

Our interpretation of the $X(3872)$ delivers a consistent explanation of its molecular and quarkonium properties. In particular, the prompt production and the radiative decays indicate that $X(3872)$ is quarkonium-like state: we understand these properties as a consequence of the $c\bar{c}$ core. At the same time, the molecular features, such as the isospin breaking transition into $J/ \psi \rho$, can be easily explained by the $DD^*$ mesonic loops dressing the $c\bar{c}$ core and appearing as an intermediate state of this channel. In addition, several predictions are made. Our model, even if rather simple, gives satisfactory results which are comparable with the experimental ones. 

\section{Introduction}
\begin{figure}[h!]
\begin{center}
\includegraphics[width=0.90 \textwidth]{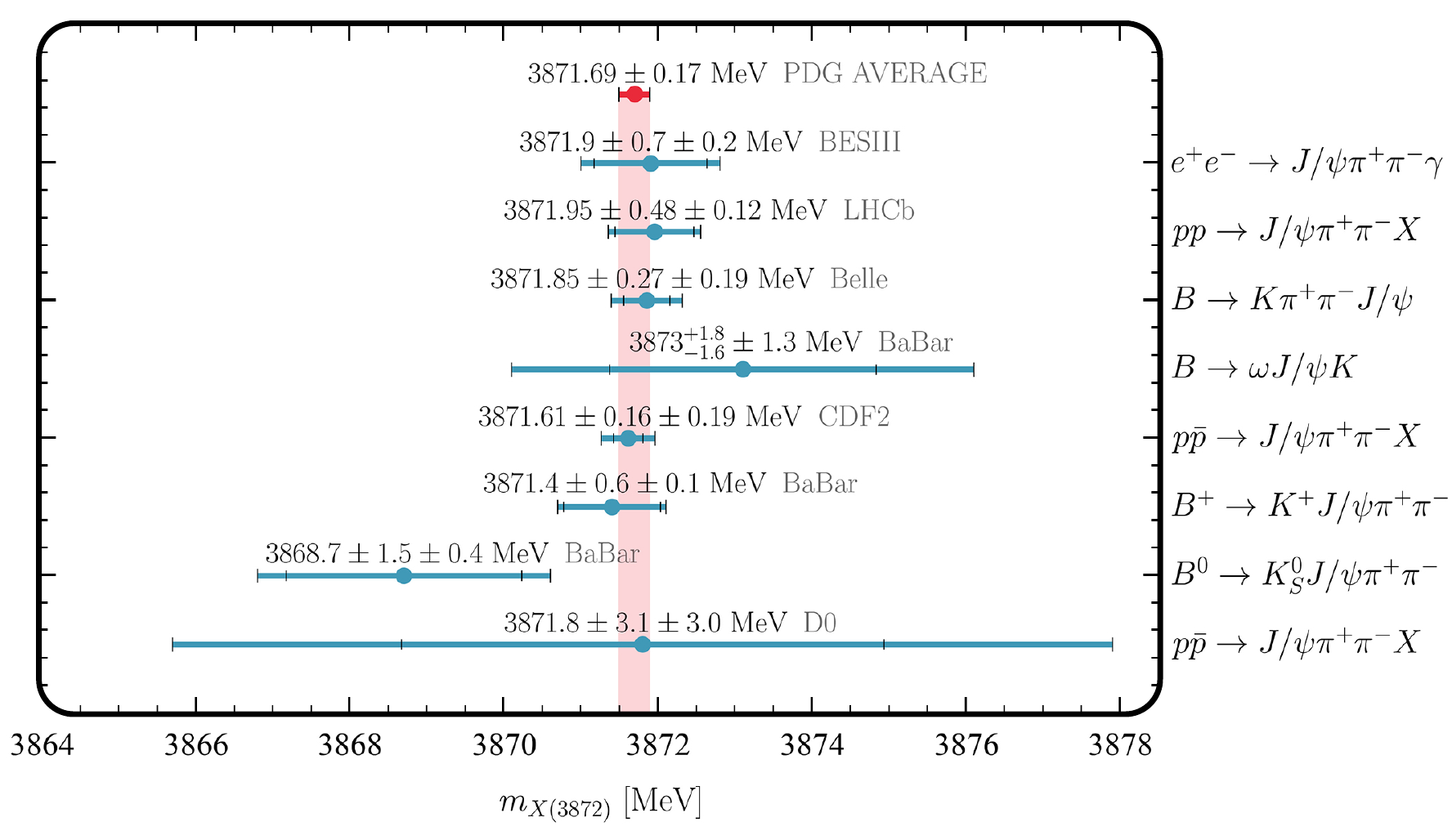}
\caption{\label{abzdrh} Comparison of the measured masses of the $X(3872)$ state used for the calculation of its average mass quoted by the PDG \cite{pdg}. }
\end{center} 
\end{figure}

In 2003 the Belle Collaboration first observed the $X(3872)$ resonance as an unexpected narrow enhancement appearing in the $B^+ \rightarrow K^+ J/ \psi \pi^+ \pi^-$ process \cite{belleX3872}. Later on, its existence has been confirmed by other collaborations in a series of experiments measuring various processes \cite{pdg}, see Figure \ref{abzdrh}. Despite huge efforts from the experimental and theoretical side, the understanding of the nature of $X(3872)$ still remains puzzling. The state $X(3872)$ has been the first resonance from the family of the so-called $X$, $Y$ and $Z$ states, which cannot be understood as conventional $q\bar{q}$ objects (for more details, see Refs. \cite{rev, rev2016, pillonireview, nielsen, Ali:2017jda, Olsen:2017bmm} and refs. therein). The resonance $X(3872)$, in PDG 2019 denoted as $\chi_{c1}(3872)$, has the average mass of $(3871.69 \pm 0.17)$ MeV, see Figure \ref{abzdrh}, where the results of different experiments are summarized. 
 Moreover, the PDG gives for the total decay width only an upper limit: 
\begin{equation}
\Gamma_{X(3872)}^{\text{PDG}}< 1.2 \text{ MeV at} \text{ 90}\% \text{ CL .} \nonumber
\end{equation} 
Accordingly, X(3872) is extremely narrow and its mass is situated very close to the $D^0D^{*0}$ threshold, but it is not yet fully clarified whether below or above it. From the predictions of the quark model there should be a $c\bar{c}$ state $\chi_{c1}(2P)$ with a mass of about $3.95$ GeV in the $J^{PC}=1^{++}$ sector (in spectroscopic notation characterized by the quantum numbers $n$ $^{2S+1}L_{J}=$ $2$ $^{3}P_{1}$) \cite{belle1, belle2, segovia3872}, however its mass is too high to be directly identified with $X(3872)$. 
 
Although the $X(3872)$ is widely studied in many experiments and intensively analyzed in theoretical models, still there is no consensus on its interpretation. Since the mass of $X(3872)$ lies nearby the $D^0D^{*0}$ threshold, it is often interpreted as a $D^0 D^{*0}$ molecule \cite{kalashnikova, Dong1, Dong2, Dong3, Gamermann1, Stapleton1, Hanhart1, Guo1}. This assignment offers an explanation of the isospin-breaking process $X(3872) \rightarrow J/ \psi \rho$. (Note, here and in the following we use $D^0D^{*0}$ for $D^0\bar{D}^{*0}+ h.c.$ in order to simplify the notation). Nevertheless, when considering the radiative decays and taking into account that $X(3872)$ was also observed in prompt production in heavy ion collisions, the charm-anticharm structure is favored \cite{Barnes1, Rogozina1, Rogozina2, Quigg1} (see Refs. \cite{Bignamini1, Esposito1, Albaladejo1, Aaboud1, EspositoGrinstein} for the related ongoing discussion). In order to unify both aspects various models were suggested \cite{Ortega12, CoitoRupp, Ferretti1, Ferretti2, Cardoso1, TakeuchiTakizawa}. These models are based on the idea that the $D^0D^{*0}$ and $c\bar{c}$ components simultaneously contribute to the $X(3872)$ wave function. Moreover, it was also proposed to interpret $X(3872)$ as a non-conventional diquark-antidiquark object \cite{Maianidiq}.

Following the idea of dynamically generated companion poles presented in previous chapters, we check if it is possible to describe the state $X(3872)$ by using a similar formalism. To this end we construct a relativistic Lagrangian containing a single axial-vector field, denoted as $\chi_{c1}^{\mu}$. This field, in the non-interacting case, corresponds to a $c\bar{c}$ seed state $\chi_{c1}(2P)\equiv \chi_{c1}$, for which a bare mass in the range from $3.9$ to $3.95$ GeV is expected. In our model $\chi_{c1}$ is strongly coupled to $D^0D^{*0}$ and $D^+D^{*-}$ channels. These decay products generate the quantum fluctuations which dress the bare $c\bar{c}$ seed state. The total decay width of $\chi_{c1}$ turns out to be about 80 MeV, consistent with the quark model predictions. Then, by studying the imaginary part of the dressed propagator, we compute the spectral function (mass distribution) of the charm-anticharm state $\chi_{c1}$. The spectral function exhibits a very peculiar shape: besides the expected (broad) peak related to the seed state, an additional and very peaked enhancement appears in the energy region nearby the $D^{*0}D^0$ threshold. In the simplest  scenario presented here, this enhancement is a \textit{virtual} state: this is a state that corresponds to a pole situated on the real energy axis just below the lowest $D^{*0}D^0$ threshold in the second Riemann sheet (II RS). Such a pole can be also interpreted as a companion pole. As one can expect, a pole of the standard $c\bar{c}$ seed state also exists and it is located above both thresholds ($D^+D^-$ and $D^0D^{*0}$) in the III RS. We find that its real part is in between 3.95 and 4 GeV, while its imaginary part is about 35 MeV. This result is compatible with the quark model predictions \cite{GodfreyIsgur, Barnes1}.

The idea presented here delivers a good explanation for both the molecular and charmonium features of the $X(3872)$. As it is shown  throughout this chapter, there is no need to consider the molecular and $q\bar{q}$ components in the Fock space separately. The result appears naturally as a consequence of mesonic quantum fluctuations which dress the bare charmonium state. It is worthwhile to emphasize that the $\chi_{c1}(2P)$ and $X(3872)$ are simultaneously described by only one spectral function, which is normalized to unity. Taking into account this normalization condition one can say that only ``one object'' is present.  Interestingly, for certain values of the parameters, only the peak nearby the $D^0 D^{*0}$ threshold is present in the spectral function and there is no peak related to the $c\bar{c}$ seed state. This may explain why the state $\chi_{c1}(2P)$ has not yet been visible in experiments.

Furthermore, we also study the strong decay of $X(3872)$ into $D^{*0}D^0$, and clearly show why this decay channel is the dominant one (the relevant decay width is estimated to be $\sim 0.5$ MeV). For completeness, we make predictions for the radiative decays into $\psi(1S)\gamma$ and $\psi(2S) \gamma$ and explain why they are consistent with the charmonium properties. Since $X(3872)$ is a part of the whole spectral function of $\chi_{c1}(2P)$, the corresponding values of the coupling constants to $\psi(1S)\gamma$ and $\psi(2S)\gamma$ are controlled by the underlying $q\bar{q}$ state (details later on). Then, the decay into $\psi(2S) \gamma$ is sizably larger than into $\psi(1S)\gamma$ and $X(3872)$ is seen in prompt production in heavy ion collisions. 

Moreover, because of dressing by the $DD^*$ mesonic loops, many properties typical for the molecular assignment also arise. The ratio 
\begin{equation}
\frac{X(3872) \rightarrow J/ \psi \rho \rightarrow J/ \psi \pi^+ \pi^-}{X(3872) \rightarrow J/ \psi \omega \rightarrow J/ \psi \pi^+ \pi^- \pi^0} \nonumber
\end{equation}
can be explained by considering the $D^+D^{*-}$ and $D^0D^{*0}$ loops and the small (but crucial) difference between them. 

Finally, the $X(3872)$ cannot exist without the $c\bar{c}$ seed state. By relying on $DD^*$ interaction strength only, and/or by setting the mass of seed state to infinity, one can observe that the peak related to $X(3872)$ fades away. This behavior is in agreement with other works on the subject \cite{CoitoRupp, Cardoso1}.

\section{Theoretical formalism}
We introduce the Lagrangian describing the decays of the state $\chi_{c1}(2P)$ into two mesonic pairs, $D^0\bar{D}^{*0}+ h.c.$ and $D^+D^{*-}+ h.c.$, respectively. Taking into account that the coupling constant, denoted here as $g_{\chi_{c1}DD^*}$ is the same for both charged and neutral channels (due to the isospin symmetry), the interaction Lagrangian has the form:
\begin{equation}
\mathcal{L}_{\chi_{c1}(2P)DD^{\ast}}=g_{\chi_{c1}DD^{\ast}}\chi_{c1,\mu
}\left[  D^{\ast0,\mu}\bar{D}^{0}+D^{\ast+,\mu}D^{-}+h.c.\right]  \text{ .}%
\end{equation}
The tree-level decay widths, calculated from the ordinary Feynmann rules, are:
\begin{align}
\Gamma_{\chi_{c1}(2P)\rightarrow D^{\ast0}\bar{D}^{0}+h.c.}(m)  &
=2\frac{k(m,m_{D^{\ast0}},m_{D^{0}})}{8\pi m^{2}}\frac{g_{\chi_{c1}DD^{\ast}%
}^{2}}{3}\left(  3+\frac{k^{2}(m,m_{D^{\ast0}},m_{D^{0}})}{m_{D^{\ast0}}^{2}%
}\right)  F_{\Lambda}(m)\text{ ,}\label{dw1}\\
\Gamma_{\chi_{c1}(2P)\rightarrow D^{\ast+}D^{-}+h.c.}(m)  &  =2\frac
{k(m,m_{D^{\ast+}},m_{D^{+}})}{8\pi m^{2}}\frac{g_{\chi_{c1}DD^{\ast}}^{2}}%
{3}\left(  3+\frac{k^{2}(m,m_{D^{\ast+}},m_{D^{+}})}{m_{D^{\ast+}}^{2}%
}\right)  F_{\Lambda}(m)\text{ .} \label{dw2}%
\end{align}
In the above equations, the quantity $m$ is the ``running'' mass of the $\chi_{c1}$ state. As usual, $k\equiv k(m)\equiv k(m, m_A, m_B)$ is the modulus of the three-momentum of one emitted meson. The mass values of the decay products, taken from the PDG \cite{pdg}, are: 
\begin{equation}
m_{D^0}=1864.83 \pm 0.05 \text{ MeV} \text{ ,}
\end{equation}
\begin{equation}
m_{D^{*0}}=2006.85 \pm 0.05 \text{ MeV} \text{ ,}
\end{equation}
\begin{equation}
m_{D^+}=m_{D^-}=1869.65 \pm 0.05 \text{ MeV} \text{ ,}
\end{equation}
\begin{equation}
 m_{D^{*+}}=m_{D^{*-}}=2010.26 \pm 0.05 \text{ MeV} \text{ .}
\end{equation}

In analogy to the previous chapters, also here we regularize the theory by using the form factor $F_{\Lambda}(k)$. We again choose the Gaussian function for the same reasons discussed previously

\begin{equation}
F_{\Lambda}(m)= e^{-\frac{2k^2(m)}{\Lambda^2}}\text{ .} \label{gaussxx}
\end{equation}
For comparison, in Appendix \ref{appX3782} we also test the dipolar form factor of Eq. (\ref{dipolarpsi4040}). As we show later on, the results are not strongly influenced by the particular choice of the form factor. The parameter  $\Lambda$ varies between $0.4$ and $0.8$ GeV ($\Lambda\simeq 0.5$ GeV for $K^*_0(700)$ and $\Lambda \simeq 0.4$ GeV for $\psi(4040)$). Here, we start with a typical value of $\Lambda=0.5$ GeV \cite{Wolkanowskija, Piotrowskapsi4040}, but it should not be regarded as a sharp value. Later on, we shall show that varying this paramater in the range from $0.4$ to $0.8$ GeV also gives qualitatively similar results. 

Next, we turn to the propagator of the $\chi_{c1}$ field, which is, in our approach, calculated at one-loop level. Its scalar part, as function of $s=m^2$, is expressed as 
 
\begin{equation}
\Delta(s)=\frac{1}{s-M_{0}^{2}+\Pi(s)}\text{ .} \label{prop}%
\end{equation}
As usual, $M_0$ corresponds to the mass of the bare $q\bar{q}$ state $\chi_{c1}(2P)$, which according to quark model is about $3.95$ GeV \cite{GodfreyIsgur}. Again, $\Pi(s)$ stands for the self-energy contribution. One can write it explicitly as the sum of the $\bar{D}^0D^{*0}$ and $D^-D^{*+}$ loops: 
\begin{equation}
\Pi(s)=\Pi_{D^{\ast0}\bar{D}^{0}+h.c.}(s)+\Pi_{D^{\ast+}D^{-}+h.c.}(s)=g_{\chi
_{c1}DD^{\ast}}^{2}\left[  \Sigma_{D^{\ast0}\bar{D}^{0}+h.c.}(s)+\Sigma
_{D^{\ast+}D^{-}+h.c.}(s)\right]\  \text{ .} \label{sigmas}%
\end{equation}
At the one-loop approximation, the functions $\Sigma_{D^{\ast+}D^{-}+h.c.}(s)$ and $\Sigma_{D^{\ast0}\bar{D}^{0}+h.c.}(s)$ entering Eq. (\ref{sigmas}) are independent of the value of the coupling constant $g_{\chi_{c1}DD^{\ast}}$. Furthermore, one can decompose $\Pi(s)$ into real and imaginary parts. The latter can be obtained from the optical theorem and it is
 
\begin{equation}
\operatorname{Im}\Pi(s)=\sqrt{s}\left[  \Gamma_{\chi_{c1}(2P)\rightarrow
D^{\ast0}\bar{D}^{0}+h.c.}(\sqrt{s})+\Gamma_{\chi_{c1}(2P)\rightarrow D^{\ast
+}D^{-}+h.c.}(\sqrt{s})\right]  \text{ }.
\end{equation}
The real part is calculated from the dispersion relation (valid only if $\sqrt{s}$ is real and exceeds the value $m_{D^{*+}}+m_{D^-}$):
\begin{align}
\operatorname{Re}\Pi(s)  &  =\frac{\mathcal{P}}{\pi}\int_{(m_{D^{\ast0}}+m_{D^{0}}%
)^{2}}^{\infty}\sqrt{s^{\prime}}\frac{\Gamma_{\chi_{c1}(2P)\rightarrow
D^{\ast0}\bar{D}^{0}+h.c.}(\sqrt{s^{\prime}})}{s^{\prime}-s}\ \mathrm{d}%
s^{\prime}\nonumber\\
&  +\frac{\mathcal{P}}{\pi}\int_{(m_{D^{*+}}+m_{D^{+}})^{2}}^{\infty}\sqrt{s^{\prime}}%
\frac{\Gamma_{\chi_{c1}(2P)\rightarrow D^{\ast+}D^{-}+h.c.}(\sqrt{s^{\prime}}%
)}{s^{\prime}-s}\ \mathrm{d}s^{\prime}.
\end{align}
For $\sqrt{s}$ smaller than $m_{D^{*0}}+m_{D^0}$ or if the imaginary part is not equal to zero, $\mathcal{P}$ is not taken and $\Pi(s=z^2)$, in the first Riemann sheet (I RS) is:
\begin{eqnarray}
\Pi(s=z^{2})&=&\frac{1}{\pi}\int_{\left(  m_{D^{\ast0}}+m_{D^{0}}\right)  ^{2}%
}^{\infty}\sqrt{s^{\prime}}\frac{\Gamma_{\chi_{c1}\rightarrow D^{\ast0}\bar
{D}^{0}+h.c.}(\sqrt{s^{\prime}})}{s^{\prime}-z^{2}}\ \mathrm{d}s^{\prime} \nonumber \\%
&+&\frac{1}{\pi}\int_{\left(  m_{D^{*+}}+m_{D^{+}}\right)  ^{2}}^{\infty}%
\sqrt{s^{\prime}}\frac{\Gamma_{\chi_{c1}\rightarrow D^{\ast+}D^{-}+h.c.}%
(\sqrt{s^{\prime}})}{s^{\prime}-z^{2}}\ \mathrm{d}s^{\prime}\text{ }.
\label{selfenergycomplex}%
\end{eqnarray}

Once more, we present the spectral function 
\begin{equation}
d_{\chi_{c1}(2P)}(m)=-\frac{2m}{\pi}\operatorname{Im}[\Delta(s=m^{2})]\text{
,} \label{dchi}%
\end{equation}
together with the crucial normalization condition 
\begin{equation}
\int_{m_{D^{\ast0}}+m_{D^{0}}}^{\infty}\mathrm{dm}\ d_{\chi_{c1}(2P)}(m)=1\text{
,} \label{norm}%
\end{equation}
which allows to verify if the numerical calculations are correct. One can interpret the quantity $\mathrm{dm}\ d_{\chi_{c1}(2P)}(m)$ as the probability that the mass of the unstable $\chi_{c1}(2P)$ state is between $m$ and $m+dm$ \cite{fggp, Giacosa:2012hd, duecan, Matthews:1959sy}.

As a next step, we introduce the radiative decays described by the following terms
\begin{equation}
\mathcal{L}_{\chi_{c1}\text{-rad}}=g_{\chi_{_{c1}}\psi(1S)\gamma}\ %
\chi_{c1,\mu}\psi(1S)_{\nu}\tilde{F}^{\mu\nu}+g_{\chi_{_{c1}}\psi
(2S)\gamma}\ \chi_{c1,\mu}\psi(2S)_{\nu}\tilde{F}^{\mu\nu}+...\ \text{,}
\label{radlag}%
\end{equation}
containing the two coupling constants $g_{\chi_{_{c1}}\psi(1S)\gamma}$ and $g_{\chi_{_{c1}}\psi(2S)\gamma}$. Their numerical values are taken from quark model of Ref. \cite{Barnes1}.

\section{Results of the model}
Let us now present the main results of our study. We start from the spectral function and poles in the complex plane. Next, we concentrate on the strong decays into the dominant channel $D^{*0}D^0$. Moreover, we discuss some important consequences of our model, such as radiative decays, prompt production, and finally, the isospin breaking decay into $J/ \psi \rho$. 
\subsection{Spectral function and poles}
Here we consider two scenarios, depending on the value of the bare mass $M_0$. We named them as `\textit{case I}' and `\textit{case II}', respectively. In both cases, the $\Lambda$ parameter is chosen to be $0.5$ GeV. Later on, we test different values of $\Lambda$, in order to check its influence on the results. The values of the masses of the decay products are taken from the PDG \cite{pdg} and are listed in Table \ref{tablemassesX3872}. 
\begin{table}[htbp] 
\centering
\renewcommand{\arraystretch}{1.35}
\begin{tabular}{c|c|c|c|c} 
\hline
\hline
State& $D^0$&$D^+$&$D^{*0}$&$D^{*+}$ \\
\hline
Value [MeV]&$1864.83 \pm 0.05$&$1869.65 \pm 0.05$&$2006.83 \pm 0.05$&$2010.26 \pm 0.05$\\
\hline
\hline
\end{tabular}
\caption{Numerical values of the masses of the decay products of $\chi_{c1}(2P)$ state, taken from PDG \cite{pdg}. We neglect the small errors in the calculations.}
\label{tablemassesX3872}
\end{table}
The relevant thresholds are: $m_{D^+}+m_{D^{*+}}=3.87985$ GeV and $m_{D^0}+m_{D^{*0}}=3.87168$ GeV, respectively. 
\begin{center}
\textbf{Case I} $\mathbf{(M_0=3.95}$ \textbf{GeV)}
\end{center}
Let us first study the case in which the bare mass $M_0$ is set to $3.95$ GeV (in close agreement with quark model predictions of Ref. \cite{GodfreyIsgur}). The coupling constant $g_{\chi_{c1}DD^*}$ is obtained by imposing the requirement:
\begin{equation}
\operatorname{Re}[\Delta^{-1}(s=m_{\ast}^{2})]=0, \text{ for }m_{\ast
}=3.874\text{ GeV ,}\label{mstar}%
\end{equation}
out of which $g_{\chi_{c1}DD^*}=9.732$ GeV. Note that the value of $m_{*}$ has been chosen to be between the $D^0D^{*0}$ and $ D^{*+}D^-$ thresholds. (Any other choice of $m_*$ within this range would not change the results significantly).

\begin{figure}[h!]
\begin{center}
\includegraphics[width=0.72 \textwidth]{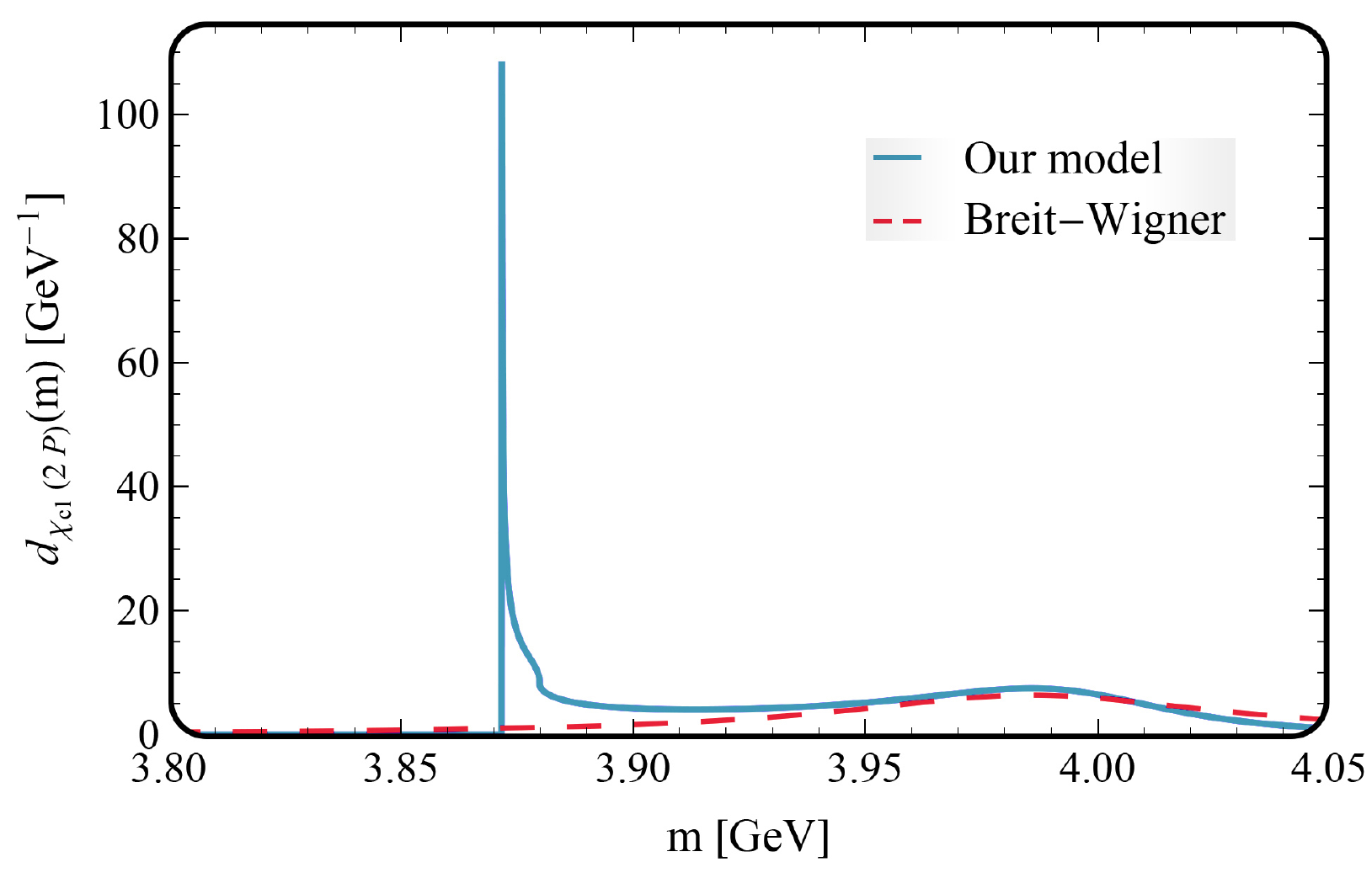}
\caption{\label{1X} The shape of the normalized spectral function $d_{\chi_{c1}(2P)}(m)$ defined in Eq. (\ref{dchi}), plotted for the case I (blue solid line) and compared to the Breit-Wigner distribution, plotted for $m_{BW}=3.986$ GeV and $\Gamma_{BW}=79.7$ MeV (red dashed line). It is visible that the spectral function of our model has two peaks. The broad one, on the right hand side, with the maximum at around $3.99$ GeV, corresponds to the $c\bar{c}$ seed state. The extremely high and narrow peak, situated close to the lowest threshold, corresponds to the $X(3872)$ state.}
\end{center}
\end{figure}

In Figure \ref{1X} we show the spectral function of Eq. (\ref{dchi}). 
It contains two peaks: the first one, extremely narrow and high, is realized very close to the lowest $D^0D^{*0}$ threshold and is assigned to the $X(3872)$ state; the second one, very broad, is located at $3.986$ GeV and corresponds to a state with a width of about 80 MeV, in agreement with the predictions of the quark model. We emphasize that both peaks come from the single seed state $\chi_{c1}(2P)$, which has been initially included in the Lagrangian. Moreover, the total spectral function of Figure \ref{1X} is normalized to unity, therefore it describes simultaneously both states $\chi_{c1}(2P)$ and the $X(3872)$. 

Let us now consider the integral  
\begin{equation}
\int_{m_{D^{\ast0}}+m_{D^{0}}}^{m_{D^{+}}+m_{D^{\ast+}}}\mathrm{dm}\ %
d_{\chi_{c1}(2P)}(m)=0.160\text{ ,}%
\end{equation}
meaning that $16\%$ of the total spectral function $d_{\chi_{c1}(2P)}$ is included in the energy region between the two thresholds. Taking into account that the decay width of $X(3872)$ is much smaller than $8$ MeV and that the experimental uncertainty of this width is around $1$ MeV, we also compute the following integral
\begin{equation}
\int_{m_{D^{\ast0}}+m_{D^{0}}}^{m_{D^{\ast0}}+m_{D^{0}}+1\text{ MeV}%
}\mathrm{dm}\ d_{\chi_{c1}(2P)}(m)=0.049\text{ ,} \label{integralthr}%
\end{equation}
according to which $X(3872)$ corresponds to $4.9\%$ of the spectral function $d_{\chi_{c1}(2P)}(m)$. Whenever referring to the peak related to the $X(3872)$ we shall keep the estimated value of $1$ MeV for the extension of the width. 

\begin{figure}[h!]
\begin{center}
\includegraphics[width=0.72 \textwidth]{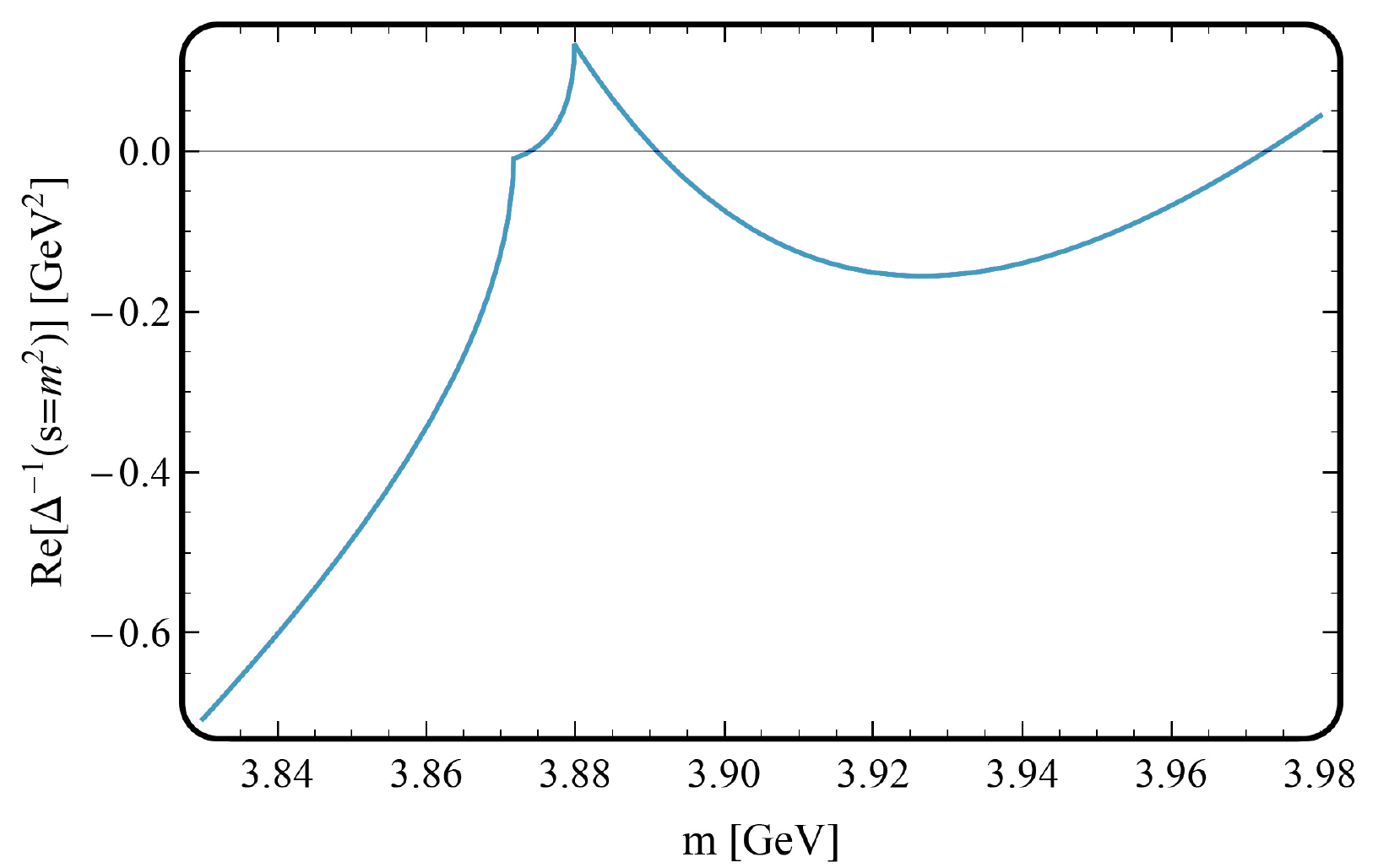}
\caption{\label{Fig2dokXf} The function $\text{Re}[\Delta^{-1}(m^2)]$ plotted for case I. One observes that three zeros are reached. The first one, on the right hand side, for $m=3.973$ GeV, corresponding to the broad peak in the spectral function, thus to the seed $c\bar{c}$ state. The second, on the left hand side, for $m_{*}=3.874$ GeV, referring to the high peak of $X(3872)$. The third zero, in the center, is not associated with any peak or pole.}
\end{center}
\end{figure}

In the next step, we investigate the function $\text{Re}[\Delta^{-1}(s=m^2)]$, which is plotted in Figure \ref{Fig2dokXf}. 
One observes that it has three zeros. The first one (on the left side) realized for $m_{*}=3.874$ GeV and corresponds to the high peak related to the $X(3872)$ state close to the $D^{*0}D^0$ threshold. The third zero (on the right side) is obtained for $m=3.973$ GeV and relates to the broad peak in the spectral function. The second zero (in the middle) for $m=3.891$ GeV is not related to any peak or state, because the derivative of the $\text{Re}[\Delta^{-1}(s=m^2)]$ is negative, see Ref. \cite{Boglioneprd} for a detailed discussion of this aspect. 

When investigating the complex plane, we found a pole of the charm-anticharm seed state corresponding to the broad peak in the spectral function. It is situated on the III RS  (above both thresholds) and has the following coordinates:
\begin{equation}
3.995-0.036i \hspace{0.25cm}\text{GeV}  \text{ .} \label{polecase1}
\end{equation}
By doubling the imaginary part one gets for the pole width the value of $72$ MeV. Moreover, we find a virtual pole on the II RS, too. It is located on the real axis, below the lowest $D^0D^{*0}$ threshold. Its coordinates are:
\begin{equation}
3.87164-i\varepsilon \hspace{0.25cm}\text{GeV}  \text{ .}
\end{equation}
This virtual pole is assigned to the $X(3872)$ state emerging as a high and narrow peak in the spectral function of Figure \ref{1X}. 

In summary, the evidences for  the $X(3872)$ state are:\newline
(i) zero of the function  $\text{Re}[\Delta^{-1}(s=m^2)]$ for $m_{*}=3.874$ GeV,\newline
(ii) existence of the virtual pole for $3.87164-i\varepsilon$ GeV on the II RS. 

Notice that the virtual pole lies very near to the $D^{*0}D^0$ threshold, namely only $0.04$ MeV below it. Indeed, the exact value of $m_{*}$ and the position of this pole may change upon varying the model parameters. However, the overall picture remains firmly stable.   

We now study the strong decay widths of $\chi_{c1}(2P)$, which can be estimated at the value of the maximum of the peak ($3.986$ GeV) as:
\begin{align}
\Gamma_{\chi_{c1}(2P)\rightarrow D^{\ast0}\bar{D}^{0}+h.c.}(3.986\text{ GeV}) &
=38.1\text{ MeV ,}\label{wch1}\\
\Gamma_{\chi_{c1}(2P)\rightarrow D^{\ast+}D^{-}+h.c.}(3.986\text{ GeV}) &
=41.6\text{ MeV .}\label{wch2}%
\end{align}
Then, the total decay width of $\chi_{c1}(2P)$ is $79.7$ MeV. We use this width to construct the Breit-Wigner function plotted in Figure \ref{1X}. Comparable results are obtained by employing the mass of the corresponding pole. 

A direct estimate of the width of the high peak corresponding to $X(3872)$ is much more complicated, since the peak is extremely narrow. Its width at half maximum has the value of about $0.1$ MeV. Even by using different sets of parameters, this value changes only slightly and never exceeds $0.5$ MeV. Therefore, as a reliable estimate of the main decay channel of $X(3872)$ we use the average value given by the following integral 
\begin{equation} 
\Gamma_{X(3872)\rightarrow D^{\ast0}\bar{D}^{0}+h.c.}^{\text{average}}%
=\int_{m_{D^{\ast0}}+m_{D^{0}}}^{m_{D^{\ast0}}+m_{D^{0}}+1\text{ MeV}%
}\mathrm{dm}\ \Gamma_{\chi_{c1}(2P)\rightarrow D^{\ast0}\bar{D}^{0}%
+h.c.}(m)d_{\chi_{c1}(2P)}(m)=0.61\text{ MeV. } \label{intwidth}%
\end{equation}
The obtained result shows that the decay $X(3872) \rightarrow \bar{D}^0D^{*0}+ h.c.$ is sizable. We recall that the width of $X(3872)$ is estimated to be smaller than $1.2$ GeV, thus our value of $0.61$ GeV looks reasonable. On the contrary, $\Gamma_{X(3872) \rightarrow D^{*+}D^-+h.c.}$ vanishes, since the mass is below the threshold. 

For completeness, in Appendix \ref{appX3782} we report the results obtained for different choices of the parameters. They are in qualitative agreement with the ones presented above and can be also understood as an estimate of the uncertainties of our study. In particular we checked how the results depend on the $\Lambda$ parameter. To this end we left $m_{*}$ unchanged by adjusting the value of the coupling constant. Moreover, we tested different values of coupling constant and fix $\Lambda$ by changing $m_{*}$. It can be observed that the smaller the coupling, the more $m_{*}$ moves toward higher $D^{*+}D^-$ threshold. At the same time, the peak associated to $X(3872)$ becomes lower and gradually disappears. For a sufficiently small $g_{\chi_{c1}DD*}$, $m_*$ exceeds the $D^{*+}D^-$ threshold and the peak of $X(3872)$ finally vanishes. In the opposite case, when the coupling constant is larger, $m_*$ tends to the lower $D^{*0}D^0$ threshold and the peak related to $X(3872)$ becomes higher. When $g_{\chi_{c1}DD^*}$ exceeds a critical value, $g_{\chi_{c1}DD^*}^{\text{critical}}=9.808$ GeV (which corresponds to $m_*$ being exactly at the $D^{*0}D^0$ threshold), a pole in the I RS appears. This means that an extra (quasi-)stable bound state arises. For such scenario, the spectral function is as follow \cite{fggp}:
\begin{equation}
d_{\chi_{c1}(2P)}(m)=Z\delta(m-m_{BS})+d_{\chi_{c1}(2P)}^{\text{above
threshold}}(m) \text{ ,}\label{realpoleds}%
\end{equation}
where the quantity $m_{BS}$ stands for the mass of this dynamically generated molecular-like bound state. Although, this bound state is strictly connected to the seed state and could not exist by itelfs. The normalization condition, given by Eq. (\ref{norm}), is still fulfilled:
\begin{equation}
Z+\int_{m_{D^{\ast0}}+m_{D^{0}}}^{\infty}\mathrm{dm}\ d_{\chi_{c1}%
(2P)}^{\text{above threshold}}(m)=1.
\end{equation}
Next, we consider a specific example in which the coupling constant exceeds the critical value. For $g_{\chi_{c1}DD^*}=10$ GeV, one obtains $m_{BS}=3.87164$ GeV (pole on the I RS) and $Z=0.0465$. The spectral function of Eq. (\ref{dchi}) and that one of Eq. (\ref{realpoleds}) have similar shapes. For this (quasi-)bound state the decays into $D^{0}D^{*0}$ and $D^-D^{*+}$ channels do not take place, but decays into the light hadrons and through radiative processes are possible. This is why the width related to this pole is very small but nonzero. Nevertheless, the peak above the $D^{*0}D^0$ threshold is still present. We stress that the amount of $d_{\chi_{c1}2P}^{\text{above threshold}}(m)$ between the two thresholds is $0.133$. However, when computing the integral from $m_{D^{*0}}+m_{D^0}$ to $m_{D^{*0}}+m_{D^0}+1 \text{ MeV}$, one has $0.035$, which is less then in the case given by Eq. (\ref{integralthr}). 
 
\begin{figure}[h!]
\begin{center}
\includegraphics[width=0.72 \textwidth]{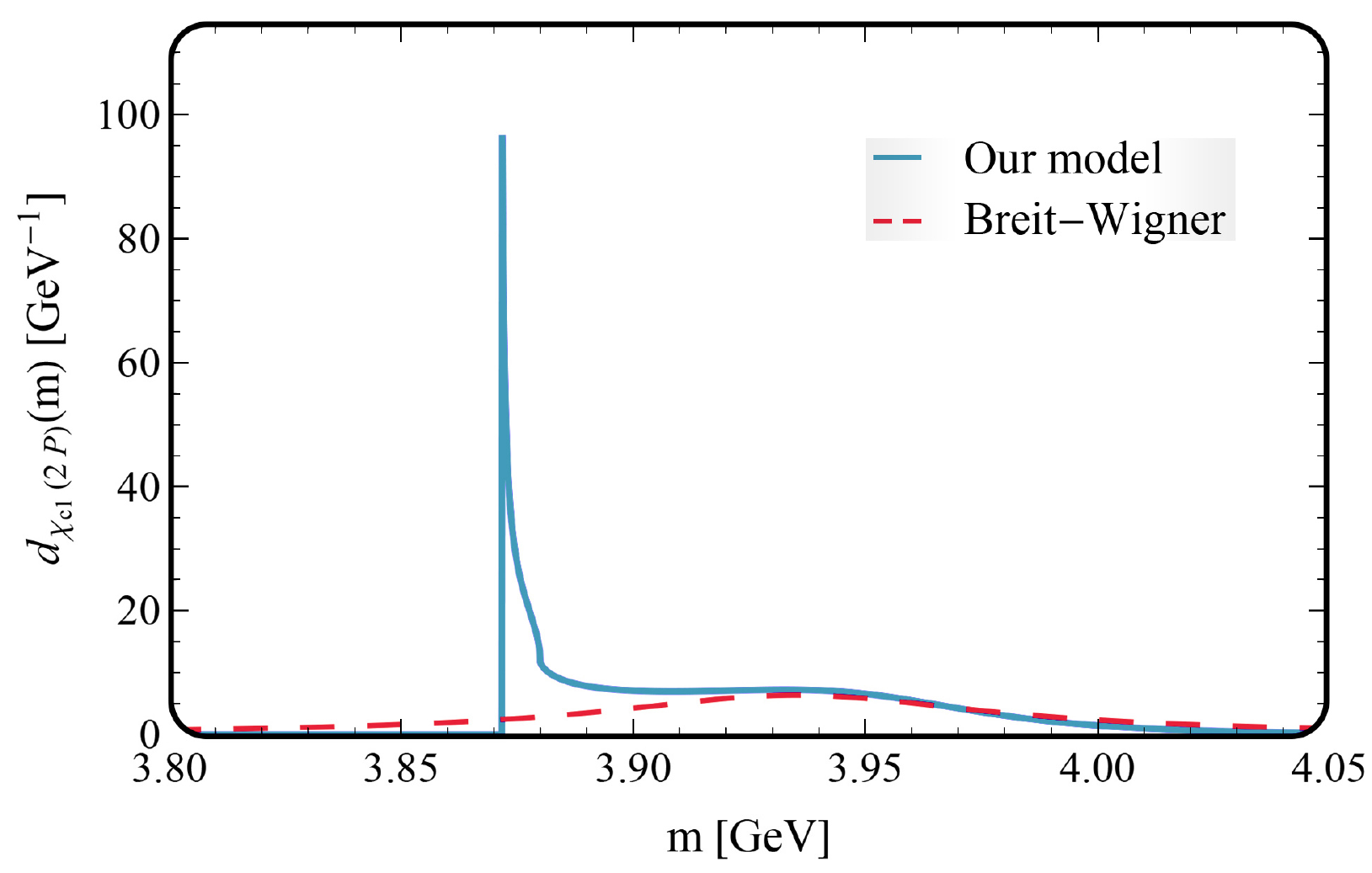}
\caption{\label{spfcase2} The same as in Figure \ref{1X} but for case II. Now the only one peak, corresponding to $X(3872)$ is present. There is no peak related to the $c\bar{c}$ seed state.}
\end{center}
\end{figure}

\begin{center}
\textbf{Case II} $\mathbf{(M_0=3.92}$ \textbf{GeV)}
\end{center}
We now consider the second scenario, in which the bare mass, $M_0$, is set to $3.92$ GeV. The choosen value is very close to the one reported in Ref. \cite{Ebertfaustov} and is only a bit smaller than the value listed in Ref. \cite{GodfreyIsgur}. Also in this case, the coupling constant is obtained by imposing the same condition as for case I, $m_{*}=3.874$ GeV, in which $g_{\chi_{c1}DD^*}=7.557$ GeV. In Figure \ref{spfcase2} we plot the spectral function, compared to standard Breit-Wigner distribution. 
One observes that the high peak nearby the $D^{*0}D^0$ threshold is still present, while the broad one, referring to the $c\bar{c}$ is not present. Interestingly, a similar behavior is reported in Refs. \cite{TakeuchiTakizawa, Braaten2007dw}. 

The integral of $d_{\chi_{c1}(2P)}(m)$ between the $D^{*0}D^0$ and the $D^{*+}D^-$ thresholds

\begin{equation}
\int_{m_{D^{\ast0}}+m_{D^{0}}}^{m_{D^{+}}+m_{D^{\ast+}}}\mathrm{dm}\ d_{\chi_{c1}%
(2P)}(m)=0.250%
\end{equation}
amounts to $25\%$ of the total spectral function (more than in case I). For what concerns the peak related to $X(3872)$, one has (using the `1 MeV width' estimate):
\begin{equation}
\int_{m_{D^{\ast0}}+m_{D^{0}}}^{m_{D^{\ast0}}+m_{D^{0}}+1\text{ MeV}}%
\mathrm{dm}\ d_{\chi_{c1}(2P)}(m)=0.067\text{ .}%
\end{equation}
This means that in this case $X(3872)$ corresponds to $6.7\%$ of the entire spectral function. 

For completeness, in Figure \ref{recase2} we show the function $\text{Re}[\Delta^{-1}(s=m^2)]$:
\begin{figure}[h!]
\begin{center}
\includegraphics[width=0.72 \textwidth]{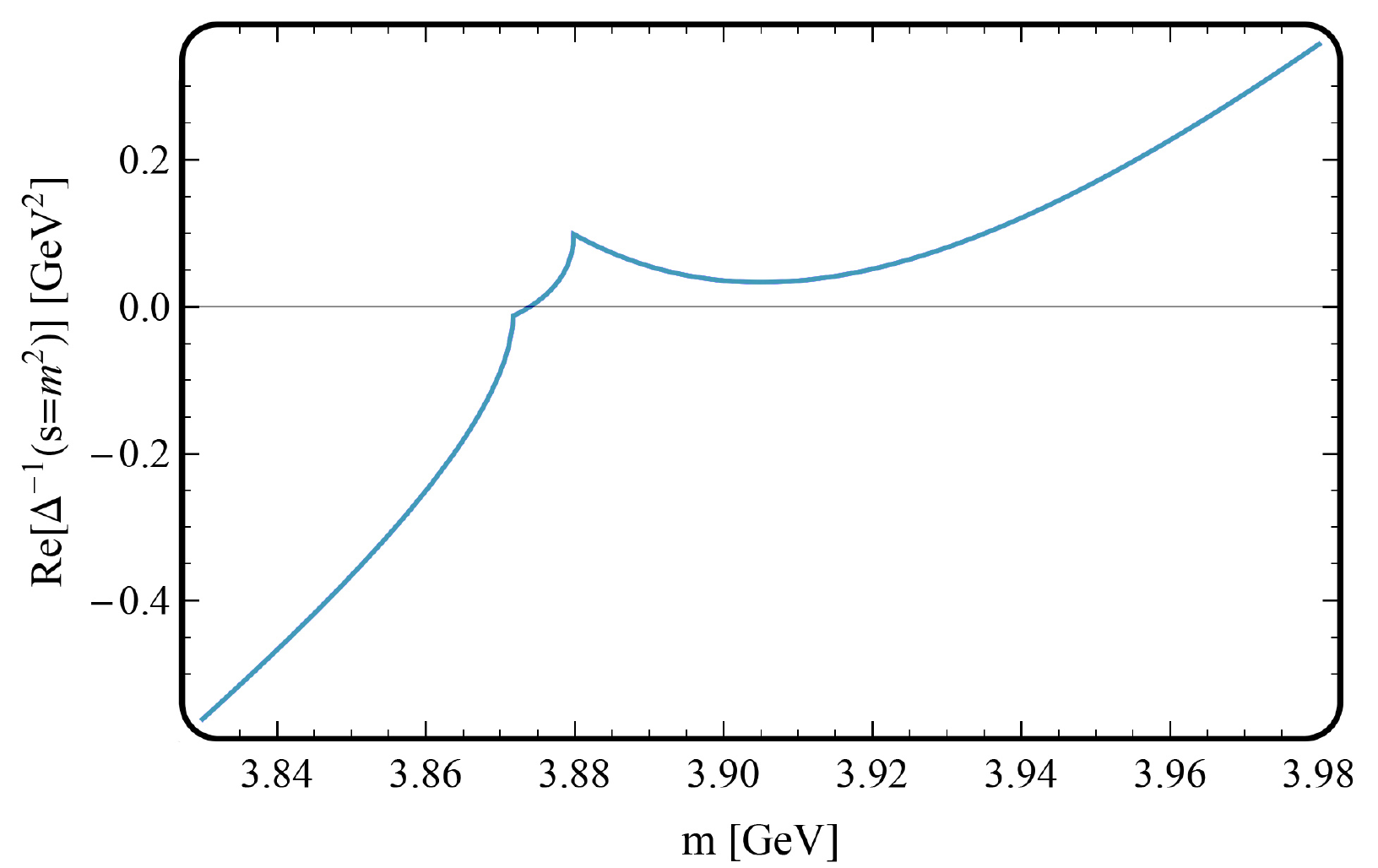}
\caption{\label{recase2} The function $\text{Re}[\Delta^{-1}(m^2)]$ plotted for case II. One observes only one zero for $m_{*}=3.874$ GeV, corresponding  to the high $X(3872)$ peak in the spectral function. The virtual companion pole is situated in the II RS.}
\end{center}
\end{figure} 
it has only one zero for $m_*=3.874$ GeV. This result is compatible with absence of the peak associated to the $c\bar{c}$ seed in the spectral function of Figure \ref{spfcase2}. Nevertheless, we have found a pole related to the seed state which is situated on the III RS and has the coordinates:
\begin{equation}
3.953-0.044i \hspace{0.25cm}\text{GeV}  \text{ .}
\label{polecase22}
\end{equation}
As one can see, it is quite similar to that one reported in Eq. (\ref{polecase1}). From Eq. (\ref{polecase22}) we conclude that the  width of the pole reads $88$ MeV. From the above considerations one may see that the determination of the poles delivers valuable information. Even if no peak is present in the spectral function, a state can still occur. Moreover, there is a virtual pole on the II RS, slightly below $D^{*0}\bar{D}^0$ threshold, realized for
\begin{equation}
3.87160-i\varepsilon \hspace{0.25cm}\text{GeV}  \text{ ,}
\label{virtpolecase22}
\end{equation}
that corresponds to $X(3872)$.
Next, we study the partial widths. Since in this case no peak is present, the on-shell mass is fixed to $\text{Re}[z_{\text{pole}}]=3.953$ GeV. We have:
\begin{align}
\Gamma_{\chi_{c1}(2P)\rightarrow D^{\ast0}\bar{D}^{0}+h.c.}(3.953)  &
=32.9\text{ MeV}\text{ },\label{sd1}\\
\Gamma_{\chi_{c1}(2P)\rightarrow D^{\ast+}D^{-}+h.c.}(3.953)  &  =35.4\text{
MeV}\text{ }. \label{sd2}%
\end{align}

The high peak associated to $X(3872)$ has about $0.91$ MeV width at half-maximum. The result
\begin{equation}
\Gamma_{X(3872)\rightarrow D^{\ast0}\bar{D}^{0}+h.c.}^{\text{average}}%
=\int_{m_{D^{\ast0}}+m_{D^{0}}}^{m_{D^{\ast0}}+m_{D^{0}}+1\text{ MeV}%
}\mathrm{dm}\ \Gamma_{\chi_{c1}(2P)\rightarrow D^{\ast0}\bar{D}^{0}%
+h.c.}(m)d_{\chi_{c1}(2P)}(m)=0.54\text{ MeV }%
\end{equation}
is in good agreement with the previous case. 

Changing the value of the parameter $\Lambda$ as well as varying the value of coupling constant $g_{\chi_{c1}DD^*}$ generate similar results, see Appendix \ref{appX3782} for more details. 
\subsection{Applications of the model}
In this subsection we study some applications of our model. First, we concentrate on radiative decays of the state $\chi_{c1}(2P)$. Then we discuss the prompt production of $X(3872)$. Finally, we describe the isospin breaking decay into $J/ \psi \rho$. In the following, we use the numerical values of the parameters of case I. 
\begin{center}
\textbf{Radiative decays}
\end{center}
We inspect closer radiative decays. We start from Eq. (\ref{radlag}) which describes the two radiative processes:
\begin{equation}
\chi_{c1}(2P) \rightarrow \psi(1S)\gamma \text{ and } \chi_{c1}(2P) \rightarrow \psi(2S)\gamma \text{ .} \nonumber
\end{equation}
This Lagrangian contains two coupling constants
\begin{equation}
g_{\chi_{c1}(2P)\psi(1S)\gamma} \text{ and } g_{\chi_{c1}(2P)\psi(2S)\gamma} \text{ ,} \nonumber
\end{equation}
which can be obtained through the quark model. From Ref. \cite{Barnes1} we have $g_{\chi_{c1}(2P)\psi(2S)\gamma}=1.737$ and $g_{\chi_{c1}(2P)\psi(1S)\gamma}=0.093$. These coupling constants scale proportionally to the overlapping area of the spatial part of the wave functions of $\chi_{c1}(2P)$ state with $\psi(1S)$ and $\psi(2S)$, respectively. Then, $g_{\chi_{c1}(2P)\psi(2S)\gamma}$ is much larger than $g_{\chi_{c1}(2P)\psi(1S)\gamma}$, because the spatial overlap $2P \leftrightarrow 2S$ is larger than $2P \leftrightarrow 1S$ (namely, for the $2P \rightarrow 1S$ transition a cancellation due to the node of $2P$ occurs). 

The theoretical expressions for the decay widths of the transitions $\chi_{c1}(2P) \rightarrow \psi(1S)\gamma$ and $\chi_{c1}(2P) \rightarrow \psi(2S)\gamma$ depending on $m$ (being the running mass of $\chi_{c1}$) are:
\begin{align}
\Gamma_{\chi_{_{c1}}(2P)\rightarrow\psi(1S)\gamma}(m)  &  =g_{\chi_{_{c1}%
}(2P)\psi(1S)\gamma}^{2}\frac{k^{3}(m,m_{\psi(1S)},0)}{8\pi m^{2}}\frac{4}%
{3}\left(  1+\frac{k^{2}(m,m_{\psi(1S)},0)}{m_{\psi(1S)}^{2}}\right)  \text{
,}\\
\Gamma_{\chi_{_{c1}}(2P)\rightarrow\psi(2S)\gamma}(m)  &  =g_{\chi_{_{c1}%
}(2P)\psi(2S)\gamma}^{2}\frac{k^{3}(m,m_{\psi(2S)},0)}{8\pi m^{2}}\frac{4}%
{3}\left(  1+\frac{k^{2}(m,m_{\psi(2S)},0)}{m_{\psi(2S)}^{2}}\right)  \text{
.}%
\end{align}
Thus, their ratio reads
\begin{equation}
\frac{\Gamma_{\chi_{_{c1}}(2P)\rightarrow\psi(2S)\gamma}(m)}{\Gamma
_{\chi_{_{c1}}(2P)\rightarrow\psi(1S)\gamma}(m)}=\left(  \frac{g_{\chi_{_{c1}%
}(2P)\psi(2S)\gamma}}{g_{\chi_{_{c1}}(2P)\psi(1S)\gamma}}\right)  ^{2}\left(
\frac{k(m,m_{\psi(2S)},0)}{k(m,m_{\psi(1S)},0)}\right)  ^{3}\left(
\frac{1+\frac{k^{2}(m,m_{\psi(2S)},0)}{m_{\psi(2S)}^{2}}}{1+\frac
{k^{2}(m,m_{\psi(1S)},0)}{m_{\psi(1S)}^{2}}}\right)  \text{ }.\label{radratio}%
\end{equation}
When considering the decays of $X(3872)$, we use $m\simeq m_{X(3872)}\simeq 3.872$ GeV and get the ratio 
\begin{equation}
\frac{\Gamma_{X(3872)\rightarrow \psi(2S)\gamma}}{\Gamma_{X(3872) \rightarrow \psi(1S)\gamma}}\simeq 5.4 \text{ .} 
\end{equation}
This result is in good agreement with the experimental value $2.6 \pm 0.4$ quoted by the PDG \cite{pdg}. For instance, the BaBar Collaboration reports for this ratio the value $3.4\pm 1.4$ \cite{Aubert:2008ae}, while the LHCb group obtained the value $2.38\pm0.64\pm0.29$ \cite{Aaij:2014ala}.

The most important aspect is that in our model the large value of the ratio $\psi(2S)\gamma$ to $\psi(1S)\gamma$ appears naturally because the standard $c\bar{c}$ component contributes mostly to these decays. Hence, this feature refers not only to the seed state but also to the $X(3872)$. On the other hand, a purely molecular state would imply that $\langle2S|r|2P\rangle$ is smaller than $\langle1S|r|2P\rangle$, at odds with the experiment. We also notice that our results for the ratio somewhat larger than the experiment can be explained by the fact that we do not include the $DD^*$ loop processes \cite{Guo1}.

In the last step we make predictions for the radiative decays. To this end, we calculate the relevant decay widths, by the following integrals (using again the `1 MeV width' estimate):
\begin{align}
\Gamma_{X(3872)\rightarrow\psi(1S)\gamma}  &  =\int_{m_{D^{\ast0}}+m_{D^{0}}%
}^{m_{D^{\ast0}}+m_{D^{0}}+1\text{ MeV}}\mathrm{dm}\ \Gamma_{\chi_{_{c1}%
}(2P)\rightarrow\psi(1S)\gamma}(m)d_{\chi_{c1}(2P)}(m)\simeq0.54\text{
keV\ ,}\label{raddec1}\\
\Gamma_{X(3872)\rightarrow\psi(2S)\gamma}  &  =\int_{m_{D^{\ast0}}+m_{D^{0}}%
}^{m_{D^{\ast0}}+m_{D^{0}}+1\text{ MeV}}\mathrm{dm}\ \Gamma_{\chi_{_{c1}%
}(2P)\rightarrow\psi(2S)\gamma}(m)d_{\chi_{c1}(2P)}(m)\simeq3.13\text{ keV\ .}
\label{raddec2}%
\end{align}
These values can be tested in future experiments. 
\begin{center}
\textbf{Prompt production}
\end{center}
Our assignment of the $X(3872)$ (as a virtual pole) can explain why this state was observed in prompt production processes. Namely, the charm-anticharm system dressed by the $D^*D$ mesons loops can be seen as a unique object represented by the spectral function of Figure \ref{1X}. This holds true even if there are two poles corresponding to two separate states. From the discussion concernig radiative decays, it follows that the value of the bare coupling constant is the same for both the broad peak associated with the seed state and for the high peak of $X(3872)$. We stress that a different issue is the role of $X(3872)$ in thermal models, where it may be subleading, see e.g. Ref. \cite{Broniowski:2015oha}.
\begin{center}
\textbf{Isospin breaking decay}
\end{center}
First, let us consider the transition $\chi_{c1}(2P) \rightarrow \psi(1S)\omega \rightarrow \psi(1S)\pi^0 \pi^+ \pi^-$ that obeys isospin symmetry. Basically, this process can be realized through two kind of mechanisms. In the first case, two glouns are emitted and then they transform into the $\omega$ meson. In the second case, the $D^*D$ loop coupled to $\psi(1S)\omega$ is involved.

Additionally, the transition $\chi_{c1}(2P) \rightarrow \psi(1S)\rho^0 \rightarrow \psi(1S)\pi^- \pi^+$ violates isospin symmetry. Moreover, the mechanism involving two-gluon emission cannot occur, since the $\rho$ cannot be created from two gluons. However, the $D^*D$ loop generates an isospin suppressed coupling of $\chi_{c1}(2P)$ to $\psi(1S)\rho$. In the following we study this issue in more details. To this end we introduce the Lagrangian which couples $D^*D$ to the $\rho^0$ and $\omega$ mesons: 
\begin{equation}
\mathcal{L}_{DD^{\ast}}=\xi_{0}D^{\ast0\mu}\bar{D}^{0}\psi(1S)^{\nu}\left[
\tilde{\omega}_{\mu\nu}+\tilde{\rho}_{\mu\nu}^{0}\right]  +\xi_{0}D^{\ast+\mu
}D^{-}\psi(1S)^{\nu}\left[  \tilde{\omega}_{\mu\nu}-\tilde{\rho}_{\mu\nu}%
^{0}\right]  +h.c.\ .\label{DDjor}%
\end{equation}
In the equation above, $\tilde{\omega}_{\mu \nu}+ \tilde{\rho}^0_{\mu \nu}\sim u\bar{u}$, thus it couples to $\bar{D}^{0}D^{*0}$, whereas $\tilde{\omega}_{\mu \nu}-\tilde{\rho}^0_{\mu \nu}\sim d\bar{d}$, thus it couples to $D^{*+}D^-$. Moreover, $\xi_0$ stands for the corresponding coupling constant. As a consequence of isospin symmetry, it is equal in both terms. The energy dependent coupling constant describing the decay of $\chi_{c1}(2P)$ into $\psi(1S)\rho$ is proportional to
\begin{equation}
\xi_{\chi_{_{c1}}(2P)\rightarrow\psi(1S)\rho}(m)=\xi_{0}g_{\chi_{c1}DD^{\ast}%
}\left[  \Sigma_{D^{\ast0}\bar{D}^{0}+h.c.}(s)-\Sigma_{D^{\ast+}D^{-}%
+h.c.}(s)\right]\ ,  \label{psi1rho}%
\end{equation}
while, for the decay into $\psi(1S)\omega$, one has ($s=m^2$):
\begin{equation}
\xi_{\chi_{_{c1}}(2P)\rightarrow\psi(1S)\omega}(m)=\xi_{0}g_{\chi_{c1}%
DD^{\ast}}\left[  \Sigma_{D^{\ast0}\bar{D}^{0}+h.c.}(s)+\Sigma_{D^{\ast+}%
D^{-}+h.c.}(s)\right]  +\lambda_{gg} \text{ .} \label{psi1omega}%
\end{equation}
Note, in Eq. (\ref{psi1omega}), the additional parameter $\lambda_{gg}$ involves the direct contribution of two gluons. 

\begin{figure}[h!]
\begin{center}
\includegraphics[width=0.72 \textwidth]{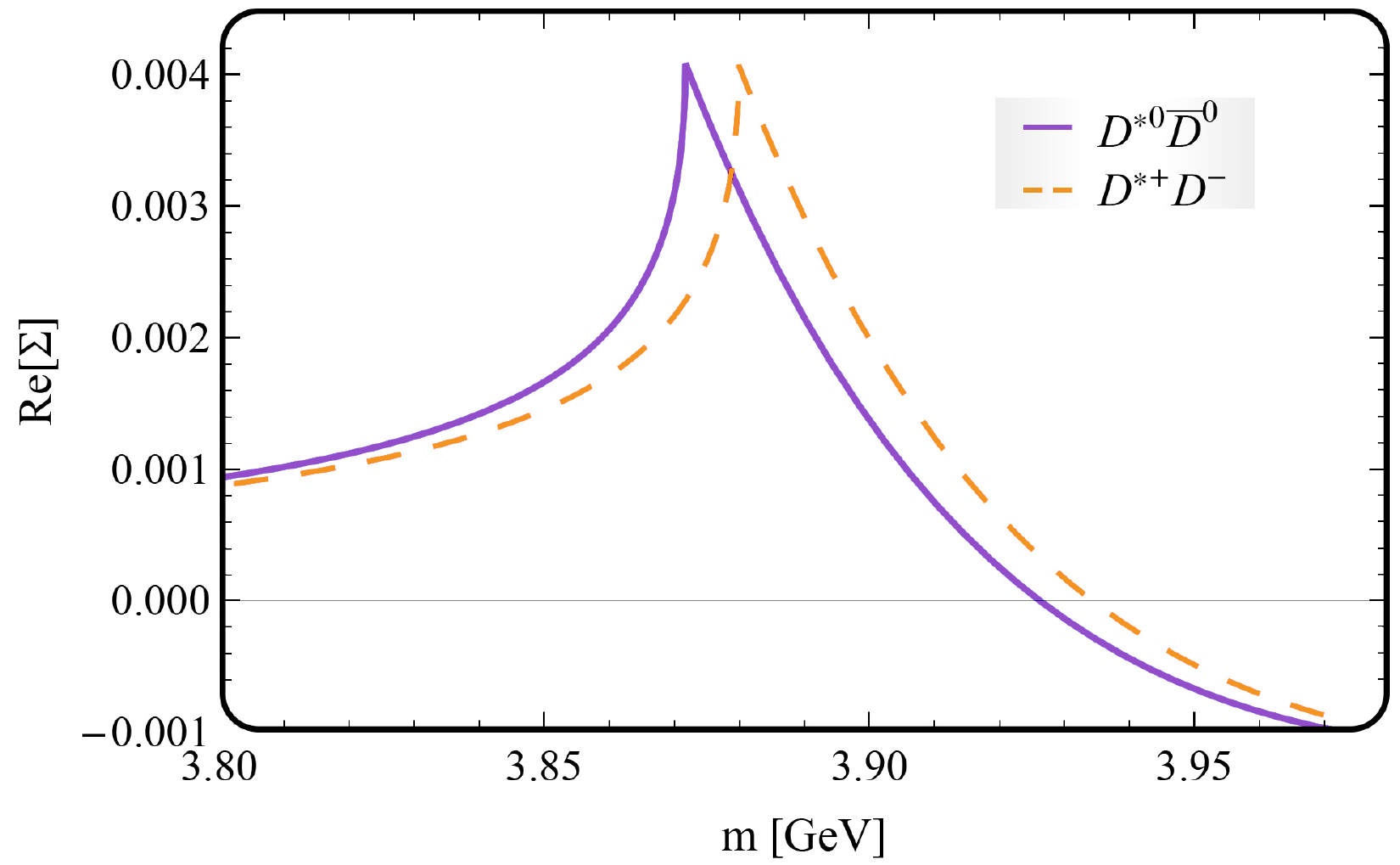}
\caption{\label{lastX} Plot of the real part of $DD^*$ loops. The solid violet line describes the loop function $\Sigma_{D^0D^{*0}}(m^2)$, while the dashed orange line describes the loop function $\Sigma_{D^+D^{*-}}(m^2)$. }
\end{center}
\end{figure}

In Figure \ref{lastX} we plot the real part of $DD^*$ loops. 
One shall notice that Eq. (\ref{psi1rho}) is not zero, since the quantum loops $\Sigma_{D^{*0}\bar{D}^0+h.c.}(s)$ and $\Sigma_{D^{*+}D^-+h.c.}(s)$ are different from each other. The reason for that is a slight difference in mass of neutral and charged $D$ and $D^*$ mesons. This, in turn, can be explained by the mass difference in between $u$ and $d$ quarks. However, for $s$ sufficiently far from the two relevant thresholds ($D^{*0}D^0$ and $D^{*+}D^-$), the loop difference entering Eq. (\ref{psi1rho}) is very small. For instance, for $m=3.986$ GeV (the mass value of the state originating from the $c\bar{c}$ seed) the coupling to $\psi(1S)\rho$ can be neglected. On the other hand, for the $X(3872)$, we cannot neglect the loop difference since it is sizably larger and rather important.  

Moreover, the sum of the loops in Eq. (\ref{psi1omega}) guarantees that the coupling $\xi_{\chi_{c1}(2P)\rightarrow \psi(1S)\omega}$  is large for the seed state as well as for $X(3872)$. As a further approximation, one may neglect the parameter $\lambda_{gg}$. This seems a valid assumption because the loops nearby the $DD^*$ thresholds are large. For $m=3.872$ GeV, referring to the $X(3872)$, the following ratio holds:

\begin{equation}
\left\vert \frac{\xi_{\chi_{_{c1}}(2P)\rightarrow\psi(1S)\omega}(m)}{\xi
_{\chi_{_{c1}}(2P)\rightarrow\psi(1S)\rho}(m)}\right\vert _{m=3.872\text{
GeV}}^{2}\simeq\left\vert \frac{\Sigma_{D^{\ast0}\bar{D}^{0}+h.c.}%
(s)+\Sigma_{D^{\ast+}D^{-}+h.c.}(s)}{\Sigma_{D^{\ast0}\bar{D}^{0}+h.c.}%
(s)-\Sigma_{D^{\ast+}D^{-}+h.c.}(s)}\right\vert _{s=3.872\text{ GeV}^{2}}%
^{2}=12.3\text{ .}%
\end{equation}
It follows that the coupling $\xi_{\chi_{_{c1}}(2P)\rightarrow\psi(1S)\rho}$ plays an important role and cannot be omitted. On the contrary, for $m=3.986$ GeV (corresponding to the broad peak), one finds the ratio $\sim 630$. This indicates that the decay into $\psi(1S)\rho$ is strongly suppressed (there is the cancellation of the loops). As a consequence, the seed state decays only into the $\psi(1S)\omega$ channel. 

The theoretical formulas for the widths of $\chi_{c1}(2P)$ decaying into the $\psi(1S)\omega$ or the $\psi(1S)\rho$ channels read:
\begin{align}
\Gamma_{\chi_{_{c1}}(2P)\rightarrow\psi(1S)\omega}(m,x)  &  =\xi_{\chi_{_{c1}%
}(2P)\rightarrow\psi(1S)\omega}^{2}V(m,x)\text{ ,}\\
\Gamma_{\chi_{_{c1}}(2P)\rightarrow\psi(1S)\rho}(m,x)  &  =\xi_{\chi_{_{c1}%
}(2P)\rightarrow\psi(1S)\rho}^{2}V(m,x)\text{ ,}%
\end{align}
where the parameters $m$ and $x$ are the running masses of $\chi_{c1}(2P)$ and $\omega$ (or $\rho$) mesons, respectively. Moreover, the quantity $V(m, x)$ reads:
\begin{equation}
V(m,x)=\frac{k}{8\pi m^{2}}\frac{1}{3m_{\psi(1S)}^{2}}\left[  4k^{4}%
+6m_{\psi(1S)}^{2}x^{2}+2k^{2}\left(  2m_{\psi(1S)}^{2}+x^{2}+2\sqrt
{k^{2}+m_{\psi(1S)}^{2}}\sqrt{k^{2}+x^{2}}\right)  \right]  \text{ ,}%
\end{equation}
where $k=k(m,m_{\psi(1S)}, x)$. When setting $m=3.872$ GeV and performing the integration over the mass of the $\rho$ meson, one can write the transition $X(3872) \rightarrow \psi(1S)\rho^0 \rightarrow \psi(1S)\pi^+\pi^-$ as:
\begin{equation}
\Gamma_{X(3872)\rightarrow\psi(1S)\pi^{+}\pi^{-}}=\left\vert \xi_{\chi_{_{c1}%
}(2P)\rightarrow\psi(1S)\rho}(m_{X(3872)})\right\vert ^{2}\int_{0}^{\infty
}\mathrm{d}x\ V(m_{X(3872)},x)d_{\rho}(x)\text{ },
\end{equation}
with $d_{\rho}(x)$ being the spectral function of the $\rho$, taken for simplicity as the relativistic Breit-Wigner distribution:
\begin{equation}
d_{\rho^{0}}(x)=N_{\rho}\frac{\theta(x-2m_{\pi^{+}})}{(x^{2}-m_{\rho^{0}}%
^{2})^{2}+\Gamma_{\rho^{0}}^{2}m_{\rho^{0}}^{2}}\ . \label{rhosf}%
\end{equation}
The numerical values of the parameters used in the above equation are taken from the PDG: $m_{\rho^0}=775.26 \pm 0.25$ MeV, $\Gamma_{\rho^0}=147.8\pm 0.9$ MeV. Moreover, the normalization factor $N_{\rho}$ is determined under the requirement that $\int\limits_0^{\infty}dxd_{\rho}(x)=1$.

Analogously, the transition $X(3872) \rightarrow \psi(1S)\omega \rightarrow \psi(1S)\pi^+\pi^-\pi^0$ can be obtained after performing the integration over the mass of the $\omega$:
\begin{equation}
\Gamma_{X(3872)\rightarrow\psi(1S)\pi^{+}\pi^{-}\pi^{0}}=\left\vert \xi
_{\chi_{_{c1}}(2P)\rightarrow\psi(1S)\omega}(m_{X(3872)})\right\vert ^{2}%
\int_{0}^{\infty}\mathrm{d}x\ V(m_{X(3872)},x)d_{\omega}(x)\ ,
\end{equation}
where the spectral function of the $\omega$ is given by the following distribution:
\begin{equation}
d_{\omega}(x)=N_{\omega}\frac{\theta(x-2m_{\pi^{+}}-m_{\pi^{0}})}%
{(x^{2}-m_{\omega}^{2})^{2}+\Gamma_{\omega}^{2}m_{\omega}^{2}}\ .
\end{equation}
The parameters used in the above equation are \cite{pdg}: $m_{\omega}=782.65\pm 0.12$ MeV, $\Gamma_{\omega}=8.49\pm0.08$ MeV and $N_{\omega}$ obtained from the normalization $\int \limits_{0}^{\infty}dx d_{\omega}(x)=1$. 

Finally, we obtain the following ratio:
\begin{equation}
\frac{\Gamma_{X(3872)\rightarrow\psi(1S)\pi^{+}\pi^{-}\pi^{0}}}{\Gamma
_{X(3872)\rightarrow\psi(1S)\pi^{+}\pi^{-}}}\simeq1.9 \text{ .} \label{ratioVVV}%
\end{equation}
This value is roughly consistent with the experimental result $0.8 \pm 0.3$. In Appendix \ref{appX3782} we report the value of the ratio found for different choices of the model parameters. 

In summary, the typically small isospin breaking describing the decay into $J/ \psi \rho$ is sizably at the value of the mass of the $X(3872)$. This is because of the significant difference between  the real parts of the charged and neutral loops. The obtained result of the ratio of Eq. (\ref{ratioVVV}) is in qualitative agreement with the experimental one, even without usage of additional parameters or any other assumption. 

However, some improvements of the model are possible. For instance, the previously neglected parameter $\lambda_{gg}$ describing the process of the direct emission of two gluons can be included into the calculations. Moreover, the obtained value of the ratio of Eq. (\ref{ratioVVV}) may be reduced by using a $\rho$ spectral function which goes beyond the standard Breit-Wigner parametrization. Nevertheless, these improvements require the use of additional parameters and are not expected to change much the overall picture. 

\section{Concluding remarks}

The experimental data deliver many puzzling features of the $X(3872)$ state. Some of them, as for instance the prompt production and radiative decays, are characteristic of a $c\bar{c}$ state. However, some other features, as its mass position nearby the $D^0D^{*0}$ threshold and the isospin breaking decay into the $J/ \psi \rho$ channel, indicate that $X(3872)$ is rather a $D^0D^{*0}$ molecule.  

In this chapter we have shown that it is possible to explain both properties by using a simple QFT model. Our relativistic Lagrangian includes originally  only a unique $c\bar{c}$ seed state assigned to the $\chi_{c1}(2P)$. The existence of the $X(3872)$ is a result of dressing this $c\bar{c}$ state by the $DD^*$ mesonic loops. Our study shows that two poles are present on the complex plane. The first one, located in the III RS, is related to the seed state, thus to $\chi_{c1}(2P)$ with a total decay width $\sim 80$ MeV. The second is a virtual pole on the II RS close to the $D^0D^{*0}$ energy threshold, which we assigned to the $X(3872)$ state. 

Moreover, we have shown that both states $\chi_{c1}(2P)$ and $X(3872)$ can be described by only one spectral function, which is normalized to unity. The $\chi_{c1}(2P)$ appears as a broad peak on the right hand side. A very interesting observation is that for some parameter choices this peak completly dissapears. That can explain why this state has not yet been observed in experiments. For what concerns the $X(3872)$, it is visible in the spectral function as an extremely high and narrow peak nearby the $D^0D^{*0}$ energy threshold. Our interpretation of the $X(3872)$ as a virtual companion pole naturally explains the puzzling properties of this state. The $c\bar{c}$ behavior found in prompt production and radiative decays is affected by the $c\bar{c}$ core. On the other hand, the molecular behavior revealed in isospin breaking decays is connected to the $DD^*$-loop mechanism. 

Moreover, our theoretical model explains not only the features of $X(3872)$ but also allows us to make some predictions. We calculated the strong decays into the $D^0D^{*0}$ channel, the radiative decays into $\psi(1S)\gamma$ and $\psi(2S)\gamma$, and the ratio of $J/ \psi \rho$ and $J/ \psi \omega$. 
 
\chapter{Conclusions}
\label{concmp}
In this thesis we have presented a detailed review of the status and interpretation of some conventional and non-conventional mesonic states with scalar and vector quantum numbers. The main results are reported in Table \ref{concl}, which summarizes the achievements of this work. 
\begin{table}[h!]
\renewcommand{\arraystretch}{1.02}
\par
\makebox[\textwidth][c] {
\begin{tabular}
[c]{c|c|c|c}\hline \hline
State(s)& Method & Interpretation & Refs. \\ \hline
$\rho(1450)$, $K^*(1410)$&QFT, tree-level,& conventional $q\bar{q}$ nonet&\cite{Piotrowskaex}, \cite{4proc}, \\ 
$\omega(1420)$, $\phi(1680)$&strong and radiative decays& $J^{PC}=1^{--}$&\cite{5proc}, \cite{6proc} \\ 
(radially excited)&& & \\ \hline 
$\rho(1700)$, $K^*(1680)$ & QFT, tree-level, & conventional $q\bar{q}$ nonet & \cite{Piotrowskaex}, \cite{4proc},\\
$\omega(1650)$, $\phi(1959)$& strong and radiative decays & predictions for $\phi(1959)$ & \cite{5proc}, \cite{6proc}\\ 
(orbitally excited)& & (not yet discovered)&\\
&&$J^{PC}=1^{--}$&\\ \hline
$K^*(892)$& QFT at the one-loop & (predom.) $u\bar{s}$, $s\bar{u}$, $d\bar{s}$, $s\bar{d}$& \cite{1proc}\\
&unitarized level, & Pole (II RS): $0.89-0.028 i$ GeV & \\
& one channel $K\pi$ & $J^{PC}=1^{--}$& \\ \hline
$K^*_0(1430)$ & QFT at the one-loop & (predom.)  $u\bar{s}$, $s\bar{u}$, $d\bar{s}$, $s\bar{d}$& \cite{Wolkanowskija}, \cite{1proc},\\
& untarized level, & Pole (II RS): $(1.413 \pm 0.02)-$& \cite{2proc}, \cite{3proc},\\
& one channel $K\pi$& $(0.127 \pm 0.003) i$ GeV & \cite{8proc}\\
&& $J^{PC}=0^{++}$& \\ \hline
$K^*_0(700)$ & QFT at the one-loop & (four quark) dynamically & \cite{Wolkanowskija}, \cite{1proc},\\
& unitarized level, & generated pole & \cite{2proc}, \cite{3proc},\\
& one channel $K\pi$ & Pole (II RS): $(0.746 \pm 0.019)-$ & \cite{8proc}\\
& & $(0.262 \pm 0.014) i$ GeV & \\
& & $J^{PC}=0^{++}$ & \\ \hline
$\psi(4040)$ & QFT at the one-loop & conventional $c\bar{c}$ state & \cite{Piotrowskapsi4040}, \cite{8proc},\\
& unitarized level, & Pole (V RS): $(4.053 \pm 0.04)-$ & \cite{7proc}\\
& five channels $DD$, $DD^*$, & $(0.040 \pm 0.010) i$ GeV & \\
& $D^*D^*$, $D_sD_s$, $D_sD_s^*$& $J^{PC}=1^{--}$ & \\ \hline
$Y(4008)$ & QFT at the one-loop & enhancement in the channel & \cite{Piotrowskapsi4040}, \cite{8proc},\\
& unitarized level, & $\psi(4040) \rightarrow DD^* \rightarrow \pi^+ \pi^- J/ \psi$ & \cite{7proc}\\
& five channels $DD$, $DD^*$, & not a real state & \\
& $D^*D^*$, $D_sD_s$, $D_sD_s^*$ & &\\ \hline
$\chi_{c1}(2P)$ & QFT at the one-loop & conventional $c\bar{c}$ state & \cite{PiotrcoitoX}, \cite{8proc}\\
& unitarized level, & (not yet found) & \\
& one channel $DD^*$ & Pole (III RS): $3.995-0.036 i$ GeV & \\
&&$J^{PC}=1^{++}$& \\ \hline
$X(3872)$ & QFT at the one-loop & (virtual) companion pole & \cite{PiotrcoitoX}, \cite{8proc}\\
& unitarized level, & Pole (II RS): $3.87164 - i \varepsilon$ GeV & \\
& one channel $DD^*$ & $J^{PC}=1^{++}$ & \\ \hline \hline
\end{tabular}
}\caption{Summary of the main results.} \label{concl}
\end{table}

We have understood that two nonets of vector mesons fit very well into the conventional quark-antiquark assignment. These are the nonets involving (predominantly) vector radial excitations with the resonances  $\{\rho(1450),$ $K^*(1410),$ $\omega(1420),$ $\phi(1680)\}$ and (predominantly) vector orbital excitations with the resonances $\{\rho(1700),$ $K^*(1680),$ $\omega(1650),$ $\phi (1959)\}$. In particular, the resonance $\phi(1959)$ should be still confirmed experimentally. Within our flavor-invariant Quantum Field Theoretical model we obtained some predictions of its mass and decays. Our (tree-level) analysis have revealed that the $\phi(1959)$ state is very broad ($\sim 430$ MeV), which explains why it is  difficult to experimentally discover it. We have shown that the main decay channels of $\phi(1959)$ are into $\bar{K}K^*(892)+h.c.$ and $\bar{K}K$ pairs. Moreover the radiative decay $\phi(1959) \rightarrow \gamma \eta$ is also possible. 
This fact implies that $\phi(1959)$ can be confirmed at the ongoing GlueX and CLAS12 experiments at Jefferson Lab, where the processes 
\begin{equation}
\gamma+p \rightarrow K^0+\bar{K}^0+p \hspace{0.3cm} \text{ and } \hspace{0.3cm} \gamma+p \rightarrow K^++K^-+p
\end{equation}
can be measured. Another possibility is the study of the reaction $e^+e^- \rightarrow K^+K^-$ at BaBar; similar processes at BESIII are also very promising. 

In this work, we have also investigated some non-conventional mesons, whose nature is not yet fully understood and needs to be clarified. In particular, we have shown that some states may emerge as  dynamically generated companion poles. This rather simple mechanism occurs when a single conventional quark-antiquark seed state couples strongly to its low-mass decay products. As a consequence, the quantum fluctuations dress the original seed state and modify its spectral function. This non-perturbative phenomenon can be investigated by using a QFT Lagrangian that couples the standard $q\bar{q}$ state to its decay products. The propagator of the unstable state is then evaluated at the one-loop resummed level. Non-conventional mesons may appear as poles in the mass distribition function of the conventional mesons originally included in the Lagrangian.  
 
 We have explored the non-conventional state $\kappa=K_0^*(700)$, which was recently updated in the summary table of the PDG implaying that a full nonet of light scalar mesons below $1$ GeV is present. (However, the annotation ``needs confirmation'' still figurates there for the $K^*_0(700)$). 
 
Our study agrees with other works on the subject by confirming the existence of the $K_0^*(700)$ state. We have proposed an independent proof for that and we conclude that $K_0^*(700)$ should be accepted in the PDG summary list as a meson. Moreover, we have proposed a physical interpretation of the nature of light $\kappa$: within our approach this state appears as a dynamically generated companion pole  of the heavier conventional $K_0^*(1430)$ resonance. The corresponding coordinates of $K_0^*(700)$
 on the complex plane are:
 \begin{equation}
 (0.746 \pm0.019)-(0.262 \pm 0.014) \text{ GeV} \hspace{1.2cm} [\kappa=K^*_0(700)]
 \end{equation}
The spectral function of the regular $K^*_0(1430)$ deviates sizably from an ordinary Breit-Wigner distribution. The light $\kappa$ is visible as its significant low-energy enhancement.
 
 Similar studies have been performed in the charmonium sector where one also observes both conventional and non-conventional mesons. The latter mesons are known under the names of $X$, $Y$ and $Z$. 
 
In this thesis we have investigated the vector state $Y(4008)$, which was observed by Belle Collaboration and whose nature still remains puzzling. Following the idea of dynamical generation we have studied the conventional $c\bar{c}$ state $\psi(4040)$ decaying into $DD$, $D^*D$, $D^*D^*$, $D_sD_s$ and $D_s^*D_s$ channels. We have observed that the spectral function deviates from the Breit-Wigner shape due to the strong enhancement close to 4 GeV. Moreover, also in this case, in addition to the pole corresponding to the $\psi(4040)$ state, a second pole appeared on the complex plane. However, this pole \textbf{cannot} be assigned to $Y(4008)$ since the imaginary part of it is by far too small when compared to experimental data. Here, a different mechanism -independent of dynamical generation- is at work. We have shown that $Y(4008)$ appears as a bump
when the $\psi(4040)$ resonance decays via $DD^*$ loop into the $\pi^+ \pi^- J/ \psi$ channel. The most important observation here is the fact that the enhancement identified with $Y(4008)$  cannot be regarded as a real resonance, but it is solely the consequence of the strong coupling of $\psi(4040)$ to the $DD^*$ loop. In general, not all bumps appering in some experimentally measured channels should be regarded as genuine states. Therefore, care is needed whenever a peak appears in a certain channel: in some cases it can correspond to an actual state, but this is defnitely not a general rule. 

The next meson that has been studied here is the axial-vector resonance $X(3872)$ (in PDG known under the name $\chi_{c1}(3872)$). This state, even if confirmed in many experiments and studied in numerous theoretical works in the field, is still not yet fully understood. We have shown that $X(3872)$ can also be explained by the mechanism of dynamical generation. However, in this case the seed state corresponding to the $c\bar{c}$ $\chi_{c1}(2P)$ meson in the quark model has not yet been discovered experimentally. We have shown that $X(3872)$ appears as the virtual companion pole of $\chi_{c1}(2P)$. An interesting observation is the fact that the spectral function of $\chi_{c1}(2P)$ state, correctly normalized to unity, represents simultaneously both states: the seed $\chi_{c1}(2P)$ and $X(3872)$. The shape of the spectral function is very specific. The $X(3872)$ corresponds to a very high and sharp peak close to the lowest $D_0D_0^*$ threshold, while $\chi_{c1}(2P)$ appears as a broad peak on the right hand side from it. Moreover, two poles are always present on the complex plane. The pole for $X(3872)$ has coordinates
\begin{equation}
3.87164-i \varepsilon  \text{ GeV } \hspace{1.2cm} [X(3872)] \text{ , }
\end{equation}
while the second one has
\begin{equation}
3.995 -0.036 \text{ GeV } \hspace{1.2cm} [\chi_{c1}(2P)] \text{ ; }
\end{equation}
the latter is in agreement with quark model predictions.
One should notice that even if the pole for the $c\bar{c}$ seed state $\chi_{c1}(2P)$ is always well-defined, it is possible that the corresponding broad peak in the spectral function fades away. This phenomenon may explain why the $\chi_{c1}(2P)$ resonance has not yet been observed experimentally. At the same time, our interpretation of the state $X(3872)$ as a virtual companion pole generated by dressing the $c\bar{c}$ seed state by $DD^*$ loops explains very well the quarkonium-like and molekular-like features of $X(3872)$. 

As a concluding remark, we hope that all the predictions presented in this work can be tested in ongoing and future experiments. Moreover, in the future one can apply the techniques used in this thesis to other interesting and not yet explained resonances, which still require a better understanding.

\begin{appendices}
\appendixpage
\noappendicestocpagenum
\addappheadtotoc
\chapter{Lagrangian for excited vector mesons} 
\label{ApA}
\section{Lagrangian form}
In this Appendix we show the explicit form of the Lagrangian presented in Chapter \ref{excitki} and described by Eq. (\ref{lagfull}).

In the following, the term $\mathcal{L}_{RPP}$ given by Eq. (\ref{apAlag1}) is related to Eqs. (\ref{term1a}) and (\ref{term2b}). Analogously, the term $\mathcal{L}_{RVP}$ given by Eq. (\ref{apAlag2}) is related to Eqs. (\ref{term3c}) and (\ref{term4d}).

The index $R$ is identified with $E$ for radially excited vector mesons and with $D$ for orbitally excited vector mesons. 
\begin{eqnarray} \label{apAlag1}
\mathcal{L}_{RPP}&=& ig_{iPP}Tr\left[\left[\partial^{\mu}P, V_{i, \mu}\right]P\right]=\frac{1}{4}ig_{iPP} \Big\{ K^{*0}_{\mu}\Big(\left(\partial^{\mu}\bar{K}^{0}\right)\pi^{0}-\bar{K}^{0}\left(\partial^{\mu} \pi^{0}\right)-\sqrt{2}\left(\partial^{\mu}K^{-}\right) \pi^{+} \nonumber \\
&+&\sqrt{2}K^{-}\left(\partial^{\mu} \pi^+\right)+\left(\partial^{\mu}\eta_{N}\right)\bar{K}^0-\eta_{N}\left(\partial^{\mu}\bar{K}^0\right)+\sqrt{2}\eta_{S}\left(\partial^{\mu}\bar{K}^0\right)-\sqrt{2}\left(\partial^{\mu} \eta_{S}\right)\bar{K}^0\Big)\nonumber \\ &+&\bar{K}^{* 0}_{\mu} \Big( K^0\left(\partial^{\mu} \pi^0\right)-\left(\partial^{\mu}K^0\right) \pi^0-\sqrt{2}K^+\left(\partial^{\mu} \pi^- \right)+ \sqrt{2}\left(\partial^{\mu} K^+\right) \pi^-+\eta_{N}\left(\partial^{\mu}K^0\right)\nonumber \\
&-& \left(\partial^{\mu}\eta_{N}\right)K^0-\sqrt{2}\eta_{S}\left(\partial^{\mu}K^0\right)+ \sqrt{2}\left(\partial^{\mu}\eta_{S}\right)K^0\Big)+ K^{*-}_{\mu}\Big(\left(\partial^{\mu}K^+\right) \pi^0-K^+\left(\partial^{\mu} \pi^{0}\right)\nonumber \\
&-& \sqrt{2}K^0\left(\partial^{\mu} \pi^{+}\right)+\sqrt{2}\left(\partial^{\mu}K^0\right) \pi^{+}+ \eta_{N} \left(\partial^{\mu}K^+\right)-\left(\partial^{\mu}\eta_{N}\right) K^+- \sqrt{2}\eta_{S}\left(\partial^{\mu}K^+\right)\nonumber \\
&+&\sqrt{2}\left(\partial^{\mu}\eta_{S}\right)K^+\Big)+K^{*+}_{\mu}\Big(K^-\left(\partial^{\mu} \pi^0\right)-\left(\partial^{\mu}K^-\right) \pi^0 -\sqrt{2}\left(\partial^{\mu}\bar{K}^0\right) \pi^- + \sqrt{2}\bar{K}^0 \left(\partial^{\mu} \pi^-\right) \nonumber \\
&+& \left(\partial^{\mu}\eta_{N}\right)K^--\eta_{N}\left(\partial^{\mu}K^-\right)+\sqrt{2}\eta_{S}\left(\partial^{\mu}K^-\right)-\sqrt{2}\left(\partial^{\mu}\eta_{S}\right)K^-\Big)+\rho_{\mu}^0 \Big(\bar{K}^0\left(\partial^{\mu}K^0\right)\nonumber \\
&-&\left(\partial^{\mu}\bar{K}^0\right) K^0 + K^+\left(\partial^{\mu}K^-\right) - \left(\partial^{\mu}K^+\right)K^-+ 2 \pi^+\left(\partial^{\mu}\pi^-\right)-2\left(\partial^{\mu}\pi^+\right) \pi^-\Big)\nonumber \\
&+& \rho^-_{\mu} \Big(\sqrt{2}K^+\left(\partial^{\mu}\bar{K}^0\right)-\sqrt{2}\left(\partial^{\mu}K^+\right)\bar{K}^0+2 \pi^0 \left(\partial^{\mu} \pi^+\right)-2 \left(\partial^{\mu}\pi^0\right)\pi^+\Big)\nonumber \\
&+&\rho^{+}_{\mu}\Big(\sqrt{2}K^0\left(\partial^{\mu}K^-\right)-\sqrt{2}\left(\partial^{\mu}K^0\right)K^-+2\left(\partial^{\mu}\pi^0\right) \pi^{-}-2\pi^0\left(\partial^{\mu}\pi^-\right)\Big)+\omega\Big(K^0\left(\partial^{\mu}\bar{K}^0\right)\nonumber \\
&-& \left(\partial^{\mu}K^0\right)\bar{K}^0 +K^+\left(\partial^{\mu}K^-\right)-\left(\partial^{\mu}K^+\right)K^-\Big)+\sqrt{2}\phi\Big(\left(\partial^{\mu}K^0\right)\bar{K}^0-K^0\left(\partial^{\mu}\bar{K}^0\right)\nonumber \\
&-&K^+\left(\partial^{\mu}K^-\right)+\left(\partial^{\mu}K^+\right)K^-\Big)\Big\}
\end{eqnarray}
\begin{eqnarray} \label{apAlag2}
\mathcal{L}_{RVP}&=&g_{iVP}Tr\left(\tilde{V}^{\mu \nu}_i\big\{V_{\mu \nu},P\big\}\right)=2g_{iVP} \varepsilon^{\mu \nu \alpha \beta} Tr \left(\left(\partial_{\alpha}V_{i, \beta}\right)\{\left(\partial_{\mu}V_{\nu}\right), P \} \right)\nonumber \\
&=& \frac{1}{2}g_{iVP} \varepsilon^{\mu \nu \alpha \beta}\Big\{\left(\partial_{\alpha}\rho^0_{i, \beta}\right)\Big(2 \pi^0\left(\partial_{\mu}\omega_{\nu}\right)+2\eta_{N}\left(\partial_{\mu}\rho^0_{\nu}\right)-\bar{K}^0\left(\partial_{\mu}K^{* 0}_{\nu}\right)-K^0\left(\partial_{\mu}\bar{K}^{* 0}_{\nu}\right)\nonumber \\
&+& K^+\left(\partial_{\mu}K^{*-}_{\nu}\right)+K^-\left(\partial_{\mu}K^{*+}_{\nu}\right)\Big)+\sqrt{2}\left(\partial_{\alpha}\rho^{-}_{i, \beta}\right)\Big(\sqrt{2} \pi^{+}\left(\partial_{\mu}\omega_{\nu}\right)+\sqrt{2}\eta_{N}\left(\partial_{\mu}\rho^{+}_{\nu}\right)\nonumber \\
&+&K^+\left(\partial_{\mu}\bar{K}^{*0}_{\nu}\right)+\bar{K}^0\left(\partial_{\mu}K^{*+}_{\nu}\right)\Big)+\sqrt{2}\left(\partial_{\alpha}\rho^{+}_{i, \beta}\right)\Big(\sqrt{2}\pi^-\left(\partial_{\mu}\omega_{\nu}\right)+\sqrt{2}\eta_{N}\left(\partial_{\mu}\rho^{-}_{\nu}\right)\nonumber \\
&+&K^-\left(\partial_{\mu}K^{*0}_{\nu}\right)+K^0\left(\partial_{\mu}K^{*-}_{\nu}\right)\Big)+\sqrt{2}\left(\partial_{\alpha}\phi_{i, \beta}\right)\Big(2\eta_{S}\left(\partial_{\mu}\phi_{\nu}\right) + \bar{K}^0\left(\partial_{\mu}K^{* 0}_{\nu}\right)\nonumber \\ &+&K^0\left(\partial_{\mu}\bar{K}_{\nu}^{*0}\right)
+ K^+\left(\partial_{\mu}K^{*-}_{\nu}\right)+K^-\left(\partial_{\mu}K^{*+}_{\nu}\right)\Big)+\left(\partial_{\alpha}\omega_{i, \beta}\right)\Big(2 \pi^0\left(\partial_{\mu}\rho^{0}_{\nu}\right)+ 2 \pi^+\left(\partial_{\mu}\rho^-_{\nu}\right)\nonumber \\
&+& 2 \pi^-\left(\partial_{\mu}\rho^{+}_{\nu}\right)+2 \eta_{N}\left(\partial_{\mu}\omega_{\nu}\right)+K^0\left(\partial_{\mu}\bar{K}^{* 0}_{\nu}\right)+ \bar{K}^0\left(\partial_{\mu}K^{* 0}_{\nu}\right)
+ K^+\left(\partial_{\mu}K^{*-}_{\nu}\right) \nonumber \\
&+& K^-\left(\partial_{\mu}K^{*+}_{\nu}\right)\Big)+\left(\partial_{\alpha}K^{* 0}_{i, \beta}\right)\Big(\bar{K}^0\left(\partial_{\mu}\omega_{\nu}\right)-\pi^0\left(\partial_{\mu}\bar{K}^{* 0}_{\nu}\right)+\sqrt{2}\pi^+\left(\partial_{\mu}K^{*-}_{\nu}\right)-\bar{K}^0\left(\partial_{\mu}\rho_{\nu}^{0}\right)\nonumber \\
&+& \sqrt{2}K^-\left(\partial_{\mu}\rho^{+}_{\nu}\right)+\eta_{N}\left(\partial_{\mu}\bar{K}^{*0}_{\nu}\right)+\sqrt{2}\eta_{S}\left(\partial_{\mu}\bar{K}^{* 0}_{\nu}\right)+\sqrt{2}\bar{K}^{0}\left(\partial_{\mu}\phi_{\nu}\right)\Big)+\left(\partial_{\alpha}\bar{K}^{* 0}_{i, \beta}\right)\Big(K^0\left(\partial_{\mu}\omega_{\nu}\right)\nonumber \\
&-&\pi^0\left(\partial_{\mu}K^{* 0}_{\nu}\right)+\sqrt{2}\pi^-\left(\partial_{\mu}K^{*+}_{\nu}\right)-K^0\left(\partial_{\mu}\rho^{0}_{\nu}\right)+\sqrt{2}K^+\left(\partial_{\mu}\rho^{-}_{\nu}\right)+\eta_{N}\left(\partial_{\mu}K^{* 0}_{\nu}\right)\nonumber \\
&+&\sqrt{2}\eta_{S}\left(\partial_{\mu}K^{* 0}_{\nu}\right)
+\sqrt{2}K^0\left(\partial_{\mu}\phi_{\nu}\right)\Big)+\left(\partial_{\alpha}K^{*-}_{i, \beta}\right)\Big(K^+\left(\partial_{\mu}\omega_{\nu}\right)+\pi^0\left(\partial_{\mu}K^{*+}_{\nu}\right)\nonumber \\
&+&\sqrt{2}\pi^{+}\left(\partial_{\mu}K_{\nu}^{* 0}\right)
+K^+\left(\partial_{\mu}\rho^0_{\nu}\right)+\sqrt{2}K^0\left(\partial_{\mu}\rho^+_{\nu}\right)+\eta_{N}\left(\partial_{\mu}K^{*+}_{\nu}\right)+\sqrt{2}\eta_{S}\left(\partial_{\mu}K^{*+}_{\nu}\right)\nonumber \\
&+&\sqrt{2}K^+\left(\partial_{\mu}\phi_{\nu}\right)\Big)+ \left(\partial_{\alpha}K^{*+}_{i, \beta}\right)\Big(K^-\left(\partial_{\mu}\omega_{\nu}\right)+\pi^0\left(\partial_{\mu}K^{*-}_{\nu}\right)+\sqrt{2}\pi^{-}\left(\partial_{\mu}\bar{K}^{* 0}_{\nu}\right)+K^-\left(\partial_{\mu}\rho^0_{\nu}\right)\nonumber \\
&+& \sqrt{2}\bar{K}^0\left(\partial_{\mu}\rho_{\nu}^-\right)+\eta_{N}\left(\partial_{\mu}K^{*-}_{\nu}\right)+\sqrt{2}\eta_{S}\left(\partial_{\mu}K^{*-}_{\nu}\right)+\sqrt{2}K^-\left(\partial_{\mu}\phi_{\nu}\right)\Big)\Big\}
\end{eqnarray}
\section{Invariance properties of the Lagrangian}
Here, we show the detailed study of the transformation properties of the nonets of ground-state pseudoscalar and vector mesons and of radially and orbitally excited vector mesons under parity ($\mathcal{P}$), charge conjugation ($\mathcal{C}$) and flavor symmetry ($U(3)_V$). Note, all vector fields $V_{\mu}, V_{D, \mu}, V_{E, \mu}$ transform in the same way.
\begin{table}[h]
\renewcommand{\arraystretch}{1.23}
\par
\makebox[\textwidth][c] { 
\par%
\begin{tabular}
[c]{|c|c|c|c|}\hline
nonet& Parity& Charge conjugation& flavor symmetry\\
\hline
$P$& $-P(t,-\vec{x})$ & $P^T$& $UPU^{\dagger}$\\
\hline
$V_{\mu}$& $V^{\mu}(t, -\vec{x})$ & $-(V_{\mu})^T$& $UV_{\mu}U^{\dagger}$\\
\hline
$\tilde{V}_{\mu \nu}$& $-\tilde{V}^{\mu \nu}$& $-(\tilde{V}_{\mu \nu})^{T}$& $U\tilde{V}_{\mu \nu}U^{\dagger}$\\
\hline
$V_{\mu \nu}$& $V^{\mu \nu}$& $-(V_{\mu \nu})^{T}$& $UV_{\mu \nu}U^{\dagger}$\\
\hline
\end{tabular}
}\caption{\label{transcpu}Transformation properties of the nonets $P$ and $V_\mu$.}%
\end{table}
\subsection{Parity}
The Lagrangian of the model presented in Chapter \ref{excitki} is invariant under the parity transformation $(\mathcal{P})$: 
\begin{eqnarray}
\mathcal{L}_{RPP}&=&ig_{RPP}Tr\left(  [\partial^{\mu}P,V_{R,\mu}]P\right)\stackrel{\mathcal{P}}{\longrightarrow}ig_{RPP}Tr\left( [\partial^{0}(-P),V_{R,0}](-P)+ [\partial_{i}(-P), V_{R}^{i}](-P)\right) \nonumber \\
&=&ig_{RPP}Tr\left([\partial^{0}P, V_{R,0}]P+[\partial^{i}P, V_{R,i}]P\right)=ig_{RPP}Tr\left([\partial^{\mu}P, V_{R,\mu}]P\right)=\mathcal{L}_{RPP}\text{ ; }
\end{eqnarray}
\begin{eqnarray}
\mathcal{L}_{RVP}&=&g_{RVP}Tr\left(\tilde{V}_{R}^{\mu \nu}\{V_{\mu \nu}, P\}\right)\stackrel{\mathcal{P}}{\longrightarrow}g_{RVP}Tr\left(-\tilde{V}_{R, \mu \nu} \{V^{\mu \nu},-P\}\right)\nonumber \\
 &=&g_{RVP}Tr\left(\tilde{V}_{R, \mu \nu} \{V^{\mu \nu},P\}\right)=g_{RVP}Tr\left(\tilde{V}_{R}^{\mu \nu} \{V_{\mu \nu}, P\}\right)=\mathcal{L}_{RVP}\text{ . }
\end{eqnarray}

\subsection{Charge conjugation}
The Lagrangian of the model presented in Chapter \ref{excitki} is invariant under the charge conjugation transformation ($\mathcal{C}$): 
\begin{eqnarray}
\mathcal{L}_{RPP}&=&ig_{RPP}Tr\left(\left[\partial^{\mu} P, V_{R,\mu}\right]P\right)\stackrel{\mathcal{C}}{\longrightarrow} ig_{RPP}Tr\left([\left(\partial^{\mu}P\right)^{T},-\left(V_{R,\mu}\right)^{T}]P^{T}\right)\nonumber \\
&=&ig_{RPP}Tr\left(\left(\left(\partial^{\mu}P\right)^{T}\left(-V_{R,\mu}\right)^{T}+V_{R,\mu}^{T}\left(\partial^{\mu}P\right)^{T}\right)P^{T}\right)\nonumber \\
&=& ig_{RPP}Tr\left(\left( \left(-V_{R,\mu} \partial^{\mu}P\right)^{T}+\left(\partial^{\mu}PV_{R,\mu}\right)^{T}\right)P^{T}\right) \nonumber \\
&=& ig_{RPP} Tr \left(\left(-PV_{R,\mu}\partial^{\mu}P\right)^{T}+\left(P\partial^{\mu}PV_{R,\mu}\right)^{T}\right)\nonumber \\
&=& ig_{RPP}Tr\left(\left(-PV_{R,\mu}\partial^{\mu}P+P\partial^{\mu}PV_{R,\mu}\right)^{T}\right)\nonumber \\
&=&ig_{RPP}Tr\left(-PV_{R,\mu}\partial^{\mu}P+P\partial^{\mu}PV_{R,\mu}\right)=ig_{RPP}Tr\left(P\left[\partial^{\mu}P,V_{R,\mu}\right]\right)\nonumber \\
&=&ig_{RPP}Tr\left(\left[\partial^{\mu}P, V_{R,\mu}\right]P\right)=\mathcal{L}_{RPP}.
\end{eqnarray}
\begin{eqnarray}
\mathcal{L}_{RVP}&=&g_{RVP}Tr\left(\tilde{V}_{R}^{\mu \nu}\left\{V_{\mu \nu},P\right\}\right)\stackrel{\mathcal{C}}{\longrightarrow}g_{RVP}Tr\left(-\left(\tilde{V}_{R}^{\mu \nu}\right)^{T}\left\{-\left(V_{\mu \nu}\right)^{T}, P^{T}\right\}\right)\nonumber \\
&=&g_{RVP}Tr\left(\left(PV_{\mu \nu}\tilde{V}_{R}^{\mu \nu}+V_{\mu \nu}P\tilde{V}_{R}^{\mu \nu}\right)^{T}\right)=g_{RVP}Tr\left(V_{\mu \nu}P \tilde{V}_{R}^{\mu \nu}+PV_{\mu \nu}\tilde{V}_{R}^{\mu \nu}\right)\nonumber \\
&=& g_{RVP}Tr\left(\tilde{V}_{R}^{\mu \nu} \left\{V_{\mu \nu}, P\right\}\right)=\mathcal{L}_{RVP}.
\end{eqnarray}
\subsection{Flavor symmetry}
The Lagrangian of the model presented in Chapter \ref{excitki} is flavor-invariant: 
\begin{eqnarray}
\mathcal{L}_{RPP}&=&ig_{RPP}Tr\left(\left[\partial^{\mu}P,V_{E}\right]P\right)\stackrel{U(3)_V}{\longrightarrow}ig_{RPP} Tr\left(\left[\partial^{\mu}UPU^{\dagger},UV_EU^{\dagger}\right]UPU^{\dagger}\right)\nonumber\\
&=&ig_{RPP}Tr\left(\left[U\partial^{\mu}PU^{\dagger}, UV_EU^{\dagger}\right]UPU^{\dagger}\right)\nonumber \\
&=& ig_{RPP}Tr\left(U\partial^{\mu}PU^{\dagger}UV_{E}U^{\dagger}UPU^{\dagger}-UV_{E}U^{\dagger}U\partial^{\mu}PU^{\dagger}UPU^{\dagger}\right)\nonumber \\
&=&ig_{RPP} Tr\left(U\partial^{\mu}PV_{E}PU^{\dagger}-UV_{E}\partial^{\mu}PPU^{\dagger}\right)\nonumber \\
&=&ig_{RPP}\left(Tr \left(U\partial^{\mu}PV_{E}PU^{\dagger}\right)- Tr\left(UV_{E}\partial^{\mu}PPU^{\dagger}\right)\right)\nonumber \\
&=& ig_{RPP} Tr\left(\left(U^{\dagger}U\partial^{\mu}PV_{E}P\right)-Tr\left(U^{\dagger}UV_{E}\partial^{\mu}PP\right)\right)\nonumber \\
&=&ig_{RPP}\left(Tr\left(\partial^{\mu}PV_{E}P\right)-Tr\left(V_{E}\partial^{\mu}PP\right)\right)\nonumber \\
&=&ig_{RPP}Tr\left(\left(\partial^{\mu}PV_{E}-V_{E}\partial^{\mu}P\right)P\right)=ig_{RPP}Tr\left(\left[\partial^{\mu}P, V_{E}\right]P\right)=\mathcal{L}_{RPP} \text{ ,}
\end{eqnarray} 

where we have used that $Tr\left[AB\right]=Tr\left[BA\right]$ and that $U^{\dagger}U=UU^{\dagger}=1.$
\newpage
Similarly:

\begin{eqnarray}
\mathcal{L}_{RVP}&=&g_{RVP}Tr\left(\tilde{V}_{R}^{\mu \nu}\left\{V^{\mu \nu},P\right\}\right)\stackrel{U(3)_V}{\longrightarrow}g_{RVP}Tr\left(U\tilde{V}_R^{\mu \nu}U^{\dagger}\left\{UV^{\mu \nu}U^{\dagger}, UPU^{\dagger}\right\}\right) \nonumber \\
&=& g_{RVP}Tr\left(U\tilde{V}_R^{\mu \nu}U^{\dagger}UV^{\mu \nu}PU^{\dagger}+U\tilde{V}_R^{\mu \nu}U^{\dagger}UPV^{\mu \nu}U^{\dagger}\right)\nonumber \\
&=& g_{RVP}\left(Tr\left(U\tilde{V}_R^{\mu \nu}V^{\mu \nu}PU^{\dagger}\right)+ Tr\left(U\tilde{V}_R^{\mu \nu}PV^{\mu \nu}U^{\dagger}\right)\right)\nonumber \\
&=& g_{RVP}\left(Tr \left(U^{\dagger}U\tilde{V}_R^{\mu \nu}V^{\mu \nu}P\right)+Tr\left(U^{\dagger}U\tilde{V}_R^{\mu \nu}PV^{\mu \nu}\right)\right)\nonumber \\
&=& g_{RVP}Tr\left(\tilde{V}_R^{\mu \nu}V^{\mu \nu}P+\tilde{V}_R^{\mu \nu}PV^{\mu \nu}\right)=g_{RVP}Tr \left(\tilde{V}_R^{\mu \nu}\left\{V^{\mu \nu}, P\right\}\right)=\mathcal{L}_{RVP}.
\end{eqnarray}
\chapter{Modification of the model parameters for~$\mathbf{\psi(4040)}$}
\label{apppsi4040}

In this appendix we show how the main results discussed in Chapter \ref{chaptpsiy} vary when changing some parameters of our model. In particular, we study different forms of the vertex function as well as various numerical values for the cutoff parameter $\Lambda$.

In Table \ref{bbhh1} we present the results for the Gaussian form factor of Eq. (\ref{gaussianpsi4040}) and higher values of the parameter $\Lambda$ (from $0.5$ to $1$ GeV).

In Table \ref{bbhh2} we present the results for the same vertex function and for lower values of $\Lambda$ parameter (from $0.38$ to $0.45$ GeV).

Similarly, in Tables \ref{bbhh3} and \ref{bbhh4} we show the results for the dipolar form factor given by Eq. (\ref{dipolarpsi4040}).

Finally, in Table \ref{partialdec4040} we present the results for partial decay widths of the resonance $\psi(4040)$ for all possible vertex functions and cutoff $\Lambda$ tested here. 

It is visible that changing the form of the vertex function does not influence significantly the overall picture. However, for what concerns the value of the parameter $\Lambda$, only the results obtained for $\Lambda \lesssim 0.6$ GeV are physically acceptable. 

 \begin{table}[h!] 
\centering

\renewcommand{\arraystretch}{1.25}
\begin{tabular}[c]{clcc}
\hline
\multicolumn{4}{c}{\textbf{Gaussian form factor}}\\
\hline
Cutoff $\Lambda$& \multicolumn{1}{c}{Parameters} & Pole for $\psi(4040)$& Pole for enhancement\\
$[$GeV$]$& \multicolumn{1}{c}{} &[GeV]& [GeV]\\
\hline
0.5&$g_{\psi DD}=19.4 \pm 4.2$&$(4.056 \pm 0.018)$&$(3.918 \pm 0.007)$\\
&$g_{\psi D^* D}=2.63 \pm 0.39$ GeV$^{-1}$&$-i(0.069 \pm 0.054)$&$-i(0.064 \pm 0.004)$\\
&$g_{\psi D^* D^*}=2.3 \pm 1.0$&&\\
&$M_{\psi}=4.04$ GeV&&\\
\hline
0.6&$g_{\psi DD}=10.3 \pm 2.7$&$(4.024 \pm 0.014)$&$(4.055 \pm 0.018)$\\
&$g_{\psi D^* D}=1.95 \pm 0.22$ GeV$^{-1}$&$-i(0.039 \pm 0.030)$&$-i(0.032 \pm 0.007)$\\
&$g_{\psi D^* D^*}=2.4 \pm 1.0$&&\\
&$M_{\psi}=4.07$ GeV&&\\
\hline
0.7&$g_{\psi DD}=6.9 \pm 1.9$&$(4.028 \pm 0.016)$&$(4.053 \pm 0.024)$\\
&$g_{\psi D^* D}=1.58 \pm 0.16$ GeV$^{-1}$&$-i(0.034 \pm 0.024)$&$-i(0.031 \pm 0.010)$\\
&$g_{\psi D^* D^*}=2.4 \pm 1.0$&&\\
&$M_{\psi}=4.09$ GeV&&\\
\hline
0.8&$g_{\psi DD}=5.2 \pm 1.5$&$(4.028 \pm 0.019)$&$(4.046 \pm 0.030)$\\
&$g_{\psi D^* D}=1.37 \pm 0.13$ GeV$^{-1}$&$-i(0.029 \pm 0.022)$&$-i(0.029 \pm 0.011)$\\
&$g_{\psi D^* D^*}=2.4 \pm 1.0$&&\\
&$M_{\psi}=4.11$ GeV&&\\
\hline
0.9&$g_{\psi DD}=4.3 \pm 1.2$&$(4.031 \pm 0.025)$&$(4.047 \pm 0.038)$\\
&$g_{\psi D^* D}=1.24 \pm 0.11$ GeV$^{-1}$&$-i(0.028 \pm 0.023)$&$-i(0.027 \pm 0.011)$\\
&$g_{\psi D^* D^*}=2.34 \pm 0.99$&&\\
&$M_{\psi}=4.14$ GeV&&\\
\hline
1.0&$g_{\psi DD}=3.8 \pm 1.1$&$(4.029 \pm 0.032)$&$(4.041 \pm 0.048)$\\
&$g_{\psi D^* D}=1.15 \pm 0.11$ GeV$^{-1}$&$-i(0.025 \pm 0.023)$&$-i(0.025 \pm 0.010)$\\
&$g_{\psi D^* D^*}=2.33 \pm 0.98$&&\\
&$M_{\psi}=4.17$ GeV&&\\
\hline

\end{tabular}
\caption{\label{bbhh1} Results of the model presented in Chapter \ref{chaptpsiy} for the Gaussian form factor and $0.5 \text{ GeV}\leqslant \Lambda \leqslant 1 \text{ GeV}$.}
\end{table} 

 \begin{table}[h!] 
\centering

\renewcommand{\arraystretch}{1.00}
\begin{tabular}[c]{clcc}
\hline
\multicolumn{4}{c}{\textbf{Gaussian form factor}}\\
\hline
Cutoff $\Lambda$& \multicolumn{1}{c}{Parameters} & Pole for $\psi(4040)$& Pole for enhancement\\
$[$GeV$]$& \multicolumn{1}{c}{} &[GeV]& [GeV]\\
\hline
0.38&$g_{\psi DD}=59.5 \pm 3.9$&$(4.050 \pm 0.002)$&$(3.938 \pm 0.003)$\\
&$g_{\psi D^* D}=3.7 \pm 1.1$ GeV$^{-1}$&$-i(0.032 \pm 0.003)$&$-i(0.015 \pm 0.002)$\\
&$g_{\psi D^* D^*}=1.44 \pm 0.78$&&\\
&$M_{\psi}=3.99$ GeV&&\\
\hline
0.39&$g_{\psi DD}=53.6 \pm 4.3$&$(4.051 \pm 0.003)$&$(3.937 \pm 0.004)$\\
&$g_{\psi D^* D}=3.6 \pm 1.0$ GeV$^{-1}$&$-i(0.034 \pm 0.004)$&$-i(0.019 \pm 0.002)$\\
&$g_{\psi D^* D^*}=1.57 \pm 0.84$&&\\
&$M_{\psi}=3.99$ GeV&&\\
\hline
0.40&$g_{\psi DD}=48.4 \pm 4.6$&$(4.051 \pm 0.003)$&$(3.936 \pm 0.005)$\\
&$g_{\psi D^* D}=3.59 \pm 0.94$ GeV$^{-1}$&$-i(0.036 \pm 0.006)$&$-i(0.022 \pm 0.001)$\\
&$g_{\psi D^* D^*}=1.69 \pm 0.88$&&\\
&$M_{\psi}=4.00$ GeV&&\\
\hline
0.41&$g_{\psi DD}=43.7 \pm 4.8$&$(4.052 \pm 0.004)$&$(3.935 \pm 0.006)$\\
&$g_{\psi D^* D}=3.51 \pm 0.87$ GeV$^{-1}$&$-i(0.038 \pm 0.008)$&$-i(0.026 \pm 0.001)$\\
&$g_{\psi D^* D^*}=1.80 \pm 0.92$&&\\
&$M_{\psi}=4.00$ GeV&&\\
\hline
0.42&$g_{\psi DD}=39.6 \pm 5.0$&$(4.053 \pm 0.004)$&$(3.934 \pm 0.006)$\\
&$g_{\psi D^* D}=3.43 \pm 0.80$ GeV$^{-1}$&$-i(0.040 \pm 0.010)$&$-i(0.030 \pm 0.001)$\\
&$g_{\psi D^* D^*}=1.90 \pm 0.95$&&\\
&$M_{\psi}=4.01$ GeV&&\\
\hline
0.43&$g_{\psi DD}=35.9 \pm 5.1$&$(4.053 \pm 0.005)$&$(3.932 \pm 0.007)$\\
&$g_{\psi D^* D}=3.34 \pm 0.73$ GeV$^{-1}$&$-i(0.042 \pm 0.013)$&$-i(0.034 \pm 0.001)$\\
&$g_{\psi D^* D^*}=1.98 \pm 0.98$&&\\
&$M_{\psi}=4.01$ GeV&&\\
\hline
0.44&$g_{\psi DD}=32.6 \pm 5.1$&$(4.054 \pm 0.005)$&$(3.930 \pm 0.007)$\\
&$g_{\psi D^* D}=3.24 \pm 0.66$ GeV$^{-1}$&$-i(0.045 \pm 0.016)$&$-i(0.038 \pm 0.001)$\\
&$g_{\psi D^* D^*}=2.06 \pm 1.0$&&\\
&$M_{\psi}=4.02$ GeV&&\\
\hline
0.45&$g_{\psi DD}=29.7 \pm 5.0$&$(4.054 \pm 0.005)$&$(3.928 \pm 0.008)$\\
&$g_{\psi D^* D}=3.13 \pm 0.60$ GeV$^{-1}$&$-i(0.048 \pm 0.020)$&$-i(0.042 \pm 0.002)$\\
&$g_{\psi D^* D^*}=2.12 \pm 1.0$&&\\
&$M_{\psi}=4.02$ GeV&&\\
\hline
\end{tabular}
\caption{\label{bbhh2} The same as Table \ref{bbhh1} but for $0.38 \text{ GeV}\leqslant \Lambda \leqslant 0.45 \text{ GeV}$.}
\end{table} 
 \begin{table}[h!] 
\centering

\renewcommand{\arraystretch}{1.25}
\begin{tabular}[c]{clcc}
\hline
\multicolumn{4}{c}{\textbf{Dipolar form factor}}\\
\hline
Cutoff $\Lambda$& \multicolumn{1}{c}{Parameters} & Pole for $\psi(4040)$& Pole for enhancement\\
$[$GeV$]$& \multicolumn{1}{c}{} &[GeV]& [GeV]\\
\hline
0.5&$g_{\psi DD}=12.5 \pm 3.0$&$(4.089 \pm 0.033)$&$(3.943 \pm 0.011)$\\
&$g_{\psi D^* D}=2.01 \pm 0.26$ GeV$^{-1}$&$-i(0.084 \pm 0.005)$&$-i(0.085 \pm 0.014)$\\
&$g_{\psi D^* D^*}=2.08 \pm 0.92$&&\\
&$M_{\psi}=4.05$ GeV&&\\
\hline
0.6&$g_{\psi DD}=7.4 \pm 2.0$&$(4.031 \pm 0.018)$&$(4.056 \pm 0.024)$\\
&$g_{\psi D^* D}=1.44 \pm 0.16$ GeV$^{-1}$&$-i(0.034 \pm 0.020)$&$-i(0.029 \pm 0.006)$\\
&$g_{\psi D^* D^*}=2.16 \pm 0.93$&&\\
&$M_{\psi}=4.08$ GeV&&\\
\hline
0.7&$g_{\psi DD}=5.1 \pm 1.4$&$(4.032 \pm 0.022)$&$(4.053 \pm 0.037)$\\
&$g_{\psi D^* D}=1.18\pm 0.12$ GeV$^{-1}$&$-i(0.032 \pm 0.021)$&$-i(0.028 \pm 0.007)$\\
&$g_{\psi D^* D^*}=2.20 \pm 0.93$&&\\
&$M_{\psi}=4.11$ GeV&&\\
\hline
0.8&$g_{\psi DD}=3.9 \pm 1.1$&$(4.032 \pm 0.028)$&$(4.046 \pm 0.047)$\\
&$g_{\psi D^* D}=1.04 \pm 0.10$ GeV$^{-1}$&$-i(0.025 \pm 0.022)$&$-i(0.026 \pm 0.010)$\\
&$g_{\psi D^* D^*}=2.22 \pm 0.94$&&\\
&$M_{\psi}=4.14$ GeV&&\\
\hline
0.9&$g_{\psi DD}=3.22 \pm 0.93$&$(4.029 \pm 0.037)$&$(4.039 \pm 0.057)$\\
&$g_{\psi D^* D}=0.97 \pm 0.09$ GeV$^{-1}$&$-i(0.023 \pm 0.024)$&$-i(0.024 \pm 0.010)$\\
&$g_{\psi D^* D^*}=2.23 \pm 0.94$&&\\
&$M_{\psi}=4.18$ GeV&&\\
\hline
1.0&$g_{\psi DD}=2.84 \pm 0.83$&$(3.936 \pm 0.008)$&$(3.890 \pm 0.048)$\\
&$g_{\psi D^* D}=0.93 \pm 0.08$ GeV$^{-1}$&$-i(0.005 \pm 0.004)$&$-i(0.003 \pm 0.005)$\\
&$g_{\psi D^* D^*}=2.24 \pm 0.94$&&\\
&$M_{\psi}=4.03$ GeV&&\\
\hline

\end{tabular}
\caption{\label{bbhh3} The same as Table \ref{bbhh1} but for the dipolar form factor.}
\end{table} 

\begin{table}[h!] 
\centering

\renewcommand{\arraystretch}{1.00}
\begin{tabular}[c]{clcc}
\hline
\multicolumn{4}{c}{\textbf{Dipolar form factor}}\\
\hline
Cutoff $\Lambda$& \multicolumn{1}{c}{Parameters} & Pole for $\psi(4040)$& Pole for enhancement\\
$[$GeV$]$& \multicolumn{1}{c}{} &[GeV]& [GeV]\\
\hline
0.38&$g_{\psi DD}=30.0 \pm 5.6$&$(4.055 \pm 0.014)$&$(3.941 \pm 0.003)$\\
&$g_{\psi D^* D}=4.03 \pm 0.70$ GeV$^{-1}$&$-i(0.045 \pm 0.013)$&$-i(0.038 \pm 0.009)$\\
&$g_{\psi D^* D^*}=1.97 \pm 0.92$&&\\
&$M_{\psi}=4.02$ GeV&&\\
\hline
0.39&$g_{\psi DD}=27.5 \pm 5.3$&$(4.056 \pm 0.017)$&$(3.941 \pm 0.003)$\\
&$g_{\psi D^* D}=3.74 \pm 0.64$ GeV$^{-1}$&$-i(0.048 \pm 0.013)$&$-i(0.042 \pm 0.009)$\\
&$g_{\psi D^* D^*}=1.97 \pm 0.92$&&\\
&$M_{\psi}=4.02$ GeV&&\\
\hline
0.40&$g_{\psi DD}=25.3 \pm 5.0$&$(4.058 \pm 0.019)$&$(3.941 \pm 0.003)$\\
&$g_{\psi D^* D}=3.48 \pm 0.58$ GeV$^{-1}$&$-i(0.051 \pm 0.012)$&$-i(0.045 \pm 0.010)$\\
&$g_{\psi D^* D^*}=1.98 \pm 0.92$&&\\
&$M_{\psi}=4.02$ GeV&&\\
\hline
0.41&$g_{\psi DD}=23.4 \pm 4.7$&$(4.061 \pm 0.021)$&$(3.942 \pm 0.004)$\\
&$g_{\psi D^* D}=3.25 \pm 0.53$ GeV$^{-1}$&$-i(0.054 \pm 0.011)$&$-i(0.049 \pm 0.010)$\\
&$g_{\psi D^* D^*}=1.99 \pm 0.91$&&\\
&$M_{\psi}=4.03$ GeV&&\\
\hline
0.42&$g_{\psi DD}=21.6 \pm 4.4$&$(4.063 \pm 0.023)$&$(3.942 \pm 0.004)$\\
&$g_{\psi D^* D}=3.05 \pm 0.49$ GeV$^{-1}$&$-i(0.057 \pm 0.010)$&$-i(0.052 \pm 0.010)$\\
&$g_{\psi D^* D^*}=2.00 \pm 0.91$&&\\
&$M_{\psi}=4.03$ GeV&&\\
\hline
0.43&$g_{\psi DD}=20.0 \pm 4.2$&$(4.066 \pm 0.024)$&$(3.942 \pm 0.005)$\\
&$g_{\psi D^* D}=2.87 \pm 0.45$ GeV$^{-1}$&$-i(0.060 \pm 0.009)$&$-i(0.056 \pm 0.011)$\\
&$g_{\psi D^* D^*}=2.00 \pm 0.91$&&\\
&$M_{\psi}=4.03$ GeV&&\\
\hline
0.44&$g_{\psi DD}=18.6 \pm 4.0$&$(4.069 \pm 0.025)$&$(3.943 \pm 0.005)$\\
&$g_{\psi D^* D}=2.71 \pm 0.41$ GeV$^{-1}$&$-i(0.063 \pm 0.008)$&$-i(0.060 \pm 0.011)$\\
&$g_{\psi D^* D^*}=2.01 \pm 0.91$&&\\
&$M_{\psi}=4.03$ GeV&&\\
\hline
0.45&$g_{\psi DD}=17.3 \pm 3.8$&$(4.072 \pm 0.027)$&$(3.943 \pm 0.006)$\\
&$g_{\psi D^* D}=2.56 \pm 0.38$ GeV$^{-1}$&$-i(0.067 \pm 0.008)$&$-i(0.064 \pm 0.011)$\\
&$g_{\psi D^* D^*}=2.03 \pm 0.91$&&\\
&$M_{\psi}=4.04$ GeV&&\\
\hline
\end{tabular}
\caption{\label{bbhh4} The same as Table \ref{bbhh2} but for the dipolar form factor.}
\end{table} 

\begin{table}[h!] 
\centering

\renewcommand{\arraystretch}{1.00}
\begin{tabular}[c]{c|c|c|c|c|c}
\cline{2-6}
 \multicolumn{1}{c}{}&\multicolumn{5}{|c}{\textbf{Partial decay width [MeV]}}\\
 \hline

\backslashbox{$\mathbf{\Lambda}$}{\textbf{Decay}} &$\mathbf{DD}$&$\mathbf{D_sD_s}$&$\mathbf{D^*D}$&$\mathbf{D^*_sD_s}$&$\mathbf{D^*D^*}$\\
\hline
\multicolumn{6}{c}{\textbf{Gaussian form factor}}\\
\hline
$0.38$ GeV& $2.90 \pm 0.38$&$62.8 \pm 8.1$&$12.1 \pm 7.0$&$0$&$2.2^{+2.4}_{-2.2}$\\
$0.39$ GeV&$3.57 \pm 0.57$&$58.9 \pm 9.4$&$14.9 \pm 8.2$&$0$&$2.7^{+2.9}_{-2.7}$\\
$0.40$ GeV&$4.26 \pm 0.81$&$55\pm 10$&$17.8\pm 9.3$&$0$&$3.2^{+3.3}_{-3.2}$\\
$0.41$ GeV&$5.0\pm 1.1$&$51 \pm 11$&$21 \pm 10$&$0$&$3.7^{+3.8}_{-3.7}$\\
$0.42$ GeV&$5.7\pm 1.4$&$46 \pm 12$&$24 \pm 11$&$0$&$4.3 \pm 4.3$\\
$0.43$ GeV&$6.3 \pm 1.8$&$42 \pm 12$&$26 \pm 12$&$0$&$4.8 \pm 4.7$\\
$0.44$ GeV&$7.0 \pm 2.2$&$39 \pm 12$&$29 \pm 12$&$0$&$5.2 \pm 5.1$\\
$0.45$ GeV&$7.6 \pm 2.5$&$35 \pm 12$&$32 \pm 12$&$0$&$5.7 \pm 5.4$\\
$0.5$ GeV&$9.8\pm 4.3$&$22.0\pm 9.6$&$41 \pm 12$&$0$&$7.4 \pm 6.6$\\
$0.6$ GeV&$11.8\pm 6.2$&$10.3\pm 5.4$&$49\pm 11$&$0$&$8.8 \pm 7.6$\\
$0.7$ GeV&$12.5\pm6.9$&$6.2\pm 3.4$&$52\pm 10$&$0$&$9.4\pm 8.0$\\
$0.8$ GeV&$12.8\pm 7.3$&$4.4 \pm 2.5$&$53 \pm 10$&$0$&$9.6 \pm 8.1$\\
$0.9$ GeV&$12.9\pm 7.4$&$3.4\pm 2.0$&$54\pm 10$&$0$&$9.7 \pm 8.2$\\
$1.0$ GeV&$13.0\pm 7.6$&$2.9\pm 1.7$&$54.3\pm 9.9$&$0$&$9.8 \pm 8.2$\\
\hline
\multicolumn{6}{c}{\textbf{Dipolar form factor}}\\

\hline
$0.38$ GeV& $8.4 \pm 3.2$&$ 30 \pm 11$&$35 \pm 12$&$0$&$6.3 \pm 5.9$\\
$0.39$ GeV&$8.6 \pm 3.3$&$29 \pm 11$&$36 \pm 12$&$0$&$6.5 \pm 6.0$\\
$0.40$ GeV&$8.8 \pm 3.5$&$28 \pm 11$&$37 \pm 12$&$0$&$6.6 \pm 6.1$\\
$0.41$ GeV&$9.0 \pm 3.6$&$27 \pm 11$&$38 \pm 12$&$0$&$6.8 \pm 6.2$\\
$0.42$ GeV&$9.2\pm 3.8$&$26\pm 10$&$38 \pm 12$&$0$&$6.9 \pm 6.3$\\
$0.43$ GeV&$9.4 \pm 3.9$&$24 \pm 10$&$39 \pm 12$&$0$&$7.1 \pm 6.4$\\
$0.44$ GeV&$9.6 \pm 4.1$&$23.3 \pm 9.9$&$40 \pm 12$&$0$&$7.2 \pm 6.5$\\
$0.45$ GeV&$9.8\pm 4.2$&$22.2 \pm 9.6$&$41 \pm 12$&$0$&$7.3 \pm 6.6$\\
$0.5$ GeV&$10.7 \pm 5.0$&$16.9 \pm 8.0$&$44\pm 12$&$0$&$8.0 \pm 7.1$\\
$0.6$ GeV&$11.9 \pm 6.3$&$9.5 \pm 5.0$&$50 \pm 11$&$0$&$8.9 \pm 7.7$\\
$0.7$ GeV&$12.6\pm 7.0$&$5.6 \pm 3.1$&$52\pm 10$&$0$&$9.4 \pm 8.0$\\
$0.8$ GeV&$12.9 \pm 7.4$&$3.7 \pm 2.1$&$54\pm 10.0$&$0$&$9.7 \pm 8.2$\\
$0.9$ GeV&$13.1 \pm 7.6$&$2.8\pm 1.6$&$54.4\pm 9.9$&$0$&$9.8 \pm 8.2$\\
$1.0$ GeV&$13.1\pm 7.7$&$2.2 \pm 1.3$&$54.8\pm 9.8 $&$0$&$9.9 \pm 8.3$\\
\hline
\end{tabular}
\caption{\label{partialdec4040} Results for the partial decay widths of $\psi(4040)$ computed for different form factors and various numerical values of the parameter $\Lambda$.}
\end{table}   

\chapter{Modification of the model parameters for~$\mathbf{X(3872)}$}
\label{appX3782}
In this appendix we discuss how the main results of our study of $X(3872)$ change upon varying the free parameters of the model presented in Chapter \ref{Xmain}. For each set of parameters we have calculated the most prominent quantities of our study, the coupling constant $g_{\chi_{c1}DD^*}$; the amount of the $\chi_{c1}(2P)$ state in between the $D^0D^{*0}$ threshold and 1 MeV above it as defined in Eq. (\ref{integralthr}); the decay width of $X(3872)$ into the main $D^{*0}\bar{D}^0$ channel as described by Eq. (\ref{intwidth}); the position of the propagator poles; the decay widths of the radiative transitions into $\psi(1S)\gamma$ and $\psi(2S)\gamma$ exspressed by Eqs. (\ref{raddec1}) and (\ref{raddec2}), the ratio given by Eq. (\ref{ratioVVV}). All the obtained results are summarized in Tables \ref{I}-\ref{VII}. 
 \begin{center}
 \textbf{Varying the coupling $\mathbf{g_{\chi_{c1}DD^*}}$}
 \end{center}
 
Let us start from testing the results in dependence of the coupling constant $g_{\chi_{c1}DD^*}$ at fixed cutoff $\Lambda=0.5$ GeV. Technically, $g_{\chi_{c1}DD^*}$ was determined by choosing different values of the mass $m_{*}$ defined in Eq. (\ref{mstar}) in the range between the $D^{0}D^{*0}$ and the $D^{-}D^{*+}$ thresholds. Notice that for $m_*$ (lowest threshold) one has the so called critical value of $g_{\chi_{c1}DD^*}$. Such cases are marked in the tables with the label ``critical''. Furthermore, one value of $m_*$ has been chosen outside the range (below $D^0D^{*0}$ energy threshold). In this case a `real' bound state (pole on the I RS) described by the spectral function of Eq. (\ref{realpoleds}) is realized. 

For the bare mass of $c\bar{c}$ state we use two numerical values, that are described in Chapter \ref{Xmain} and correspond to case I and case II, respectively. In table \ref{I} we set $M_0=3.95$ GeV and in Table  \ref{III} $M_0=3.92$ GeV. For completeness, we also test two types of form factors, i.e. the Gaussian one of Eq. (\ref{gaussxx}) and the dipolar function
\begin{equation}
F_{\Lambda}(m)\equiv F_{\Lambda}^{\text{dipolar}}(m)=\left(1+\frac{k^4(m)}{\Lambda^4}\right)^{-2} \text{ .}
\end{equation}
As a general observation, the modification of the constant $g_{\chi_{c1} DD^*}$ and the use of different form factors (either Gaussian or dipolar) do not change much the results. The overall picture is quite stable. One observes that, for each case tested here, a significant enhancement appears nearby the lowest $D^0D^{*0}$ threshold. Moreover, two propagator poles are always present in the complex plane. One of them refers to the seed $c\bar{c}$ state $\chi_{c1}(2P)$, while the second to the $X(3872)$ state. For clarity, in all tables the seed pole is indicated by $\bullet$, while the pole of $X(3872)$ by $\blacklozenge$. We recall that the pole assigned to $X(3872)$ is virtual only if the value of $g_{\chi_{c1} DD^*}$ is subcritical. Such a pole is located in the II RS. Otherwise, when $g_{\chi_{c1} DD^*}$ exceeds the critical value, a real bound-state is realized: in this case there is a pole on the I RS. 

Quite interestingly, the state $X(3872)$ dissapears if the value of the coupling constant $g_{\chi_{c1} DD^*}$ is sufficiently small. Such behavior is consistent with our interpretation of $X(3872)$ as emergent from the nearby charmonium state and its $DD^*$ dressing.
\label{xrozneg}
\begin{table}[h!]
\centering
\renewcommand{\arraystretch}{0.95}\label{tableggauss}
\begin{tabular}[c]{ccclcccc}
\hline
\hline
\multicolumn{8}{c}{\textbf{Parameters: \hspace{0.5cm} $\mathbf{M_0=3.95}$ \textbf{GeV,} \hspace{0.5cm}  $\mathbf{\Lambda=0.5}$ GeV,\hspace{0.5cm} $\mathbf{m_* \neq}$ \textbf{const.}}}\\
\hline
\multicolumn{8}{c}{\textbf{GAUSSIAN FORM FACTOR}}\\
\hline
$g_{\chi_{c1}DD^*}$ & Eq. (\ref{integralthr}) &Eq. (\ref{intwidth})& \hspace{0.7cm}Pole positions & RS & Eq. (\ref{raddec1}) & Eq. (\ref{raddec2}) & Eq. (\ref{ratioVVV})\\

[GeV]& &[MeV]&\hspace{1.2cm} [GeV] & &[keV]& [keV] & \\
\hline
\hline
9.808&0.057&0.636&$\bullet  \hspace{0.5cm} 3.9961-0.0359$ $i$& III &0.628&3.64&1.92\\
(critical)&&&$\blacklozenge \hspace{0.45cm} 3.8717-i\varepsilon$&II&&&\\
\hline
9.732&0.049&0.607&$\bullet  \hspace{0.5cm} 3.9954-0.0357$ $i$& III &0.539&3.13&1.92\\
(case I)&&&$\blacklozenge \hspace{0.45cm} 3.8716-i\varepsilon$& II &&&\\
\hline
9.500&0.029&0.408&$\bullet  \hspace{0.5cm} 3.9933-0.0354$ $i$& III &0.323&1.88&1.92\\
&&&$\blacklozenge \hspace{0.45cm} 3.8715-i\varepsilon$& II &&&\\
\hline
9.300&0.019&0.263&$\bullet  \hspace{0.5cm} 3.9915-0.0350$ $i$& III &0.206&1.20&1.92\\
&&&$\blacklozenge \hspace{0.45cm} 3.8710-i\varepsilon$&II&&&\\
\hline
9.000&0.010&0.136&$\bullet  \hspace{0.5cm} 3.9887-0.0344$ $i$&III&0.110&0.64&1.92\\
&&&$\blacklozenge \hspace{0.45cm} 3.8699-i\varepsilon$&II&&&\\
\hline
8.800&0.007&0.091&$\bullet  \hspace{0.5cm} 3.9869-0.0339$ $i$&III&0.076&0.44&1.92\\
&&&$\blacklozenge \hspace{0.45cm} 3.8689-i\varepsilon$&II&&&\\
\hline
8.000&0.002&0.024&$\bullet  \hspace{0.5cm} 3.9796-0.0316$ $i$&III&0.024&0.14&1.92\\
&&&$\blacklozenge \hspace{0.45cm} 3.8609-i\varepsilon$&II&&&\\
\hline
10.000&0.035&0.505&$\bullet  \hspace{0.5cm} 3.9978-0.0361$ $i$&III&0.387&2.25&1.92\\
&&&$\blacklozenge \hspace{0.45cm} 3.8716-i\varepsilon$&I&&&\\ 
\hline
\multicolumn{8}{c}{\textbf{DIPOLAR FORM FACTOR}}\\
\hline
8.339&0.078&0.630&$\bullet  \hspace{0.5cm} 4.0075-0.0390$ $i$&III&0.856&4.97&2.87\\
(critical)&&&$\blacklozenge \hspace{0.45cm} 3.8717-i\varepsilon$&II&&&\\
\hline
8.179&0.047&0.481&$\bullet  \hspace{0.5cm} 4.006-0.0389$ $i$&III&0.520&3.02&2.87\\
&&&$\blacklozenge \hspace{0.45cm} 3.8715-i\varepsilon$&II&&&\\
\hline
7.800&0.013&0.138&$\bullet  \hspace{0.5cm} 4.001-0.0385$ $i$&III&0.146&0.85&2.87\\
&&&$\blacklozenge \hspace{0.45cm} 3.8675-i\varepsilon$&II&&&\\
\hline
7.500&0.060&0.059&$\bullet  \hspace{0.5cm} 3.9980-0.0381$ $i$&III&0.066&0.38&2.87\\
&&&$\blacklozenge \hspace{0.45cm} 3.8616-i\varepsilon$&II&&&\\
\hline
7.200&0.0032&0.029&$\bullet  \hspace{0.5cm} 3.9945-0.0376$ $i$&III&0.35&0.21&2.87\\
&&&$\blacklozenge \hspace{0.45cm} 3.8628-i\varepsilon$&II&&&\\
\hline
7.000&0.0022&0.020&$\bullet  \hspace{0.5cm} 3.9921-0.0373$ $i$&III&0.025&0.15&2.87\\
&&&$\blacklozenge \hspace{0.45cm} 3.8707-i\varepsilon$&II&&&\\
\hline
6.500&0.0011&0.008&$\bullet  \hspace{0.5cm} 3.9861-0.0361$ $i$&III&0.012&0.072&2.87\\
&&&$\blacklozenge \hspace{0.45cm} 3.8658-i\varepsilon$&II&&&\\
\hline
8.500&0.039&0.417&$\bullet  \hspace{0.5cm} 4.009-0.0391$ $i$&III&0.426&2.48&2.87\\
&&&$\blacklozenge \hspace{0.45cm} 3.8716-i\varepsilon$&I&&&\\
\hline
\hline
\end{tabular}
\caption{\label{I} Results of the model obtained by changing the value of the coupling constant $g_{\chi_{c1}DD^*}$ and for two types of form factors (Gaussian and dipolar). The other fixed parameters are $M_0=3.95$ GeV and $\Lambda=0.5$ GeV.  The symbol $(\bullet)$ indicates the pole of the seed state, while $(\blacklozenge)$ the pole of the state $X(3872)$.}
\end{table}

\begin{table}[h!]
\centering
\renewcommand{\arraystretch}{0.95}
\begin{tabular}[c]{ccclcccc}
\hline
\hline
\multicolumn{8}{c}{\textbf{Parameters: \hspace{0.5cm} $\mathbf{M_0=3.92}$ \textbf{GeV,} \hspace{0.5cm} $\mathbf{\Lambda=0.5}$ GeV, \hspace{0.5cm} $\mathbf{m_* \neq}$ \textbf{const.}}}\\
\hline
\multicolumn{8}{c}{\textbf{GAUSSIAN FORM FACTOR}}\\
\hline
$g_{\chi_{c1}DD^*}$ &Eq. (\ref{integralthr}) &Eq. (\ref{intwidth})& \hspace{0.7cm}Pole positions &RS& Eq. (\ref{raddec1}) & Eq. (\ref{raddec2}) & Eq. (\ref{ratioVVV})\\

[GeV]&&[MeV]&\hspace{1.2cm} [GeV] & & [keV]& [keV] & \\
\hline
\hline
7.689&0.092&0.634&$\bullet  \hspace{0.5cm} 3.9547-0.0444$ $i$&III&1.02&5.91&1.92\\
(critical)&&&$\blacklozenge \hspace{0.45cm} 3.8717-i\varepsilon$&II&&&\\
\hline
7.557&0.067&0.544&$\bullet  \hspace{0.5cm} 3.9531-0.0440$ $i$&III&0.74&4.27&1.92\\
(case II)&&&$\blacklozenge \hspace{0.45cm} 3.8716-i\varepsilon$&II&&&\\
\hline
7.400&0.043&0.373&$\bullet  \hspace{0.5cm} 3.9513-0.0435$ $i$&III&0.48&2.77&1.92\\
&&&$\blacklozenge \hspace{0.45cm} 3.8713-i\varepsilon$&II&&&\\
\hline
7.300&0.033&0.283&$\bullet  \hspace{0.5cm} 3.9501-0.0432$ $i$&III&0.36&2.08&1.92\\
&&&$\blacklozenge \hspace{0.45cm} 3.8710-i\varepsilon$&II&&&\\
\hline
7.000&0.015&0.123&$\bullet  \hspace{0.5cm} 3.9465-0.0420$ $i$&III&0.16&0.95&1.92\\
&&&$\blacklozenge \hspace{0.45cm} 3.8689-i\varepsilon$&II&&&\\
\hline
6.800&0.009&0.075&$\bullet  \hspace{0.5cm} 3.9441-0.0412$ $i$&III&0.10&0.60&1.92\\
&&&$\blacklozenge \hspace{0.45cm} 3.8664-i\varepsilon$&II&&&\\
\hline
6.600&0.006&0.048&$\bullet  \hspace{0.5cm} 3.9417-0.0402$ $i$&III&0.07&0.41&1.92\\
&&&$\blacklozenge \hspace{0.45cm} 3.8629-i\varepsilon$&II&&&\\
\hline
8.000&0.034&0.340&$\bullet  \hspace{0.5cm} 3.9582-0.0452$ $i$&III&0.38&2.19&1.92\\
&&&$\blacklozenge \hspace{0.45cm} 3.8714-i\varepsilon$&I&&&\\
\hline
\multicolumn{8}{c}{\textbf{DIPOLAR FORM FACTOR}}\\
\hline
6.537&0.125&0.62&$\bullet  \hspace{0.5cm} 3.9700-0.0498$ $i$&III&1.38&8.00&2.87\\
(critical)&&&$\blacklozenge \hspace{0.45cm} 3.8717-i\varepsilon$&II&&&\\
\hline
6.351&0.062&0.40&$\bullet  \hspace{0.5cm} 3.9674-0.0498$ $i$&III&0.68&3.94&2.87\\
&&&$\blacklozenge \hspace{0.45cm} 3.8712-i\varepsilon$&II&&&\\
\hline
6.000&0.015&0.094&$\bullet  \hspace{0.5cm} 3.9624-0.0498$ $i$&III&0.16&0.96&2.87\\
&&&$\blacklozenge \hspace{0.45cm} 3.8678-i\varepsilon$&II&&&\\
\hline
5.800&0.081&0.048&$\bullet  \hspace{0.5cm} 3.95934-0.0497$ $i$&III&0.09&0.52&2.87\\
&&&$\blacklozenge \hspace{0.45cm} 3.8682-i\varepsilon$&II&&&\\
\hline
5.500&0.0039&0.021&$\bullet  \hspace{0.5cm} 3.9546-0.0495$ $i$&III&0.043&0.25&2.87\\
&&&$\blacklozenge \hspace{0.45cm} 3.8688-i\varepsilon$&II&&&\\
\hline
5.200&0.0022&0.011&$\bullet  \hspace{0.5cm} 3.9495-0.0492$ $i$&III&0.024&0.14&2.87\\
&&&$\blacklozenge \hspace{0.45cm} 3.8694-i\varepsilon$&II&&&\\
\hline
4.500&0.0008&0.0029&$\bullet  \hspace{0.5cm} 3.9355-0.0477$ $i$&III&0.0088&0.051&2.87\\
&&&$\blacklozenge \hspace{0.45cm} 3.8960-i\varepsilon$&II&&&\\
\hline
6.800&0.032&0.23&$\bullet  \hspace{0.5cm} 3.9735-0.0498$ $i$&III&0.35&2.03&2.87\\
&&&$\blacklozenge \hspace{0.45cm} 3.8713-i\varepsilon$&I&&&\\
\hline
\hline
\end{tabular}
\caption{\label{III} Similar to Table \ref{I}, but for $M_0=3.92$ GeV.}%
\end{table}
\begin{center}
\textbf{Testing the $\mathbf{\Lambda}$ parameter}
\end{center}

We also study the changes of the results in dependence of the cutoff parameter $\Lambda$. To this end, we have repeated the calculations for some values of $\Lambda$ in the range $0.4$-$0.8$ GeV. We consider two values for the bare mass of the seed, $M_0=3.95$ GeV and $M_0=3.92$ GeV, as well as two types of vertex functions (Gaussian and dipolar one). The numerical value of the constant $g_{\chi_{c1}DD^*}$ was obtained for $m_{*}=3.874$ GeV, just as in Chapter \ref{Xmain}. One should recall that, although $m_*$ is fixed, the value of $g_{\chi_{c1}DD^*}$ changes since $\Lambda$ changes. 

One observes that for a quite wide interval of $\Lambda$, the main results are qualitatively consistent with the ones presented in Chapter \ref{Xmain}. For each $\Lambda$ we have found two propagator poles. Nevertheless, care is needed for $\Lambda=0.8$ GeV (or larger), since it generates a too large decay width of the seed state in comparison to the quark model predictions. For instance, for $M_0=3.95$ GeV the decay width is about $225$ MeV for the Gaussian vertex function and about $292$ MeV for the dipolar one. When using $M_0=3.92$ GeV we obtain similar large values. In consequence, such a broad state would be not easy to observe in experiments. We thus conclude that $\Lambda=0.8$ GeV represents an upper limit for the cutoff value.

\begin{table}[h!]
\small{
\centering
\renewcommand{\arraystretch}{0.95}
\begin{tabular}[c]{cccclcccc}
\hline
\multicolumn{9}{c}{\textbf{Parameters:  \hspace{0.5cm} $\mathbf{M_0=3.95}$ \textbf{GeV,}
\hspace{0.5cm} $\mathbf{\Lambda\neq}$ \textbf{const.}, \hspace{0.5cm} $\mathbf{m_*=3.874}$ \textbf{GeV.} }}\\
\hline
\multicolumn{9}{c}{\textbf{GAUSSIAN FORM FACTOR}}\\
\hline
$\Lambda$&$g_{\chi_{c1}DD^*}$ & Eq. (\ref{integralthr}) &Eq. (\ref{intwidth})& \hspace{0.7cm}Pole positions &RS& Eq. (\ref{raddec1})  & Eq. (\ref{raddec2}) & Eq. (\ref{ratioVVV})\\

[GeV]&[GeV]&&[MeV]&\hspace{1.2cm} [GeV] && [keV]& [keV] & \\
\hline
\hline
0.4&11.259&0.040&0.629&$\bullet  \hspace{0.5cm} 3.9861-0.0171$ $i$&III&0.444&2.57&1.32\\
&&&&$\blacklozenge \hspace{0.45cm} 3.8717-i\varepsilon$&II&&&\\
\hline
0.42&10.897&0.045&0.636&$\bullet  \hspace{0.5cm} 3.9883-0.0204$ $i$&III&0.499&2.89&1.43\\
&&&&$\blacklozenge \hspace{0.45cm} 3.8717-i\varepsilon$&II&&&\\
\hline
0.45&10.413&0.048&0.632&$\bullet  \hspace{0.5cm} 3.9913-0.0258$ $i$&III&0.528&3.07&1.61\\
&&&&$\blacklozenge \hspace{0.45cm} 3.8717-i\varepsilon$&II&&&\\
\hline
0.5&9.732&0.049&0.607&$\bullet  \hspace{0.5cm} 3.9954-0.0357$ $i$&III&0.539&3.13&1.92\\
&(case I)&&&$\blacklozenge \hspace{0.45cm} 3.8716-i\varepsilon$&II&&&\\
\hline
0.55&9.169&0.050&0.577&$\bullet  \hspace{0.5cm} 3.9983-0.0468$ $i$&III&0.551&3.20&2.27\\
&&&&$\blacklozenge \hspace{0.45cm} 3.8716-i\varepsilon$&II&&&\\
\hline
0.6&8.694&0.051&0.549&$\bullet  \hspace{0.5cm} 3.9998-0.0588$ $i$&III&0.562&3.26&2.65\\
&&&&$\blacklozenge \hspace{0.45cm} 3.8716-i\varepsilon$&II&&&\\
\hline
0.7&7.930&0.053&0.497&$\bullet  \hspace{0.5cm} 3.9983-0.0848$ $i$&III&0.582&3.38&3.49\\
&&&&$\blacklozenge \hspace{0.45cm} 3.8715-i\varepsilon$&II&&&\\
\hline
0.8&7.338&0.054&0.454&$\bullet  \hspace{0.5cm} 3.9899-0.1123$ $i$&III&0.600&3.49&4.45\\
&&&&$\blacklozenge \hspace{0.45cm} 3.8715-i\varepsilon$&II&&&\\
\hline
\multicolumn{9}{c}{\textbf{DIPOLAR FORM FACTOR}}\\
\hline
0.4&9.369&0.0440&0.552&$\bullet  \hspace{0.5cm} 3.9909-0.0192$ $i$&III&0.485&2.82&1.90\\
&&&&$\blacklozenge \hspace{0.45cm} 3.8716-i\varepsilon$&II&&&\\
\hline
0.42&9.090&0.0447&0.536&$\bullet  \hspace{0.5cm} 3.9939-0.0226$ $i$&III&0.493&2.86&2.08\\
&&&&$\blacklozenge \hspace{0.45cm} 3.8716-i\varepsilon$&II&&&\\
\hline
0.45&8.714&0.0457&0.514&$\bullet  \hspace{0.5cm} 3.9984-0.0282$ $i$&III&0.503&2.92&2.36\\
&&&&$\blacklozenge \hspace{0.45cm} 3.8716-i\varepsilon$&II&&&\\
\hline
0.5&8.179&0.0472&0.481&$\bullet  \hspace{0.5cm} 4.0057-0.0389$ $i$&III&0.520&3.02&2.87\\
&&&&$\blacklozenge \hspace{0.45cm} 3.8715-i\varepsilon$&II&&&\\
\hline
0.55&7.732&0.0486&0.452&$\bullet  \hspace{0.5cm} 4.0126-0.0515$ $i$&III&0.536&3.12&3.43\\
&&&&$\blacklozenge \hspace{0.45cm} 3.8714-i\varepsilon$&II&&&\\
\hline
0.6&7.351&0.0499&0.427&$\bullet  \hspace{0.5cm} 4.0190-0.0660$ $i$&III&0.550&3.20&4.04\\
&&&&$\blacklozenge \hspace{0.45cm} 3.8714-i\varepsilon$&II&&&\\
\hline
0.7&6.732&0.0522&0.384&$\bullet  \hspace{0.5cm} 4.0298-0.1013$ $i$&III&0.575&3.34&5.42\\
&&&&$\blacklozenge \hspace{0.45cm} 3.8711-i\varepsilon$&II&&&\\
\hline
0.8&6.249&0.0539&0.348&$\bullet  \hspace{0.5cm} 4.0376-0.1458$ $i$&III&0.595&3.46&7.00\\
&&&&$\blacklozenge \hspace{0.45cm} 3.8706-i\varepsilon$&II&&&\\
\hline
\end{tabular}
}
\caption{\label{V} Results of the model obtained by changing the value of the cutoff parameter $\Lambda$ and for two types of form factors (Gaussian and dipolar). The other fixed parameters are $M_0=3.95$ GeV and $m_{*}=3.874$ GeV. The symbol $(\bullet)$ indicates the pole of the seed state, while $(\blacklozenge)$ the pole of the state $X(3872)$.}%

\end{table}
\begin{table}[h!]
\centering
\small{
\renewcommand{\arraystretch}{1.0}
\begin{tabular}[c]{cccclcccc}
\hline
\multicolumn{9}{c}{\textbf{Parameters:  \hspace{0.5cm} $\mathbf{M_0=3.92}$ \textbf{GeV,}\hspace{0.5cm} $\mathbf{\Lambda\neq}$ \textbf{const.}, \hspace{0.5cm} $\mathbf{m_*=3.874}$ \textbf{GeV.} }}\\
\hline
\multicolumn{9}{c}{\textbf{GAUSSIAN FORM FACTOR}}\\
\hline
$\Lambda$&$g_{\chi_{c1}DD^*}$ & Eq. (\ref{integralthr})&Eq. (\ref{intwidth})& \hspace{0.7cm}Pole positions &RS& Eq. (\ref{raddec1}) & Eq. (\ref{raddec2}) & Eq. (\ref{ratioVVV})\\

[GeV]&[GeV]&&[MeV]&\hspace{1.2cm} [GeV] && [keV]& [keV] & \\
\hline
\hline
0.4&8.743&0.0662&0.626&$\bullet  \hspace{0.5cm} 3.9507-0.0246$ $i$&III&0.729&4.23&1.32\\
&&&&$\blacklozenge \hspace{0.45cm} 3.8717-i\varepsilon$&II&&\\
\hline
0.42&8.461&0.0663&0.612&$\bullet  \hspace{0.5cm} 3.9517-0.0282$ $i$&III&0.730&4.24&1.43\\
&&&&$\blacklozenge \hspace{0.45cm} 3.8717-i\varepsilon$&II&&&\\
\hline
0.45&8.086&0.0664&0.587&$\bullet  \hspace{0.5cm} 3.9527-0.0339$ $i$&III&0.732&4.25&1.61\\
&&&&$\blacklozenge \hspace{0.45cm} 3.8716-i\varepsilon$&II&&&\\
\hline
0.5&7.557&0.0667&0.544&$\bullet  \hspace{0.5cm} 3.9531-0.0440$ $i$&III&0.735&4.27&1.92\\
(case II)&&&&$\blacklozenge \hspace{0.45cm} 3.8716-i\varepsilon$&II&&&\\
\hline
0.55&7.120&0.0670&0.504&$\bullet  \hspace{0.5cm} 3.9519-0.0548$ $i$&III&0.738&4.29&2.27\\
&&&&$\blacklozenge \hspace{0.45cm} 3.8716-i\varepsilon$&II&&&\\
\hline
0.6&6.751&0.0672&0.468&$\bullet  \hspace{0.5cm} 3.9490-0.0659$ $i$&III&0.741&4.31&2.65\\
&&&&$\blacklozenge \hspace{0.45cm} 3.8715-i\varepsilon$&II&&&\\
\hline
0.7&6.157&0.0675&0.408&$\bullet  \hspace{0.5cm} 3.9373-0.0883$ $i$&III&0.744&4.33&3.49\\
&&&&$\blacklozenge \hspace{0.45cm} 3.8713-i\varepsilon$&II&&&\\
\hline
0.8&5.698&0.0674&0.359&$\bullet  \hspace{0.5cm} 3.9172-0.1087$ $i$&III&0.744&4.33&4.45\\
&&&&$\blacklozenge \hspace{0.45cm} 3.8708-i\varepsilon$&II&&&\\
\hline
\multicolumn{9}{c}{\textbf{DIPOLAR FORM FACTOR}}\\
\hline
0.4&7.275&0.060&0.483&$\bullet  \hspace{0.5cm} 3.9572-0.0265$ $i$&III&0.660&3.84&1.90\\
&&&&$\blacklozenge \hspace{0.45cm} 3.8715-i\varepsilon$&II&&&\\
\hline
0.42&7.058&0.060&0.463&$\bullet  \hspace{0.5cm} 3.9594-0.0305$ $i$&III&0.664&3.86&2.08\\
&&&&$\blacklozenge \hspace{0.45cm} 3.8715-i\varepsilon$&II&&&\\
\hline
0.45&6.766&0.061&0.437&$\bullet  \hspace{0.5cm} 3.9626-0.0371$ $i$&III&0.670&3.90&2.36\\
&&&&$\blacklozenge \hspace{0.45cm} 3.8714-i\varepsilon$&II&&&\\
\hline
0.5&6.351&0.062&0.399&$\bullet  \hspace{0.5cm} 3.9674-0.0498$ $i$&III&0.678&3.94&2.87\\
&&&&$\blacklozenge \hspace{0.45cm} 3.8712-i\varepsilon$&II&&&\\
\hline
0.55&6.004&0.062&0.366&$\bullet  \hspace{0.5cm} 3.9715-0.0647$ $i$&III&0.684&3.98&3.43\\
&&&&$\blacklozenge \hspace{0.45cm} 3.8709-i\varepsilon$&II&&&\\
\hline
0.6&5.708&0.062&0.338&$\bullet  \hspace{0.5cm} 3.9748-0.0819$ $i$&III&0.688&4.00&4.04\\
&&&&$\blacklozenge \hspace{0.45cm} 3.8703-i\varepsilon$&II&&&\\
\hline
0.7&5.228&0.0632&0.291&$\bullet  \hspace{0.5cm} 3.9790-0.1244$ $i$&III&0.692&4.03&5.42\\
&&&&$\blacklozenge \hspace{0.45cm} 3.8698-i\varepsilon$&II&&&\\
\hline
0.8&4.852&0.063&0.254&$\bullet  \hspace{0.5cm} 3.9813-0.1793$ $i$&III&0.690&4.01&7.00\\
&&&&$\blacklozenge \hspace{0.45cm} 3.8704-i\varepsilon$&II&&&\\
\hline
\end{tabular}
}
\caption{\label{VII} Similar to Table \ref{V}, but for $M_0=3.92$ GeV.}%
\end{table}

\end{appendices}


\begin{thebibliography}{99}                                                                                               %
 \bibitem{GodfreyIsgur}
  S.~Godfrey and N.~Isgur,
  ``Mesons in a Relativized Quark Model with Chromodynamics,''
  Phys.\ Rev.\ D {\bf 32} (1985) 189.

 \bibitem {pdg}M.~Tanabashi \textit{et al.} (Particle Data Group), Phys.
Rev. D \textbf{98}, 030001 (2018).

\bibitem{Klempt}
E.~Klempt,
``Glueballs, hybrids, pentaquarks: Introduction to hadron spectroscopy and review of selected topics,'' Contribution to: 18th Annual Hampton University Graduate Studies (HUGS at JLab 2003)
[arXiv:hep-ph/0404270 [hep-ph]].

\bibitem{Friendly}
R.~Klauber, ``Student Friendly Quantum Field Theory ,'' Sandtrove Press (2013).


\bibitem{nowe1}
C.~Alexandrou,
``Hadron Structure in Lattice QCD,''
Prog. Part. Nucl. Phys. \textbf{67} (2012), 101-116
[arXiv:1111.5960 [hep-lat]].

\bibitem{Politzer}
H.~D.~Politzer,
``The dilemma of attribution,''
Proc. Nat. Acad. Sci. \textbf{102} (2005), 7789-7793

\bibitem{peskinQFT}
M.~E.~Peskin and D.~V.~Schroeder,
``An Introduction to Quantum Field Theory ,'' The Perseus Books Group (1995).

\bibitem{PiotrcoitoX}
F.~Giacosa, M.~Piotrowska and S.~Coito,
  ``$X(3872)$ as virtual companion pole of the charm-anticharm state $\chi_{c1}(2P)$,''
  Int.\ J.\ Mod.\ Phys.\ A {\bf 34} (2019) no.29,  1950173
  [arXiv:1903.06926 [hep-ph]].

\bibitem{adfran}
  A.~Koenigstein and F.~Giacosa,
  ``Phenomenology of pseudotensor mesons and the pseudotensor glueball,''
  Eur.\ Phys.\ J.\ A {\bf 52} (2016) no.12,  356
  [arXiv:1608.08777 [hep-ph]].
  
\bibitem{kloemixang}
  G.~Amelino-Camelia {\it et al.},
  ``Physics with the KLOE-2 experiment at the upgraded DA$\phi$NE,''
  Eur.\ Phys.\ J.\ C {\bf 68} (2010) 619
 [arXiv:1003.3868 [hep-ex]]. 
  
\bibitem {dick}
D.~Parganlija, P.~Kovacs, G.~Wolf, F.~Giacosa and D.~H.~Rischke, ``Meson
vacuum phenomenology in a three-flavor linear sigma model with (axial-)vector
mesons,'' Phys.\ Rev.\ D \textbf{87} (2013) no.1, 014011
[arXiv:1208.0585 [hep-ph]].

\bibitem {bass}
S.~D.~Bass and A.~W.~Thomas, \textquotedblleft eta bound states in nuclei: A
Probe of flavor-singlet dynamics,\textquotedblright\ Phys.\ Lett.\ B
\textbf{634} (2006) 368 
[hep-ph/0507024]. 

\bibitem {feldmann}T.~Feldmann, P.~Kroll and B.~Stech, \textquotedblleft
Mixing and decay constants of pseudoscalar mesons,\textquotedblright%
\ Phys.\ Rev.\ D \textbf{58} (1998) 114006 
[hep-ph/9802409].

\bibitem {qqpdg}Quark model, Standard Model and related topics, Reviews,
Tables and Plots of the PDG \cite{pdg}.

 \bibitem{Prelovsekex}
  S.~Prelovsek, L.~Leskovec, C.~B.~Lang and D.~Mohler,
  ``K $\pi$ Scattering and the K* Decay width from Lattice QCD,''
  Phys.\ Rev.\ D {\bf 88} (2013) no.5,  054508
  [arXiv:1307.0736 [hep-lat]].
  
  \bibitem{Dudekex}
  J.~J.~Dudek, R.~G.~Edwards, B.~Joo, M.~J.~Peardon, D.~G.~Richards and C.~E.~Thomas,
  ``Isoscalar meson spectroscopy from lattice QCD,''
  Phys.\ Rev.\ D {\bf 83} (2011) 111502
  [arXiv:1102.4299 [hep-lat]].
  
  \bibitem{Dudekex2}
  J.~J.~Dudek {\it et al.} [Hadron Spectrum Collaboration],
  ``Toward the excited isoscalar meson spectrum from lattice QCD,''
  Phys.\ Rev.\ D {\bf 88} (2013) no.9,  094505
  [arXiv:1309.2608 [hep-lat]].

  \bibitem{OConnellpearce}
H.~B.~O'Connell, B.~Pearce, A.~W.~Thomas and A.~G.~Williams,
``$\rho - \omega$ mixing, vector meson dominance and the pion form-factor,''
Prog. Part. Nucl. Phys. \textbf{39} (1997), 201-252
[arXiv:hep-ph/9501251 [hep-ph]].

\bibitem{fggp}
  F.~Giacosa and G.~Pagliara,
  ``On the spectral functions of scalar mesons,''
  Phys.\ Rev.\ C {\bf 76} (2007) 065204
 [arXiv:0707.3594 [hep-ph]].
 
 
\bibitem {Aston84}D.~Aston \textit{et al.}, ``Observation of Two Nonleading
Strangeness 1 Vector Mesons,'' Phys.\ Lett.\ \textbf{149B} (1984) 258.

\bibitem {Aston88}D.~Aston \textit{et al.}, ``A Study of K- pi+ Scattering in
the Reaction K- p ---> K- pi+ n at 11-GeV/c,'' Nucl.\ Phys.\ B \textbf{296}
(1988) 493. 

\bibitem {Donnachie91}A.~Donnachie and A.~B.~Clegg, ``The Decays of the
rho-prime(1) and omega-prime(1) mesons,'' Z.\ Phys.\ C \textbf{51} (1991) 689.

\bibitem {Clegg94}A.~B.~Clegg and A.~Donnachie, ``Higher vector meson states
produced in electron - positron annihilation,'' Z.\ Phys.\ C \textbf{62}
(1994) 455. 

\bibitem {Buon82}J.~Buon, D.~Bisello, J.~C.~Bizot, A.~Cordier, B.~Delcourt,
F.~Mane and J.~Layssac, ``Interpretation of Dm1 Results on $e^{+} e^{-}$
Annihilation Into Exclusive Channels Between 1.4-{GeV} and 1.9-{GeV} With a
$\rho^{\prime}\omega^{\prime}\phi^{\prime}$ Model,''
Phys.\ Lett.\ \textbf{118B} (1982) 221.

\bibitem {Aulchenko15}V.~M.~Aulchenko \textit{et al.} [SND Collaboration],
``Measurement of the $e^{+}e^{-} \to\eta\pi^{+}\pi^{-}$ cross section in the
center-of-mass energy range 1.22-2.00 GeV with the SND detector at the
VEPP-2000 collider,'' Phys.\ Rev.\ D \textbf{91} (2015) no.5, 052013
[arXiv:1412.1971 [hep-ex]].

\bibitem {Fukui91}S.~Fukui \textit{et al.}, ``Study of omega pi0 system in the
pi- p charge exchange reaction at 8.95-GeV/c,'' Phys.\ Lett.\ B \textbf{257}
(1991) 241. 

\bibitem {Coan04}
T.~E.~Coan \textit{et al.} [CLEO Collaboration], \textquotedblleft Wess-Zumino
current and the structure of the decay tau- ---%
$>$
K- K+ pi- nu(tau),\textquotedblright\ Phys.\ Rev.\ Lett.\ \textbf{92} (2004)
232001
[hep-ex/0401005].

\bibitem {Achasov14}
M.~N.~Achasov \textit{et al.}, ``Study of the process $e^{+}e^{-}\to\eta
\gamma$ in the center-of-mass energy range 1.07--2.00 GeV,'' Phys.\ Rev.\ D
\textbf{90} (2014) no.3, 032002 
[arXiv:1312.7078 [hep-ex]].

\bibitem {Achasov16B}M.~N.~Achasov \textit{et al.}, ``Measurement of the
$e^{+}e^{-} \to\omega\eta$ cross section below $\sqrt{s}=2$ GeV,''
Phys.\ Rev.\ D \textbf{94} (2016) no.9, 092002 
[arXiv:1607.00371 [hep-ex]].

\bibitem {Aulchenko15A}V.M. Aulchenko \textit{et al.} ``Study of the
$e^{+}e^{-} \rightarrow\pi^{+} \pi^{-}\pi^{0}$ process in the energy range
$1.05-2.00$ GeV,'' J.Exp.Theor.Phys. \textbf{121} (2015) no.1, 27-34,
Zh.Eksp.Teor.Fiz. \textbf{148} (2015) no.1, 34-41

\bibitem {Aubert08S}B.~Aubert \textit{et al.} [BaBar Collaboration],
``Measurements of $e^{+} e^{-} \to K^{+} K^{-} \eta$, $K^{+} K^{-} \pi^{0}$
and $K^{0}_{s} K^{\pm}\pi^{\mp}$ cross- sections using initial state radiation
events,'' Phys.\ Rev.\ D \textbf{77} (2008) 092002
[arXiv:0710.4451 [hep-ex]].

\bibitem {Alavi-harati 02B}
A.~Alavi-Harati \textit{et al.} [KTeV Collaboration], ``Search for the K(L)
---> pi0 pi0 e+ e- decay in the KTeV experiment,''
Phys.\ Rev.\ Lett.\ \textbf{89} (2002) 211801
[hep-ex/0210056].

\bibitem {Akhmetshin01B}R.~R.~Akhmetshin \textit{et al.} [CMD-2
Collaboration], ``Study of the process e+ e- ---> eta gamma in center-of-mass
energy range 600-MeV to 1380-MeV at CMD-2,'' Phys.\ Lett.\ B \textbf{509}
(2001) 217
[hep-ex/0103043].


\bibitem {Diekmann88}B.~Diekmann, ``Spectroscopy of Mesons Containing Light
Quarks ($u$, $d$, $s$) or Gluons,'' Phys.\ Rept.\ \textbf{159} (1988) 99.


\bibitem {Kurdadze83}L.~M.~Kurdadze \textit{et al.}, ``Measuring Of Pion
Form-factor Within The Region S**(1/2) From 640-mev To 1400-mev,'' JETP
Lett.\ \textbf{37} (1983) 733 [Pisma Zh.\ Eksp.\ Teor.\ Fiz.\ \textbf{37}
(1983) 613].

\bibitem {Akhmetshin05}R.~R.~Akhmetshin \textit{et al.} [CMD-2 Collaboration],
``Study of the processes e+ e- ---> eta gamma, pi0 gamma ---> 3 gamma in the
c.m. energy range 600-MeV to 1380-MeV at CMD-2,'' Phys.\ Lett.\ B \textbf{605}
(2005) 26 
[hep-ex/0409030].

\bibitem {Henner02}V.~K.~Henner, T.~S.~Belozerova, V.~G.~Solovev and
P.~G.~Frick, ``Application of wavelet analysis to the spectrum of omega'
states and ratio R(e+ e-),'' Eur.\ Phys.\ J.\ C \textbf{26} (2002) 3.

\bibitem {Becker79}
H.~Becker \textit{et al.} [CERN-Cracow-Munich Collaboration],
\textquotedblleft A Model Independent Partial Wave Analysis of the pi+ pi-
System Produced at Low Four Momentum Transfer in the Reaction pi- p
(Polarized) ---%
$>$
pi+ pi- n at 17.2-GeV/c,\textquotedblright\ Nucl.\ Phys.\ B \textbf{151}
(1979) 46. 

\bibitem {Martin78C}A.~D.~Martin and M.~R.~Pennington, ``How Imposing
Analyticity on a pi pi Phase Shift Analysis Can Reveal New Solutions, Explore
Experimental Structures and Investigate the Possibility of New Resonances,''
Annals Phys.\ \textbf{114} (1978) 1.



\bibitem {Froggatt77}C.~D.~Froggatt and J.~L.~Petersen, ``Phase Shift Analysis
of pi+ pi- Scattering Between 1.0-GeV and 1.8-GeV Based on Fixed Momentum
Transfer Analyticity. 2.,'' Nucl.\ Phys.\ B \textbf{129} (1977) 89.



\bibitem {Hyams73}B.~Hyams \textit{et al.}, ``$\pi\pi$ Phase Shift Analysis
from 600-MeV to 1900-MeV,'' Nucl.\ Phys.\ B \textbf{64} (1973) 134.




\bibitem {Delcourt81B}A.~Cordier, D.~Bisello, J.~C.~Bizot, J.~Buon,
B.~Delcourt, L.~Fayard and F.~Mane, ``Study of the $e^{+} e^{-} \to\pi^{+}
\pi^{-} \pi^{+} \pi^{-}$ Reaction in the 1.4-{GeV} to 2.18-{GeV} Energy
Range,'' Phys.\ Lett.\ \textbf{109B} (1982) 129.
 B.~Delcourt \textit{et al.}, ``$e^{+}%
\setminus e^{-}$ Annihilation at DCI with the Magnetic Detector DM1 for $1.4 <
\sqrt{s} > 2.2$ GeV,'' eConf C \textbf{810824} (1981) 205.




\bibitem {Aston80}D.~Aston \textit{et al.} [Bonn-CERN-Ecole
Poly-Glasgow-Lancaster-Manchester-Orsay-Paris-Rutherford-Sheffield
Collaboration], ``Observation of the $\rho^{-}$prime (1600) in the Channel
$\gamma p \to\pi^{+} \pi^{-} p$,'' Phys.\ Lett.\ \textbf{92B} (1980) 215.

\bibitem {Bizot80}J.~C.~Bizot \textit{et al.}, \textquotedblleft Observation
of a $\phi(1.65)$ vector meson in $e^{+}e^{-}$ annihilation at
DCI,\textquotedblright\ AIP Conf.\ Proc.\ \textbf{68} (1981) 546.

\bibitem {Delcourt82}B.~Delcourt, D.~Bisello, J.~C.~Bizot, J.~Buon, A.~Cordier
and F.~Mane, ``Study of the Reactions $e^{+} e^{-} \to\rho\eta$, $\rho\pi$,
$\phi\pi$ and $\phi\eta$ for Total Energy Ranges Between 1.4-{GeV} and
2.18-{GeV},'' Phys.\ Lett.\ \textbf{113B} (1982) 93 Erratum:
[Phys.\ Lett.\ \textbf{115B} (1982) 503].

\bibitem {Antonelli88}A.~Antonelli \textit{et al.} [DM2 Collaboration],
``Measurement of the Reaction $e^{+} e^{-} \to\eta\pi^{+} \pi^{-}$ in the
Center-of-mass Energy Interval 1350-{MeV} to 2400-{MeV},'' Phys.\ Lett.\ B
\textbf{212} (1988) 133. 

\bibitem {Aston87}D.~Aston \textit{et al.}, ``The Strange Meson Resonances
Observed in the Reaction $K^{-} p \to\bar{K}0 \pi^{+} \pi^{-} n$ at
11-{GeV}/$c$,'' Nucl.\ Phys.\ B \textbf{292} (1987) 693.

\bibitem {Achasov03D}
M.~N.~Achasov \textit{et al.}, \textquotedblleft Study of the process
$e^{+}e^{-}\rightarrow\pi^{+}\pi^{-}\pi^{0}$ in the energy region $\sqrt{s}$
below 0.98-GeV,\textquotedblright\ Phys.\ Rev.\ D \textbf{68} (2003) 052006
[hep-ex/0305049].

\bibitem {Aubert06D}
B.~Aubert \textit{et al.} [BaBar Collaboration],
Phys.\ Rev.\ D \textbf{73} (2006) 052003 
[hep-ex/0602006].

\bibitem{GallasGiacosalinear}
S.~Gallas, F.~Giacosa and D.~H.~Rischke,
``Vacuum phenomenology of the chiral partner of the nucleon in a linear sigma model with vector mesons,''
Phys. Rev. D \textbf{82} (2010), 014004
[arXiv:0907.5084 [hep-ph]].

\bibitem{sigmaolgiam}
L.~Olbrich, M.~Zétényi, F.~Giacosa and D.~H.~Rischke,
``Three-flavor chiral effective model with four baryonic multiplets within the mirror assignment,''
Phys. Rev. D \textbf{93} (2016) no.3, 034021
[arXiv:1511.05035 [hep-ph]].

\bibitem{Leesbabar}
J.~Lees \textit{et al.} [BaBar],
``Precision measurement of the $e^+e^- \rightarrow K^+K^-(\gamma)$ cross section with the initial-state radiation method at BABAR,''
Phys. Rev. D \textbf{88} (2013) no.3, 032013
[arXiv:1306.3600 [hep-ex]].

\bibitem{Sauli}
V.~Sauli,
``Hadronic Vacuum Polarization in $e^{+}e^{-}\to \mu ^{+}\mu ^{-}$ Process Below 3 GeV,''
Acta Phys. Polon. Supp. \textbf{10} (2017) no.4, 1159-1164
[arXiv:1704.01887 [hep-ph]].

\bibitem{oneloopapprox}
  J.~Schneitzer, T.~Wolkanowski and F.~Giacosa,
  ``The role of the next-to-leading order triangle-shaped diagram in two-body hadronic decays,''
  Nucl.\ Phys.\ B {\bf 888} (2014) 287
  [arXiv:1407.7414 [hep-ph]].

\bibitem{Mrowczynski}
S.~Mrówczyński, ``ABC kwantowej teorii pola,'' Wydawnictwo Uniwersytetu Jana Kochanowskiego w Kielcach (2016).

 \bibitem {3p0old}
E.~S.~Ackleh, T.~Barnes and E.~S.~Swanson, \textquotedblleft On the mechanism
of open flavor strong decays,'Phys.\ Rev.\ D \textbf{54} (1996) 6811
[hep-ph/9604355].

\bibitem {3p0new}
Z.~G.~Luo, X.~L.~Chen and X.~Liu, \textquotedblleft B(s1)(5830) and
B*(s2)(5840),\textquotedblright\ Phys.\ Rev.\ D \textbf{79} (2009) 074020
[arXiv:0901.0505 [hep-ph]].

\bibitem{nonlocalmilena}
  M.~Soltysiak and F.~Giacosa,
  ``A covariant nonlocal Lagrangian for the description of the scalar kaonic sector,''
  Acta Phys.\ Polon.\ Supp.\  {\bf 9} (2016) 467
  [arXiv:1607.01593 [hep-ph]].

\bibitem{complexmmpp1}
L. Ahlfors, ``Complex Analysis, 3 ed.'', McGraw-Hill, 1979.

\bibitem{complexmmpp2}
Ablowitz and Fokas, ``Complex Variables: Introduction and Applications,'' Cambridge, 2003  

\bibitem{a0980revisited}
  T.~Wolkanowski, F.~Giacosa and D.~H.~Rischke,
  ``$a_{0}(980)$ revisited,''
  Phys.\ Rev.\ D {\bf 93} (2016) no.1,  014002
  [arXiv:1508.00372 [hep-ph]].  
    

\bibitem{lebed}
R.~F.~Lebed,
``Phenomenology of large N(c) QCD,''
Czech.\ J.\ Phys.\ \textbf{49} (1999) 1273 
[nucl-th/9810080].

\bibitem{largencwitten}
  E.~Witten,
  ``Baryons in the 1/n Expansion,''
  Nucl.\ Phys.\ B {\bf 160} (1979) 57.

\bibitem{pdgold}
C.~Patrignani \textit{et al.} (Particle Data Group), Chin. Phys. C, \textbf{40}, 100001 (2016) and 2017 update.


\bibitem{Ishida97B}
  S.~Ishida, M.~Ishida, T.~Ishida, K.~Takamatsu and T.~Tsuru,
  ``Analysis of K $\pi$ scattering phase shift and existence of $\kappa$(900) particle,''
  Prog.\ Theor.\ Phys.\  {\bf 98} (1997) 621
  [hep-ph/9705437]. 
 
 \bibitem{Descotes}
  S.~Descotes-Genon and B.~Moussallam,
  ``The $K^*_0 (800)$ scalar resonance from Roy-Steiner representations of $\pi$K scattering,''
  Eur.\ Phys.\ J.\ C {\bf 48} (2006) 553
 [hep-ph/0607133].
 
 \bibitem{Zhou06}
  Z.~Y.~Zhou and H.~Q.~Zheng,
  ``An improved study of the kappa resonance and the non-exotic $s$ wave $\pi K$ scatterings up to $\sqrt{s}=2.1$GeV of LASS data,''
  Nucl.\ Phys.\ A {\bf 775} (2006) 212
  [hep-ph/0603062].
  
  \bibitem{Pelaez17}
  J.~R.~Peláez, A.~Rodas and J.~Ruiz de Elvira,
  ``Strange resonance poles from $K\pi $ scattering below 1.8 GeV,''
  Eur.\ Phys.\ J.\ C {\bf 77} (2017) no.2,  91
  [arXiv:1612.07966 [hep-ph]].
  
  \bibitem{Bugg10}
  D.~V.~Bugg,
  ``An Update on the Kappa,''
  Phys.\ Rev.\ D {\bf 81} (2010) 014002
 [arXiv:0906.3992 [hep-ph]].

 
 \bibitem{Jaffe}
  R.~L.~Jaffe,
  ``Multi-Quark Hadrons. 1. The Phenomenology of (2 Quark 2 anti-Quark) Mesons,''
  Phys.\ Rev.\ D {\bf 15} (1977) 267.
 
 \bibitem{Jaffe2005}
  R.~L.~Jaffe,
  ``Exotica,''
  Phys.\ Rept.\  {\bf 409} (2005) 1
 [hep-ph/0409065].
 
 \bibitem{Maiani2004}
  L.~Maiani, F.~Piccinini, A.~D.~Polosa and V.~Riquer,
  ``A New look at scalar mesons,''
  Phys.\ Rev.\ Lett.\  {\bf 93} (2004) 212002
 [hep-ph/0407017].
 
 \bibitem{Giacosa2006}
  F.~Giacosa,
  ``Strong and electromagnetic decays of the light scalar mesons interpreted as tetraquark states,''
  Phys.\ Rev.\ D {\bf 74} (2006) 014028
  [hep-ph/0605191].
  
  \bibitem{GiacosaPagliara}
  F.~Giacosa and G.~Pagliara,
  ``Decay of light scalar mesons into vector-photon and into pseudoscalar mesons,''
  Nucl.\ Phys.\ A {\bf 833} (2010) 138
  [arXiv:0905.3706 [hep-ph]].
  
  \bibitem{FariborzJora}
  A.~H.~Fariborz, R.~Jora and J.~Schechter,
  ``Toy model for two chiral nonets,''
  Phys.\ Rev.\ D {\bf 72} (2005) 034001
 [hep-ph/0506170].
 
 \bibitem{Fariborz:2003}
  A.~H.~Fariborz,
  ``Isosinglet scalar mesons below 2-GeV and the scalar glueball mass,''
  Int.\ J.\ Mod.\ Phys.\ A {\bf 19} (2004) 2095
  [hep-ph/0302133].
  
  \bibitem{FariborzAzizi}
  A.~H.~Fariborz, A.~Azizi and A.~Asrar,
  ``Proximity of f$_0$(1500) and f$_0$(1710) to the scalar glueball,''
  Phys.\ Rev.\ D {\bf 92} (2015) no.11,  113003
  [arXiv:1511.02449 [hep-ph]].
  
  \bibitem{Napsuciale}
  M.~Napsuciale and S.~Rodriguez,
  ``A Chiral model for anti-q q and anti-qq qq mesons,''
  Phys.\ Rev.\ D {\bf 70} (2004) 094043
  [hep-ph/0407037].
  
 \bibitem{CloseTornqvist}
  F.~E.~Close and N.~A.~Tornqvist,
  ``Scalar mesons above and below 1-GeV,''
  J.\ Phys.\ G {\bf 28} (2002) R249
  [hep-ph/0204205]. 
  
  \bibitem{Pelaez:2004xp}
  J.~R.~Pelaez,
  ``Light scalars as tetraquarks or two-meson states from large N(c) and unitarized chiral perturbation theory,''
  Mod.\ Phys.\ Lett.\ A {\bf 19} (2004) 2879
  [hep-ph/0411107].
  
  \bibitem{OllerOset}
  J.~A.~Oller and E.~Oset,
  ``N/D description of two meson amplitudes and chiral symmetry,''
  Phys.\ Rev.\ D {\bf 60} (1999) 074023
 [hep-ph/9809337].
 \bibitem{lightestnonet}
  E.~van Beveren, D.~V.~Bugg, F.~Kleefeld and G.~Rupp,
  ``The Nature of sigma, kappa, a(0)(980) and f(0)(980),''
  Phys.\ Lett.\ B {\bf 641} (2006) 265
  [hep-ph/0606022].
  
  \bibitem{JaminOller}
  M.~Jamin, J.~A.~Oller and A.~Pich,
  ``S wave K pi scattering in chiral perturbation theory with resonances,''
  Nucl.\ Phys.\ B {\bf 587} (2000) 331
  [hep-ph/0006045].
  
 \bibitem{AlbaladejoOller}
  M.~Albaladejo and J.~A.~Oller,
  ``Identification of a Scalar Glueball,''
  Phys.\ Rev.\ Lett.\  {\bf 101} (2008) 252002
  [arXiv:0801.4929 [hep-ph]]. 
  
  \bibitem{PelaezRiosprl}
  J.~R.~Pelaez and G.~Rios,
  ``Nature of the f$_0$(600) from its N(c) dependence at two loops in unitarized Chiral Perturbation Theory,''
  Phys.\ Rev.\ Lett.\  {\bf 97} (2006) 242002
 [hep-ph/0610397].
 
 \bibitem{Pelaez:2003dy}
  J.~R.~Pelaez,
  ``On the Nature of light scalar mesons from their large N(c) behavior,''
  Phys.\ Rev.\ Lett.\  {\bf 92} (2004) 102001
  [hep-ph/0309292].
  
  \bibitem{MorganPennington}
  D.~Morgan and M.~R.~Pennington,
  ``New data on the K anti-K threshold region and the nature of the f0 (S*),''
  Phys.\ Rev.\ D {\bf 48} (1993) 1185.
 
 \bibitem{vanBeveren:1986ea}
  E.~van Beveren, T.~A.~Rijken, K.~Metzger, C.~Dullemond, G.~Rupp and J.~E.~Ribeiro,
  ``A Low Lying Scalar Meson Nonet in a Unitarized Meson Model,''
  Z.\ Phys.\ C {\bf 30} (1986) 615
  [arXiv:0710.4067 [hep-ph]].
  
  \bibitem{Tornqvistzp}
  N.~A.~Tornqvist,
  ``Understanding the scalar meson q anti-q nonet,''
  Z.\ Phys.\ C {\bf 68} (1995) 647
 [hep-ph/9504372].
 
 \bibitem{TornqvistRoos}
  N.~A.~Tornqvist and M.~Roos,
  ``Resurrection of the sigma meson,''
  Phys.\ Rev.\ Lett.\  {\bf 76} (1996) 1575
 [hep-ph/9511210].
 
 \bibitem{BoglionePenningtonprl}
  M.~Boglione and M.~R.~Pennington,
  ``Unquenching the scalar glueball,''
  Phys.\ Rev.\ Lett.\  {\bf 79} (1997) 1998
 [hep-ph/9703257].
 
 \bibitem{Boglioneprd}
  M.~Boglione and M.~R.~Pennington,
  ``Dynamical generation of scalar mesons,''
  Phys.\ Rev.\ D {\bf 65} (2002) 114010
  [hep-ph/0203149].
  
  \bibitem{OllerOsetnp}
  J.~A.~Oller and E.~Oset,
  ``Chiral symmetry amplitudes in the S wave isoscalar and isovector channels and the $\sigma$, f$_0$(980), a$_0$(980) scalar mesons,''
  Nucl.\ Phys.\ A {\bf 620} (1997) 438
   Erratum: [Nucl.\ Phys.\ A {\bf 652} (1999) 407]
  [hep-ph/9702314].
  \bibitem{OllerOsetPelaezprl}
  J.~A.~Oller, E.~Oset and J.~R.~Pelaez,
  ``Nonperturbative approach to effective chiral Lagrangians and meson interactions,''
  Phys.\ Rev.\ Lett.\  {\bf 80} (1998) 3452
  [hep-ph/9803242].
  
  \bibitem{OllOse}
  J.~A.~Oller, E.~Oset and J.~R.~Pelaez,
  ``Meson meson interaction in a nonperturbative chiral approach,''
  Phys.\ Rev.\ D {\bf 59} (1999) 074001
   Erratum: [Phys.\ Rev.\ D {\bf 60} (1999) 099906]
   Erratum: [Phys.\ Rev.\ D {\bf 75} (2007) 099903]
  [hep-ph/9804209].
  
\bibitem{Aston}
  D.~Aston {\it et al.},
  ``A Study of K- pi+ Scattering in the Reaction K- p ---> K- pi+ n at 11-GeV/c,''
  Nucl.\ Phys.\ B {\bf 296} (1988) 493.
 
\bibitem{Guokappa}
  F.~K.~Guo, R.~G.~Ping, P.~N.~Shen, H.~C.~Chiang and B.~S.~Zou,
  ``S wave K$\pi$ scattering and effects of $\kappa$ in J/$\psi \rightarrow \bar{K}^{*0}(892) K^+ \pi^-$,''
  Nucl.\ Phys.\ A {\bf 773} (2006) 78
 [hep-ph/0509050].
 
 
 \bibitem{Ablikimkappa}
  M.~Ablikim {\it et al.} [BES Collaboration],
  ``Evidence for $\kappa$ meson production in J/$\psi \rightarrow \bar{K}^{*0}(892) K^+ \pi^-$ process,''
  Phys.\ Lett.\ B {\bf 633} (2006) 681
  [hep-ex/0506055].
  
  \bibitem{evidenceskappa}
  M.~Ablikim {\it et al.} [BES Collaboration],
  ``Observation of charged $\kappa$ in $J/\psi \rightarrow K^*(892)^{-+}K_s\pi^{+-}$, $K^*(892)^{-+} \rightarrow K_s\pi^{-+}$ at BESII,''
  Phys.\ Lett.\ B {\bf 698} (2011) 183
  [arXiv:1008.4489 [hep-ex]].
  
  \bibitem{lattice}
  J.~J.~Dudek {\it et al.} [Hadron Spectrum Collaboration],
  ``Resonances in coupled $\pi K -\eta K$ scattering from quantum chromodynamics,''
  Phys.\ Rev.\ Lett.\  {\bf 113} (2014) no.18,  182001
 [arXiv:1406.4158 [hep-ph]].  
  
\bibitem{gluint1}
C.~Amsler and F.~E.~Close,
``Evidence for a scalar glueball,''
Phys. Lett. B \textbf{353} (1995), 385-390
[arXiv:hep-ph/9505219 [hep-ph]].

\bibitem{gluint2}
J.~Sexton, A.~Vaccarino and D.~Weingarten,
``Numerical evidence for the observation of a scalar glueball,''
Phys. Rev. Lett. \textbf{75} (1995), 4563-4566
[arXiv:hep-lat/9510022 [hep-lat]].



\bibitem{gluint3}
M.~Strohmeier-Presicek, T.~Gutsche, R.~Vinh Mau and A.~Faessler,
``Glueball quarkonia content and decay of scalar - isoscalar mesons,''
Phys. Rev. D \textbf{60} (1999), 054010
[arXiv:hep-ph/9904461 [hep-ph]].

\bibitem{gluint4}
F.~Giacosa, T.~Gutsche, V.~E.~Lyubovitskij and A.~Faessler,
``Scalar nonet quarkonia and the scalar glueball: Mixing and decays in an effective chiral approach,''
Phys. Rev. D \textbf{72} (2005), 094006
[arXiv:hep-ph/0509247 [hep-ph]].



\bibitem{gluint5}
F.~Brünner, D.~Parganlija and A.~Rebhan,
``Glueball Decay Rates in the Witten-Sakai-Sugimoto Model,''
Phys. Rev. D \textbf{91} (2015) no.10, 106002
[arXiv:1501.07906 [hep-ph]].



\bibitem{gluint6}
S.~Janowski, D.~Parganlija, F.~Giacosa and D.~H.~Rischke,
``The Glueball in a Chiral Linear Sigma Model with Vector Mesons,''
Phys. Rev. D \textbf{84} (2011), 054007
[arXiv:1103.3238 [hep-ph]].
  
\bibitem{Gasser}
  J.~Gasser and H.~Leutwyler,
  ``Chiral Perturbation Theory to One Loop,''
  Annals Phys.\  {\bf 158} (1984) 142.
  
  \bibitem{EckerGasser}
  G.~Ecker, J.~Gasser, A.~Pich and E.~de Rafael,
  ``The Role of Resonances in Chiral Perturbation Theory,''
  Nucl.\ Phys.\ B {\bf 321} (1989) 311.
  
  \bibitem{Scherer}
  S.~Scherer,
  ``Introduction to chiral perturbation theory,''
  Adv.\ Nucl.\ Phys.\  {\bf 27} (2003) 277
 [hep-ph/0210398].    
  
  \bibitem{KoRudaz}
  P.~Ko and S.~Rudaz,
  ``Phenomenology of scalar and vector mesons in the linear sigma model,''
  Phys.\ Rev.\ D {\bf 50} (1994) 6877.
  
  \bibitem{urban}
  M.~Urban, M.~Buballa and J.~Wambach,
  ``Vector and axial vector correlators in a chirally symmetric model,''
  Nucl.\ Phys.\ A {\bf 697} (2002) 338
  [hep-ph/0102260].
  
  \bibitem{ParganlijaKovacs}
  D.~Parganlija, P.~Kovacs, G.~Wolf, F.~Giacosa and D.~H.~Rischke,
  ``Meson vacuum phenomenology in a three-flavor linear sigma model with (axial-)vector mesons,''
  Phys.\ Rev.\ D {\bf 87} (2013) no.1,  014011
  [arXiv:1208.0585 [hep-ph]].
  
  
 
 \bibitem{AmslerCloseglue}
  C.~Amsler and F.~E.~Close,
  ``Is f0 (1500) a scalar glueball?,''
  Phys.\ Rev.\ D {\bf 53} (1996) 295
  [hep-ph/9507326].
  
  \bibitem{CloseKirk}
  F.~E.~Close and A.~Kirk,
  ``Scalar glueball q anti-q mixing above 1-GeV and implications for lattice QCD,''
  Eur.\ Phys.\ J.\ C {\bf 21} (2001) 531
  [hep-ph/0103173].
  
  
  \bibitem{Giacosagutsche}
  F.~Giacosa, T.~Gutsche, V.~E.~Lyubovitskij and A.~Faessler,
  ``Scalar meson and glueball decays within a effective chiral approach,''
  Phys.\ Lett.\ B {\bf 622} (2005) 277
  [hep-ph/0504033].
 
  
  
  \bibitem{Giacosa:2012hd}
  F.~Giacosa and G.~Pagliara,
  ``Spectral function of a scalar boson coupled to fermions,''
  Phys.\ Rev.\ D {\bf 88} (2013) no.2,  025010
[arXiv:1210.4192 [hep-ph]].

\bibitem{PDG2014kin}
K.A. Olive et al. (Particle Data Group), Kinematics, Chin. Phys. C, 38, 090001 (2014).


 \bibitem{Ruppschannel}
  G.~Rupp, E.~van Beveren and M.~D.~Scadron,
  ``Comment on `Intrinsic and dynamically generated scalar meson states',''
  Phys.\ Rev.\ D {\bf 65} (2002) 078501
 [hep-ph/0104087].
 
 \bibitem{Haradaschannel}
  M.~Harada, F.~Sannino and J.~Schechter,
  ``Comment on `Confirmation of the sigma meson',''
  Phys.\ Rev.\ Lett.\  {\bf 78} (1997) 1603
 [hep-ph/9609428].
 
 \bibitem{Isgurschannel}
  N.~Isgur and J.~Speth,
  ``Comment on `Confirmation of the sigma meson',''
  Phys.\ Rev.\ Lett.\  {\bf 77} (1996) 2332.

\bibitem {pealaezrev4}
J.~R.~Pelaez, \textquotedblleft From controversy to precision on the sigma
meson: a review on the status of the non-ordinary $f_{0}(500)$
resonance,\textquotedblright\ Phys.\ Rept.\ \textbf{658} (2016) 1
[arXiv:1510.00653 [hep-ph]].

\bibitem{Zhou:2010ra}
  Z.~Y.~Zhou and Z.~Xiao,
  ``The Origin of light $0^{+}$ scalar resonances,''
  Phys.\ Rev.\ D {\bf 83} (2011) 014010
  [arXiv:1007.2072 [hep-ph]].
  
  \bibitem{Guo:2011pat}
  Z.~H.~Guo and J.~A.~Oller,
  ``Resonances from meson-meson scattering in U(3) CHPT,''
  Phys.\ Rev.\ D {\bf 84} (2011) 034005
  [arXiv:1104.2849 [hep-ph]].
  
  \bibitem{Guo:2012yt}
  Z.~H.~Guo, J.~A.~Oller and J.~Ruiz de Elvira,
  ``Chiral dynamics in form factors, spectral-function sum rules, meson-meson scattering and semi-local duality,''
  Phys.\ Rev.\ D {\bf 86} (2012) 054006
  [arXiv:1206.4163 [hep-ph]].

\bibitem{taylormmpp3}
J. R. Taylor, ``An Introduction to Error Analysis: The Study of Uncertainties in Physical Measurements 2nd Edition,'' University Science Books, 1997,  NY 10012.


\bibitem {rev}
N.~Brambilla \textit{et al.}, ``Heavy quarkonium: progress, puzzles, and opportunities,'' Eur.\ Phys.\ J.\ C \textbf{71} (2011) 1534
[arXiv:1010.5827 [hep-ph]].

\bibitem {rev2016}H.~X.~Chen, W.~Chen, X.~Liu and S.~L.~Zhu, \textquotedblleft
The hidden-charm pentaquark and tetraquark states,\textquotedblright%
\ Phys.\ Rept.\ \textbf{639} (2016) 1
[arXiv:1601.02092 [hep-ph]].


\bibitem {pillonireview}
A.~Esposito, A.~L.~Guerrieri, F.~Piccinini, A.~Pilloni and A.~D.~Polosa,
\textquotedblleft Four-Quark Hadrons: an Updated Review,\textquotedblright%
\ Int.\ J.\ Mod.\ Phys.\ A \textbf{30} (2015) 1530002
[arXiv:1411.5997 [hep-ph]].

\bibitem {nielsen}M.~Nielsen, F.~S.~Navarra and S.~H.~Lee, \textquotedblleft
New Charmonium States in QCD Sum Rules: A Concise Review,\textquotedblright%
\ Phys.\ Rept.\ \textbf{497} (2010) 41
[arXiv:0911.1958 [hep-ph]].

\bibitem {eichten4040}
E.~Eichten, K.~Gottfried, T.~Kinoshita, K.~D.~Lane and T.~M.~Yan,
\textquotedblleft Charmonium: The Model\textquotedblright, Phys.\ Rev.\ D
\textbf{17} (1978) 3090 Erratum: [Phys.\ Rev.\ D \textbf{21} (1980) 313].

\bibitem{Eichten:1979mp}
E.~Eichten, K.~Gottfried, T.~Kinoshita, K.~D.~Lane and T.~M.~Yan,
\textquotedblleft Charmonium: Comparison with Experiment\textquotedblright,
Phys.\ Rev.\ D \textbf{21} (1980) 203. 

\bibitem {segoviarevmp}
J.~Segovia, D.~R.~Entem, F.~Fernandez and E.~Hernandez, \textquotedblleft
Constituent quark model description of charmonium
phenomenology,\textquotedblright\ Int.\ J.\ Mod.\ Phys.\ E \textbf{22} (2013)
1330026 
[arXiv:1309.6926 [hep-ph]].

\bibitem {belle1}C.~Z.~Yuan \textit{et al.} [Belle Collaboration],
\textquotedblleft Measurement of e+ e- ---%
$>$
pi+ pi- J/psi cross-section via initial state radiation at
Belle,\textquotedblright\ Phys.\ Rev.\ Lett.\ \textbf{99} (2007) 182004 
  [arXiv:0707.2541 [hep-ex]].

\bibitem {belle2}Z.~Q.~Liu \textit{et al.} [Belle Collaboration],
\textquotedblleft Study of $e^{+}e$ and Observation of a Charged
Charmoniumlike State at Belle,\textquotedblright%
\ Phys.\ Rev.\ Lett.\ \textbf{110} (2013) 252002
[arXiv:1304.0121 [hep-ex]].

\bibitem {besiii}M.~Ablikim \textit{et al.} [BESIII Collaboration],
\textquotedblleft Precise measurement of the $e^{+}e^{-}\rightarrow\pi^{+}%
\pi^{-}J/\psi$ cross section at center-of-mass energies from 3.77 to 4.60
GeV,\textquotedblright\ Phys.\ Rev.\ Lett.\ \textbf{118} (2017) no.9, 092001
[arXiv:1611.01317 [hep-ex]].

\bibitem {babar4008}B.~Aubert \textit{et al.} [BaBar Collaboration],
\textquotedblleft Exclusive Initial-State-Radiation Production of the D
anti-D, D* anti-D*, and D* anti-D* Systems,\textquotedblright\ Phys.\ Rev.\ D
\textbf{79} (2009) 092001
[arXiv:0903.1597 [hep-ex]].

\bibitem {molecular}X.~Liu, ``Understanding the newly observed Y(4008) by
Belle,'' Eur.\ Phys.\ J.\ C \textbf{54} (2008) 471
[arXiv:0708.4167 [hep-ph]].


\bibitem {molecular2}W.~Xie, L.~Q.~Mo, P.~Wang and S.~R.~Cotanch,
\textquotedblleft Coulomb gauge model for hidden charm
tetraquarks,\textquotedblright\ Phys.\ Lett.\ B \textbf{725} (2013) 148
[arXiv:1302.5737 [hep-ph]].

\bibitem {tetraquark1}L.~Maiani, F.~Piccinini, A.~D.~Polosa and V.~Riquer,
``The Z(4430) and a New Paradigm for Spin Interactions in Tetraquarks,''
Phys.\ Rev.\ D \textbf{89} (2014) 114010
[arXiv:1405.1551 [hep-ph]].


\bibitem {tetraquark2}P.~Zhou, C.~R.~Deng and J.~L.~Ping, ``Identification of
Y (4008), Y (4140), Y (4260), and Y (4360) as Tetraquark States,''
Chin.\ Phys.\ Lett.\ \textbf{32} (2015) no.10, 101201.

\bibitem {interference}D.~Y.~Chen, X.~Liu, X.~Q.~Li and H.~W.~Ke, ``Unified
Fano-like interference picture for charmoniumlike states Y(4008), Y(4260) and
Y(4360),'' Phys.\ Rev.\ D \textbf{93} (2016) 014011
[arXiv:1512.04157 [hep-ph]].

\bibitem {charmstate1}B.~Q.~Li and K.~T.~Chao, \textquotedblleft Higher
Charmonia and X,Y,Z states with Screened Potential,\textquotedblright%
\ Phys.\ Rev.\ D \textbf{79} (2009) 094004
[arXiv:0903.5506 [hep-ph]].


\bibitem {charmstate2}L.~J.~Chen, D.~D.~Ye and A.~Zhang, \textquotedblleft Is
$Y(4008)$ possibly a $1^{--}\psi(3^{3}S_{1})$ state?,\textquotedblright%
\ Eur.\ Phys.\ J.\ C \textbf{74} (2014) no.8, 3031
[arXiv:1402.5470 [hep-ph]].

\bibitem{scoito3770}
S.~Coito and F.~Giacosa,
``Line-shape and poles of the $\psi(3770)$,''
Nucl.\ Phys.\ A {\bf 981} (2019) 38
 [arXiv:1712.00969 [hep-ph]].

 \bibitem {duecan}
F.~Giacosa, \textquotedblleft Non-exponential decay in quantum field theory
and in quantum mechanics: the case of two (or more) decay channels,\textquotedblright\ Found.\ Phys.\ \textbf{42} (2012) 1262
[arXiv:1110.5923 [nucl-th]].

\bibitem {babarpsi4040}
B.~Aubert \textit{et al.} [BaBar Collaboration],
\textquotedblleft Exclusive Initial-State-Radiation Production of the D
anti-D, D* anti-D*, and D* anti-D* Systems,\textquotedblright\ Phys.\ Rev.\ D
\textbf{79} (2009) 092001
[arXiv:0903.1597 [hep-ex]].

\bibitem {wangpsi4040}X.~L.~Wang \textit{et al.} [Belle Collaboration],
\textquotedblleft Observation of $\psi(4040)$ and $\psi(4160)$ decay into
\^{I}\textperiodcentered J/$\psi$,\textquotedblright\ Phys.\ Rev.\ D
\textbf{87} (2013) no.5, 051101
[arXiv:1210.7550 [hep-ex]].


\bibitem {belle3psi4040}M.~N.~Anwar, Y.~Lu and B.~S.~Zou, \textquotedblleft Modeling
Charmonium-$\eta$ Decays of $J^{PC}=1^{--}$ Higher
Charmonia,\textquotedblright\ Phys.\ Rev.\ D \textbf{95} (2017) no.11, 114031
[arXiv:1612.05396 [hep-ph]].



\bibitem {cleopsi4040}T.~E.~Coan \textit{et al.} [CLEO Collaboration],
\textquotedblleft Charmonium decays of Y(4260), psi(4160) and
psi(4040),\textquotedblright\ Phys.\ Rev.\ Lett.\ \textbf{96} (2006) 162003
  [hep-ex/0602034].

\bibitem{nonorthm}G. V. Baryshevskii, V. I. Lyuboshitz, M .I. Podgorerskii,
Nonorthogonal quasisationary states, Soviet Physics JETP, Vol. 30, no. 1,
January 1970.


\bibitem {achasovmmk}
N.N.~Achasov and G.N.~Shestakov, \textit{Line shape of
$\psi(3770)$ in $e^{+}e^{-}\rightarrow D\bar{D}$},
{Phys.~Rev.~D
\textbf{86}, 114013 (2012)}
[arXiv:1208.4240 [hep-ph]].


\bibitem{belleX3872}
S.K.~Choi \textit{et al.} (Belle Collaboration),
``Observation of a narrow charmonium-like state in exclusive $B^\pm\to K^\pm \pi^+ \pi^- J/\psi$ decays,''
\ Phys.\ Rev.\ Lett.\ \textbf{91} 262001 (2003).
[hep-ex/0309032].

\bibitem{Ali:2017jda}
  A.~Ali, J.~S.~Lange and S.~Stone,
  ``Exotics: Heavy Pentaquarks and Tetraquarks,''
  Prog.\ Part.\ Nucl.\ Phys.\  {\bf 97} (2017) 123
  [arXiv:1706.00610 [hep-ph]].
  
\bibitem{Olsen:2017bmm}
  S.~L.~Olsen, T.~Skwarnicki and D.~Zieminska,
  ``Nonstandard heavy mesons and baryons: Experimental evidence,''
  Rev.\ Mod.\ Phys.\  {\bf 90} (2018) no.1,  015003
  [arXiv:1708.04012 [hep-ph]].

 \bibitem {segovia3872}
P.~G.~Ortega, J.~Segovia, D.~R.~Entem and F.~Fern\'{a}ndez, ``Charmonium
resonances in the 3.9 GeV/$c^{2}$ energy region and the $X(3915)/X(3930)$
puzzle,'' Phys.\ Lett.\ B \textbf{778} (2018) 1
[arXiv:1706.02639 [hep-ph]].


\bibitem{kalashnikova}
Y.~S.~Kalashnikova and A.~V.~Nefediev,
``X(3872) in the molecular model,''
Phys. Usp. \textbf{62} (2019) no.6, 568-595
[arXiv:1811.01324 [hep-ph]].

\bibitem{Dong1}
  Y.~b.~Dong, A.~Faessler, T.~Gutsche and V.~E.~Lyubovitskij,
 ``Estimate for the $X(3872) \rightarrow \gamma J/\psi$ decay width,''
  Phys.\ Rev.\ D {\bf 77} (2008) 094013
  [arXiv:0802.3610 [hep-ph]].
  
 \bibitem{Dong2}
  Y.~Dong, A.~Faessler, T.~Gutsche, S.~Kovalenko and V.~E.~Lyubovitskij,
 ``X(3872) as a hadronic molecule and its decays to charmonium states and pions,''
  Phys.\ Rev.\ D {\bf 79} (2009) 094013
  [arXiv:0903.5416 [hep-ph]].
  
  \bibitem{Dong3}
  Y.~Dong, A.~Faessler, T.~Gutsche and V.~E.~Lyubovitskij,
 ``$J/\psi \gamma$ and $\psi(2S) \gamma$ decay modes of the X(3872),''
  J.\ Phys.\ G {\bf 38} (2011) 015001
  [arXiv:0909.0380 [hep-ph]].
  
 \bibitem{Gamermann1}
  D.~Gamermann, J.~Nieves, E.~Oset and E.~Ruiz Arriola, ``Couplings in coupled channels versus wave functions: application to the X(3872) resonance,'' Phys.\ Rev.\ D {\bf 81} (2010) 014029
  [arXiv:0911.4407 [hep-ph]].
  
  \bibitem{Stapleton1}
  E.~Braaten and J.~Stapleton,
 ``Analysis of $J/\psi \pi^+ \pi^-$ and $D^0 \bar{D}^0 \pi^0$ Decays of the X(3872),''
  Phys.\ Rev.\ D {\bf 81} (2010) 014019
  [arXiv:0907.3167 [hep-ph]].
  
  \bibitem{Hanhart1}
  C.~Hanhart, Y.~S.~Kalashnikova, A.~E.~Kudryavtsev and A.~V.~Nefediev,
 ``Reconciling the X(3872) with the near-threshold enhancement in the $D^0 \bar{D}^{*0}$ final state,''
  Phys.\ Rev.\ D {\bf 76} (2007) 034007
  [arXiv:0704.0605 [hep-ph]].
  
 \bibitem{Guo1}
  F.~K.~Guo, C.~Hanhart, Y.~S.~Kalashnikova, U.~G.~Meißner and A.~V.~Nefediev,
  ``What can radiative decays of the X(3872) teach us about its nature?,''
  Phys.\ Lett.\ B {\bf 742} (2015) 394
  [arXiv:1410.6712 [hep-ph]].
  
  \bibitem{Barnes1}
  T.~Barnes and S.~Godfrey,
  ``Charmonium options for the X(3872),''
  Phys.\ Rev.\ D {\bf 69} (2004) 054008
  [hep-ph/0311162].
  
  \bibitem{Rogozina1}
  N.~N.~Achasov and E.~V.~Rogozina,
  ``X(3872), $I^G(J^{PC})=0^+(1^{++})$, as the $\chi_{1c}(2P)$ charmonium,''
  Mod.\ Phys.\ Lett.\ A {\bf 30} (2015) no.33,  1550181
  [arXiv:1501.03583 [hep-ph]].
  
  \bibitem{Rogozina2}
  N.~N.~Achasov and E.~V.~Rogozina,
  ``Towards the nature of $\mathrm{X(3872)}$ resonance,''
  J.\ Univ.\ Sci.\ Tech.\ China {\bf 46} (2016) no.7,  574
  [arXiv:1510.07251 [hep-ph]].
  
  \bibitem{Quigg1}
  E.~J.~Eichten, K.~Lane and C.~Quigg,
  ``Charmonium levels near threshold and the narrow state $X(3872) \to \pi^{+}\pi^{-}J/\psi$,''
  Phys.\ Rev.\ D {\bf 69} (2004) 094019
  [hep-ph/0401210].
  
  \bibitem{Bignamini1}
  C.~Bignamini, B.~Grinstein, F.~Piccinini, A.~D.~Polosa and C.~Sabelli,
 ``Is the X(3872) Production Cross Section at Tevatron Compatible with a Hadron Molecule Interpretation?,''
  Phys.\ Rev.\ Lett.\  {\bf 103} (2009) 162001
  [arXiv:0906.0882 [hep-ph]].
  
  \bibitem{Esposito1}
  A.~Esposito, A.~L.~Guerrieri, L.~Maiani, F.~Piccinini, A.~Pilloni, A.~D.~Polosa and V.~Riquer,
  ``Observation of light nuclei at ALICE and the X(3872) conundrum,''
  Phys.\ Rev.\ D {\bf 92} (2015) no.3,  034028
  [arXiv:1508.00295 [hep-ph]].
  
  \bibitem{Albaladejo1}
  M.~Albaladejo, F.~K.~Guo, C.~Hanhart, U.~G.~Meißner, J.~Nieves, A.~Nogga and Z.~Yang,
  ``Note on X(3872) production at hadron colliders and its molecular structure,''
  Chin.\ Phys.\ C {\bf 41} (2017) no.12,  121001
  [arXiv:1709.09101 [hep-ph]].
  
  \bibitem{Aaboud1}
  M.~Aaboud {\it et al.} [ATLAS Collaboration],
  ``Measurements of $\psi(2S)$ and $X(3872) \to J/\psi\pi^+\pi^-$ production in $pp$ collisions at $\sqrt{s} = 8$ TeV with the ATLAS detector,''
  JHEP {\bf 1701} (2017) 117
  [arXiv:1610.09303 [hep-ex]].
  
  \bibitem{EspositoGrinstein}
  A.~Esposito, B.~Grinstein, L.~Maiani, F.~Piccinini, A.~Pilloni, A.~D.~Polosa and V.~Riquer,
  ``Comment on ‘Note on X(3872) production at hadron colliders and its molecular structure’,''
  Chin.\ Phys.\ C {\bf 42} (2018) no.11,  114107
  [arXiv:1709.09631 [hep-ph]].
  
  \bibitem{Ortega12}
  P.~G.~Ortega, J.~Segovia, D.~R.~Entem and F.~Fernandez,
 ``Coupled channel approach to the structure of the X(3872),''
  Phys.\ Rev.\ D {\bf 81} (2010) 054023
  [arXiv:0907.3997 [hep-ph]].
  
  \bibitem{CoitoRupp}
  S.~Coito, G.~Rupp and E.~van Beveren,
  ``X(3872) is not a true molecule,''
  Eur.\ Phys.\ J.\ C {\bf 73} (2013) no.3,  2351
  [arXiv:1212.0648 [hep-ph]].
  
 \bibitem{Ferretti1}
  J.~Ferretti, G.~Galatà and E.~Santopinto,
  ``Interpretation of the X(3872) as a charmonium state plus an extra component due to the coupling to the meson-meson continuum,''
  Phys.\ Rev.\ C {\bf 88} (2013) no.1,  015207
  [arXiv:1302.6857 [hep-ph]].
  
  \bibitem{Ferretti2}
  J.~Ferretti, G.~Galatà and E.~Santopinto,
  ``Quark structure of the $X(3872)$ and $\chi_b(3P)$ resonances,''
  Phys.\ Rev.\ D {\bf 90} (2014) no.5,  054010
  [arXiv:1401.4431 [nucl-th]].
  
  \bibitem{Cardoso1}
  M.~Cardoso, G.~Rupp and E.~van Beveren,
  ``Unquenched quark-model calculation of $X(3872)$ electromagnetic decays,''
  Eur.\ Phys.\ J.\ C {\bf 75} (2015) no.1,  26
  [arXiv:1411.1654 [hep-ph]].

\bibitem{TakeuchiTakizawa}
  S.~Takeuchi, K.~Shimizu and M.~Takizawa,
  ``On the origin of the narrow peak and the isospin symmetry breaking of the $X$(3872),''
  PTEP {\bf 2014} (2014) no.12,  123D01
   Erratum: [PTEP {\bf 2015} (2015) no.7,  079203]
  [arXiv:1408.0973 [hep-ph]].
  
  \bibitem{Maianidiq}
  L.~Maiani, F.~Piccinini, A.~D.~Polosa and V.~Riquer,
  ``Diquark-antidiquarks with hidden or open charm and the nature of X(3872),''
  Phys.\ Rev.\ D {\bf 71} (2005) 014028
  [hep-ph/0412098].
  
\bibitem{Wolkanowskija}
  T.~Wolkanowski, M.~Sołtysiak and F.~Giacosa,
  Nucl.\ Phys.\ B {\bf 909} (2016) 418
  [arXiv:1512.01071 [hep-ph]].  
  
  \bibitem{Piotrowskapsi4040}
  M.~Piotrowska, F.~Giacosa and P.~Kovacs,
  ``Can the $\psi(4040)$ explain the peak associated with $Y(4008)$?,''
  Eur.\ Phys.\ J.\ C {\bf 79} (2019) no.2,  98
  [arXiv:1810.03495 [hep-ph]].
  
  \bibitem{Matthews:1959sy}
  P.~T.~Matthews and A.~Salam,
  ``Relativistic theory of unstable particles. 2,''
  Phys.\ Rev.\  {\bf 115} (1959) 1079.
  
  \bibitem{Ebertfaustov}
  D.~Ebert, R.~N.~Faustov and V.~O.~Galkin,
  ``Spectroscopy and Regge trajectories of heavy quarkonia and $B_c$ mesons,''
  Eur.\ Phys.\ J.\ C {\bf 71} (2011) 1825
  [arXiv:1111.0454 [hep-ph]].
  
  \bibitem{Braaten2007dw}
  E.~Braaten and M.~Lu,
  ``Line shapes of the X(3872),''
  Phys.\ Rev.\ D {\bf 76} (2007) 094028
  [arXiv:0709.2697 [hep-ph]].
  
  \bibitem{Aubert:2008ae}
  B.~Aubert {\it et al.} [BaBar Collaboration],
  ``Evidence for $X(3872) \to \psi_{2S} \gamma$ in $B^\pm \to X_{3872} K^\pm$ decays, and a study of $B \to c \bar{c} \gamma K$,''
  Phys.\ Rev.\ Lett.\  {\bf 102} (2009) 132001
  [arXiv:0809.0042 [hep-ex]].
  
  \bibitem{Aaij:2014ala}
  R.~Aaij {\it et al.} [LHCb Collaboration],
  ``Evidence for the decay $X(3872)\rightarrow\psi(2S)\gamma$,''
  Nucl.\ Phys.\ B {\bf 886} (2014) 665
  [arXiv:1404.0275 [hep-ex]].
  
  \bibitem{Broniowski:2015oha}
  W.~Broniowski, F.~Giacosa and V.~Begun,
  ``Cancellation of the $\sigma$ meson in thermal models,''
  Phys.\ Rev.\ C {\bf 92} (2015) no.3,  034905
  [arXiv:1506.01260 [nucl-th]].
  


\bibitem{Piotrowskaex}
M.~Piotrowska, C.~Reisinger and F.~Giacosa,
``Strong and radiative decays of excited vector mesons and predictions for a new $\phi(1930)$ resonance,''
Phys. Rev. D \textbf{96} (2017) no.5, 054033
[arXiv:1708.02593 [hep-ph]].

\bibitem{4proc}
M.~Piotrowska and F.~Giacosa,
``Strong decays of excited vector mesons,''
Acta Phys. Polon. Supp. \textbf{10} (2017), 1015
[arXiv:1708.03175 [hep-ph]].

\bibitem{5proc}
M.~Piotrowska and F.~Giacosa,
``A study of the excited radial vector meson $\rho$,''
PoS Hadron2017 (2018), 237
[arXiv:1712.05617 [hep-ph]].

\bibitem{6proc}
M.~Piotrowska and F.~Giacosa,
``Excited vector mesons: phenomenology and predictions for a yet unknown vector $s\bar{s}$ state with a mass of about 1.93 GeV,''
EPJ Web Conf. \textbf{182} (2018), 02097
[arXiv:1712.01087 [hep-ph]].

\bibitem{1proc}
M.~Soltysiak, T.~Wolkanowski and F.~Giacosa,
``Large-$N_{c}$ pole trajectories of the vector kaon $K^{\ast}(892) $ and of the scalar kaons $K_{0}^{\ast}(800)$ and $K_{0}^{\ast}(1430)$,''
Acta Phys. Polon. Supp. \textbf{9} (2016), 321
[arXiv:1604.01636 [hep-ph]].

\bibitem{2proc}
M.~Soltysiak, T.~Wolkanowski and F.~Giacosa,
``A study of the resonances $K_{0}^{*}(800)$ and $K_{0}^{*}(1430)$,''
J. Phys. Conf. Ser. \textbf{742} (2016) no.1, 012014
[arXiv:1606.02970 [hep-ph]].

\bibitem{3proc}
M.~Soltysiak and F.~Giacosa,
``A covariant nonlocal Lagrangian for the description of the scalar kaonic sector,''
Acta Phys. Polon. Supp. \textbf{9} (2016), 467-472
[arXiv:1607.01593 [hep-ph]].

\bibitem{8proc}
M.~Piotrowska,
``Study of some (non-)conventional mesons in the framework of effective models,''
[arXiv:2004.09970 [hep-ph]].

\bibitem{7proc}
M.~Piotrowska and F.~Giacosa,
``A study of the vector meson $\psi(4040)$,''
EPJ Web Conf. \textbf{199} (2019), 04013
[arXiv:1810.12702 [hep-ph]].




  

\end{thebibliography}
\end{document}